\documentclass[superscriptaddress,11pt]{article}
\usepackage{authblk}
\usepackage{array}
\usepackage{type1cm}
\usepackage{lettrine}
\usepackage{graphicx}
\usepackage{geometry}
\usepackage{psfrag}
\usepackage{amsmath}
\usepackage{amsfonts}
\usepackage{appendix}
\usepackage[margin=10pt,font=footnotesize,labelfont=sc,format=hang,labelformat=parens]%
{caption}
\usepackage{subcaption}
\usepackage{amssymb}%
\providecommand{\U}[1]{\protect\rule{.1in}{.1in}}
\numberwithin{equation}{section}
\def\be{\begin{equation}}
\def\ee{\end{equation}}
\def\ba{\begin{eqnarray}}
\def\ea{\end{eqnarray}}
\def\bi{\begin{itemize}}
\def\ei{\end{itemize}}

\def\bra{\langle}
\def\ket{\rangle}
\def\phir{\phi_{ref}}
\def\phic{\phi_{con}}
\def\sref{s^{ref}}
\def\scon{s^{con}}
\def\bd{\bar{\delta}}

\begin{document}

\title{The constraint algebra in Smolins'  $G\rightarrow 0$ limit of 4d Euclidean Gravity}

\author{Madhavan Varadarajan}
\affil{Raman Research Institute\\Bangalore-560 080, India}

\maketitle

\begin{abstract}
Smolin's  generally covariant $G_{\mathrm{Newton}}\rightarrow0$ limit of 4d Euclidean gravity 
is 
a useful toy model for the study of the constraint algebra in Loop Quantum Gravity. In particular, the commutator between its Hamiltonian constraints
has a metric dependent structure function.
While a prior LQG like construction  of  non-trivial anomaly free constraint  commutators for the model exists, 
that work suffers from two defects. First, Smolin's remarks on the inability of the quantum dynamics to generate propagation effects
apply. Second, 
the construction only
yields the action of a single Hamiltonian constraint  together with  the action of its commutator through a continuum 
limit of corresponding discrete approximants; the continuum limit of a product of 2 or more constraints does not exist.
Here, we incorporate changes in the quantum dynamics
through structural modifications in the choice of discrete approximants to the quantum Hamiltonian constraint.
The new structure is  motivated by that  responsible for propagation in an LQG like quantization of Paramaterized Field Theory
and significantly alters the space of physical states. 
We study the off shell constraint algebra of the model in the context of
these structural
changes and show that the continuum limit action of multiple products of Hamiltonian constraints is (a) supported on an
appropriate domain of states (b) yields anomaly free commutators between pairs of Hamiltonian constraints
and (c) is diffeomorphism covariant.
Many of our considerations seem robust enough to be applied to the setting of 4d Euclidean gravity. \\

\end{abstract}

\section{\label{sec1}Introduction}

The construction of a physically viable quantum dynamics for Loop Quantum Gravity 
(see for e.g. \cite{aabook,aajurekreview,ttbook,gpbook,apbook} and the references therein)
constitutes a key open problem.
Two desirable features of such a  dynamics are its compatibility with general covariance  and its ability to 
propagate perturbations \cite{leeprop}. 
Here, we focus on the issue of general covariance in 
the context of Smolin's novel weak coupling limit of Euclidean gravity \cite{leeG}. 
General covariance is expected to be encoded in a representation 
of the algebra of Hamiltonian and spatial diffeomorphism constraints \cite{hkt}. 
Accordingly, we construct 
a domain of quantum states for the model together with the action of constraint operator products thereon in such a way that 
the resulting algebra of constraints exhibits anomaly free constraint commutators.
The model shares several structural aspects
with canonical General Relativity and we  expect our considerations here to serve as essential inputs in the 
construction of a generally covariant dynamics for LQG.

On the other hand, propagation properties of quantum dynamics in LQG like quantizations seem to be 
related to  certain structural properties of the Hamiltonian constraint \cite{proppft}. While we defer an analysis
of propagation properties of the dynamics of  this model to future work \cite{u13prop}, we note that the general structural
properties  believed to be connected  with propagation effects in our study of Parameterised Field Theory \cite{proppft}  play a key role in our demonstration of an anomaly
free constraint algebra here.

We initiated our study of the quantum constraint algebra of the model in \cite{mect,diffcov}.
The phase space of the system consists of a triplet of abelian connections and conjugate electric fields, its dynamics is 
driven by Hamiltonian and diffeomorphism constraints with a Poisson Bracket algebra isomorphic to that of (Euclidean) gravity and
the LQG like quantum theory supports a representation of operators consisting of holonomies of connections around spatial loops and 
electric fluxes through spatial surfaces.
While the quantum theory supports a unitary representation of spatial diffeomorphisms, the action of the Hamiltonian constraint 
operator is defined in an indirect manner via a continuum limit of appropriate discrete approximants. The reason, as in LQG, is as follows. The classical
constraint depends on the curvature of the connection. While the classical curvature can be defined via a `shrinking loop' limit of an  approximant constructed out of  classical holonomies, the corresponding
quantum holonomy operator limit does not exist because the background independent quantum theory cannot distinguish between a bigger loop and its smaller shrinking versions. 
However, following \cite{ttrick},  it {\em is} nevertheless possible to construct a classical approximant to the Hamiltonian constraint through a suitable  
conglomeration of such discrete approximants in such a way that the limit of the action of the corresponding conglomeration of operators  can be defined despite individual operator limits being ill defined.
Since the limit involves shrinking of `discrete regulating labels' such as loops and graphs, it is referred to as a `continuum limit' and the approximants are referred to as discrete approximants.

The work in Reference \cite{mect}
constructs the continuum limit of the action of a single Hamiltonian constraint and an anomaly free continuum limit action of the commutator between two Hamiltonian constraints
from suitably defined discrete approximants. The work in Reference \cite{diffcov} improves upon the single Hamiltonian constraint action so as to render it spatially covariant thus ensuring an
anomaly free commutator of the single Hamiltonian constraint action with the spatial diffeomorphism constraint. This is achieved while maintaining the anomaly free nature of the commutator between 
a pair of Hamiltonian constraints.
It is important to note that the work in \cite{mect,diffcov} constructs the continuum limit of a discrete approximant to  the commutator between a pair of Hamiltonian constraints rather than 
the commutator between continuum limit products.
More in detail, the product of the action of 2 discrete approximant single Hamiltonian constraints is constructed, the commutator of this product is
evaluated {\em first} and {\em then} the continuum limit is taken. Instead, a better implementation of the commutator  between the quantum constraints would be to
{\em first} take the continuum limit of the product of a pair of discrete single Hamiltonian constraint actions and then take the commutator of this product.
However it turns out that with the  choice of discrete approximants used in \cite{mect,diffcov}, while the continuum limit of the discrete commutator action is well defined, the 
limit of the discrete product action is not. This is because  certain terms with  divergent continuum limits in the discrete product action drop out when commutation is performed before continuum limit.

Here we significantly improve upon the analysis of  \cite{mect,diffcov} as follows.  We construct the continuum limit action of multiple products of Hamiltonian constraints each such constraint 
smeared by a `$c$- number' lapse i.e. we are 
able to compute the action of a string of Hamiltonian constraint operators ${\hat C}(N_1)..{\hat C}(N_m)$
\footnote{\label{fnk}Specifically, we are able to define the action of upto $k-1$ products of these constraints where we use the $C^k$ semianalytic category and $k$ can be chosen to arbitrarily large.
Note this is similar to the fact that for $C^k$ vector fields one can only define upto $k$ nested commutators and this is the analog of the Lie algebra for the group of $C^{k+1}$ diffeomorphsims.  
}. From this action we can compute the action of the operator obtained
by replacing, in this operator string, any number of  pairs of succesive smeared Hamiltonian constraint operators by their commutators i.e we can compute actions of  operator products of the type
\ba
{\hat C}(N_1)..{\hat C}(N_{i_1-1})[{\hat C}(N_{i_1}), {\hat C}(N_{i_{1}+1})]{\hat C}(N_{i_1+2})..{\hat C}(N_{i_2-1})[{\hat C}(N_{i_2}), {\hat C}(N_{i_2+1})]{\hat C}(N_{i_2+2})...
\nonumber \\
....{\hat C}(N_{i_j-1})[{\hat C}(N_{i_j}), {\hat C}(N_{i_j+1})]{\hat C}(N_{i_j+2})..{\hat C}(N_m).
\label{eqn1}
\ea
We show that each of the commutators in this string is anomaly free in the sense that each can be replaced by the operator correspondent of the corresponding  classical Poisson bracket
(this operator correspondent, as in General Relativity,  is itself {\em not} a Hamiltonian constraint smeared by a $c$- number lapse because of the occurrence of structure functions in the 
Poisson bracket algebra).
We are also able to show that the continuum limit action of multiple  products of smeared Hamiltonian constraints is diffeomorphism covariant and that the group of finite spatial diffeomorphisms 
is implemented in an anomaly free manner. This is almost but not quite the same as what is conventionally referred to as the implementation of the constraint algebra without anomalies in that 
we do not concern ourselves with higher order commutators of the type $[....[[{\hat C(N_1), {\hat C}(N_2)],{\hat C}(N_3}],...., {\hat C}(N_j)]$. We shall return to this point in the concluding section
of this work. Till then we shall refer to our results as an {\em anomaly free single commutator implementation of the constraint algebra}.

While  our basic strategy  is the same as in References \cite{mect, diffcov} (referred to here on as P1, P2 respectively),
its implementation here is more complex than in those works.
A brief summary of the strategy, as implemented here, follows.
As in P1,P2 we deal with the Hamiltonian constraint of density $4/3$ smeared with a density weight $-1/3$ lapse as this seems essential for  {\em nontrivial}
anomaly free commutators (see, for e.g. section 9  in \cite{pftham} and Chapter 2 of \cite{apbook}). 
For reasons explained above, we first define the action of suitable discrete approximants to this constraint and then take the continuum limit. 
As for LQG \cite{rsloop,ttrick}, the  action of these discrete approximants on a charge network state
\footnote{Charge network states are  the abelian analog of the Spin Network basis states of LQG \cite{spinnet} each such state being labelled by a spatial graph whose edges are labelled by integer valued `charges'.}
receives contributions only from vertices of the charge net. As in P1,P2, we confine our attention to the case of chargenets with a single contributing vertex. Since the lapse function has a non-trivial density 
weight the action of a discrete approximant to the constraint (henceforth referred to as the {\em discrete  action of the constraint}) can only be computed with the aid of a coordinate patch around the contributing vertex. 
This action on such a chargenet state generates deformations of the state and the `size' of these deformations is measured, in a precise sense,
by the coordinate patch  associated with the chargenet being acted upon.  The continuum limit action then involves shrinking the size of these deformations away.
Thus, the constraint action depends on a choice of `regulating' coordinate patches, one for (the contributing vertex of) each charge net.
%

While the discrete action
is defined on any charge network state, 
the continuum limit of this discrete action can only be defined on distributional states which are non-normalizable infinite sums over charge network states and which lie in the algebraic dual to the finite
span of charge networks states.
\footnote{The algebraic dual comprises of linear mappings from this finite span to the complex numbers; its elements may be thought of as (in geneneral non-normalizable) sums of charge network bras.}
In this work, as in P1,P2 we restrict attention to the case where the coefficents in these sums are non-vanishing only for `single vertex' charge nets of the type described above.
The coefficients in this sum
are determined  through the specification of a density weighted function and a Riemmanian metric on the 3d Cauchy slice. This is in contrast to the 
specification of the scalar `vertex smooth' function \cite{donjurek}  of P1, P2. Due to the density weight of the function and the tensorial nature of the metric,  the evaluation of these
coefficients also requires a choice of coordinate patches at vertices of the charge network states they multiply. We choose these coordinate patches used to evaluate these coefficients to be the same as the regulating coordinate
patches chosen above to 
define the discrete action of the Hamiltonian constraint.
This choice of coordinate patches then  allows the coefficients to be evaluated and, consequently, the distributional states which support the continuum limit constraint action to be specified.
It is on this set of  distributional states that anomaly freedom is verified. Each such state will be called an `anomaly free state' and the set of states will be referred to as the 
`anomaly free domain'.

The requirement of anomaly free single commutators  is phrased in terms of 
an identity (\ref{key}) discovered in P1 which expresses the Poisson bracket between a pair of classical Hamiltonian constraints in terms of Poisson brackets between certain phase space
functions known as Electric Diffeomorphism constraints (this name derives from their construction as smearings of  the diffeomorphism constraint with Electric field dependent vector fields).
Anomaly freedom is the requirement that this identity holds 
between the commutator between a pair of Hamiltonian constraints and the (continuum limit of the) corresponding  electric 
diffeomorphism commutators.  Since the electric fields in quantum theory are not smooth, the deformations corresponding to electric diffeomorphisms are `singular' versions of 
smooth diffeomorphisms, and, hence, {\em distinct} from the latter. This enables us to focus first on the  construction of an anomaly free single commutator implementation of the algebra of Hamiltonian constraints
and analyse spatial diffeomorphism covariance of our constructions in a second step as follows.

Classical
diffeomorphism  covariance is encoded in the Poisson brackets between the diffeomorphism constraint and the Hamiltonian constraint  and between the diffeomorphism constraints
themselves. The diffeomorphism constraint generates the action of infinitesmal diffeomorphisms on the connection and electric fields. In contrast, in LQG like representations the 
natural operators are those which implement {\em finite} diffeomorphisms. 
It is possible to encode the content of the Poisson brackets involving the diffeomorphism constraint in terms of  the action of finite diffeomorphisms. The Poisson bracket between the 
diffeomorphism constraints is encoded in the requirement that the group of finite diffeomorphisms connected to identity is represented faithfully.  
The Poisson brackets between the diffeomorphism constraint and the Hamiltonian constraint are 
encoded in the requirement that the action of the Hamiltonian constraint be appropriately diffeomorphism covariant (see equation (\ref{r-3})).
Since LQG like representations provide a unitary representation of the group of finite diffeomorphisms, we need concentrate only on the diffeomorphism covariance
of the Hamiltonian constraint action on states in the anomaly free domain.
It is here that 
the  {\em metric dependence} of states in the anomaly free domain allows, relative to P2,  a 
qualitatively {\em new} mechanism for the implementation of diffeomorphism covariance of the  continuum limit action of the Hamiltonian constraint.

Recall that this continuum limit action arises as the limit of the action of discrete approximants to the constraint. Also recall that
this  discrete  action underlying the continuum limit action  requires, for its definition, the choice of a  regulating coordinate patch around the contributing vertex of the charge net being acted upon.
Hitherto (see P2), these coordinate patches (and hence the corresponding discrete deformations generated by the  discrete approximant to the constraint)  were chosen once and for all independent of the choice of the anomaly free state. 
The new ingredient in this work is to tie the choice of
these  structures to the metric label of the state as follows.
Smooth diffeomorphisms are represented unitarily on the space of charge network states. Hence they have a well defined dual action on any anomaly free state. Consider one such state with metric label $h_{ab}$.
Then it turns out that the dual action of a finite diffeomorphism $\phi$ on this state maps the state to a new anomaly free state with metric label $\phi^*h_{ab}$  which is the push forward of $h_{ab}$ by $\phi$.
Let the choice of coordinate patch around the contributing vertex $v$  of the charge net state $c$ when the anomaly free state has metric label $h_{ab}$ be $\{x\}$.
Similar to the case of LQG spin nets, the unitary action of the diffeomorphism $\phi$ on $c$ yields the chargenet $c_{\phi}$ with contributing vertex $\phi(v)$.
Then the idea is to choose  the coordinate patch around the contributing vertex of the charge net state $c_{\phi}$ when the anomaly free state has metric label $\phi^*h_{ab}$ to be $\phi^*\{x\}$.

As we shall see in the main body of the paper, tying the choice of regulating coordinate patches to the metric label of the state in this `diffeomorphism covariant' manner results in 
an elegant and immediate implementation of diffeomorphism covariance of the continuum limit action of the Hamiltonian constraint.  To summarise: we have a tight formalism wherein the label of the anomaly free distributional state dictates the
choice of discrete approximant to the Hamiltonian constraint which in turn defines a discrete action whose continuum limit is  diffeomorphism covariant.
This  implementation of diffeomorphism covariance seems to us to be a robust and beautiful
phenomenon with possible applicability to full blown LQG.
This concludes our summary of the strategy employed in this paper. 

Our considerations in the main body of the paper are based on the contents of P1 and P2. 
While we shall aim at a self contained presentation, the reader interested in a complete understanding is urged to establish some familiarity with P1, P2
especially sections 2, 4, 5 and Appendix C4  of P1 and sections 3.2 and 3.3 and 5.5 of P2.
The reader interested in only a birds eye view of  our results may peruse sections \ref{sec2}, \ref{sec3},  \ref{secresults} and \ref{sec10}.. 
Before we proceed to a description of the layout of the paper, we note that this  model was first studied in an LQG representation in \cite{3du13} wherein the authors focussed on the case of 3 dimensions. 
The model was studied in 4d in \cite{mect,diffcov}.
An attempt was made to apply  the lessons learnt from these studies, together with a remarkable identity discovered by Ashtekar \cite{aapvt} (see also \cite{al} where this identity is reproduced) 
and  earlier pioneering work by Bruegman \cite{bernie}, to    4d Euclidean gravity in \cite{al}.

The layout of the paper is as follows. 
In Section \ref{sec2} we briefly review the model and the derivation of the discrete approximants used in P2.
In Section \ref{sec3} we briefly review the structural lessons learnt from the study of propagation in Parameterised Field Theory \cite{proppft} and  show how to incorporate
these lessons into a modified choice of discrete approximants for the action of the Hamiltonian and the electric diffeomorphism constraint on a certain restricted class of states.
The modifications, though seemingly minor,  are responsible for an anomaly free single commutator  implementation of the constraint algebra.
Due to the nature of the modifications it turns out that the set of restricted states considered in section \ref{sec3} are not large enough for our purposes because the action of the constraints maps
these states out of this set. Hence it is necessary to define the discrete constraint action on a slightly larger set. We develop this  for  a restricted  class of elements of this larger set in section \ref{secgr}  and lift 
this restriction in section \ref{secneg}, 
wherein  we display our  detailed choice for the action on elements of  this larger set (called the Ket Set in section \ref{sec4}).
%
%
%
%
%

In  section \ref{sec4} we construct the discrete action of products of constraint operators. This action derives from multiple applications of actions each of the type specified in section \ref{secneg}.
The specification in section \ref{secneg} on elements of the Ket Set is  not complete in that the coordinate patches underlying the constraint action remain unspecified. In section \ref{sec4} we remedy this and
provide a complete construction of the  action corresponding to  discrete  approximants to  products of constraints on elements of the Ket Set. Finally,
we also indicate as to how the constraints act on states outside this larger set. It turns out that for our purposes, this action on the complement of this set does not need to
be specified in great detail; any action which maps the complement to itself suffices.

In section \ref{sec5} we construct the anomaly free domain of quantum states. 
As mentioned earlier the quantum states in the anomaly free domain are obtained as  non-normalizable sums over kinematic states  with certain coefficients. 
Since it is mathematically more precise to think of these states as residing in a dual space, the sum is over `bras' rather than kets.
The set of bras being summed over is referred to as the  Bra Set. As in P1, P2, for simplicity, we restrict attention to a Bra set in which  each bra has a single nontrivial vertex at which the 
constraints act.  These bras are `bra' correspondents of states of the type encountered in section \ref{secneg}.
Every  state in the anomaly free domain is labelled by a density weighted function 
and a Riemmanian metric on the Cauchy slice. 
The coefficient which multiplies  a bra in the bra set is  evaluated from the structure of the graph underlying the bra together with the density weighted function and metric associated with the anomaly free state.
As mentioned earlier, the continuum limit action of discrete approximant is defined through the contraction  of the discrete deformations generated by the approximant. 
The  dual action of the discrete approximant  on an anomaly free state transfers this contraction  behaviour to the  contraction behaviour of coefficients which characterise the anomaly free state.
We analyse this behavior in section \ref{sec6} and Appendices \ref{aconb}, \ref{acong} as a necessary prerequisite to the computation of the continuum limit action.
In section \ref{sec7} we evaluate  the   continuum  limit action of a product of 2 Hamiltonian constraints on an anomaly free state.  This defines the action of its commutator.
Next, we compute the continuum limit action of the appropriate commutator between 2 electric diffeomorphism constraints and demonstrate equality with the Hamiltonian constraint
commutator, thus showing that the action of a product of 2 Hamiltonian constraints is well defined and anomaly free.
In section \ref{sec8} we extend this result to the action of  higher order products of constraints so as to show that the commutators in (\ref{eqn1}) are anomaly free. 
In Section \ref{sec9} we show that the action of the constraint products of section \ref{sec8} is also diffeomorphism covariant. We briefly summarise and display our results in section \ref{secresults}. 
Section \ref{sec10} is devoted to discussion.
\\

\noindent{\bf Notation and Conventions}:  We set the speed light to be unity  but retain factors of $\hbar$. 
The analog of spin net states in LQG are called charge network states here. We refer to a charge network state as $c$ or $|c\ket$ depending  on our convenience, even changing from one   to the 
other in the course of a single calculation. The symbol $c$ is used for the charge network label (see section \ref{sec2} ) underlying a charge net state.
We  work with the $C^k$ semianalytic category \cite{lost,ttbook}. Due to the finite number of English alphabets, the letter $k$ may occassionaly refer to objects other than the differentiability degree; however the context
should make the usage clear. The Cauchy slice $\Sigma$ is semianalytic, oriented, connected and compact without boundary. All semianalytic charts used are right handed.
The pushforward action of a $C^k$ semianalytic diffeomorphism $\phi$ is denoted by $\phi^*$ and its pull back action by $\phi_*$ so that $\phi^*\phi_*= {\bf 1}$.

\section{\label{sec2} Review of Essential Background from P1,P2}
Almost all the material below is contained in P1. The only part of P2 we allude to is in the choice
of conical deformations at the end of section \ref{sec2.3} below. The only new material not from P1,P2 is in the last two paragraphs ofsection \ref{sec2.2} 
wherein we describe our choice of the inverse metric determinant operator.
\subsection{\label{sec2.1}Classical description of the model} 
The phase space variables $(A_a^i, E^a_i, i=1,2,3)$ are a triplet of  $U(1)$  connections and conjugate density weight one electric fields
on the Cauchy slice $\Sigma$
so that the phase space is that of a $U(1)^3$ gauge theory. 
We define the  density weight 2 contravariant metric  $qq^{ab}:= \sum_i E^a_iE^b_i$, $q$ being the determinant of the corresponding covariant metric $q_{ab}$.
The phase space functions:
\begin{align}
G[\Lambda]  &  =\int\mathrm{d}^{3}x~\Lambda^{i}\partial_{a}E_{i}^{a}\\
D[\vec{N}]  &  =\int\mathrm{d}^{3}x~N^{a}\left(  E_{i}^{b}F_{ab}^{i}-A_{a}%
^{i}\partial_{b}E_{i}^{b}\right) \label{defclassd}\\
H[N]  &  =\tfrac{1}{2}\int\mathrm{d}^{3}x~{N}q^{-1/3}\epsilon^{ijk}E_{i}^{a}E_{j}%
^{b}F_{ab}^{k}, \label{defclassh}
\end{align}
are the Gauss law, diffeomorphism, and Hamiltonian constraints of the theory,
and where $F_{ab}^{i}:=\partial_{a}{A}_{b}^{i}-\partial_{b}{A}_{a}^{i}$. 
The Poisson brackets between the constraints are: 
\begin{align}
\{G[\Lambda],G[\Lambda^{\prime}]\}  &  =\{G[\Lambda],H[N]\}=0\\
\{D[\vec{N}],G[\Lambda]\}  &  =G[\pounds _{\vec{N}}\Lambda]\\
\{D[\vec{N}],D[\vec{M}]\}  &  =D[\pounds _{\vec{N}}\vec{M}] \label{classdd}\\
\{D[\vec{N}],H[N]\}  &  =H[\pounds _{\vec{N}}N]\label{classdh}\\
\{H[N],H[M]\}  &  =D[\vec{\omega}]+G[A\cdot\vec{\omega}],\qquad\omega
^{a}:=q^{-2/3}E_{i}^{a}E_{i}^{b}\left(  M\partial_{b}N-N\partial_{b}M\right) \label{classhh}
\end{align}
The last Poisson bracket (between the Hamiltonian constraints) exhibits
structure functions just as in gravity. 

It is useful to define the Electric Shifts $N^a_i$ by
\be
N^a_i = NE^a_i q^{-1/3}
\label{defes}
\ee
and the Electric Diffeomorphism Constraints $D({\vec N}_i)$ by 
\be
D[\vec{N}_i]   =\int\mathrm{d}^{3}x~N_i^{a}  E_{j}^{b}F_{ab}^{j}
\ee
Assuming the Gauss Law constraint is satisfied, a key identity derived in P1 is:
\be 
\{H[N],H[M]\}  = (-3) \sum_{i=1}^3\{D[{\vec N}_i],D[{\vec M}_i]\} 
\label{key}
\ee

\subsection{\label{sec2.2}Quantum Kinematics}
The basic functions of interest are $U(1)^3$  holonomies associated with  oriented closed graphs  colored by representations of $U(1)^3$ and electric fluxes through surfaces.
Colored graphs are labelled by charge network labels. A charge network label $c$ is the collection $(\gamma, {\vec {q_I}}, I=1,,N )$ where $\gamma$ is an oriented graph with $N$ edges, the 
$I$th edge $e_I$ colored with a triplet of $U(1)$ charges $(q^1_I,q^2_I, q^3_I)\equiv {\vec q_I}$. The holonomy associated with $c$ is $h_c$,
\be
h_{c}:=\prod_{I=1}^{N}\; 
\mathrm{e}^{\mathrm{i}\kappa\gamma q_I^{j}\int_{e_{I}}A_{a}%
^{j}\mathrm{d}x^{a}}.
\ee
Here $\kappa$ is a fixed constant with dimensions $ML^{-1}$ and $\gamma$ is a dimensionless Immirzi parameter. In what follows we shall choose units such
that $\kappa \gamma =1$.

$h_c$ is $U(1)^3$ gauge invariant if the total $U(1)^3$ charge flowing into every vertex is the same as that flowing out of the vertex,  where `into' and `out of' corresponds to whether the 
edge in question is incoming or outgoing at the vertex. In the rest of this paper we restrict attention exclusively to gauge invariant charge net labels.
The gauge invariant electric flux through a surface $S$ is $E_i (S)$, 
\begin{equation}
{E}_{i}(S):=\int_{S}\eta_{abc}{E}_{i}^{a}.
\end{equation}
where $\eta_{abc}$ is the coordinate 3- form.  The holonomy flux Poisson bracket algebra is closed and represented on the space of charge network states. Each charge network state $\vert c\ket$
is labelled by a charge network label $c$. Holonomies act by multiplication and electric flux operators count the discrete electric flux corresponding to the  weighted sum of the charge carried
by edges of $c$ which intersect $S_i$ with the weights being ${\pm 1},0$ depending on the orientation and placement of the intersecting edges relative to the (oriented) surface $S$.

Next consider the Electric shift operator 
\be
{\hat N^a_i} = N{\hat E^a_i q^{-1/3}}
\label{defqes}
\ee
corresponding to the classical expression (\ref{defes}). It turns out that this operator only has a nontrivial action at
vertices of chargenet states and to compute its explicit action  we need a regulating coordinate patch at the vertex in question (see P1). The final expression for the operator action at a vertex
$v$ of the chargent $|c\ket$ is:
\be
{\hat N^a_i}(v) |c\ket=  N^a_i (v) |c\ket:= \sum_{I_v} N^{a}_{I_v i} |c\ket, \;\;\; N^{a}_{I_v i} := \frac{3}{4\pi} N(x(v)) \nu_v^{-2/3}q^i_{I_v}{\hat e}^a_{I_v} .
\label{qsev}
\ee
Here $I_v$ refers to the $I_v$th edge at $v$, and ${\hat e}^a_{I_v}$  to the unit $I_v$th edge tangent vector, unit with respect to the coordinates $\{x\}$ at $v$ and $N(x(v))$ denotes the 
evaluation of the density weighted lapse $N$ at $v$ in this coordinate system. $\nu^{-2/3}_v$ is proportional to the eigen value of the ${\hat q}^{-1/3}$ operator  in equation (\ref{defqes}).
 Specifically, a regulated version of this operator acting  at the vertex $v$ of the charge net state $|c\ket$ can be defined. It has the eigen value
$\nu^{-2/3} \epsilon^2$ where $\epsilon^3$ is the coordinate size of a small regulating region around $v$ so that ${\hat q}^{-1/3}(v) |c\ket := (\nu_v^{-2/3}  \epsilon^2)|c\ket$.
In  P1 this regulated version of ${\hat q}^{-1/3}$  is defined through a Thiemann trick \cite{ttrick,ttbook}.

In this work we use a slightly different definition of ${\hat q}^{-1/3}$ as follows. 
From P1, we have that the regulated metric determinant operator ${\hat q}$ acts at $v$ as ${\hat q}(v) = \epsilon^{-6}{\hat q}_{loc}(v) |c\ket$ where, again, $\epsilon^3$ is the coordinate size of a small regulating region around $v$
and where the operator ${\hat q}_{loc}(v)$ is defined through:
\begin{equation}
\hat{q}_{loc}(v)|c\rangle=\tfrac{1}{48}\hbar^{3}%
\big(|{\textstyle\sum\nolimits_{IJK}}
\epsilon^{IJK}\epsilon_{ijk}q_{I}^{i}q_{J}^{j}q_{K}^{k}|\big)|c\rangle =: \hbar^3(\nu_v)^2 |c\rangle 
\label{evq}
\end{equation}
where each of the three sums (over $I,J,K$) extends over the valence of $v,$
with $I,J,K$ labeling (outgoing) edges $e_{I},e_{J},e_{K}$ emanating from $v.$
$\epsilon^{IJK}=0,+1,-1$ depending on whether the tangents of $e_{I}%
,e_{J},e_{K}$ are linearly dependent, define a right-handed frame (with
respect to the orientation of the underlying manifold), or define a
left-handed frame, respectively. 
We define ${\hat q}^{-1/3}(v)$ by spectral decomposition of ${\hat q}(v)$ on states with non-zero eigenvalues for $\hat{q}_{loc}(v)$ so that on such states $\nu_v^{-2/3}$ is given 
by the $-2/3$rd power of $\nu_v$ in (\ref{evq}). The vertex $v$ for such states
will be referred to as a {\em nondegenerate} vertex.
\footnote{It turns out that this notion of non-degeneracy is appropriate for the `GR' vertices of P1, P2 and section \ref{sec3}. We shall enounter a different type of 
vertex in section \ref{secgr} of this work called a `CGR vertex and will discuss the notion of non-degeneracy
for such a vertex in section \ref{secgr.1}}
On the zero eigen value subspace we define it through the Thiemann trick employed in P1. The result pertinent to the rest of this work is that for
the type of zero eigen value states  of ${\hat q}_{loc}$ enountered in this work, the Thiemann trick returns a vanishing eigen value for ${\hat q}^{-1/3}(v)$.
This is similar to the definitions of inverse metric operators  employed in the Loop Quantum Cosmology context of References \cite{edparam}.

\subsection{\label{sec2.3}Discrete Hamiltonian Constraint from P1}
The action of the discrete approximant to the Hamiltonian constraint operator of P1 is motivated through the following heuristics.
Give a charge net label define the charge net coordinate $c^{ai}(x)$:
\begin{equation}
c_{}^{ai}(x)=c_{}^{ai}(x;\{e_{I}\},\{q_{I}\})=\sum_{I=1}^{M}\mathrm{i}%
q_{I}^{i}\int\mathrm{d}t_{I}~\delta^{(3)}(e_{I}(t_{I}),x)\dot
{e}_{I}^{a}(t_{I}). \label{chrgcoord}%
\end{equation}
The associated holonomy $h_c$ can then be written as $h_c= \exp\left(  \int\mathrm{d}^{3}x~c_{i}^{a}A_{a}^{i}\right)$. A charge net state can be thought of heuristically as a wave function of
the connection $c(A)= h_c(A)$. Holonomy operators then act by multiplication and the electric field operator by functional differentiation so that ${\hat E}^{a}_i (x) = -i\hbar \frac{\delta}{\delta A_a^i(x)}$.

The Hamiltonian constraint in terms of the Electric Shift is:
\begin{eqnarray}
H[N]&=& \tfrac{1}{2}\int_{\Sigma}\mathrm{d}^{3}x~\epsilon^{ijk}N^a_iF_{ab}^{k}%
E_{j}^{b}  + \tfrac{1}{2}\int_{\Sigma}\mathrm{d}^{3}x~N^a_iF_{ab}^{i}%
E_{i}^{b}  \nonumber\\
&=& \tfrac{1}{2}\int_{\Sigma}\mathrm{d}^{3}x~\left(
-\epsilon^{ijk}(\pounds _{\vec{N}_{j}}A_{b}^{k})E_{i}^{b}+%
{\textstyle\sum\nolimits_{i}}
(\pounds _{\vec{N}_{i}}A_{b}^{i})E_{i}^{b}\right) 
\label{hamconst1}
\end{eqnarray}
Here the second term  on the right hand side of the first line vanishes classically and the second line is obtained using the identity 
$N_{i}^{a}F_{ab}^{k}=\pounds _{\vec{N}_{i}}A_{b}^{k}- \partial_{b}(N_{i}%
^{c}A_{c}^{i})$.

The quantum analog of (\ref{hamconst1}) acts on a charge net wave function. For simplicity restrict attention to charge nets with a single non-degenerate vertex. 
The electric shift is then replaced by its operator analog (\ref{defqes}) which is, in turn, replaced by its 
eigen value $N^a_i (v)$ (\ref{qsev}) to yield:
\be
\hat{C}[N]c(A) =-\frac{\hbar}{2\mathrm{i}}c(A)\int_{\Sigma}\mathrm{d}^{3}x~A_{a}%
^{i}\left(  \epsilon^{ijk}\pounds _{\vec{N}_{j}}c_{k}^{a}+\pounds _{\vec
{N}_{i}}c_{i}^{a}\right)  \label{heurcn}
\ee
We refer to $N^a_i (v)$ as the quantum shift. While $N^a_i (v)$ is non zero only at the point $v$ on the Cauchy slice $\Sigma$ we shall think of some regulated version thereof which 
is of small compact support $\Delta_{\delta}(v)$ of coordinate size $\delta^3$ about $v$ (in the coordinates we used to define the quantum shift). Expanding the quantum shift into its edge components (\ref{qsev})
yields:
\begin{equation}
\hat{C}[N]c(A)=\sum_{I_v}
-\frac{\hbar}{2\mathrm{i}}c(A)\int
_{\triangle_{\delta(v)}}\mathrm{d}^{3}x~A_{a}^{i}\left(  \epsilon
^{ijk}\pounds _{\vec{N}_{j}^{I_{v}}}c_{k}^{a}+\pounds _{\vec{N}_{i}^{I_{v}}%
}c_{i}^{a}\right)
\label{edgecomp}
\end{equation}
Next, we approximate the Lie derivative by the difference of a small diffeomorphism and the identity as follows:
\begin{equation}
(\pounds _{\vec{N}_{i}^{I}}c_{j}^{a})A_{a}^{k}=-\frac{3}{4\pi}N(x(v))\nu
_{v}^{-2/3}q_{I_{v}}^{i}\frac{\varphi(\vec{{\hat{e}}}_{I},\delta)^{\ast}%
c_{j}^{a}A_{a}^{k}-c_{j}^{a}A_{a}^{k}}{\delta}+O(\delta). \label{liee}%
\end{equation}
where we imagine extending the edge tangents $\vec{{\hat{e}}}_{I}$ to  $\Delta_{\delta}(v)$ in some smooth compactly supported way  and define
$\varphi(\vec{{\hat{e}}}_{I},\delta)$ to be the finite diffeomorphism corresponding to translation by an affine amount $\delta$  along this edge tangent vector field. Using the 
replacement (\ref{liee}) and using the compact support property of the edge tangent vector field to replace the integration domain $\Delta_{\delta}(v)$  by $\Sigma$ yields:
\begin{equation}
\hat{C}[N]c(A)=\frac{1}{\delta}\frac{\hbar}{2\mathrm{i}%
}c(A)\frac{3}{4\pi}N(x(v))\nu_{v}^{-2/3}\sum_{I_v}q_{I_{v}}^{i}\int_{\Sigma}%
\mathrm{d}^{3}x~\left[  \cdots\right]  _{\delta}^{I_{v},i}+O(\delta), 
\label{qoutsidep1}
\end{equation}
\begin{align}
\left[  \cdots\right]  _{\delta}^{I_{v},1}  &  =\left[  (\varphi c_{2}%
^{a})A_{a}^{3}-c_{2}^{a}A_{a}^{3}\right]  +\left[  (\varphi\bar{c}_{3}%
^{a})A_{a}^{2}-\bar{c}_{3}^{a}A_{a}^{2}\right]  +\left[  (\varphi c_{1}%
^{a})A_{a}^{1}-c_{1}^{a}A_{a}^{1}\right] \nonumber\\
\left[  \cdots\right]  _{\delta}^{I_{v},2}  &  =\left[  (\varphi c_{3}%
^{a})A_{a}^{1}-c_{3}^{a}A_{a}^{1}\right]  +\left[  (\varphi\bar{c}_{1}%
^{a})A_{a}^{3}-\bar{c}_{1}^{a}A_{a}^{3}\right]  +\left[  (\varphi c_{2}%
^{a})A_{a}^{2}-c_{2}^{a}A_{a}^{2}\right] \nonumber\\
\left[  \cdots\right]  _{\delta}^{I_{v},3}  &  =\left[  (\varphi c_{1}%
^{a})A_{a}^{2}-c_{1}^{a}A_{a}^{2}\right]  +\left[  (\varphi\bar{c}_{2}%
^{a})A_{a}^{1}-\bar{c}_{2}^{a}A_{a}^{1}\right]  +\left[  (\varphi c_{3}%
^{a})A_{a}^{3}-c_{3}^{a}A_{a}^{3}\right]  , \label{postlien3}%
\end{align}
where we have written ${\bar c}^a_i\equiv -c^a_i$ and where we have suppressed the edge label $I_v$ and set  $\varphi c_{j}^{a}\equiv\varphi(\vec{{\hat{e}}}_{I_{v}},\delta)^{\ast
}c_{j}^{a}$.

The   integral in (\ref{qoutsidep1}) is of order $\delta$ and we approximate by its exponential minus the identity to get our final expression:
\be
\hat{C}[N]c(A) =\frac{\hbar}{2\mathrm{i}}c(A)\frac{3}{4\pi}N(x(v))\nu
_{v}^{-2/3}\sum_{I_{v}}\sum_{i}q_{I_{v}}^{i}\frac{\mathrm{e}^{\int_{\Sigma
}\left[  \cdots\right]  _{\delta}^{I_{v},i}}-1}{\delta}+O(\delta).\label{p1cnf1}
\ee
For each fixed $(I_v,i)$ the  exponential term is a  product of edge holonomies corresponding to the chargenet labels specified through  (\ref{postlien3}). This product
may be written as $h^{-1}_{c_{(i, flip})}h_{c_{(i, flip, I_v, \delta})}$ where 
$c_{(i, flip, I_v, \delta)}$ is the deformation of $c_{(i, flip)}$ by $\varphi(\vec{{\hat{e}}}_{I},\delta)$ and $c_{i, flip}$ has the same graph as $c$ but `flipped' charges. To see what these
charges are,  fix $i=1$ and some edge $I_v$ corresponding to the 
the first line of (\ref{postlien3}).  In $c_{(1,flip)}$, the  
connection $A_a^3$ corresponding to the 3rd copy of $U(1)$ is multiplied by the charge net $c_2^a$ corresponding to the second copy of $U(1)$. This implies that in
the holonomy $h_{c_{(1,flip)}}$ the charge label  in the 3rd copy of $U(1)$ for any edge is exactly the charge label in the second copy of $U(1)^3$ of the same edge in $c$ i.e. in obvious notation
$q^3|_{c_{(1,flip)}} = q^2|_{c}$ where we have suppresses the edge label. A similar analysis for all the remaining terms in (\ref{postlien3}) indicates that  charges  $^{(i)}\!q^{j}, j=1,2,3$  on any edge of  $c_{(i,flip)}$
are given by the following `$i$- flipping' of the charges on the same edge of $c$.
\begin{equation}
\left.  ^{(i)}\!q^{j}\right.  =\delta^{ij}q^{j}-%
{\textstyle\sum\nolimits_{k}}
\epsilon^{ijk}q^{k} \label{defchrgeflip}%
\end{equation}
The exact nature of the deformed chargenet $c_{(i, flip, I_v, \delta)}$  depends on the definition of the deformation. Since the deformation is of compact support around $v$, the combination
$h^{-1}_{c_{(i, flip)}}h_{c_{(i, flip, I_v, \delta})}$ is just identity except for a small region around $v$. From (\ref{p1cnf1}), this term multiplies $c(A)$. We call the resulting chargenet
as $c_{(i,I_v,\delta )}$. Our final expression as derived in P1 for the discrete approximant to  the Hamiltonian constraint then reads:
\be
\hat{C}[N]_{\delta}c(A) =\frac{\hbar}{2\mathrm{i}}\frac{3}{4\pi}N(x(v))\nu_{v}^{-2/3}\sum_{I_{v}}\sum_{i}q_{I_{v}}^{i}
\frac{c_{(i,I_v,\delta)}- c}{\delta}
\label{hamfinalp1}
\ee
An identical analysis for the action of the electric diffeomorphism constraint yields the result:
\be
\hat{D}_{\delta}[\vec{N}_{i}]c   =\frac{\hbar}{\mathrm{i}}\frac{3}{4\pi}%
N(x(v))\nu_{v}^{-2/3}\sum_{I_{v}}q_{I_{v}}^{i}\frac{1}{\delta
}(c_{(I_{v},\delta)}-c)
\label{dnfinalp1}
\ee
where $c_{(I_{v},\delta )}$ is obtained from $c$ only by deformation without any charge flipping so that 
\begin{equation}
(c_{({I_{v},\delta})})_{i}^{a}(x):=\varphi(\vec{{\hat{e}}}_{I_v}%
,\delta)^{\ast}c_{i}^{a}(x).
\label{defd1cdef}
\end{equation}

It remains to specify the deformation $\varphi(\vec{{\hat{e}}}_{I_v}, \delta)$. From the discussion above this deformation must distort the graph underling $c$ in the vicinity of its vertex $v$ in such a way that its vertex
is displaced by a coordinate distance $\delta$ along the  $I_v$th edge direction to leading order in $\delta$. Due to the vanishing of the  quantum shift except at $v$, this regulated deformation is visualised to
adbruptly pull the vertex structure at $v$ in the direction of the $I_v$ the edge.
In P1 this was achieved by moving the vertex `almost' along the edge 
by an amount $\delta$ but not exactly along it so that the displaced vertex lay in a $\delta^q, q>1$ vicinity  of the edge. The edges connected to the original vertex $v$ were then pulled along the direction of
the displaced vertex. Due to the `abrupt' pulling the original edges developed certain kinks signalling the point from which they were suddenly pulled. The reader is urged to consult the figures in P1 detailing this.
The final picture of the distortion is one in which the off-edge displaced vertex is connected to a kink on the $I_v$th edge by an edge which `almost' coincides with the original $I_v$th, and is connected to the 
kinks on the remaining edges by edges which point  `almost' exactly opposite to the $I_v$th one, the structure in the vicinity of the displaced vertex resembling (and in P2 being exactly that of) the latter set of edges
lying along a `downward' cone with the former edge being upward along the cone axis.
This completes our summary of discrete constraint action as developed in P1, P2.

\section{\label{sec3} Modified Discrete Constraint Action}

In section \ref{sec3.1} we recall some of the structures responsible for propagation in Parameterized field theory \cite{proppft}, discuss their analogs in the context of the $U(1)^3$ model studied here and 
argue that constraint actions in P1,P2 do not display these 
structural analogs. 

In section \ref{sec3.2}  we indicate how these structural features can be incorporated into a modified constraint action which we display in equations (\ref{hamfinal}),(\ref{dnfinal}). 
We shall focus on the case in which the chargenet being acted upon has a single GR vertex where (as in P1,P2) a GR vertex is defined as one which has valence greater than 3 and at which 
no triple of edge tangents is linearly dependent. In addition we shall restrict attention to {\em linear} GR vertices; a vertex will be said to be linear iff there exists a neighbourhood of the vertex equipped
with a coordinate patch such that the entire set of edges at this vertex in this neighbourhood are straight lines in this coordinate patch. 
\footnote{\label{fn6}A further technicality which may be ignored for now is that we also restrict the chargenets here to be `primordial' in the language of section \ref{sec4.2}.}
The constraints generate displacements and deformations of the vertex structure around the linear GR vertex. The deformed vertex structure take the form of a cone, this conical structure being defined
in terms of the coordinates associated with the linear structure of the GR vertex. For pedagogical reasons we shall focus on `downward' conical deformations in this section. It turns out that it is also necessary to
consider `upward' conical deformations and that the choice of upward or downward conicality is linked to the positivity properties of the edge charge labels at the GR vertex.  A complete treatment
will be presented in section \ref{secneg}.

In section \ref{sec3.3}  we show the existence of an alternate  choice of charge flips to that defined by equation (\ref{defchrgeflip}); as we shall see later  both choices of flips
are needed to obtain the crucial `minus' sign on the right hand side of (\ref{key}). In section \ref{sec3.4} we summarise our results.
We remind the reader that as mentioned in section 2, all charge nets encountered in the remainder of the paper are $U(1)^3$ gauge invariant.




\subsection{\label{sec3.1}Structures responsible for Propagation}
Our comments in this section will be very brief as our main focus in this work is the construction of an anomaly free constraint algebra rather than an analysis of propagation. We intend to analyse the issue of propagation 
in this model in future work \cite{u13prop}.

Smolin \cite{leeprop} argued that LQG methods necessarily yield discrete constraint actions whose repeated application on spin network states create nested structures around the original vertices of the spin net. These nested
deformations are created independently for each different vertex. As a result, a deformation near one vertex cannot have any bearing on  that near another vertex and in this sense no information can propagate from
the vicinity of one of the orginal vertices  of the spin net  to another. In Reference \cite{proppft}, we showed that while Smolin's observations are indeed valid, propagation should be viewed as a property 
of physical states lying in the kernel of the constraints rather than as a property of repeated actions of the discrete approximants to the constraint on kinematical states.
Propagation can be viewed in terms of the structure of a given physical state as follows. A physical state is a (in general, kinematically non-normalizable) sum of kinematic states. 
We may then view the physical state as one which encodes propagation effects if kinematic states in this sum are related by propagation \cite{proppft}.
Since physical states are solutions of the quantum constraints, their structure depends on that of the constraints which in turn derives from the structure of the chosen discrete approximants.
It was argued in Reference \cite{proppft} that one of the features responsible for propagation was
 the `$\frac{{\hat O} -{\bf 1}}{\delta}$' of these discrete approximants,  where ${\hat O}$ is some kinematic operator which has a finite well defined action on any spin net state.
Roughly speaking, this structure together with requirement that a continuum limit exist,  ensures that the sum over kinematic states which represents any physical state must have a structure such that 
if the `offspring' state ${\hat O}|s\ket$ is in this sum then the `parent' state $|s\ket$ must also be in the sum.
While at first sight, equations (\ref{hamfinalp1}), (\ref{dnfinalp1}) seem to have this structure, a more careful perusal of these equations shows that due to gauge invariance 
$\sum_{I_{v}}q_{I_{v}}^{i} =0$ so that the `$-{\bf 1}$' term is absent.

Secondly, in the simple context of \cite{proppft} the analog of spin network states live on 1 dimensional graphs so that any two succcesive vertices are connected by an edge. It is this connection
which provides a path for putative propagation effects i.e.  a deformation from one vertex can  putativetly propagate to another along this `conducting' edge.
In contrast (\ref{hamfinalp1}), (\ref{dnfinalp1}) generate deformations which move {\em off} the  edges of the graph (see the material at the end of the section \ref{sec2.3})
and this feature is preserved by
repeated actions of the type  (\ref{hamfinalp1}), (\ref{dnfinalp1}).

In view of these remarks we shall modify the discrete action (\ref{hamfinalp1}), (\ref{dnfinalp1}) so that:\\
\noindent (i) there is a non-trivial $-{\bf 1}$ term in the expression for the discrete constraint action.\\
\noindent (ii) the displaced vertex   $\varphi(\vec{{\hat{e}}}_{I_v},\delta)\cdot v$ is along the $I_v$th edge of the graph rather than off it.

\subsection{\label{sec3.2}Modified Action for linear GR Vertices}

We  implement (i) in section \ref{sec3.2.1} and  (ii) in section \ref{sec3.2.2}.
As mentioned above we shall restrict our considerations to the context of linear GR vertices. 
Recall that a linear vertex is one equipped with a coordinate patch in its neighborhood with respect to which the edges at the vertex in this neighbourhood
appear as straight lines. The vertex will be said to be linear with respect to such a coordinate patch. In what follows the coordinate patch used to specify the deformations generated by constraints is assumed 
to be one with respect to which the vertex is linear.  The detailed choice of these coordinates will be discussed in section \ref{sec4}.



\subsubsection{\label{sec3.2.1}Addressing the $-{\bf 1}$ issue}

We refer the reader to equation (\ref{liee}). 
Let us  scale the (regulated, compact supported in $\triangle_{\delta(v)}$)
 vector field $\vec{{\hat{e}}}_{I_v}$   by its charge label $q^i_{I_v}$ and define $\varphi(q^i_I\vec{{\hat{e}}}_{I_v},\delta)$ 
 to be the small diffeomorphism generated 
by the resulting vector field $q^i_I\vec{{\hat{e}}}_{I_v}$. If we use this diffeomorphism to approximate the Lie derivative on the left hand side of (\ref{liee}), we obtain the equation:
\begin{equation}
(\pounds _{\vec{N}_{i}^{I}}c_{j}^{a})A_{a}^{k}=-\frac{3}{4\pi}N(x(v))\nu
_{v}^{-2/3}  \frac{\varphi(q_{I_{v}}^{i}\vec{{\hat{e}}}_{I},\delta)^{\ast}%
c_{j}^{a}A_{a}^{k}-c_{j}^{a}A_{a}^{k}}{\delta}+O(\delta). \label{lieeq}%
\end{equation}
Using equation (\ref{lieeq}) as our starting point instead of equation (\ref{liee}) and repeating the subsequent argumentation and steps of section \ref{sec2.3}, 
we see that the $q^i_{I}$ factor in (\ref{p1cnf1}) now disappears by virtue of the replacement of $\varphi(\vec{{\hat{e}}}_{I},\delta)$
by $\varphi(q^i_I\vec{{\hat{e}}}_{I},\delta)$. As a result, the holonomy $h_{ c_{ (i, flip, I_v, \delta ) }  }$  is replaced by $h_{ c_{(i, flip, q^i_{I_v}, I_v, \delta) }      }$, 
where $c_{(i, flip, q^i_{I_v}, I_v, \delta)}$ is the image of of $c_{(i, flip)}$ by 
$\varphi(q_{I_{v}}^{i}\vec{{\hat{e}}}_{I},\delta)^{\ast}$:
\be
(c_{(i, flip, q^i_{I_v}, I_v, \delta})^a_j (x):= \varphi(q^i_I\vec{{\hat{e}}}_{I},\delta)^{\ast} (c_{(i, flip)})_j^a(x)
\label{ciflipq}
\ee
Consequently, the deformed charge net $c_{(i,I_v,\delta)}$ in (\ref{hamfinalp1}) is replaced by the chargenet
$c_{(i,q^i_{I_v},I_v,\delta)}$ which is obtained by the action of the holonomy  $h^{-1}_{c_{(i, flip})}h_{c_{(i, flip, q^i_{I_v}, I_v, \delta})}$ on $c$. This 
leads us to
the constraint action:
\be
\hat{C}[N]_{\delta}c(A) =\frac{\hbar}{2\mathrm{i}}\frac{3}{4\pi}N(x(v))\nu_{v}^{-2/3}\sum_{I_{v}}\sum_{i}  
\frac{c_{(i,q^i_{I_v},I_v,\delta)}- c}{\delta}
\label{hamfinalp1q}
\ee

An identical analysis for the action of the electric diffeomorphism constraint yields the result:
\be
\hat{D}_{\delta}[\vec{N}_{i}]c   =\frac{\hbar}{\mathrm{i}}\frac{3}{4\pi}%
N(x(v))\nu_{v}^{-2/3}\sum_{I_{v}}    \frac{1}{\delta
}(   c_{   (  q^i_{I_v},I_{v},\delta   )  }  -   c   ).
\label{dnfinalp1q}
\ee
where $c_{   (   q^i_{I_v}, I_{v},\delta   )  }   $ is obtained from $c$ only by the action of   $\varphi(q^i_I\vec{{\hat{e}}}_{I},\delta)$ without any charge flipping so that 
\begin{equation}
(c_{   (   q^i_{I_v},I_{v},\delta   )    }   )_{j}^{a}(x):=\varphi(q^i_{I_v}\vec{{\hat{e}}}_{I_v}%
,\delta)^{\ast}c_{j}^{a}(x).
\label{defd1cdefq}
\end{equation}
Clearly this addresses issue (i) of section \ref{sec3.1}.

\subsubsection{\label{sec3.2.2}Addressing the `conducting' edge issue}

Instead of the off edge placement of the displaced vertex by $\varphi(\vec{{\hat{e}}}_{I_v},\delta)$  as in P1, we
place the vertex on the 
edge $e_{I_v}$.  
In view of the considerations of section \ref{sec3.2.1}, the action of 
$\varphi(q^i_{I_v}\vec{{\hat{e}}}_{I_v},\delta)$ is defined to displace the vertex $v$ by a coordinate distance $q^{i}_{I_v}$ along the $I_v$th edge.
Denote the displaced vertex by $v_{q^i_{I_v}, I_v, \delta}$.
The remaining edges $e_{J_v\neq I_v}$ are dragged along in the direction of the $I_v$th edge so as to form a `downward pointing cone' in the vicinity of the cone vertex at $v_{q^i_{I_v}, I_v, \delta}$
where `upward' refers to the  direction  of the edge $e_{I_v}$ and where, as in, P1, P2, all edges at $v_{q^i_{I_v}, I_v, \delta}$ are taken to point outwards from $v_{q^i_{I_v}, I_v, \delta}$.
These remaining 
edges develop kinks at the points ${\tilde v}_{J_v}$ at which the edge tangents are discontinuous. As in P1, P2 we refer to these kinks as  $C^0$ kinks (for a formal definition see Appendix \ref{adefkink}.

An explicit construction of the relevant deformation is provided in Appendix \ref{acone} where the linear GR condition is used.
\footnote{\label{fnsec3.2.2} More precisely, as we shall see in section \ref{sec5}, the deformation constructed in Appendix \ref{acone} is diffeomorphic to that discussed here. Hence all diffeomormphism
invariant properties of the latter are identical to that of the former.}
The deformations based on the construction of  Appendix \ref{acone} are displayed in Figure \ref{gr}.  We shall summarize the content of this figure in section \ref{sec3.4}.

The downward conical deformations of Appendix \ref{acone} displace the vertex $v$ `upward' along the $I_v$th edge. This is clearly appropriate only if $q^i_{I_v}$ is {\em positive}.
If $q^i_{I_v}$ is negative it is necessary to consider deformations which displace $v$ in the {\em opposite} direction. This, in turn, requires the further construction of an {\em extension} of the 
edge $I_v$ together with an `upward' conical deformation of the vertex structure around $v$. We shall defer a discussion of such upward conical deformations and graph extensions to section \ref{secneg}
in the interests of pedagogy. Hence  the deformations described above are only valid for deformations along edges for which the charges labels are {\em positive}. 

In view of the discussion in section \ref{sec3.1}, we refer to the edge along which the vertex is displaced in the deformed charge net as the {\em conducting} edge in the deformed charge net. The remaining edges
at the displaced vertex in the deformed charge net which connect the displaced vertex with  $C^0$ kinks will be called {\em non-conducting} edges.
In the case of Hamiltonian constraint type deformations  the conducting edge at the displaced vertex  of the deformed charge net
$c_{(i,q^i_{I_v},I_v,\delta)}$
splits into 2 parts, a `lower' conducting edge which connects the displaced vertex with the vertex $v$ (i.e. with the vertex of $c$) and  
an upper part beyond the displaced vertex.


\subsection{\label{sec3.3}Charge Flips}

Note that in section \ref{sec2.3} we could equally have started with a minus sign in  front of the second term in (\ref{hamconst1}) since that term is non-vanishing.  Let us do this.
This leads to the replacement of  equation (\ref{heurcn}) by
\be
\hat{C}[N]c(A) =\frac{\hbar}{2\mathrm{i}}c(A)\int_{\Sigma}\mathrm{d}^{3}x~A_{a}%
^{i}\left( - \epsilon^{ijk}\pounds _{\vec{N}_{j}}c_{k}^{a}+\pounds _{\vec
{N}_{i}}c_{i}^{a}\right) . \label{heurcn-}
\ee
Repeating the subsequent argumentation, we are lead  to define the charge net $c(-i,flip)$  instead of $c(i,flip)$, with `$-i$ flipped' charges $\left.  ^{(-i)}\!q^{j}\right.$ instead of
the `$i$ flipped' charges of equation (\ref{defchrgeflip}), with these $-i$ flipped' charges defined as:
\begin{equation}
\left.  ^{(-i)}\!q^{j}\right.  =\delta^{ij}q^{j} +%
{\textstyle\sum\nolimits_{k}}
\epsilon^{ijk}q^{k} \label{defchrgeflip-}%
\end{equation}
The exponential term in equation (\ref{p1cnf1}) is then replaced, in obvious notation, by \\
$h^{-1}_{c_{(-i, flip})}h_{c_{(-i, flip, I_v, \delta})}$ and
we are lead to, instead of equation (\ref{hamfinalp1}),  the  expression:
\be
\hat{C}[N]_{\delta}c(A) = -\frac{\hbar}{2\mathrm{i}}\frac{3}{4\pi}N(x(v))\nu_{v}^{-2/3}\sum_{I_{v}}\sum_{i}q_{I_{v}}^{i}
\frac{c_{(-i,I_v,\delta)}- c}{\delta}
\label{hamfinalp1-}
\ee
where $c_{(-i,I_v,\delta)}$  is exactly the same as $c_{(i,I_v,\delta)}$ of (\ref{hamfinalp1}) except that the  `$i$ flipped charges' of 
equation (\ref{defchrgeflip}) are replaced by their `$-i$ flipped' version in equation (\ref{defchrgeflip-}).
Repeating the considerations of section \ref{sec3.2.1} we are lead to the final equation:
\be
\hat{C}[N]_{\delta}c(A) =-\frac{\hbar}{2\mathrm{i}}\frac{3}{4\pi}N(x(v))\nu_{v}^{-2/3}\sum_{I_{v}}\sum_{i}  
\frac{c_{(-i,q^i_{I_v},I_v,\delta)}- c}{\delta}
\label{hamfinalp1q-}
\ee
where, once again in obvious notation, $c_{(-i,q^i_{I_v},I_v,\delta)}$ is exactly the same as 
$c_{(i,q^i_{I_v},I_v,\delta)}$ except that the role of $i$-flipping is replaced by that of $-i$- flipping. 

To summarise: We are able to generate an overall  minus sign in the expression (\ref{hamfinalp1q-}) relative to (\ref{hamfinalp1q}) by changing the charge flip from a  $i$- flip to a -$i$ flip.
Putting everything together (and using the notation  $c_{(+i,q^i_{I_v},I_v,\delta)} \equiv c_{(i,q^i_{I_v},I_v,\delta)}$   we are lead to 2 possible discrete actions of the Hamiltonian constraint:
\be
\hat{C}[N]_{\delta}c(A) =\pm\frac{\hbar}{2\mathrm{i}}\frac{3}{4\pi}N(x(v))\nu_{v}^{-2/3}\sum_{I_{v}}\sum_{i}  
\frac{c_{(\pm i,q^i_{I_v},I_v,\delta)}- c}{\delta} .
\label{hamfinal}
\ee
As no charge flipping is involved, the expression for the electric diffeomorphism constraint remains the same:
\be
\hat{D}_{\delta}[\vec{N}_{i}]c   =\frac{\hbar}{\mathrm{i}}\frac{3}{4\pi}%
N(x(v))\nu_{v}^{-2/3}\sum_{I_{v}}   \frac{1}{\delta
}(   c_{   (  q^i_{I_v},I_{v},\delta   )  }  -   c   ).
\label{dnfinal}
\ee
In view of the considerations of section \ref{sec3.2.2} the deformations in equations (\ref{hamfinal}), (\ref{dnfinal}) are of the `on edge, conical type'. We slighty abuse
notation and continue to use the notation 
$\varphi(q^i_I\vec{{\hat{e}}}_{I},\delta)$ of section \ref{sec3.2.1} for the deformation map corresponding to the modified deformations of section \ref{sec3.2.2}.
In section \ref{sec5} we shall find it necessary to use both the versions of discrete Hamiltonian action described in (\ref{hamfinal}).

Finally, as emphaisised in section \ref{sec3.2}, the deformations along the $I_v$th edge constructed therein are valid only if $q^i_{I_v}>0$. For $q^i_{I_v}<0$, we shall define the deformed states
$c_{(\pm i,q^i_{I_v},I_v,\delta)}, c_{   (  q^i_{I_v},I_{v},\delta   )  }$ in equations (\ref{hamfinal}), (\ref{dnfinal}), in section \ref{secneg}. 

\subsection{\label{sec3.4}Summary}

\begin{figure}
  \begin{subfigure}[h]{0.3\textwidth}
    \includegraphics[width=\textwidth]{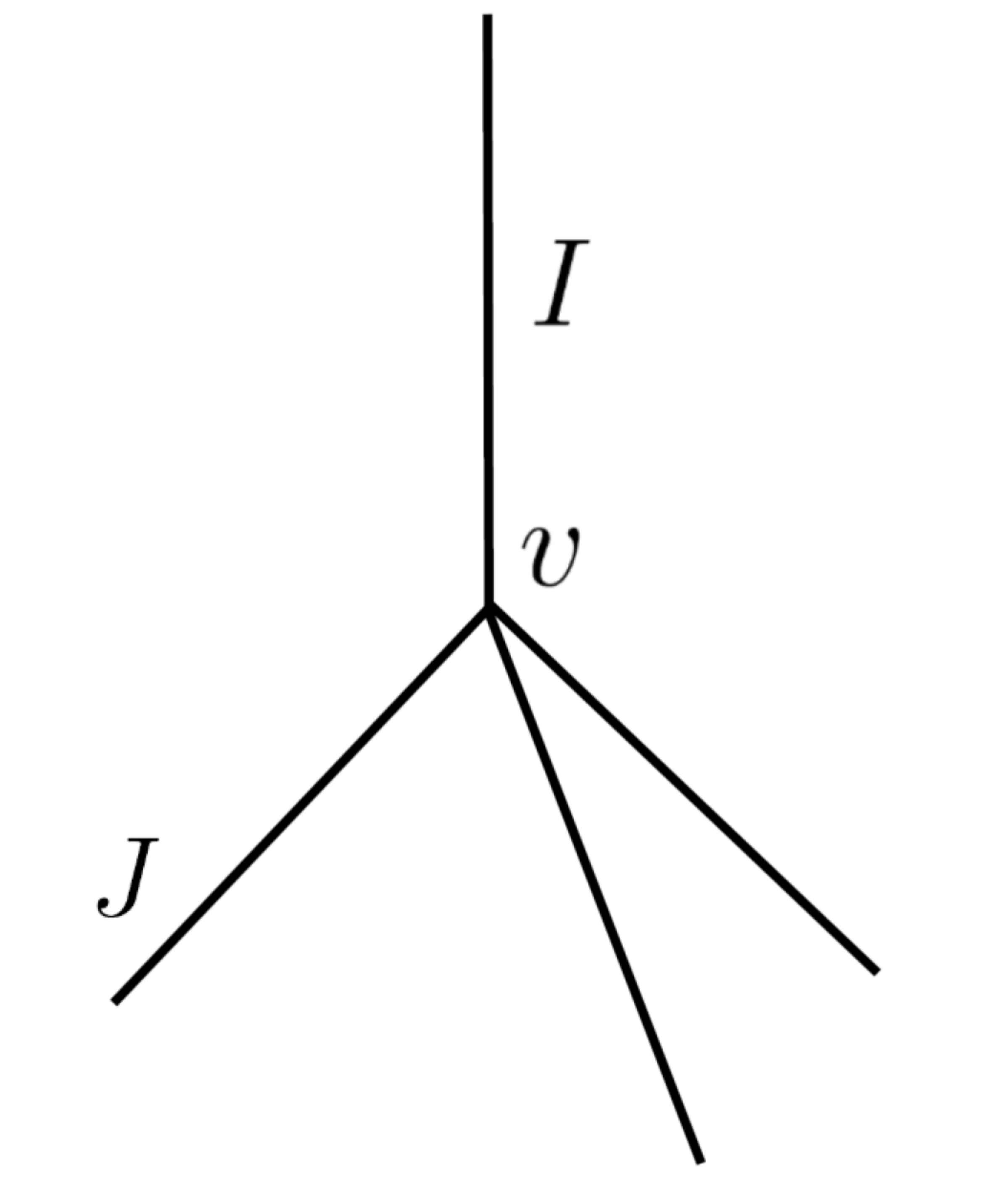}
    \caption{}
 \label{gra}
  \end{subfigure}
  \begin{subfigure}[h]{0.3\textwidth}
    \includegraphics[width=\textwidth]{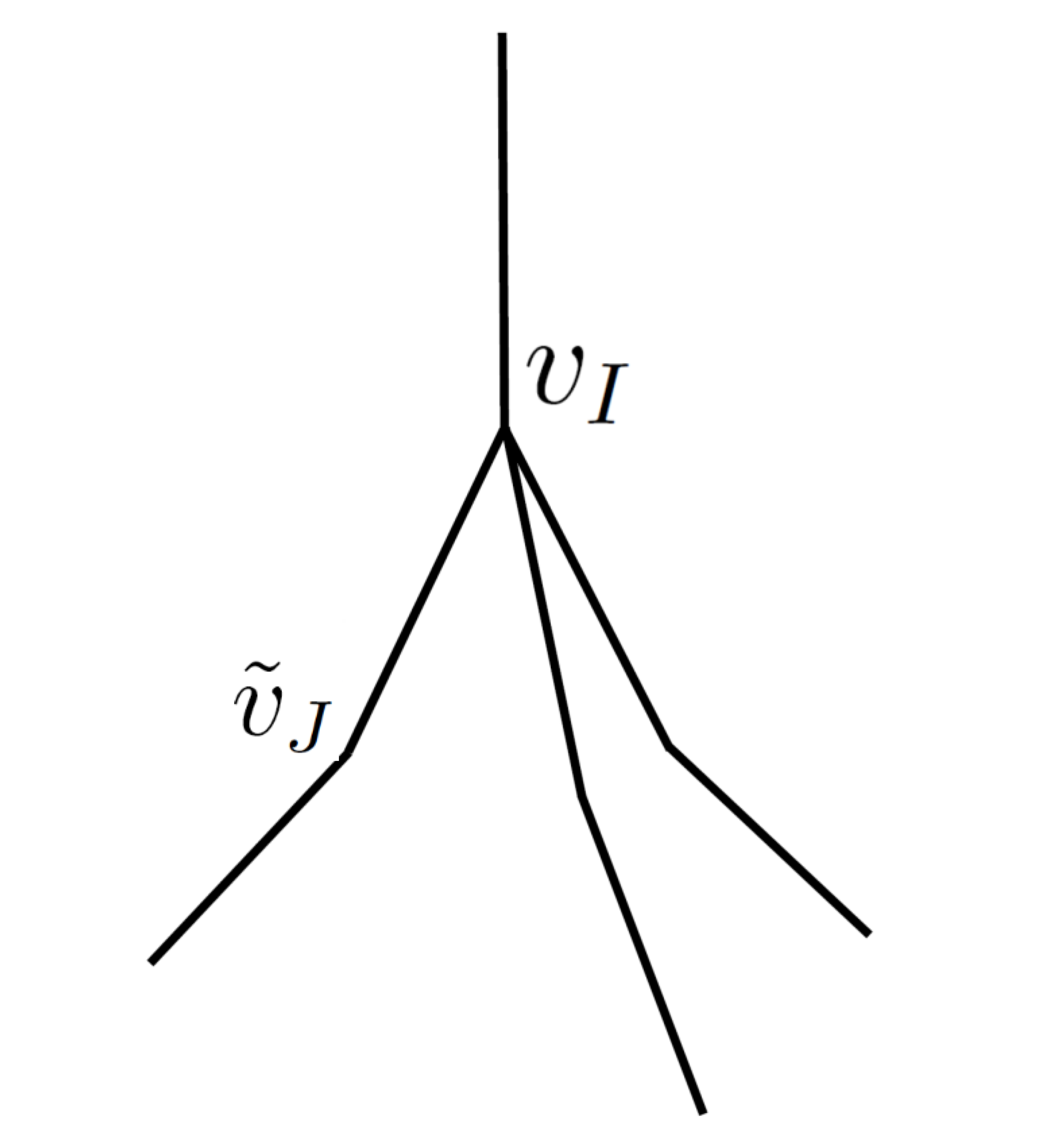}
    \caption{}
   \label{grb}
  \end{subfigure}
\begin{subfigure}[h]{0.3\textwidth}
    \includegraphics[width=\textwidth]{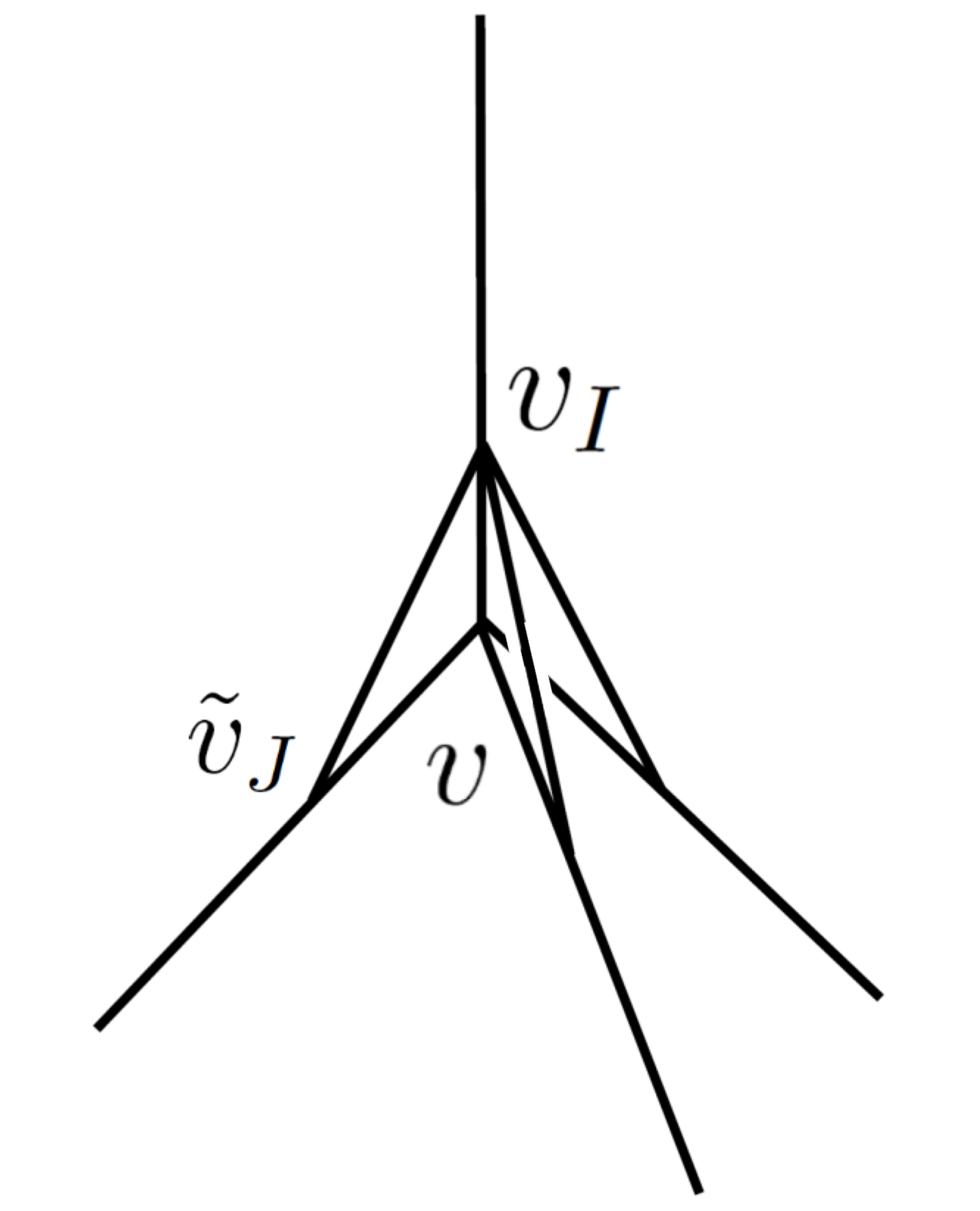}
    \caption{}
   \label{grc}
  \end{subfigure}
  \caption{ Fig \ref{gra} shows an undeformed GR vertex $v$ of a chargenet $c$  with its $I$th and $J$th edges as labelled. The vertex is deformed along its $I$th edge in Fig \ref{grb} wherein the displaced
vertex $v_I$ and the $C^0$ kink, ${\tilde v}_J$ on the $J$th edge are labelled. Fig \ref{grc} shows the result of a Hamiltonian type deformation 
obtained by multiplying the chargenet holonomies obtained by 
coloring  the edges of Fig \ref{grb}  by 
flipped images of charges  on their counterparts in $c$ ,  Fig \ref{gra}  by negative of these flipped charges and Fig \ref{gra} by the charges on $c$. If the edges of Fig \ref{grb} are colored by the charges on 
their counterparts in $c$ then one obtains an electric diffemorphism deformation.
}%
\label{gr}%
\end{figure}

For the case that $q^i_{I_v}>0$, we  display the deformed charge net   $c_{(\pm i,q^i_{I_v},I_v,\delta)}$ of (\ref{hamfinal}) in Figure \ref{grc}.  This charge net can be visualised as the product of following three holonomies:\\
\noindent (i) a holonomy labelled by the deformed chargenet colored with flipped charges, $h_{c_{(-i, flip, I_v, \delta})}$, shown in Figure \ref{gra}.\\
\noindent (ii) a holonomy labelled by an undeformed chargenet based on the same graph (see Fig \ref{gra} as $c$ and colored with  the negative of the flipped charges  $h^{-1}_{c_{(-i, flip})}$, the negative sign coming from the inverse.\\
\noindent (iii) the original chargenet holonomy based on the graph shown in Fig \ref{gra}.\\
As a result, the charge carried by the undeformed counterparts of the non-conducting edges at $v$ in  $c_{(\pm i,q^i_{I_v},I_v,\delta)}$ (namely the edges which connect $v$ to the $C^0$ kinks)  have vanishing $i$th component. 
By gauge invariance the 
charge
along the (lower) conducting  edge passing through $v$ in $c_{(\pm i,q^i_{I_v},I_v,\delta)}$  also has vanishing $i$th component. It is then straightforward to see that, similar to  P1,P2,  the vertex $v$ is {\em degenerate} in 
$c_{(\pm i,q^i_{I_v},I_v,\delta)}$.  Also note that each non-conducting edge in (i) carries flipped versions of the charges carried by its undeformed counterpart in $c$. Hence,  using gauge invariance
at the displaced vertex in  $c_{(\pm i,q^i_{I_v},I_v,\delta)}$, we have the following Remark:\\

\noindent{\em Remark 0}: The {\em difference} between the outgoing and incoming charges along the conducting edge at the deformed vertex in  $c_{(\pm i,q^i_{I_v},I_v,\delta)}$ 
is  the $\pm i$-flipped version of the charge along the $I_v$th edge  in $c$.
\\
Finally, recall that vertex structure in a sufficiently small vicinity of the displaced vertex when viewed in terms of the coordinates associated with the linear vertex $v$ in $c$ takes the following form.
All edges are straight lines. 
The conducting edge in $c_{(\pm i,q^i_{I_v},I_v,\delta)}$ is
split  into two parts by the displaced vertex. The remaining (non-conducting)  edges at the displaced vertex form a `downward' cone.
With respect to the `downward' direction of the  cone the conducting edge splits into an upper conducting edge and a lower conducting edge.

The deformed charge net  $c_{   (  q^i_{I_v},I_{v},\delta   )  } $ of (\ref{dnfinal}) is based on the same  deformed graph as that in (i) above; the only difference
is that the charge labels are unflipped i.e. each deformed edge in $c_{   (  q^i_{I_v},I_{v},\delta   )  } $ has  the   same charge as its undeformed counterpart in  $c$.

\section{\label{secgr}Modified action: Linear CGR vertices}

In the last section we have restricted attention to linear GR  vertices. The action of the Hamiltonian constraint (\ref{hamfinal}) displaces such a vertex along a conducting edge
so that  the conducting edge splits into an incoming and outgoing part at the displaced vertex and  the incoming and outgoing conducting edge tangents comprise a linearly dependent pair at the displaced vertex (see Fig \ref{grc}).
Hence  any triple of edge tangents which contains the incoming and outgoing conducting edge tangents is no longer linearly independent and the displaced vertex is not strictly GR.
Due to the  role played by the conducting edge in altering the (linear) GR structure of
such a vertex, we shall call it a (linear) Conducting Edge-Altered GR vertex or a CGR vertex. 
\footnote{\label{fndefinvgr}Note that the transition from a GR vertex to a CGR vertex by the Hamiltonian constraint action  is {\em not} generated by the action of the deformation map $\varphi(q^i_{I_v}\vec{{\hat{e}}}_{I_v},\delta)$. Indeed, the graph underlying the deformed
chargenet created by the action of the deformation map on $c$ displays a single GR vertex as shown in Figure \ref{grb}. Rather, the CGR property stems from the fact that $c_{(\pm i,q^i_{I_v},I_v,\delta)}$
is constructed not only from the deformed charge net of Fig \ref{grb} but also the undeformed ones based on the graph shown in  Fig \ref{gra}. 
Indeed, the electric diffeomorphism constraint action (\ref{dnfinal}) retains the GR nature of the vertex acted upon as displayed in Figure \ref{grb}.}

In section \ref{secgr.1} we  isolate the structure in the vicinity of such a vertex, discuss it in detail and 
define modified discrete constraint actions for states with such a vertex. As in the previous section the coordinates with respect to which the deformations generated by these constraints actions are defined will be assumed to be 
ones with respect to which the vertex is linear. The detailed choice of these coordinates will be discussed in section \ref{sec4}.
In section \ref{secgr.3} we define a single  notation which succintly describes the deformed states produced by the modified constraint actions both for the GR and the CGR cases. 
\subsection{\label{secgr.1} Linear CGR vertices:Definition and Constraint action}
From section \ref{sec3.4}, we define a (linear) CGR vertex as follows. A vertex $v$ of a charge net $c$  will be said to be linear CGR if:\\
(i) There exists a coordinate patch around $v$ such that all edges at $v$ are straight lines.\\
(ii) The union of 2 of the edges at $v$ form a single straight line so that $v$ splits this  straight line into 2 parts\\
(iii) The set of remaining edges  together with any one of the two edges in (i) constitute a GR vertex in the following sense. Consider, at $v$, the set of out going edge tangents
to each of the remaining edges together with the outgoing edge tangent to one of the two edges in (i). Then any triple of elements of this set is linearly dependent.

We shall call the edges other than those in (ii)  as non-conducting  in $c$ and the two edges in (ii) as upper and lower conducting edges in $c$ and refer to the union of the conducting edges
as the conducting line in $c$.
\footnote{Here we assume that we are given a specification of which of the two edges is upper and which is lower; how this specification arises will be discussed in section \ref{secneg}.}
Let the  upper conducting edge and the non-conducting edges be assigned an outward pointing orientation from $v$ in $c$ and let the lower conducting
edge be assigned an incoming orientation at $v$ in $c$ so that the conducting line acquires a natural well defined orientation induced from the conducting edges.
Let the number of non-conducting edges be $N-1$. Hence there are $N+1$ edges at $v$ but these edges define only  $N$ distinct oriented straight lines passing through $v$ in $c$, one of them being the 
conducting line and the remaining $N-1$ being the non-conducting edges.
Let $J_v=1,..,N$ be an index which numbers these straight lines. Let the conducting line be the $K_v$th one. It follows that  the non-conducting edges are assigned indices $\{J_v, J_v\neq K_v\}$.
Denote such a non-conducting edge by $e_{J_v}$ for some  $J_v\neq K_v$ and its outgoing charge by $q^i_{J_v}$. Denote the upper conducting edge
with outward orientation 
by $e_{K_v,out}$,  the lower conducting edge with incoming orientation by $e_{K_v,in}$ and their respective outgoing and incoming charges by $q^i_{K_v,out}$ and $q^i_{K_v,in}$. 

We turn now to a derivation of modified constraint actions on a state $c$ with a  linear CGR vertex using the notation discussed above.
We shall convert the situation into one in which the lower conducting edge is absent at $v$ and the upper conducting edge acquires a charge $q^i_{K_v,out}-q^i_{K_v,in}$. The vertex $v$ then becomes
GR and we may then use the deformations described in Appendix \ref{acone1}. In this section we shall restrict attention to the case where the `net' conducting charge $q^i_{K_v,out}-q^i_{K_v,in}$ is {\em positive}.
This restriction is for pedagogical reasons which are identical to those which underlie the applicability of the `downward conical' deformations  of section \ref{sec3.2} to the case of `$q^i_{I_v}>0$' (see the discussion at the end
of section \ref{sec3.2}). 
The general case involving charges with no
positivity restrictions together with the consideration of `upward conical deformations' will be discussed in section \ref{secneg}.

\begin{figure}[h]
\centering
\includegraphics[width=12cm]{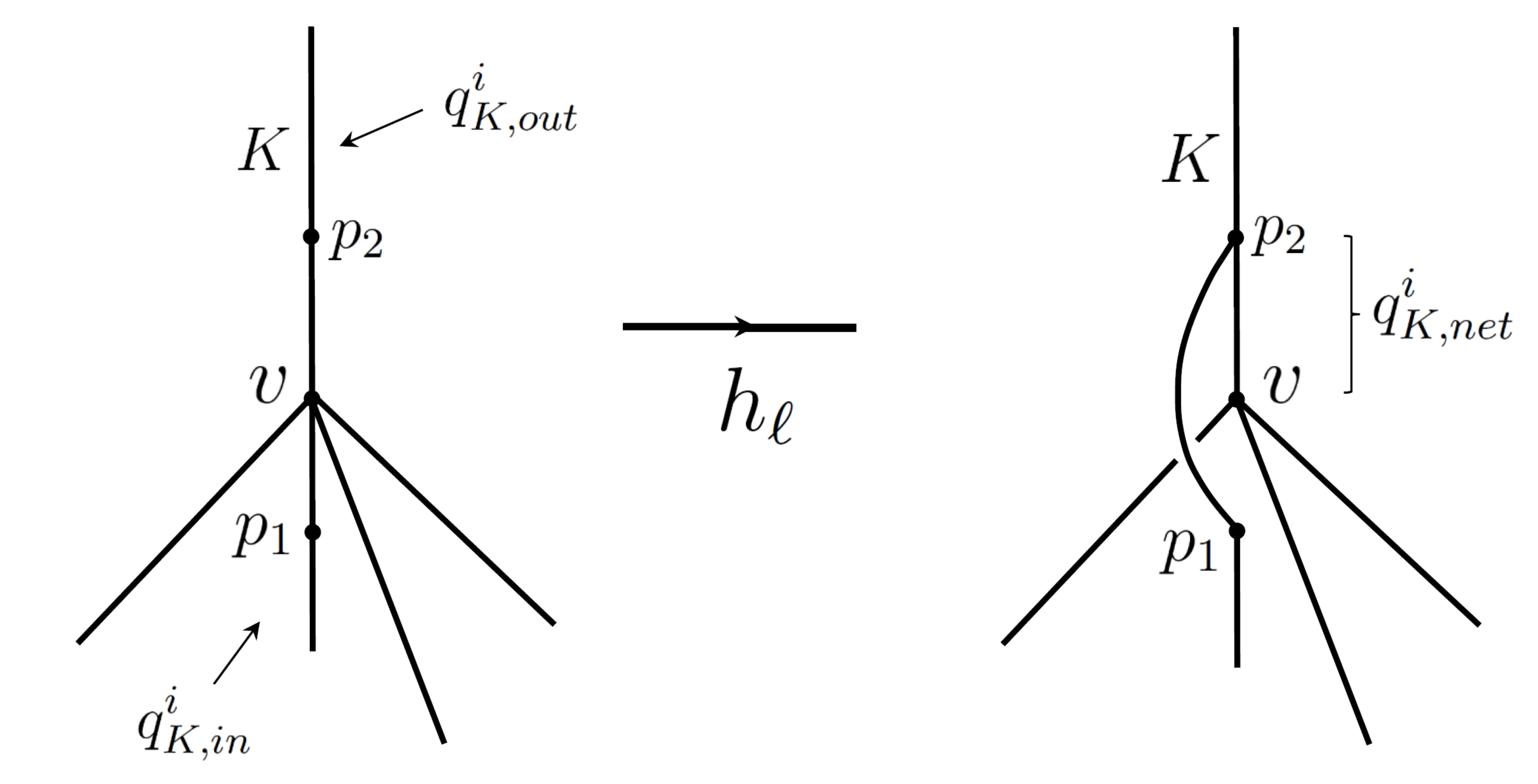}
\caption{The figure on the left shows the 
vertex structure at the CGR vertex $v$. The conducting edges are  the $K$th ones. The effect of multiplication by the intervening holonomy $h_l$ on this vertex structure is shown on the right. 
The lower conducting edge at $v$ is removed and the upper conducting edge is charged with the net conducting charge.
}%
\label{cgrint}%
\end{figure}

We are interested in the discrete action of the constraints at small enough discretization parameter $\delta$ where $\delta$ is measured by the coordinate system in (i).
Consider a loop $l$  made up of two edges $l_1, l_2$ so that $l= l_1\circ l_2$.
Let  $l_1$ be  a segment of the conducting line running between  two of its points $p_1$ and $p_2$  equidistant from $v$, where $p_1$ is  below $v$ and $p_2$ is above $v$. Let $p_1$ and $p_2$ be chosen
such that the coordinate length of $l_1$ is $C\delta, C>16q_{max}$
\footnote{See (a)- (c), section \ref{secneg1.2}  for the reason for this choice of $C$.}
where
%
%
\be 
q_{max} = \max_{(i=1,2,3),(I_v=1,..,N)} |q^i_{I_v}|.
\label{defqmax}
\ee
Further, let  $l_1$ be oriented so as to run from $p_1$ to $p_2$. Let $l_2$ be a semicircular arc connecting $p_2$ with $p_1$ such that its diameter is $C\delta$. 
Let $l$ lie in a coordinate plane $P_l$ such that no non-conducting edge lies in $P_l$. 
Define the holonomy $h_l$ to run along $l$ with charge equal to $-q_{K_v,in}$ i.e. $h_l$ is charged with the negative of the incoming charge at $v$ carried by the incoming lower conducting edge.
Note that for any smooth connection $A^j_a$,
\be
h_l := \exp i(-\sum_{j=1}^3q^j_{K_v,in} \int_{l} A_a^jdx^a) \sim 1 + O(\delta^2).
\label{defhl}
\ee
Since the classical holonomy $h_l$ is unity to order $\delta^2$  multiplication of an approximant to a constraint  by $h_l$  continues to yield an acceptable approximant.
Accordingly, we first multiply $c$ by $h_l$.
Clearly, this yields the chargenet $c_l$ in which, as mentioned above, the lower conducting edge of $c$ is absent from $p_1$ to $v$, the
upper conducting edge acquires a charge $q^i_{K_v,out}-q^i_{K_v,in}$ between $v$ and $p_2$ and the nonconducting edges are untouched. As shown in Fig \ref{cgrint}, the  vertex $v$  in $c_l$ then becomes
GR and we may then act on the result by the discrete approximant to the constraint of interest as in section \ref{sec3}, the vertex structure  deformations
of $c_l$ being constructed 
along the lines described in Appendix \ref{acone1}.

We act on the result by ${\hat h}^{-1}_l$. Since the deformation of Appendix \ref{acone1} is confined to within a ball of radius $2q_{max}\delta$ about $v$ (see (\ref{defqmax}) for the definition of $q_{max}$), 
the semicircular arc $l_2$ does not touch the 
deformed structures, and due to its placement does not touch the undeformed structure (for small enough $\delta$) except at $p_1,p_2$. Hence the action of ${\hat h}^{-1}_l$ simply
removes the `extra segment' $l_2$ from the chargenets generated hitherto and restores the missing part of the conducting line, so that we have:
\begin{eqnarray}
\hat{C}[N]_{\delta}c(A) &=& \pm{\hat h}^{-1}_{l}
\frac{\hbar}{2\mathrm{i}}\frac{3}{4\pi}N(x(v))\nu_{v}^{-2/3} \sum_{I_{v}}\sum_{i}  
\frac{c_{l(\pm i,q^i_{I_v},I_v,\delta)}- c_l}{\delta} \nonumber \\
&=& \pm\frac{\hbar}{2\mathrm{i}}\frac{3}{4\pi}N(x(v))\nu_{v}^{-2/3} \big( \sum_{I_{v}\neq K_v}\sum_{i}  
\frac{c_{(\pm i,q^i_{I_v},I_v,\delta)}- c}{\delta} \nonumber \\
& + & \sum_i\frac{  c_{ (\pm i,\;q^i_{K_v, out}- q^i_{K_v, in},\;K_v,\delta)  }- c}{\delta} \big)
\label{hamfinalc}
\end{eqnarray}
In the second and third lines we have used $\nu_v$ to denote the volume eigen value of $c_l$ at its GR vertex $v$. 
Note that this is {\em not} the same as the volume eigen value for $c$. 
\footnote{From  (\ref{evq}), it follows that the volume eigen value is sensitive only to the structure of $c$ in a small vicinity of $v$. If we replace this structure by one which has
identical colored non-conducting edges, no lower conducting edge and an upper conducting edge which has charge $q^i_{K_v, out} + q^i_{K_v, in}$, the volume eigen value for this structure is the
same as that for $c$. This differs from that for $c_l$ because the vertex structure there has the upper conducting edge charge as $q^i_{K_v, out} - q^i_{K_v, in}$.}
The fact that a non-trivial constraint action is only possible if $v$ is non-degenerate
in $c_l$ (rather than in $c$) suggests that we define our notion of non-degeneracy for a CGR vertex  to be tied to that of the corresponding GR vertex obtained by modifying the CGR one through the intervention of the holonomy $h_l$.
We shall formalise this definition in sections \ref{secgr.1a} and \ref{secneg}.
%

\begin{figure}
\begin{subfigure}[h]{0.3\textwidth}
\includegraphics[width=\textwidth]{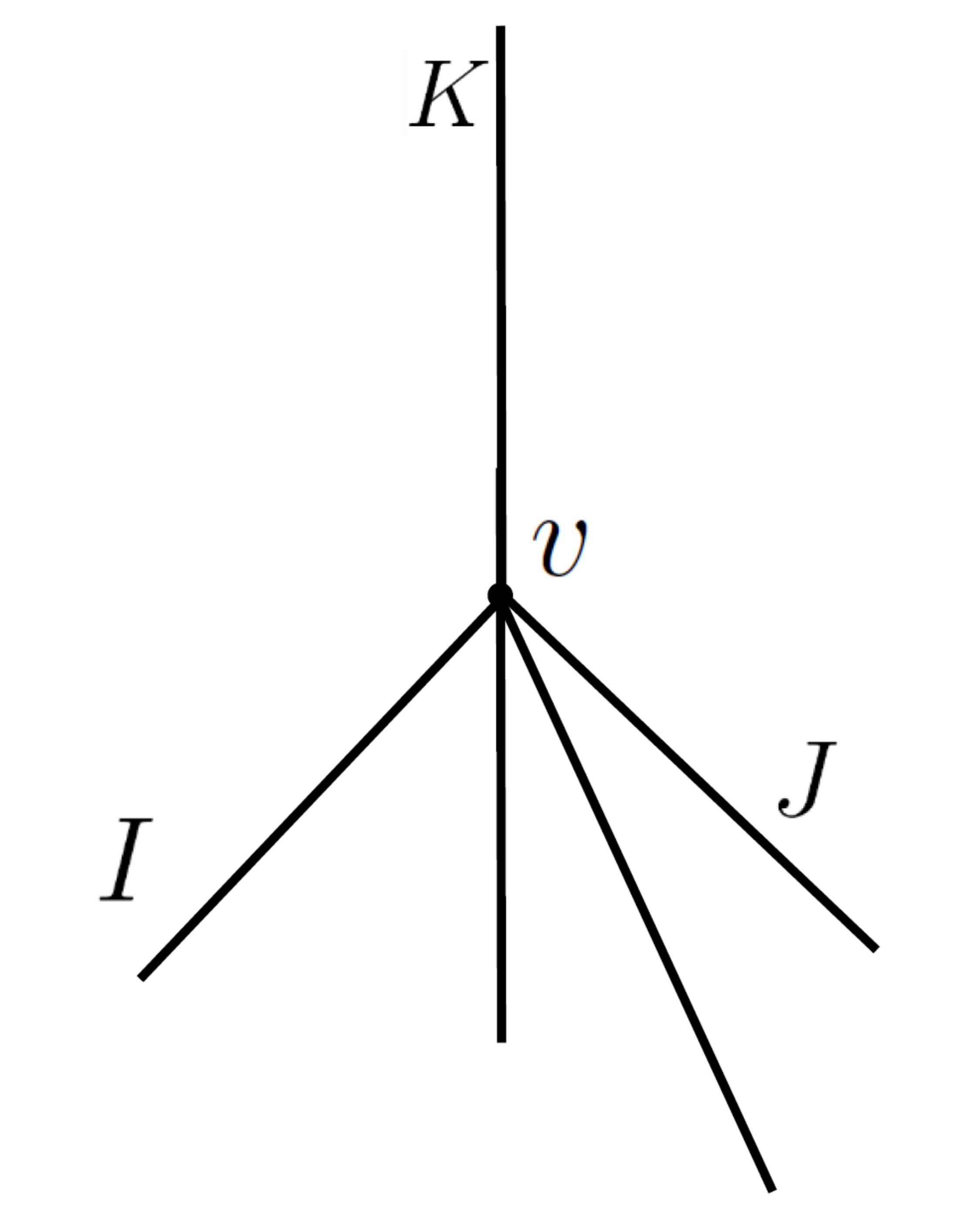}
\caption{}
    \label{cgrknia}
  \end{subfigure}
  \begin{subfigure}[h]{0.3\textwidth}
    \includegraphics[width=\textwidth]{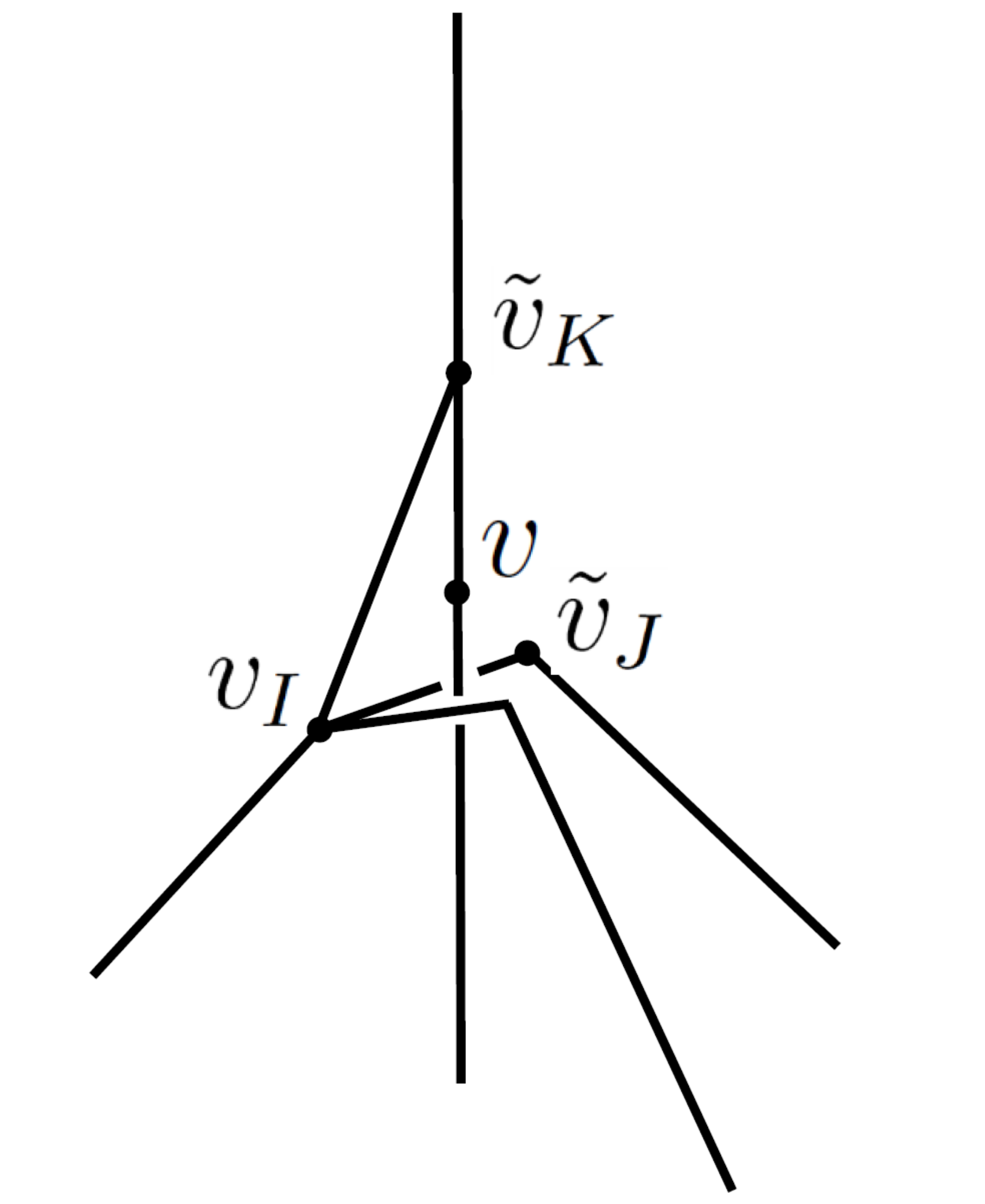}
    \caption{}
    \label{cgrknib}
  \end{subfigure}
\begin{subfigure}[h]{0.3\textwidth}
    \includegraphics[width=\textwidth]{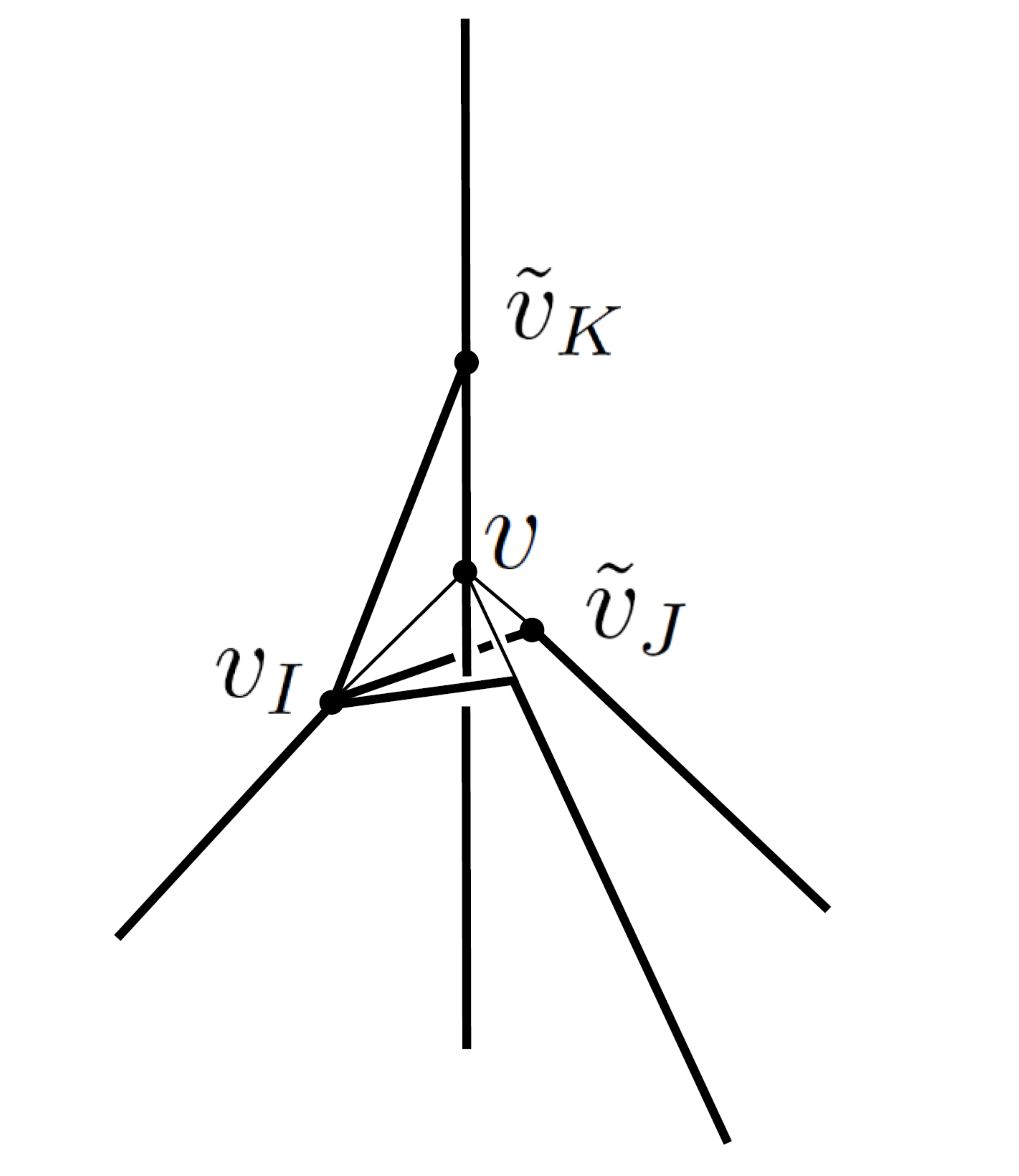}
    \caption{}
    \label{cgrknic}
  \end{subfigure}
  \caption{Fig \ref{cgrknia} shows an undeformed CGR vertex $v$ of a chargenet $c$  with its $K$th conducting edge and $I$th and  $J$th non-conducting edges as labelled. In  Fig \ref{cgrknib}  the vertex  structure of Fig \ref{cgrknia} is deformed
along its $I$th edge and the displaced
vertex $v_I$ and the $C^0$ kinks  ${\tilde v}_J$, ${\tilde v}_K$ on the $J$th, $K$th edges are as labelled. 
Fig \ref{cgrknic} shows the result of a  Hamiltonian type deformation. To obtain this result:(i)in Fig \ref{cgrknib}    
color the edge from  $v_I$   to ${\tilde v}_K$  with the  flipped image of the {\em net} conducting charge in $c$, that from
$v$ to ${\tilde v}_K$ with the flipped image of the lower conducting charge at $c$ and the remaining edges  with the  
flipped images of the charges  on their undeformed counterparts in $c$
(ii) color the edges of  Fig \ref{cgrknia} by the negative of the  
flipped charges on $c$ (iii) color  the edges of Fig \ref{cgrknia}  by the charges on $c$ (iv) multiply the holonomies corresponding to (i),(ii), (iii).  In Fig \ref{cgrknib}, if the edge from  $v_I$   to ${\tilde v}_K$
is colored with the {\em net} conducting charge in $c$, that from $v$ to ${\tilde v}_K$ by the lower conducting charge in $c$ and
the remaining edges by the charges on  their counterparts in $c$ one obtains the result of an electric diffemorphism deformation.
}%
\label{cgrkni}%
\end{figure}

The deformed chargenet $c_{(\pm i,q^i_{I_v},I_v,\delta)}$ for $I_v\neq K_v$  and for the case $q^i_{I_v}>0$
\footnote{\label{fnposcgr}
We will tackle the $q^i_{I_v}<0$ case in 
section \ref{secneg}. Hence the deformed chargenets  $c_{l(\pm i,q^i_{I_v},I_v,\delta)}$ for $q^i_{I_v}<0$ will be constructed in detail only in that section.}
is shown in Figure \ref{cgrknic}.

It may be viewed as the product of 3 holonomies:
one which is deformed and has flipped charges as shown in Fig \ref{cgrknib}, a second which is based on the undeformed graph of Fig \ref{cgrknia} with negative of the flipped charges  and the last which is just the holonomy 
corresponding to $c$.  Due to the  deformations of the GR vertex structure  of $c_l$, each of the edges of $c_{l(\pm i,q^i_{I_v},I_v,\delta)}$ at its nondegenerate vertex
other than the $I_v$th one  meet their undeformed counterparts in  $C^0$ kinks. Since there is no lower conducting 
edge at the vertex $v$ of $c_{l(\pm i,q^i_{I_v},I_v,\delta)}$, the subsequent multiplication by ${\hat h}^{-1}_{l}$ results in a restoration of this `missing' part of $e_{K_v,in}$ without any further kink.
Thus the deformed graph structure underlying   $c_{(\pm i,q^i_{I_v},I_v,\delta)}$ obtained by     first intervening with ${\hat h}_l$ then deforming the resulting GR structure and finally intervening with    ${\hat h}^{-1}_{l}$  is to  
(besides generating the 
the displaced vertex and its attendant vertex structure)  deform  the graph underlying $c$ so as to generate  a  $C^0$ kink on  each  non-conducting edge of $c$ other than the 
$I_v$th  one and to generate  a single $C^0$ kink on the conducting line of $c$, this kink lying on the upper conducting edge of $c$ with the lower conducting edge having no kink.

Note that the lower conducting edge of $c$ between $p_1$ and $v$ does not intersect the deformed edges of $c_{l(\pm i,q^i_{I_v},I_v,\delta)}$. To see this proceed as follows. Note that 
the deformation  in Appendix \ref{acone} is constructed first out of straight lines and then the straight lines at the displaced vertex are `conically' deformed in a sufficiently small neighbourhood
of the  displaced vertex. Clearly this neighbourhood can always be chosen to be small enough that the lower conducting edge is in its complement. Hence if we show that if this edge does not 
intersect the initial construction of the deformation in terms of exclusively straight lines, it does not intersect their conical deformation. For the initial part of the construction in Appendix \ref{acone1} 
(a)-(c) below hold:\\
(a) Consider the deformation of the upper conducting edge in $c$ which connects a kink vertex on the upper conducting edge in $c$ to the displaced vertex 
in $c_{l(\pm i,q^i_{I_v},I_v,\delta)}$ which lies along the $I_v$th edge of $c$ at a position distinct from $v$. This deformed edge cannot intersect the lower conducting edge  because 2 distinct straight lines can intersect at most at a single point.\\
(b) Clearly the lower conducting edge of $c$ does not intersect the $I_v$th (upper conducting and lower conducting) edge in $c_{l(\pm i,q^i_{I_v},I_v,\delta)}$ except at $v$, 
once again because 2 distinct straight lines can intersect at most at a single point.\\
(c) Consider the $J_v$th non-conducting edge in $c$ with $J_v\neq I_v$. Its deformation connects a kink vertex on the $J_v$th edge to the displaced vertex.  
From Appendix \ref{acone1} this deformed edge lies in a plane containing the $I_v$th and the $J_v$th edges. The lower conducting edge can only intersect this plane at $v$ by virtue of the fact that $v$ is CGR in $c$.\\
From (a)-(c) it follows as claimed that the lower conducting edge between $p_1$ and $v$ does not intersect the deformed edges of $c_{l(\pm i,q^i_{I_v},I_v,\delta)}$.  It then follows that the 
multiplication by ${\hat h}^{-1}_{l}$  in equation (\ref{hamfinalc}) simply restores this part of the lower conducting edge without creating any more intersections.

\begin{figure}
  \begin{subfigure}[h]{0.3\textwidth}
    \includegraphics[width=\textwidth]{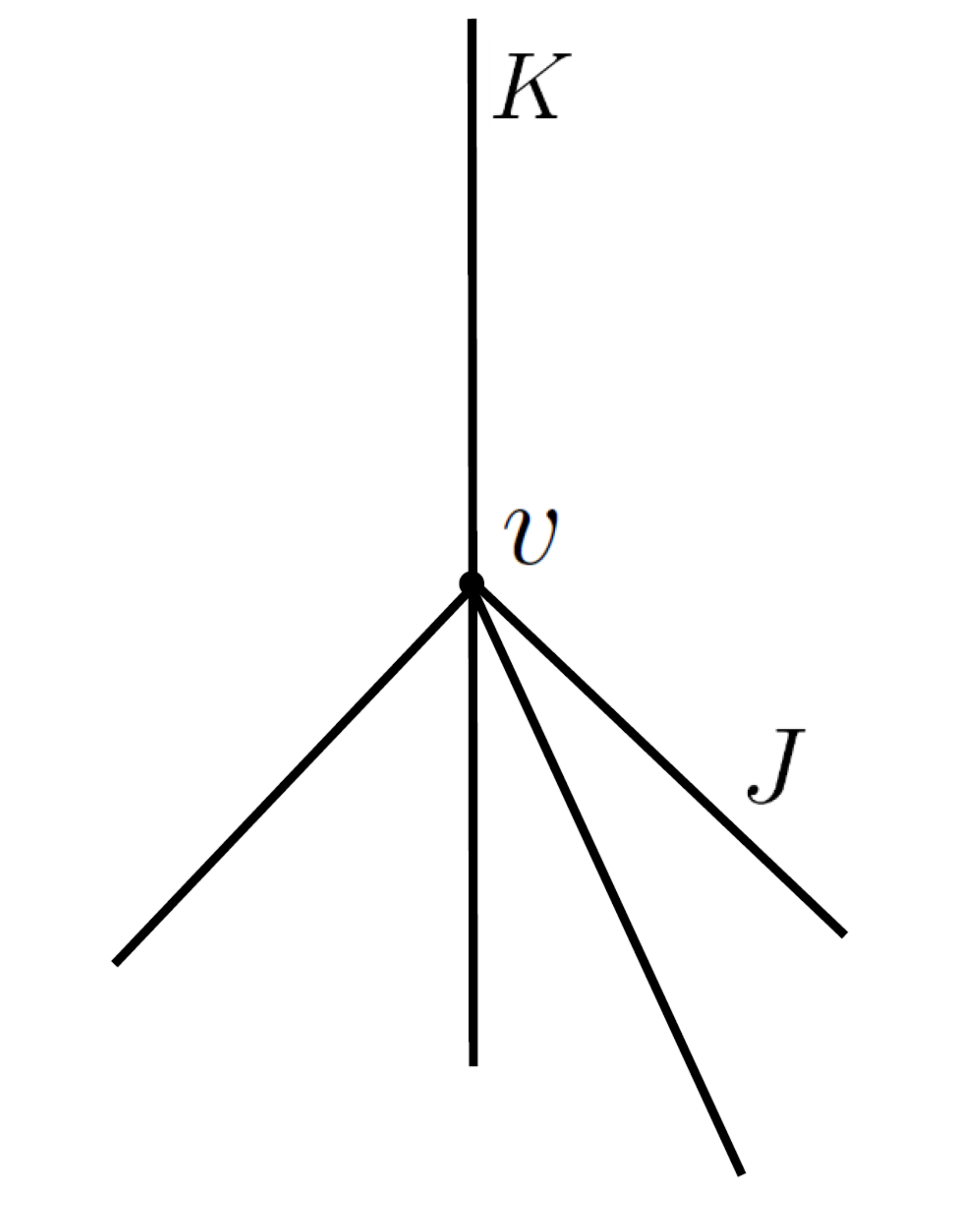}
    \caption{}
    \label{cgrk=ia}
  \end{subfigure}
  \begin{subfigure}[h]{0.3\textwidth}
    \includegraphics[width=\textwidth]{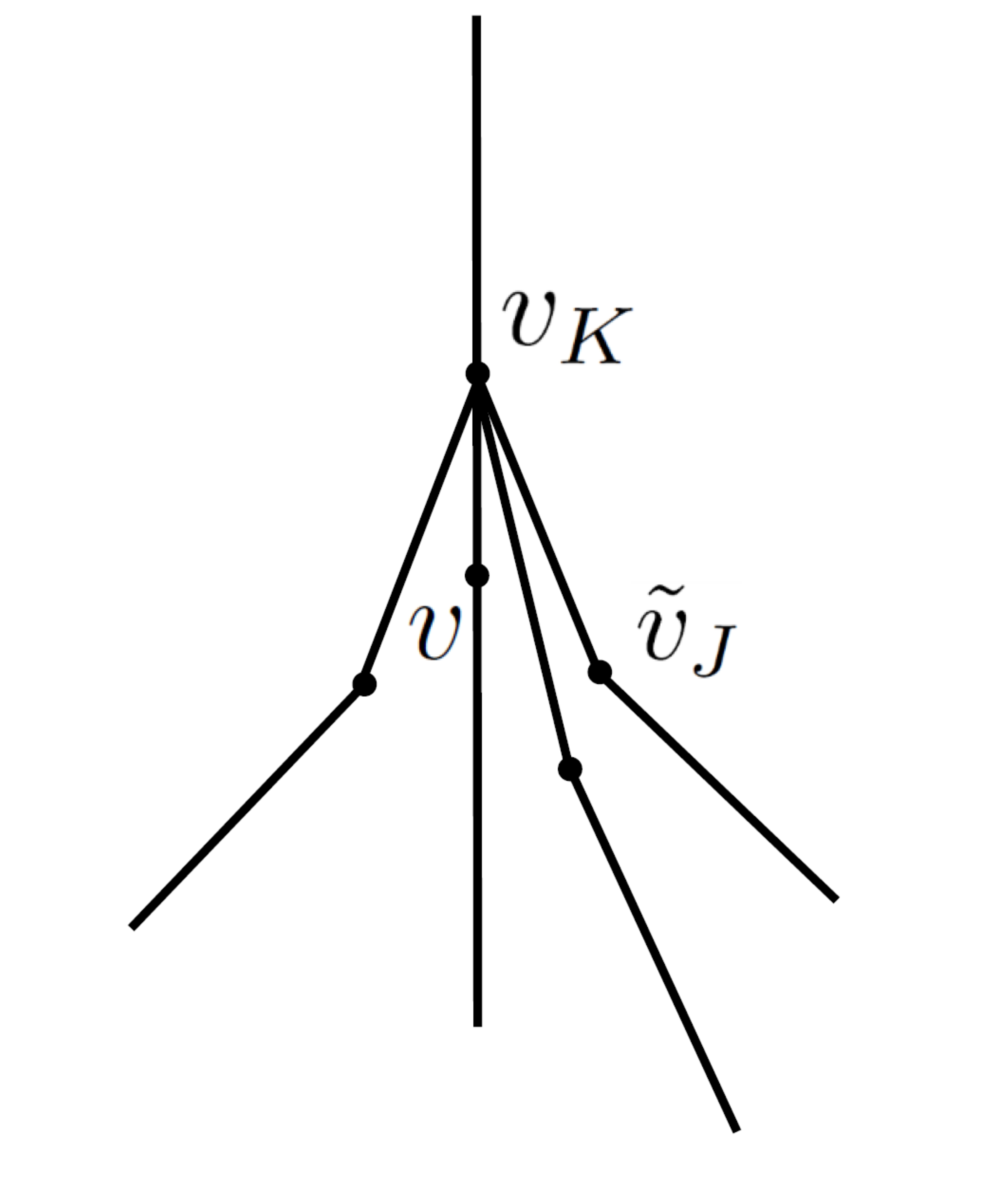}
    \caption{}
    \label{cgrk=ib}
  \end{subfigure}
\begin{subfigure}[h]{0.3\textwidth}
    \includegraphics[width=\textwidth]{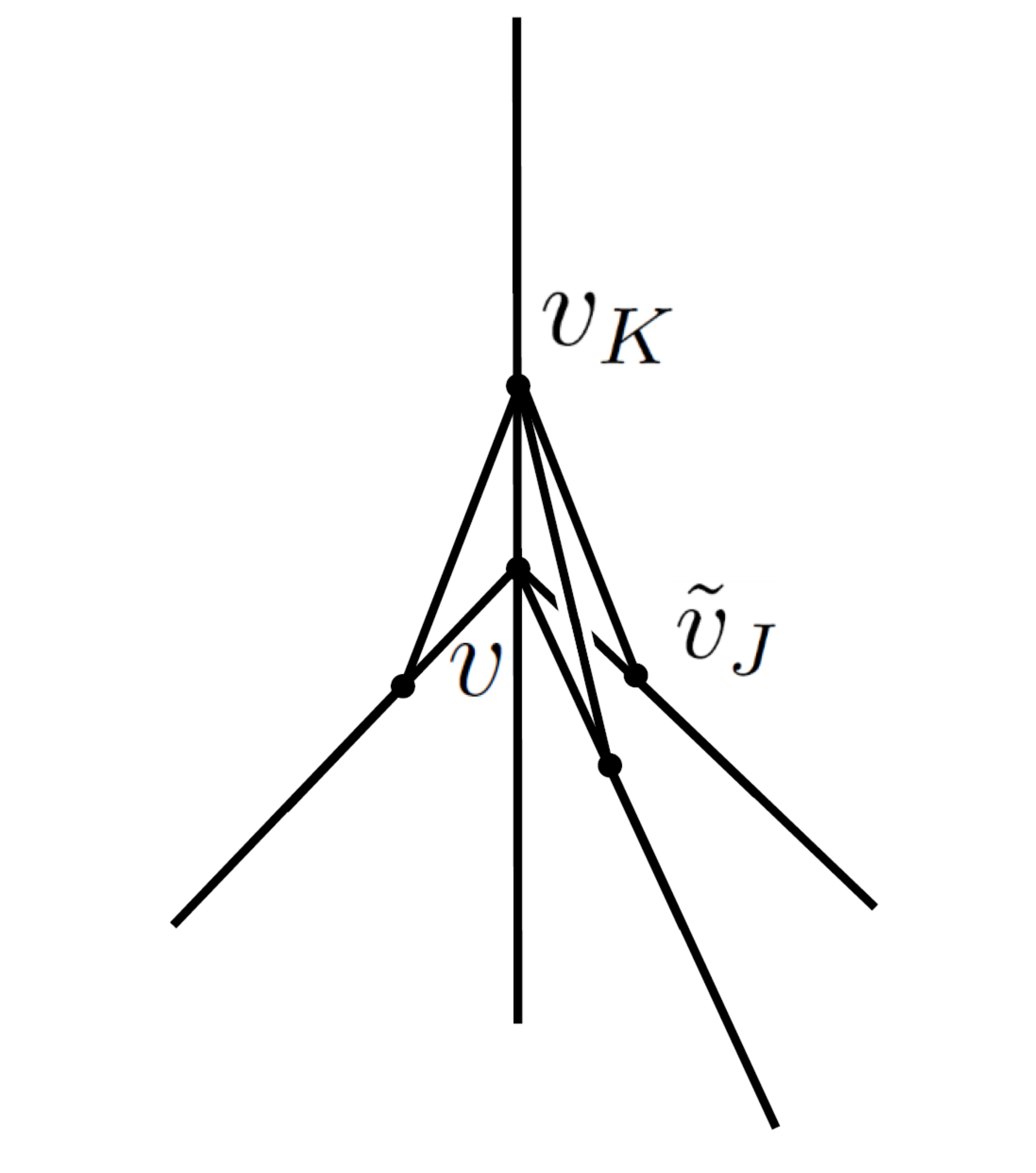}
    \caption{}
    \label{cgrk=ic}
  \end{subfigure}
  \caption{ In  Fig \ref{cgrk=ib}  the vertex  structure of Fig \ref{cgrk=ia} is deformed
along its $K$th edge  and the displaced
vertex $v_K$ and the $C^0$ kink ${\tilde v}_J$ on the $J$th edge are as labelled. Fig \ref{cgrk=ic} shows the result of a Hamiltonian type deformation  obtained by multiplying the 3 chargenet holonomies obtained by 
coloring  the edges of Fig \ref{cgrk=ib} by the  flipped images of the charges  on their counterparts in $c$ ,  the edges of Fig \ref{cgrk=ia} by the negative of these  flipped charges and the edges of Fig \ref{cgrk=ia} 
by the charges on $c$. If the edges of Fig \ref{cgrk=ib} are colored by the charges on 
their counterparts in $c$ then one obtains an electric diffemorphism deformation.
}%
\label{cgrk=i}%
\end{figure}

For the case that $I_v=K_v$,
the deformed chargenet $c_{ (\pm i,\;q^i_{K_v, out}- q^i_{K_v, in},\;K_v,\delta)  }$  is displayed in Figure \ref{cgrk=ic}. 
This chargenet can be thought of as the product of 
3 holonomies (see Figures \ref{cgrk=ia}, \ref{cgrk=ib}).
Once again it is easy to see that the deformed edges of $c_{l(\pm i,q^i_{K_v},K_v,\delta)}$ do not intersect the lower conducting edge in $c$ from the fact that 2 distinct lines can intersect at most at a point.
Hence once again the   multiplication by ${\hat h}^{-1}_{l}$ simply restores this part of the lower conducting edge without creating any more intersections.


Similarly, we have 
\ba
\hat{D}_{\delta}[\vec{N}_{i}]c  & =& {\hat h}_{l,{\vec q}_{K, in}}\frac{\hbar}{\mathrm{i}}\frac{3}{4\pi}%
N(x(v))\nu_{v}^{-2/3}   (\sum_{I_{v}}     \frac{1}{\delta
}(   c_{ l  (  q^i_{I_v},I_{v},\delta   )  }  -   c_l  )\nonumber\\
&=& \frac{\hbar}{\mathrm{i}}\frac{3}{4\pi}%
N(x(v))\nu_{v}^{-2/3}   \big( \sum_{I_{v}\neq K_v}  \frac{1}{\delta
}(   c_{  (  q^i_{I_v},I_{v},\delta   )  }  -   c  )\nonumber\\
&+& \frac{1}{\delta}(   c_{  (  q^i_{K_v,out}- q^i_{K_v,in} ,\; K_{v},\delta   )  }  -   c  )\big). 
\label{dnfinalc}
\ea
The charge net which is obtained through a  deformation of $c$  along an edge which is non-conducting in $c$ looks identical to that in Figure \ref{cgrknib} except that the charge labels are identical to their counterparts in $c$.
\footnote{Here and below, similar to Footnote \ref{fnposcgr}, our comments only apply to those deformed chargenets $c_{  (  q^i_{I_v},I_{v},\delta   )  }$  for which $q^i_{I_v}>0$. The deformed chargenets in (\ref{dnfinalc}))  for which this
condition does not apply will be defined in section \ref{secneg}.} Similarly, 
the charge net which is obtained through a  deformation of $c$  along an edge which is conducting in $c$ looks identical to that in Figure \ref{cgrk=ib} except that the charge labels are identical to their counterparts in $c$

\subsection{\label{secgr.1a}The net conducting charge: Remarks}

We define the the difference between the outgoing upper and incoming lower conducting charges at a CGR vertex to be the {\em net conducting charge} at that vertex.
The following remarks highlight the significance of this difference of conducting charges.

In the case of the action of the Hamiltonian constraint (\ref{hamfinalc}) we have that:\\
\noindent {\em Remark 1}: The deformed
$K_v$th edge in $c_{(\pm i,q^i_{I_v},I_v,\delta)}$ carries the {\em difference} between  the flipped   charges of the outgoing upper and incoming lower conducting edges in $c$.\\

\noindent {\em Remark 2}: The displaced vertex in the deformed chargenet $c_{ (\pm i,\;q^i_{K_v, out}- q^i_{K_v, in},\;K_v,\delta)  }$  is displaced by an amount 
$|q^i_{K_v, out}- q^i_{K_v, in}| \delta$ from $v$.\\

\noindent{\em Remark 3}: The difference between the charges on the outgoing upper and incoming lower conducting edges  at the non-degenerate vertex of $c_{ (\pm i,\;q^i_{K_v, out}- q^i_{K_v, in},\;K_v,\delta)  }$ 
is the $\pm i$-flipped image of the difference between the charges on the outgoing upper and  incoming lower conducting edges at the non-degenerate vertex of $c$.
\\

In the case of the Electric diffeomorphism constraint action (\ref{dnfinalc}),   we have that:\\

\noindent {\em Remark 4}: The deformed
$K_v$th edge in  $c_{  (  q^i_{I_v},I_{v}\neq K_v,\delta   )}$   carries the {\em difference} between  the    charges of the upper and lower conducting edges in $c$.\\
\\
\noindent {\em Remark 5}: The displaced vertex in the deformed chargenet  $c_{  (  q^i_{K_v,out}- q^i_{K_v,in} ,\; K_{v},\delta   )  }$  is displaced by an amount 
$(q^i_{K_v, out}- q^i_{K_v, in}) \delta$ from $v$.\\

\noindent{\em Remark 6}: The difference between the charges on the outgoing upper and incoming lower conducting edges  at the non-degenerate vertex of  $c_{  (  q^i_{K_v,out}- q^i_{K_v,in} ,\; K_{v},\delta   )  }$ 
is equal to the difference between the charges on the outgoing upper and incoming lower conducting edges at the non-degenerate vertex of $c$.
\\

\noindent {\em Remark 7}: Were it not for the intervention 
by the holonomy around the small loop $l$, this difference in   Remarks (2) and (5) would be replaced by the {\em sum} because the heuristics of sections \ref{sec2} and \ref{sec3.2} indicate a displacement 
of the vertex by $\delta (q^i_{K_v,out} {\vec{\hat e}}_{K_v,out} +  q^i_{K_v,in} {\vec{\hat e}}_{K_v,in})$ with the outgoing upper conducting edge tangent ${\vec{\hat e}}_{K_v,out}$ being equal to the ingoing
lower conducting edge tangent ${\vec{\hat e}}_{K_v,in}$. As will be apparent in sections \ref{sec8}, \ref{sec9} this `difference of charges associated with the conducting edge' plays a key role in anomaly freedom.  

As we have noted in section \ref{secgr.1}, 
we may obtain this intervention for the Hamiltonian constraint by starting from (\ref{hamconst1}) and putting in factors of the holonomy around $l$ and its inverse and then proceeding along the lines of 
the subsequent heuristics of section \ref{sec2.3}. Since `classically', the holonomy and its inverse cancel (and since, furthermore,  the classical holonomy is unity to higher order terms in $\delta$ than the leading order 
required by the putative approximant), the intervention leads to an equally acceptable discrete action. Similar heuristics hold for the electric diffeomorphism constraint.

\subsection{\label{secgr.1b} Nondegeneracy of CGR vertices}
From Figures \ref{cgrkni}, \ref{cgrk=i} and our discussion above it follows that the displaced vertices in the deformed chargenets generated by (\ref{hamfinalc}) and (\ref{dnfinalc}) are CGR or GR.
\footnote{\label{fncgrtogr}Note that in Figure \ref{cgrk=ic}, the displaced vertex is generically CGR; however it is possible for the charge values to conspire so that the  charge at the lower conducting edge at the 
displaced vertex vanishes in which case the displaced vertex would be GR.}
While the notion of nondegeneracy of a GR vertex is just the non-vanishing of the volume eigen value at the vertex, in the case of a CGR vertex,
the action of the constraints (\ref{hamfinalc}), (\ref{dnfinalc}) is sensitive to the non-degeneracy of the (GR)  vertex in  $c_l$ rather than than the (CGR) vertex in $c$. Accordingly, we define the notion of non-degeneracy of a CGR vertex as follows:

\noindent{\em Definition 1: Nondegeneracy of a CGR vertex:} A CGR vertex of a charge net $c$ will be said  to be non-degenerate iff the corresponding GR vertex in the charge net $c_l$ is non-degenerate. If the vertex in $c_l$ is degenerate
we shall say that the CGR vertex in $c$ is degenerate.
\footnote{This notion of (non)degeneracy requires the intervention by $h_l$, which in turn is fixed by the specification of which part of the conducting edge is upper and which is lower. A unique specification will be
given in section \ref{secneg}. Such a specification then makes the notion of (non)degeneracy of a CGR vertex a well defined one.}


With the definition of nondegeneracy above, the original `parent' CGR vertex $v$  is degenerate in the deformed chargenets 
generated by (\ref{hamfinalc}).  To see this, recall that the deformed chargenets
$c_{(\pm i,q^i_{I_v},I_v \neq K_v,\delta)}$,  $c_{  (  q^i_{K_v,out}- q^i_{K_v,in} ,\; K_{v},\delta   )  }$ in that equation are obtained 
from the action of $h_l^{-1}$ on $c_{l(\pm i,q^i_{I_v},I_v \neq K_v,\delta)}$,  $c_{l (  q^i_{K_v,out}- q^i_{K_v,in} ,\; K_{v},\delta   )  }$. The latter are obtained by the Hamiltonian constraint action on $c_l$
at its GR vertex 
and hence, as noted in section \ref{sec3.4},   the charges on the edges at the vertex $v$ in these deformed and `$i$- flipped chargenets have vanishing $i$th component.
In particular the edges in $c_{l(\pm i,q^i_{I_v},I_v \neq K_v,\delta)}$,  $c_{l (  q^i_{K_v,out}- q^i_{K_v,in} ,\; K_{v},\delta   )  }$ which connect $v$ to the $C^0$ kinks
have charges with vanishing $i$th component.
Since the action of $h_l^{-1}$ does not affect the charges on the edges at $v$ which connect $v$ to the $C^0$ kinks, this is also true for these edges in the chargenets 
$c_{(\pm i,q^i_{I_v},I_v \neq K_v,\delta)}$,  $c_{  (  q^i_{K_v,out}- q^i_{K_v,in} ,\; K_{v},\delta   )  }$. Gauge invariance implies that the {\em net} conducting charge at $v$ in these chargenets
also has vanishing $i$th component. Now, {\em independent} of which part of the conducting edge at $v$ we assign as upper/lower, it is straightforward to check that the appropriate intervention on 
$c_{(\pm i,q^i_{I_v},I_v \neq K_v,\delta)}$,  $c_{  (  q^i_{K_v,out}- q^i_{K_v,in} ,\; K_{v},\delta   )  }$
yields chargenets each of which has the left over upper conducting edge  at the (now GR) vertex $v$  colored with the net conducting charge at $v$.  
The other edges at $v$ retain their charges so that all the edge charges at $v$ now have vanishing $i$th componet
which implies that the volume eigen value after the intervention vanishes. Hence using the  definition of nondegeneracy above, we see that the CGR vertex $v$ in 
$c_{(\pm i,q^i_{I_v},I_v \neq K_v,\delta)}$ and in  $c_{  (  q^i_{K_v,out}- q^i_{K_v,in} ,\; K_{v},\delta   )  }$ is degenerate.

In the case of deformations generated by (\ref{dnfinalc}), the vertex $v$ is  bivalent in the deformed chargenets 
$c_{  (  q^i_{I_v},I_{v}\neq K_v,\delta   )}$ , $c_{  (  q^i_{K_v,out}- q^i_{K_v,in} ,\; K_{v},\delta   )  }$  and hence degenerate.



\subsection{\label{secgr.3} Convenient Notation}

Given a charge net $c$ with a single nondegenerate  linear GR or CGR vertex $v$ , its deformations by the discrete action of the Hamiltonian constraint in equations (\ref{hamfinal}), (\ref{hamfinalc}) can be specified through:
\footnote{While we have only explicitly defined deformed chargenets for deformations along edges of $c$ which have positive charges, it turns out that the specifications below also extend to the general case tackled in section \ref{secneg}.}
\noindent (a) the edge $e_{I_v}$ along which the deformation occurs and 
its associated charge label. If $v$ is GR this is just  $q^i_{I_v}$ and the specification is denoted by $(I_v, q^i_{I_v})$. If $v$ is CGR and the deformation is 
along the conducting line in $c$ the appropriate conducting line index $K_v$ must be specified together with the 
difference between the upper and lower conducting edge charges $q^i_{K_v,out}-q^i_{K_v, in}$. If $v$ is CGR but the deformation is along an edge $e_{I_v}$, $I_v\neq K_v$, the specification is, as for the 
GR case, $(I_v, q^i_{I_v})$.
\\
\noindent (b) the charge flip involved  which is specified by a sign $\pm$  and a $U(1)^3$ index $i$ (which is the same as that of the charge labels in (a)).
\\
\noindent (c) the coordinate patch around $v$ and the nature of the deformation it specifies including the size of the deformation parameter $\delta$ measured by it.

In section \ref{sec6} we will see that the coordinate patch is uniquely specified for every $c$  as is the nature of the deformation given the value of the deformation parameter $\delta$ and the information in (a),(b).
The information in (a), (b) is known given the charge net label $c$ (which includes all its edges and charges), 
the deformation edge/line index $I_v$, the $U(1)^3$ index $i$ and a parameter $\beta$ which takes values $+1$ or  $-1$ corresponding to a  $+i$ or $-i$ charge flip.
Hence, suppressing the (unique) specification of the coordinate patch associated with $c$, we denote the deformed chargenets 
$c_{(\pm i,q^i_{I_v},I_v,\delta)}$  in (\ref{hamfinal}), (\ref{hamfinalc}) and 
$c_{ (\pm i,\;q^i_{K_v, out}- q^i_{K_v, in},\;K_v,\delta)  }$ in (\ref{hamfinalc}) by the symbol $c_{(i,I,\beta, \delta)}$ where we have suppressed the `$v$' subscript as we shall need this notation only for 
states with a single nondegenerate (linear GR or CGR) vertex.

Similarly we denote the chargenets
$c_{  (  q^i_{I_v},I_{v},\delta   )  }$ in (\ref{dnfinal}), (\ref{dnfinalc}) and 
$ c_{  (  q^i_{K_v,out}- q^i_{K_v,in} ,\; K_{v},\delta   )  } $ in (\ref{dnfinalc}) 
by the symbol $c_{(i,I,0, \delta)}$ where `$0$' signifies that the deformation is of the electric deformation type.
By allowing $\beta$ to range over $0$ in addition to $\pm1$, we refer to the deformed chargenets in (\ref{hamfinal}), (\ref{hamfinalc}),
(\ref{dnfinal}), (\ref{dnfinalc}) by the single symbol $c_{(i,I,\beta, \delta)}$ and  say that $c_{(i,I,\beta, \delta)}$ is the $(i,I,\beta, \delta)$- deformed child of the parent $c$.
In terms of this notation, equations (\ref{hamfinal}), (\ref{hamfinalc}) take the form:
\be
\hat{C}[N]_{\delta}c(A) = \beta\frac{\hbar}{2\mathrm{i}}\frac{3}{4\pi}N(x(v))\nu_{v}^{-2/3}\sum_{I}\sum_{i}  
\frac{c_{(i,I,\beta, \delta)}- c}{\delta} ,
\label{ham}
\ee
with $\beta=+1$ or $\beta =-1$,
and equations (\ref{dnfinal}), (\ref{dnfinalc}) take the form:
\be
\hat{D}_{\delta}[\vec{N}_{i}]c   =\frac{\hbar}{\mathrm{i}}\frac{3}{4\pi}%
N(x(v))\nu_{v}^{-2/3}\sum_{I}   \frac{1}{\delta
}(   c_{   (  i, I, \beta=0, \delta   )  }  -   c   ).
\label{dn}
\ee

\section{\label{secneg}Linear GR and CGR vertices: the general case}

In sections \ref{sec3} and \ref{secgr} the explicit `downward conical' deformations considered  were applicable only for those   outgoing edges at the vertex of interest 
which had charges with certain positivity properties.
The positivity property for GR vertices 
was that the outgoing charge had to be positive and  for CGR vertices
that the outgoing charge for a  non-conducting edge had to be positive  and that the outgoing net conducting charge had to be  positive. The associated `downward' conicality of the deformation was defined with respect an assignation of 
`upward direction', this direction coinciding with the outgoing edge direction for GR vertices
\footnote{This choice of upward direction made in section \ref{sec3},
even with the positivity restrictions therein,   coincides with the choice outlined in this section only for special cases of GR vertices, an example being those which are `primordial' in the language of section \ref{sec4}.
We had pointed out this further restriction of the considerations of section \ref{sec3}  to such vertices in Footnote \ref{fn6}.}
and being arbitrarily prescribed for the CGR case.
Here we shall lift the positivity
restrictions on charges and also remove the arbitrariness in the definition of upward and downward directions in the CGR case. 
In what follows  we shall, as in sections \ref{sec3} and \ref{secgr},  appeal to the constructions of Appendix \ref{acone1}. However, in addition, we shall 
also find it necessary to embellish these constructions with an appropriate placement of kinks through the constructions of Appendix \ref{acone2}. 


We proceed as follows. First in section \ref{secneg1} we formalise the definitions of upward and downward conical deformations for GR and CGR vertices.
As we shall see, these deformations will be defined to be downward or upward conical with respect to an edge orientation determined by the kink structure in the vicinity of the 
vertex rather than with respect to the outward pointing edge tangent.
Next, in sections \ref{secneg2} and \ref{secneg3}  we tie
the choice of downward or upward conical deformation  for GR and CGR vertices to the sign of the charge labels on the edges at the vertex, with the definition of upward and downward fixed by the kink structure
in the vicinity of the vertex as in section \ref{secneg1}.
The intricacy of these choices plays a key role  in the emergence of anomaly free commutators in the continuum limit. Had we not been guided by the anomaly free requirement, it would have 
been difficult to home in on these choices.
In sections \ref{secneg2} and \ref{secneg3} we also show how  each of these choices 
is implemented through a  corresponding choice of discrete approximants to the action
of the Hamiltonian and electric  diffeomorphism constraints. We summarise our results in section \ref{secneg4}. 
In what follows we use  the notion of a $C^{m}$   kink  $m=0,1,2$  as defined in Appendix \ref{adefkink}.

\subsection{\label{secneg1}Upward and downward conically  deformed states}

\subsubsection{\label{secneg1.1}Linear GR vertex}


Let $v$ be a linear GR  vertex of the charge net $c$. Let the coordinates around $v$ with respect to which $v$ is linear be $\{x\}$.  
In this section we shall construct upward and downward conically deformed states
obtained by subjecting the graph underlying $c$ to upward and downward
conical deformations. These deformed states are the analogs of  the deformed chargenets depicted in Figure \ref{gr}.

A conical deformation of $c$ along the edge $e_I$ at the vertex $v$ of $c$ is one in which the deformed state  $c_I$ has 
a vertex $v_I$ displaced with respect to $v$ along the  straight line determined by $e_I$, deformations of the edges $e_{J\neq I}$ which connect 
the edges $e_J$ in $c$ to $v_I$, these deformations being straight lines in the vicinity of $v_I$ which form a regular cone around the line joing $v$ to $v_I$.
To characterise the conical deformation as downward or upward it is necessary to specify which direction is up.
Accordingly,  let  ${\vec V}_I$ be a tangent vector at $v$ which points either parallel to the outward pointing edge tangent to the edge $e_I$ or antiparallel
to the outward pointing  edge tangent to the edge $e_I$. Given a choice of ${\vec V}_I$, the direction along ${\vec V}_I$  is defined to be upward and 
the direction opposite  to that of ${\vec V}_I$ is defined to be downward.
A conical deformation of $c$ at $v$ will be called {\em downward} with respect to ${\vec V}_I$ if:\\
\noindent (a) the deformed edges (other than the $I$th one) form a  
downward cone around the upward direction defined by ${\vec V}_I$ so that the angle between this upward axis and any such edge as measured by $\{x\}$ is 
greater than $\frac{\pi}{2}$, and\\
\noindent (b) there is a specific kink structure in the vicinity of the displaced vertex in  the deformed state which is consistent with the choice of ${\vec V}_I$ in a sense
which we shall describe as we go along.

In particular,  if ${\vec V}_I$ is specified as being parallel to the outward pointing edge tangent ${\vec {\hat e}}_I$ at $v$ in $c$ then the deformations described in 
section \ref{sec3} are {\em downward} pointing 
because the cone is downward pointing.
In addition we use the construction of Appendix \ref{acone2} to place kinks around the displaced vertex $v_I$ as follows.
Using the terminology of section \ref{secgr.1},  the displaced vertex $v_I$ lies on the conducting line passing through $v$. We place a $C^2$ kink at a point $v_{I,2}$ on this conducting line `beyond' $v_I$ so that the part of the 
conducting line {\em from}  $v_I$ {\em to} $v_{I,2}$ is oriented {\em parallel} to ${\vec V}_I$. We also place a $C^1$ kink at a point $v_{I,1}$ on the part of the conducting line between $v$ and $v_I$ so that the part of the 
conducting line {\em from} $v_I$ {\em to} $v_{I,1}$ is oriented {\em anti- parallel} to ${\vec V}_I$. It follows that the upward direction ${\vec V}_I$ can be {\em inferred} from the position of these kinks from the orientation
of the straight lines (with respect to $\{x\}$) from the displaced vertex $v_I$ to these kinks. This is what we mean by the consistency of the kink placement with the specification of the choice of ${\vec V}_I$ in (b).


Similarly an {\em upward} conical deformation of $c$ at $v$ with respect to ${\vec V}_I$ is a conical deformation in which the 
deformed edges (other than the $I$th one) point upwards so that the angle between any such edge and ${\vec V}_I$ is acute and such that there is an appropriately defined
kink structure which is consistent with the choice of ${\vec V}_I$. As an example of an upward conical deformation,
%
%
consider the case where, once again,  ${\vec V}_I$ is specified as being parallel to the outgoing edge tangent ${\vec {\hat e}}_I$ at $v$ in $c$. We define the {\em upward}
conical deformation of $c$ along $e_I$ at $v$ as follows. 
First we describe the deformation of the graph underlying $c$ so as to obtain the analog of Fig \ref{grb}.
Recall that $v$ is linear with respect to $\{x\}$. Extend the (straight line) edge $e_I$ linearly past $v$ in the ingoing direction
opposite to ${\vec V}_I$. Let the extension, $e^{(-,\tau)}_I$ be of coordinate length $\tau$ with $\tau$ small enough that $e^{(-,\tau)}_I$ does not intersect any part of $c$ other than $v$.
\footnote{That such a small enough extension exists, follows from the linear GR nature of the vertex; the linear GR property  implies that the edges $e_{J\neq I}$ of $c$ in the vicinity of their vertex $v$ 
are straight lines, none of which are parallel to $e_I$.}  Let us consider the  altered vertex structure at $v$ when we include this extension as an edge at $v$. Clearly, 
the addition of this edge to the existing set of edges at $v$ converts $v$ into a linear CGR vertex.
The deformation of this CGR vertex structure  is similar to that  for CGR vertices  in section \ref{secgr} with $e^{(-,\tau)}_I$ playing the role of the upper conducting edge, and is as follows.
We (a) displace the vertex $v$ by an amount $\epsilon = \frac{\tau}{2}$  along $e^{(-,\tau)}_I$ to the point $v_I$, (b) connect $v_I$ to the edges $e_{J\neq I}$ at the 
$C^0$ kinks ${\tilde v}_J$  by straight lines as described in Appendix \ref{acone1} and section \ref{secgr.1}, (c) deform the resulting vertex structure in a small enough vicinity of $v_I$ 
along the lines of Appendix \ref{acone1} so as to obtain a regular conical structure in this vicinity. 
The deformed graph is then obtained by removing the parts of the edges of the original graph between $v$ and the $C^0$ kinks $\{{\tilde v}_J\}$ as well as the part of the extension $e^{(-,\tau)}_I$
beyond $v_I$ so that $v_I$ is now a GR vertex.  We emphasize here that the deformation detailed through (a) to (c) does not require any holonomy intervention of the sort provided by $h_l$ and
its inverse in section \ref{secgr}. That (a)-(c) can be implemented without the creation of any further unwanted intersections follows from an argumentation similar to that in section \ref{secgr.1}
using the properties of straight lines and the small compactly supported nature of the transformations of the type detailed in Appendix \ref{acone1} which render the conical structure regular.

\begin{figure}
  \begin{subfigure}[h]{0.3\textwidth}
    \includegraphics[width=\textwidth]{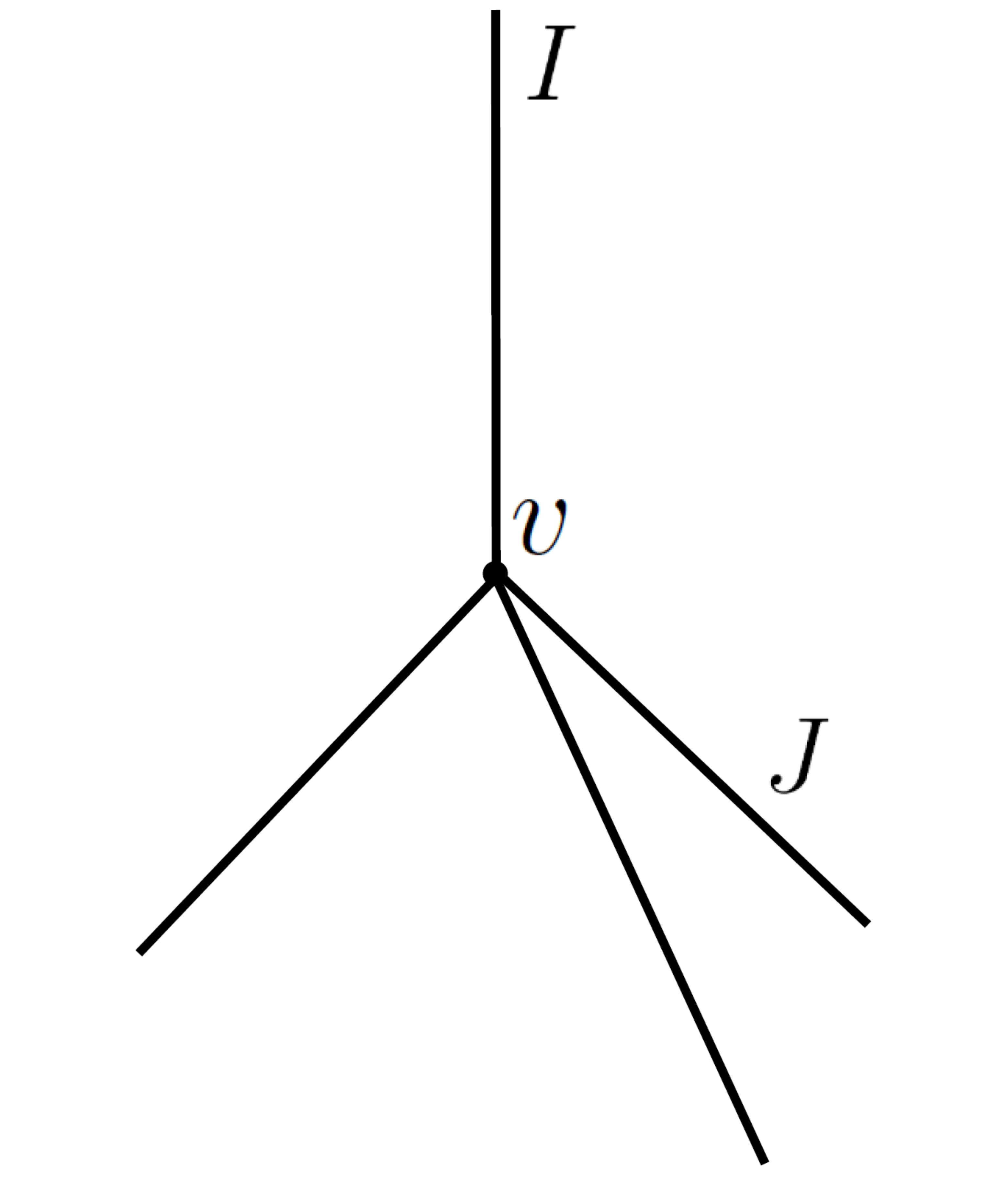}
    \caption{}
 \label{grexta}
  \end{subfigure}
  \begin{subfigure}[h]{0.3\textwidth}
    \includegraphics[width=\textwidth]{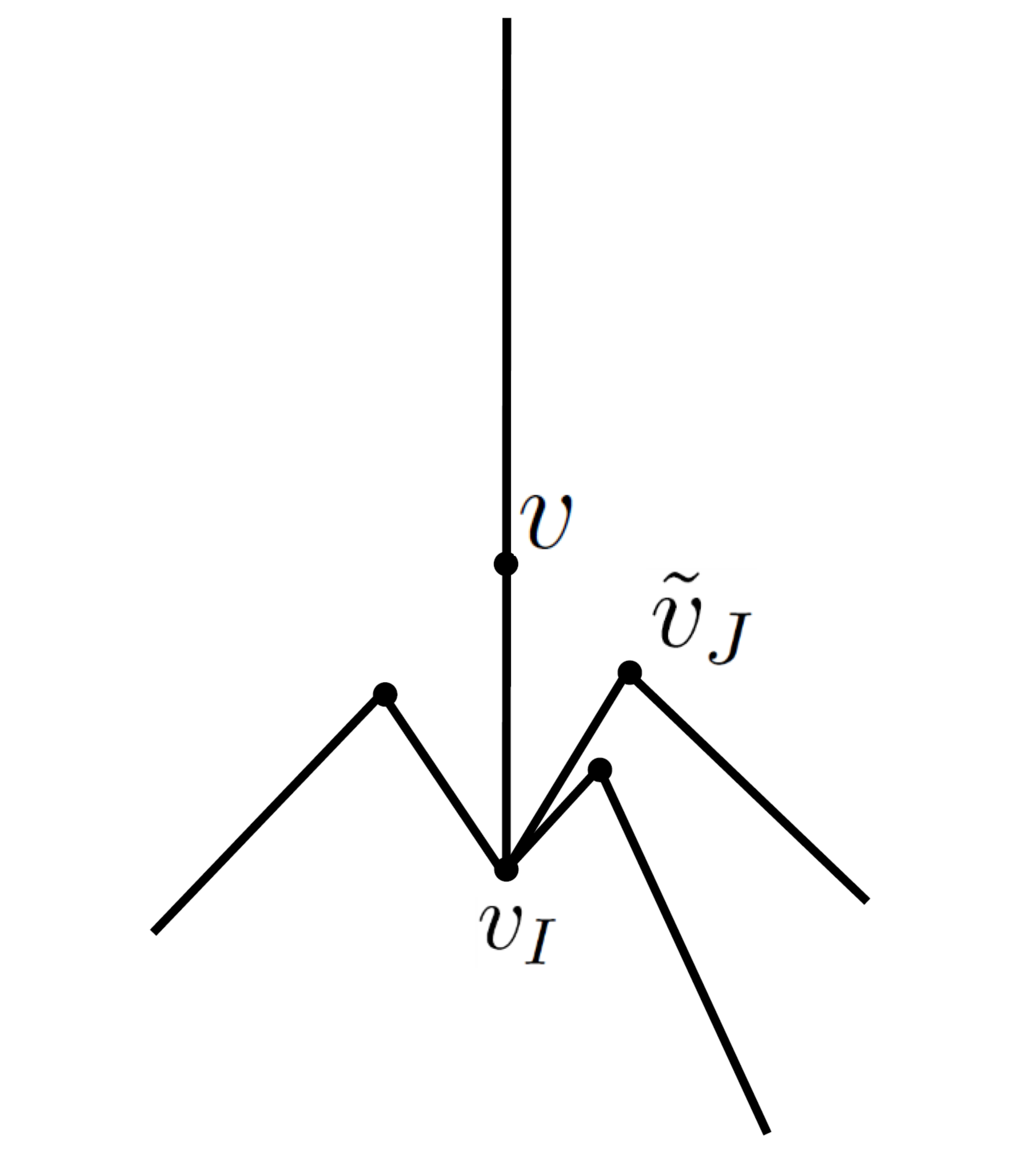}
    \caption{}
   \label{grextb}
  \end{subfigure}
\begin{subfigure}[h]{0.3\textwidth}
    \includegraphics[width=\textwidth]{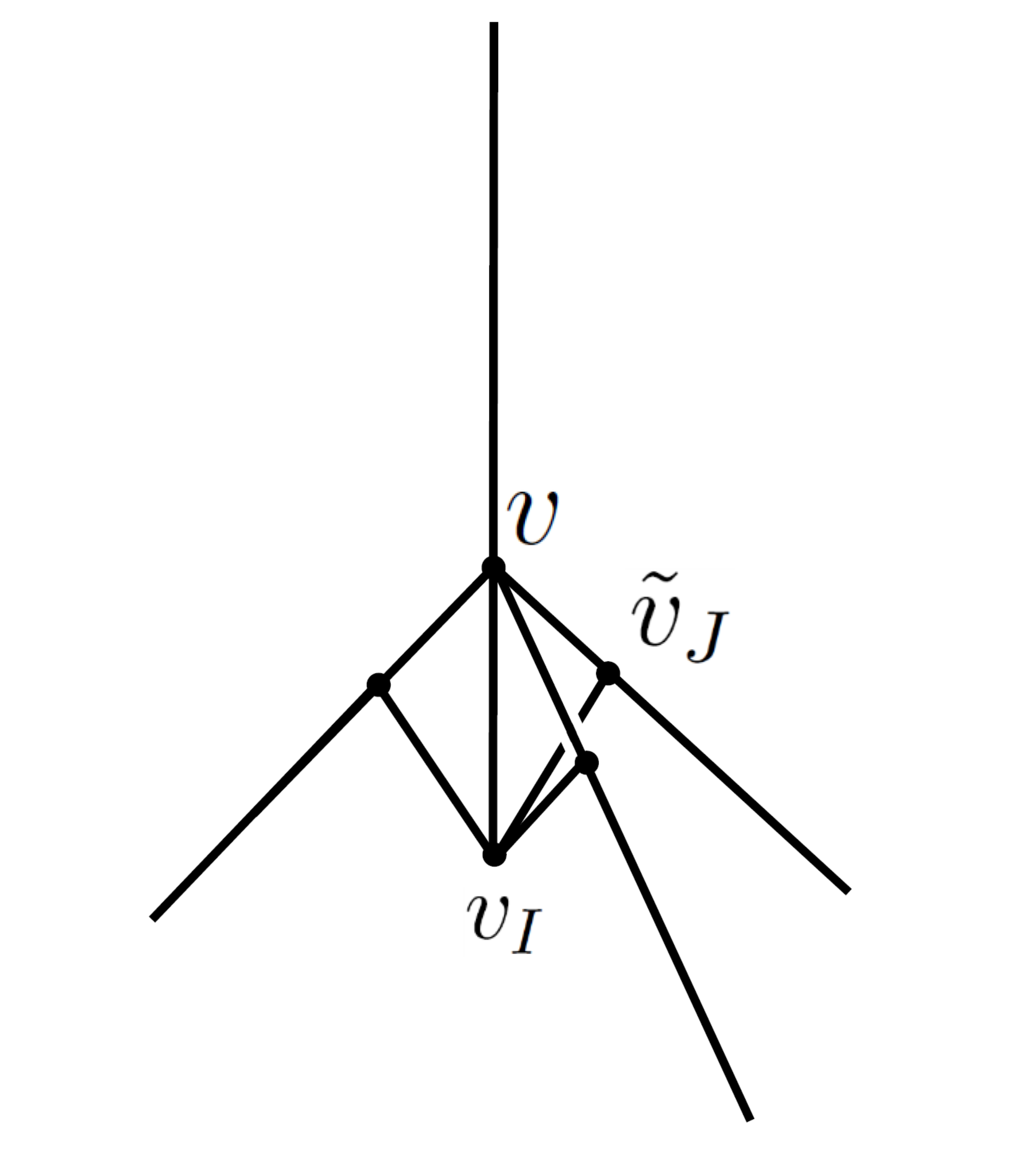}
    \caption{}
   \label{grextc}
  \end{subfigure}
  \caption{ Fig \ref{grexta} shows an undefromed GR vertex $v$ of a chargenet $c$  with its $I$th and $J$th edges as labelled. The $I$th edge is extended beyond $v$ and the vertex is displaced along  this extended edge in Fig \ref{grextb} wherein the displaced
vertex $v_I$ and the $C^0$ kink, ${\tilde v}_J$ on the $J$th edge are labelled. Fig \ref{grextc} shows the result of a Hamiltonian type deformation $(i, I,\beta, \delta)$  obtained by multiplying the chargenet holonomies obtained by 
coloring  the edges of Fig \ref{grextb}  by $(\beta, i)$ flipped images of charges  on their counterparts in $c$ ,  Fig \ref{grexta}  by negative of these $(\beta, i)$ flipped charges and Fig \ref{grexta} by the charges on $c$. If the edges of Fig \ref{grextb} are colored by the charges on 
their counterparts in $c$ then one obtains an electric diffemorphism deformation.
}%
\label{grext}%
\end{figure}

Next,  if the deformation is  of the `Hamiltonian constraint' type, the deformed graph is colored with appropriate $(\beta, i)$- flipped charges and 
the displacement $\epsilon$ of the displaced vertex $v_I$ is chosen to be $|q_I^i|\delta$
where $\delta$ is  the discretization parameter associated with the Hamiltonian constraint action  and  $q_I^i$ is the charge of the outgoing edge $e_I$ in $c$ at $v$. 
The  holonomy corresponding to this deformed charge net is 
multiplied
by 
the inverse charge net holonomy with $(\beta, i)$ flipped charges on the graph underlying $c$ together with the holonomy corresponding to $c$. The product of these
three yield a deformed chargenet generated by the Hamiltonian constraint. We show this in Figure \ref{grext}.

If the deformed charge net is  generated by the electric diffeomorphism constraint at discretization  parameter value $\delta$, its edges  bear
the same charges as their  counterparts in $c$ and we have, once again, that 
$\epsilon = |q_I^i|\delta$. The graph underlying the deformed chargenet is the one shown in Figure \ref{grextb}.

Finally, we apply a construction of the type detailed in Appendix \ref{acone2} so as to introduce a $C^2$ kink at a point $v_{I,2}$ between $v_I$ and $v$
on the remaining part of $e^{(-,\tau)}_I$.
From the arguments of section \ref{secgr.1}  and Appendix \ref{acone}, it follows that the deformed structure  does not intersect $c$ except at the points $\{v, {\tilde v}_J, J\neq I\}$
and that the deformed edges form an {\em upward}  cone with respect to the specified upward direction ${\vec V}_I$. Further, the kink structure in the vicinity of $v_I$ is, once again,
such that the oriented line from $v_I$ to the  $C^2$ kink $v_{I,2}$ is in the direction of ${\vec V}_I$
Note that in this case there is no lower conducting edge `beyond' $v_I$ and hence no $C^1$ kink placement.

Next consider the case where ${\vec V}_I$ is antiparallel to the outgoing edge tangent ${\vec {\hat e}}_I$ at $v$ in $c$. The {\em downward} conical deformation 
of $c$ along $e_I$ at $v$  with respect to this choice of ${\vec V}_I$ is exactly the same as the {\em upward} conical deformation with the opposite choice of direction of  ${\vec V}_I$
which we sketched immediately above, except  that the $C^2$ kink is replaced by a $C^1$ kink 
 so that, once again, this placement is consistent with ${\vec V}_I$ in the sense that the oriented line from $v_I$ to the  $C^1$ kink $v_{I,1}$ is in the direction opposite to that of ${\vec V}_I$.

Finally consider the case where ${\vec V}_I$ is antiparallel to the outgoing edge tangent ${\vec {\hat e}}_I$ at $v$ in $c$ and   conical deformation is {\em upward}
of $c$ along $e_I$ at $v$  with respect to this choice of ${\vec V}_I$. This  is exactly the same as the {\em downward} conical deformation with the opposite choice of direction of  ${\vec V}_I$
 which we discussed as our first example (and which we have encountered in section \ref{sec3}),
 except for the placement of the kinks. In this case, relative to our first example,  the location of the $C^2, C^1$ kinks are interchanged 
 so that once again, this placement is consistent with ${\vec V}_I$. Thus the oriented line from $v_I$ to the  $C^2$ kink $v_{I,2}$ is in the direction of ${\vec V}_I$ where as that from $v_I$ to the  $C^1$ kink $v_{I,1}$ is 
 in the direction opposite to ${\vec V}_I$ and  $v_{I,2}$ is placed between $v$ and $v_I$    whereas $v_{I,1}$ is placed on the other side of  $v_I$ on $e_I$.



\subsubsection{\label{secneg1.2}Linear CGR vertex}

We extend the considerations of 
section \ref{secneg1.1} to the case where $v$ is a linear CGR vertex of $c$ with linear coordinate patch $\{x\}$ and conducting line $e_K$.
Let the prescribed upward direction for the deformation along any nonconducting  edge $e_I$ be ${\vec V}_I$ and let 
the prescribed upward direction  for the deformation along the conducting line be ${\vec V}_K$.  

Recall from section \ref{secgr} that the deformations  of the CGR vertex constructed there involved the conversion of this vertex
to a GR one through the intervention of the holonomy $h_l$. The loop $l$ 
has a part which runs along the conducting line at $v$ in the direction of its upper conducting edge .
Here we use exactly the same intervention 
with this straight line part of $l$  oriented along the direction  ${\vec V}_K$ i.e we use  ${\vec V}_K$ to identify the upper and lower conducting edges.

Accordingly, let the net conducting charge at $v$ (namely the {\em sum} of the  {\em outgoing} charges along the two edges at $v$ which comprise the conducting line through $v$) be $q_{ K, net}^i$:
\be
q^i_{K,net}:= q^{i}_{K,1} + q^{i}_{K,2}
\label{defqnet}
\ee
where both $e_{K,1}, e_{K,2}$ are taken to be outward pointing at $v$ in $c$ so that $q^{i}_{K,1},  q^{i}_{K,2}$ are the outward edge charges.
\footnote{Note that this is exactly the same as the {\em difference} between the outgoing and {\em incoming} charges which we used in section \ref{secgr}.}
Without loss of generality, let us designate the outward pointing edge $e_{K,2}$ to be parallel to ${\vec V}_K$.
Let the intervening holonomy $h_l$  run around the loop $l$ with $l$  constructed as in section \ref{secgr}. Let the orientation of $l$  be such that the straight line part of 
$l$ runs upward (i.e. in the direction parallel to ${\vec V}_K$). Let $l$ with this orientation be charged with $q_{K,1}^i$.
Multiplication by $h_l$ converts the CGR vertex into a GR vertex and the resulting chargenet is called, as in section \ref{secgr}, $c_l$. Note that the $K$th edge of $c_l$ has charge $q^i_{K,net}$.
\footnote{As in Definition 1, section \ref{secgr.1b}, the notion of degeneracy of the CGR vertex in $c$ relevant to the action of the constraints is that of the corresponding GR vertex in $c_l$.}

Since the non-conducting edges are unaffected by this intervention, we assign the $I$th edge of $c_l$ ($I\neq K$) the same upward direction ${\vec V}_I$ as for the same edge in  $c$. Similarly for the $K$th edge of $c_l$
we assign the same upward direction ${\vec V}_K$  as for the $K$th (i.e. conducting) line of $c$ so that ${\vec V}_K$  is parallel to the outgoing $K$th edge of $c_l$ at $v$.
 Thus the assignments $\{{\vec V}_I\}$ for the edges at $v$ in $c$, induce (the same) assignments for the corresponding 
edges in $c_l$. The upward and downward conical deformations of this GR  vertex  along the $I$th edge  of $c_l$ with respect to ${\vec V}_I$ are then constructed as in section \ref{secneg1.1} except for the placement of the kinks.
Note that the deformations are small enough that they are restricted to a coordinate ball whose diameter is  smaller than the length of the straight line part of $l$ and is also small enough that the 
ball doesnt intersect the curved part of $l$. To see this recall that:\\
\noindent (a)for downward deformations, replacing $\delta$ in Appendix \ref{acone1} by $|q^i_I|\delta $ for $I\neq K$ and by $|q^i_{K,net}|\delta$ for $I=K$,   the deformation is confined to within  ball of size 
$2|q^i_I|\delta$ around $v$ for $I\neq K$ and within a ball of size  $2|q^i_{K,net}|\delta$ for $I=K$. \\
\noindent (b) for upward deformations also (a) is true; this follows from the construction of such deformations as detailed in section \ref{secneg1.1}. Further the length of the extension $e^{-,\tau}_I$ of the graph underlying the 
single GR vertex state $c_l$ (see section \ref{secneg1.1})
is chosen to be twice  that of the displacement of the vertex $v$ to its displaced position so that $\tau = 2|q^i_I|\delta, I\neq K $ and $2|q^i_{K,net}|\delta$ for $I=K$.\\
\noindent (c) the length $|l_1|$ is chosen to be larger than  $16q_{max}$  (see (\ref{defqmax}) so that $|l_1| > \tau$.

In the case of Hamiltonian deformations,  the colorings of the deformed graph and its multiplication by the two graph holonomies based on the undeformed graph underlying $c_l$ 
are as in section \ref{secneg1.1}. For the electric diffeomorphism case as well we follow section \ref{secneg1.1} applied to $c_l$ instead of $c$.

Subsequent to this, as in section \ref{secgr} we multiply the result by the inverse holonomy $h^{-1}_l$ which removes the curved part of $l$ from the deformed chargenets 
$c_{l(i,I,\beta,\delta)}$.
Finally we use constructions similar to that in Step 2 of Appendix \ref{acone}
to place a $C^1$ kink  and $C^2$ kink around the displaced vertex so that this placement is consistent with ${\vec V}_I$ in the sense described in section \ref{secneg1.1}.
Thus the straight line from the displaced vertex $v_I$ to the $C^2$ kink is parallel to ${\vec V}_I$ and that from $v_I$ to the $C^1$ kink is opposite to 
${\vec V}_I$.

This completes our discussion of the linear CGR vertex case.

\subsection{\label{secneg2}Choices of Deformation: Linear GR vertex}

\subsubsection{\label{secneg2.1}Choice of Conical Deformation Type}


Let $v$ be a nondegenerate linear GR vertex of $c$. We are interested in making a choice of upward or downward deformation at $v$ when the deformation is specified as $(i,I, \beta, \delta)$
where similar to section \ref{secgr.3},  $\beta \neq 0$ specifies a deformation with $(\beta,i)$ flipped charges along the edge $e_I$ with parameter  $\delta$ and where  $\beta=0$ specifies a deformation 
with unflipped charges   along $e_I$ with parameter $\delta$.

Let the outgoing tangent at $v$ along $e_I$ be ${\vec {\hat e}}_I$.
We define the {\em nearest} vertex   on $e_I$ to be the first $C^0, C^1$ or $C^2$ vertex  which is encountered on $e_I$ as $e_I$ is traversed in the outward direction from $v$ in $c$.
From our considerations in sections 3, \ref{secgr} and \ref{secneg1}, in the  $C^1, C^2$ cases the vertex is bivalent and  in the $C^0$ case the vertex can be bi or trivalent.

In all cases of interest, if the outgoing charge $q^i_I>0$ the deformation is chosen to {\em downward} conical and if $q^i_I<0$ the deformation is chosen to be {\em upward} conical. In both case the displaced vertex is at a distance 
$|q^i_I|\delta $ from $v$.
It turns out that for future purposes, only the following cases are of interest:

\noindent (1) The nearest vertex  is $C^0$:  Then ${\vec V}_I$ is chosen parallel to ${\vec {\hat e}}_I$.
\\
\noindent (2) The nearest vertex is $C^1$:  ${\vec V}_I$  is chosen antiparallel to ${\vec {\hat e}}_I$.
\\
\noindent (3) The nearest vertex  is $C^2$:  ${\vec V}_I$  is chosen parallel to ${\vec {\hat e}}_I$.
\\
\noindent (4) There is no nearest vertex:  ${\vec V}_I$  is chosen parallel to ${\vec {\hat e}}_I$.

\subsubsection{\label{secneg2.2}Choice of Discrete Approximant to Constraint}
In this section we describe the choice of discrete approximants to the constraints for which the ensuing discrete  action implements (1)- (4) of section \ref{secneg2.1}.

In cases (1), (3), (4) of section \ref{secneg2.1}  the heuristics of section 2 and 3 can be repeated to conclude that these deformations are generated by the diffeomorphsim
$\varphi(q^i_I\vec{{\hat{e}}}_{I_v},\delta)$ of section \ref{sec3.2.1} because ${\vec V}_I$  is in the direction of ${\vec {\hat e}}_I$ and, from the initial part of section \ref{secneg2.1},  the positive or negative character of 
$q^i_I$ then dictates whether the displaced vertex is displaced in the direction of ${\vec {\hat e}}_I$ or opposite to it. If the displacement is in the direction 
of ${\vec {\hat e}}_I$  then the deformation corresponding to equation (\ref{defd1cdefq}) is downward conical and if the displacement is in the direction opposite to ${\vec {\hat e}}_I$ the deformation is defined to be  upward conical.

In  all these three cases, in accordance with the heuristics of sections \ref{sec2} and \ref{sec3},  if the deformation is generated by the Hamiltonian constraint, the deformed graph is colored with appropriate $(\beta, i)$- flipped charges 
and the displacement $\epsilon$ of the displaced vertex $v_I$ in section \ref{secneg1.1} is chosen to be $|q_I^i|\delta$,
where $\delta$ is  the discretization parameter associated with the Hamiltonian constraint action.
The  holonomy corresponding to this deformed charge net is 
multiplied, as in Figures \ref{grc} and \ref{grextc} by the inverse charge net holonomy with $(\beta, i)$ flipped charges on the graph underlying $c$ together with the holonomy corresponding to $c$. The product of these
three yield a deformed chargenet generated by the Hamiltonian constraint. Any deformed charge net generated by the electric diffeomorphism constraint at discretization  parameter value $\delta$ bears
the same charges on each of its edges as  on the counterpart of this edge  in $c$ (see Figs \ref{grb}, \ref{grextb}) and we have that 
$\epsilon = |q_I^i|\delta$.  
Finally, using the constructions of Appendix \ref{acone2}, $C^1$ or $C^2$ kinks are placed at appropriate positions around the displaced  vertex $v_I$ in a manner consistent with the specification of ${\vec V}_I$ at $v$
in the sense described in section \ref{secneg1.1}.
We use the notation of section \ref{secgr.3} to denote the deformed charge nets generated in this way by $c_{(i,I,\beta, \delta)}$ and $c_{   (  i, I, \beta=0, \delta   )}$.

In case (2) of section \ref{secneg2.1}, the vertex displacement corresponds to that generated by 
$\varphi(-q^i_I\vec{{\hat{e}}}_{I_v},\delta)$ due to the fact that ${\vec V}_I$  is opposite to ${\vec {\hat e}}_I$. In order to remove this conflict with the considerations of section \ref{sec3.2.1} (see equation (\ref{defd1cdefq})), it is necessary 
to introduce an intervention of the type
used in section \ref{secgr}. Accordingly,  we first multiply the state $c$ by a holonomy  $h_{\bar l}$
around a loop  ${\bar l}$  made up of two edges ${\bar l}_{1}, {\bar l}_{2}$ so that ${\bar l}= {\bar l}_1\circ {\bar l}_2$.
Let  ${\bar l}_1$   run from  ${\bar p}_1$  to  ${\bar p}_2$. Here ${\bar p}_1$, ${\bar p}_2$ are  equidistant from $v$, with ${\bar p}_1$ on the linear extension of $e_I$ past $v$ and ${\bar p}_2$ on $e_I$. Let ${\bar p}_1$ and ${\bar p}_2$ be chosen
such that the coordinate length of $l_1$ is $C\delta, C= 8q_{max}$. Let ${\bar l}_2$ be a semicircular arc connecting ${\bar p}_2$ with ${\bar p}_1$ such that its diameter is $C\delta$. 
Let ${\bar l}$ lie in a coordinate plane $P_{\bar l}$ such that no non-conducting edge lies in $P_{\bar l}$. 
Define the holonomy $h_{\bar l}$ to run along ${\bar l}$ with charge equal to $-q^i_I$.
Multiplication of $c$ by this holonomy 
yields the state $c_{\bar l}$ with a GR vertex. The $I$th outgoing edge of $c_{\bar l}$ has outgoing charge $q^i_I$ and the 
outgoing tangent to this edge is parallel to ${\vec V}_I$. We now act with an approximant of the type underlying the action 
of section \ref{sec3.2.1} on  $c_{\bar l}$.
As discussed in the first paragraph of this section,  the deformation generated by this approximant  is upward (or downward) with respect to ${\vec V}_I$ if $q^i_I$ is negative (or positive). At this stage we  
refrain from  placing any $C^1$ or $C^2$ kinks. 
Next, we multiply the result by the  inverse holonomy $h^{-1}_{\bar l}$.
\footnote{Note that we have chosen the size of the loop ${\bar l}$ slightly smaller than that of $l$ in section \ref{secgr}. Nevertheless, ${\bar l}$ is still
large enough that an arguementation similar to (a)- (c) of section \ref{secneg1.2} shows that no unwanted intersections ensue due to this intervention.} 
Finally we place a  $C^1$ or a  $C^2$ kink  between the displaced vertex and $v$ in a manner consistent with ${\vec V}_I$, this placement being achieved through multiplication by a holonomy  which is classically close to identity similar to that 
employed in Step 2 of Appendix \ref{acone}.
Clearly the end result is equivalent  to deforming $c$ as indicated in section \ref{secneg2.1}. 
It turns out that for future purposes the situation of interest in this case (i.e. Case (2)), is one in which the other edges at $c$ conform to Case (1). Hence in this situation, the action of the constraints needs no further intervention 
beyond that of  $h_{\bar l}$ and its inverse.

\subsection{\label{secneg3}Choices of Deformation: Linear CGR vertex}

\subsubsection{\label{secneg3.1} Choice of Conical Deformation type}
Let $v$ be a linear GR vertex of $c$. Let the deformation of interest be  $(i,I, \beta, \delta)$.

Let the conducting edge in $c$ be $e_K$ so that $v$ seperates $e_K$ into two parts  $e_{K,1}$ and $e_{K,2}$. Let us first consider the case where $I=K$ so that the deformation is along the conducting edge.
We first need to determine the vector ${\vec V}_K$.
It turns out that the cases of interest are such that $e_{K,1}$ has a nearest kink which  is $C^1$ and $e_{K,2}$ has a nearest kink which is $C^2$ or {\em vice versa}.
In each case we apply the appropriate criteria (i.e. one of (2),(3))  of section \ref{secneg2.1} to either the edge $e_{K,1}$ oriented in the outgoing direction from $v$  or to the edge $e_{K,2}$, also oriented in the outgoing direction
from $v$) to obtain ${\vec V}_K$. It is easy to check that irrespective of whether the criteria are applied to  $e_{K,1}$ or to  $e_{K,2}$, the same choice of ${\vec V}_K$ ensues.
Next, we base our choice of upward or downward deformation with respect to ${\vec V}_K$ on the sign of the net conducting charge $q^i_{K,net}$ (see (\ref{defqnet})).
If $q^i_{K,net}>0$ we choose the deformation 
$(i,I=K, \beta, \delta)$
of $c$ to be downward with respect to ${\vec V}_K$
and if $q^i_{K,net}<0$ we choose this deformation of $c$ to be upward  with respect to ${\vec V}_K$. The deformations corresponding to these choices are constructed as in section \ref{secneg1.2}.

Next consider the case where $I\neq K$. It turns out that the case of interest is then such that $e_I$ has a nearest kink which is $C^0$. In this case we apply criterion (1) of section \ref{secneg2.1} i.e. we choose
${\vec V}_I$ to be along the outgoing edge direction. We then choose the deformation to be upward with respect to ${\vec V}_I$ if the outgoing charge $q^i_I>0$ and downward if $q^i_I<0$.
The deformation is then implemented as in section \ref{secneg1.2}.

\subsubsection{\label{secneg3.2}Choice of Discrete Approximant to Constraint}
The  choice of discrete approximants which implement  the  choices described in section \ref{secneg3.1} is then as follows.
First, as in section \ref{secneg1.2}, we apply the intervention $h_l$ with $l$ chosen in accord with ${\vec V}_K$ as described in that section.
For $l$ of small enough area the classical holonomy $h_l$ is a good approximant to identity and for small enough $\delta$, the straight line part of $l$ does not 
overlap with any nearest kinks on $e_K$. The intervention  yields the state $c_l$ with a GR vertex at $v$. 

We then use the appropriate choice of approximant detailed in section \ref{secneg2.2} to generate the chosen (upward or downward) deformation  of $c_l$  (according to the assignment $\{{\vec V}_I\}$ induced from $c$ to $c_l$ as 
explained in section \ref{secneg1.2})
\footnote{Note that  no edge of $c_l$  satisfies criterion (2). The only possibility is an edge along the conducting line in $c$; however only the {\em upper} conducting edge is retained in $c_l$, its outward orientation
coinciding with the upward direction.  
}
except that we refrain from placing the 
desired $C^2,C^1$ kinks i.e we do not implement the analog of step 2, Appendix \ref{acone}. Since this placement is implemented   via multiplication by a holonomy whose classical correspondent is a good approximant to 
the identity, the postponement of this implementation does not affect the viability of the approximant used.
We then multiply the resulting deformed charge net by the inverse holonomy $h^{-1}_l$.

Finally  
we use the analog of Step 2, Appendix \ref{acone} to place kinks consistent with
the choice of $\{{\vec V}_I\}$ 
Accordingly, when $I=K$, the conducting line of the deformed chargenet is also labelled by $K$ and we place $C^2,C^1$ kinks consistent with the specification of ${\vec V}_K$ for $c$. When 
$I\neq K$, the conducting line in the deformed chargenet is along the $I$th non-conducting edge (or its extension)  of the undeformed charge net $c$ and we place $C^2, C^1$ kinks around the 
displaced vertex in a manner consistent with the specification of ${\vec V}_I$ at $v$ in  $c$

\subsection{\label{secneg4a}(Non)degeneracy of Vertex types}

Given a GR vertex, constraint operators act nontrivially at this vertex only if it is non-degenerate, its nondegeneracy being defined as the non-vanishing of its volume eigenvalue $\nu$ (\ref{evq}).
At a CGR vertex,  the action of a constraint is sensitive to the (non)degeneracy of the same (but now GR) vertex in its image by intervention described in section \ref{secneg1.2} It is useful to formalise this
notion of degeneracy as a definition identical  to Definition 1, section \ref{secgr.1b}. Before doing it so it is useful to catalog the kinds of vertices which are generated by the deformations of GR and CGR vertices described in 
sections \ref{secneg1}- \ref{secneg3} with a view to analysing their  possible non-degeneracy. Since the $C^1,C^2$ vertices are always bivalent and hence degenerate, and since their placement does not affect the vertex
structures at other vertices, we need only analyse the vertex types generated prior to their placement. 

An exhaustive analysis of such vertex structures is provided in Appendix \ref{acolor}, the catalog of vertex types  being those encountered in Cases 1a, 1b, 2a.1, 2a.2, 2b.1, 2b.2, 3 therein. Figures pertinent to Cases 1a,1b
are Figures \ref{gr}, \ref{grext} and to Cases 2a.1, 2a.2 are Figures \ref{cgrkni}, \ref{cgrk=i}. 
Figures \ref{cgrextkni}, \ref{cgrextk=i}, pertinent to Cases 2b.1, 2b.2  are displayed below. 
\footnote{These figures are schematic and show the edge intersection structure at vertices of interest. They do not faithfully reproduce the deformations of section \ref{acone2} which result in
regular conicality of the deformed vertex, nor do they show the $C^1,C^2$ kinks.}
From the disussion in  Appendix \ref{acolor},  the figures for Case 3 may be obtained by setting the upper conducting charge equal to zero in Figures \ref{cgrkni}, \ref{cgrk=i}. 
Figures \ref{cgrextkni}, \ref{cgrextk=i}.

As discussed in Appendix \ref{acolor}, and as seen in the relevant figures, Cases 1a, 1b, 2a.1, 2a.2 and 2b.2
do not present any new potentially non-degenerate vertices of types other than GR and CGR.  However as seen in Figure \ref{cgrextkni} and discussed in Appendix \ref{acolor}, Case 2b.1 presents 2 new vertex types, both associated
with with parental vertices in deformed children. These are the 4 valent vertex of Fig \ref{cgrextknib} and the $N+2$ valent vertex of Fig \ref{cgrextknic}. The former is a planar vertex and hence degenerate. The latter is a
{\em linear doubly CGR vertex} where we define such a vertex as follows.\\

\noindent{\em Definition 3: Linear Doubly CGR Vertex}: A  $N+2$ valent vertex $v$ of a charge net $c$  will be said to be linear doubly CGR if:\\
(i) There exists a coordinate patch around $v$ such that in a small enough neighbourhood of $v$ all edges at $v$ are straight lines.\\
(ii) There are 2 sets of 2 edges such that the union of the 2 edges in each set forms a straight line so that $v$ splits this line into 2 parts and such that the 2 straight lines corresponding to each of these 2  sets
have a single isolated intersection at $v$. Each of these lines will be called conducting lines, each conducting line consisting of a pair of conducting edges.\\
(iii) The set of the remaining $N-2$ edges (called non-conducting edges)  together with any one of the two edges in each pair of (ii)  constitute a GR vertex in the following sense. Consider, at $v$, the set of out going edge tangents
to each of the remaining edges together with each of the outgoing edge tangents to one of the two edges in each  pair in  (ii). Then any triple of elements of this set is linearly dependent.
\\

We now formalise the definition of (non)degeneracy of CGR and doubly CGR vertices.\\

\noindent{\em Definition 4: Nondegeneracy of a CGR vertex:} A CGR vertex of a charge net $c$ will be said  to be non-degenerate iff the corresponding GR vertex in the charge net $c_l$ is non-degenerate. If the vertex in $c_l$ is degenerate
we shall say that the CGR vertex in $c$ is degenerate.  

This  definition provides a unique definition of (non)-degeneracy for the kind of CGR vertices we encounter. These vertices correspond to the following two cases.
In the first case the CGR vertex in the state $c$ is generated through a conical deformation of a parent state $c_{p}$ as specified in sections \ref{secneg1} -\ref{secneg3}.
In this case, the choice of `upward' and `downward'
directions at its displaced vertex and hence the choice of any intervention if required, is uniquely defined and Definition 3 may be applied unambiguously to this vertex.
The second case corresponds to
a conical deformation of the parent state $c_p$ at its vertex $v_p$ such that this vertex is  CGR in $c$ and $c_p$; here we are interested in the application of Definition 3 to this vertex in $c$.
In this case we interpret Definition 3 applied to the vertex $v_p$ in $c$ to mean that the degeneracy of this vertex is well defined iff it is {\em independent} of which part of the edge passing through $v_p$ in $c$ is chosen to be 
upper and lower. Since in our considerations, such a state  $c$ is obtained through a Hamiltonian constraint type $\beta, i$ flipped deformation of $c_p$, it follows from Appendix \ref{acolor} that the net charges at $v_p$ in $c$ have vanishing $i$th component so that 
$v_p$ is degenerate independent of this choice and hence independent of the corresponding choice of intervention.

Next, note that a doubly CGR vertex can be rendered GR through 2 holonomy interventions $h_{l^i}, i=1,2$  with $l^i$ chosen to be `semicircular' with the straight line parts of $l^i$ being along the $i$th conducting line 
defined by the $i$th set of  edges in (ii), Definition 4. These interventions leave the $N-2$ edges in (iii), Definition 4, unaffected and remove one of the conducting edges from each conducting line in (ii). The remaining 
conducting edge in each line is colored with the net conducting charge corresponding to that conducting line. For our purposes the following definition suffices:\\

\noindent{\em Definition 5: Degeneracy of a Doubly CGR Vertex}: A doubly CGR vertex will be said to be degenerate if the GR vertex obtained by any choice of interventions is degenerate.

Since the edges in the parental vertex of the deformed charge net discussed above and in (2b.1), Appendix \ref{acolor} are such that the non-conducting charges and the net conducting charges all have vanishing 
$i$th component, this doubly CGR vertex is degenerate.

\begin{figure}
\begin{subfigure}[h]{0.3\textwidth}
 \includegraphics[width=\textwidth]{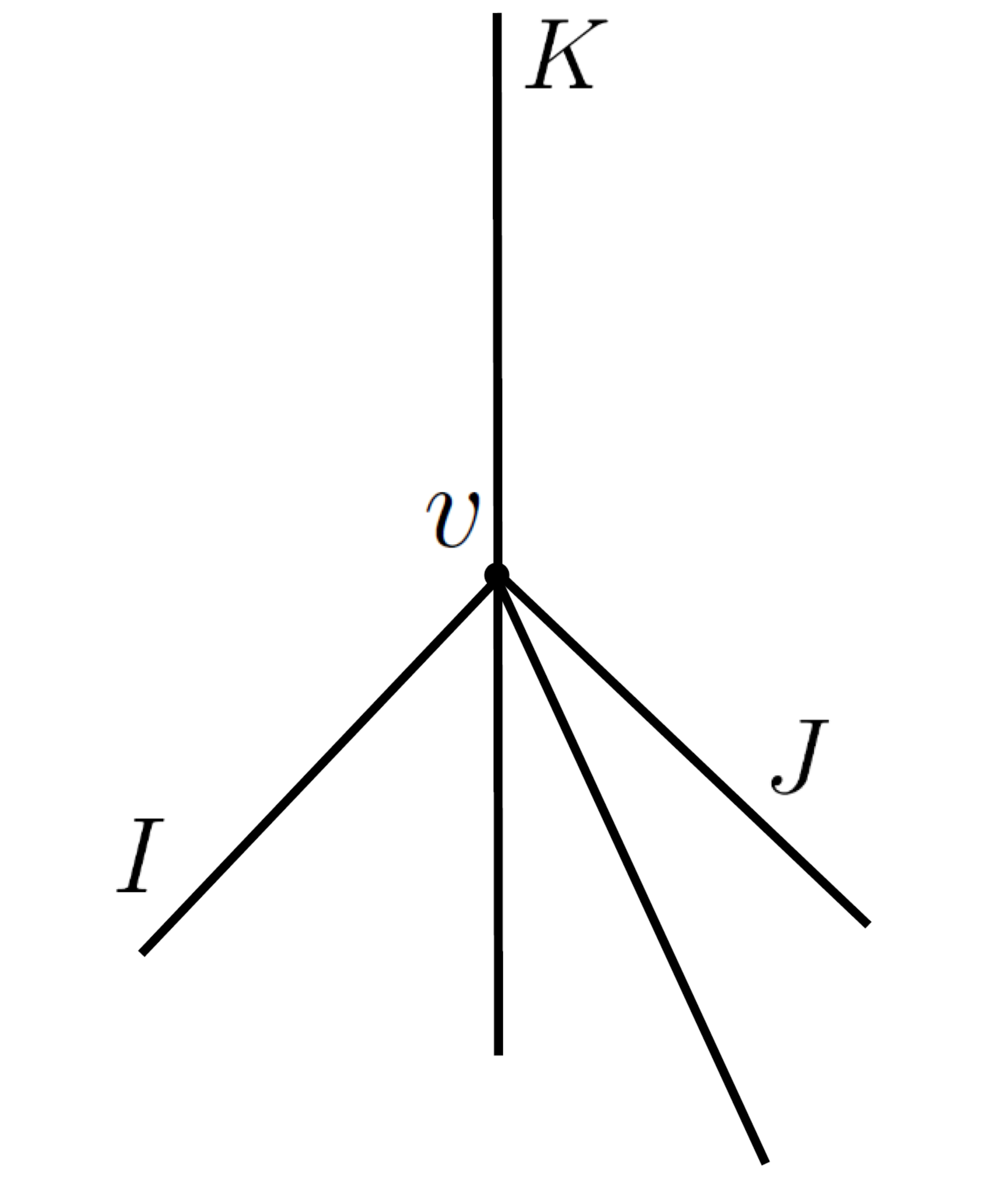}
  \caption{}
   \label{cgrextknia}
 \end{subfigure}
  \begin{subfigure}[h]{0.3\textwidth}
    \includegraphics[width=\textwidth]{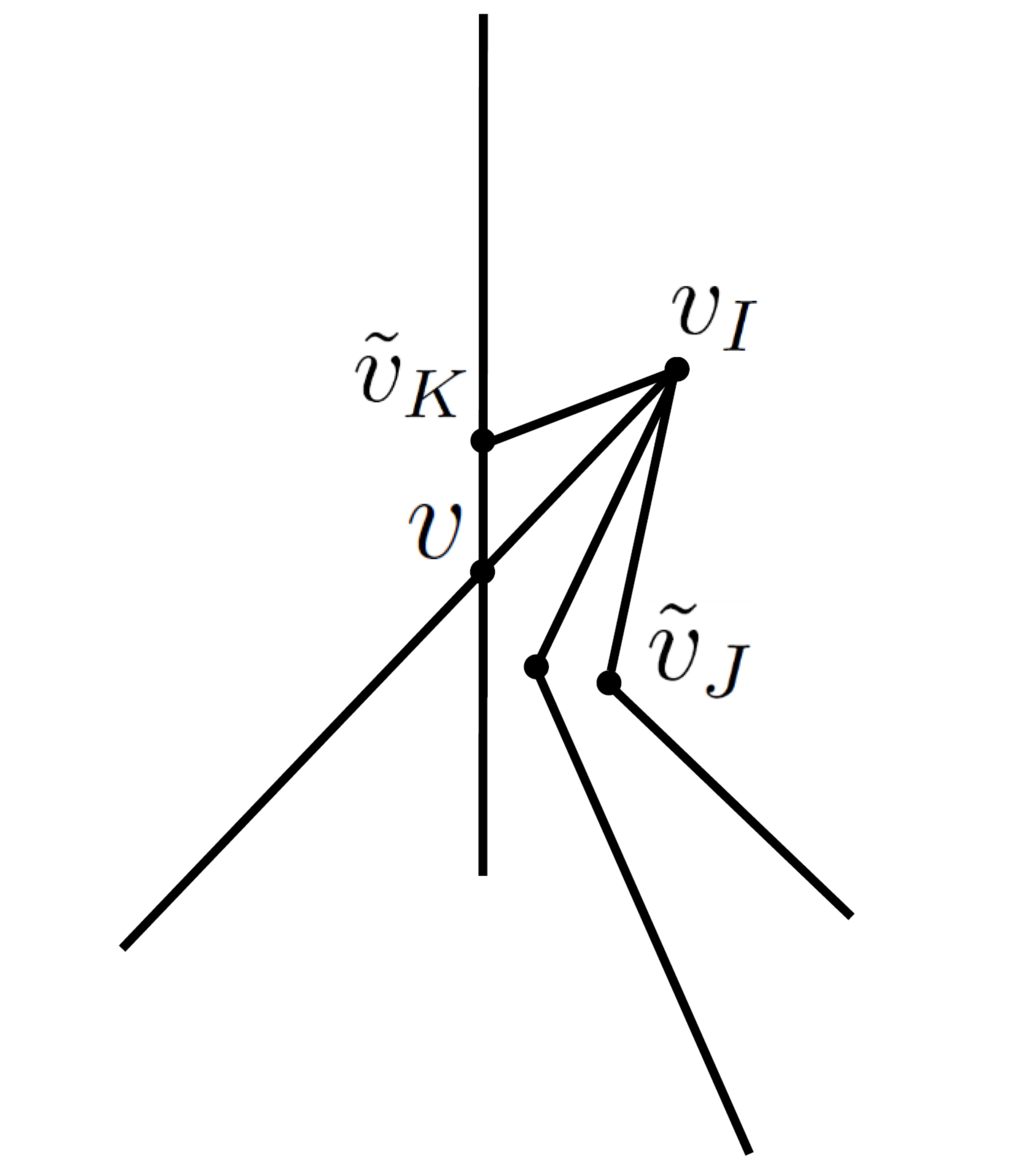}
    \caption{}
    \label{cgrextknib}
  \end{subfigure}
\begin{subfigure}[h]{0.3\textwidth}
    \includegraphics[width=\textwidth]{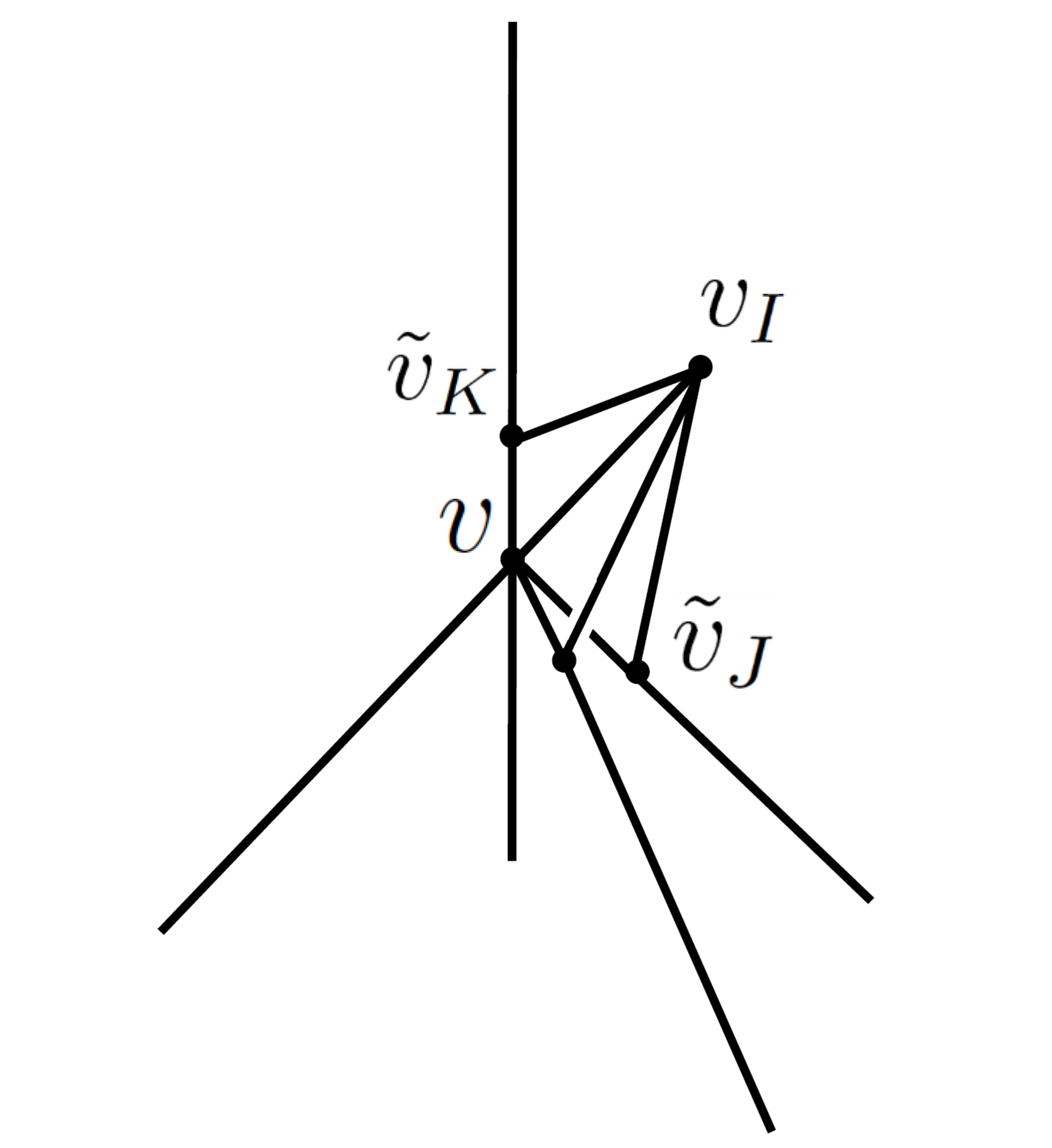}
    \caption{}
    \label{cgrextknic}
  \end{subfigure}
  \caption{Fig \ref{cgrextknia} shows an undeformed CGR vertex $v$ of a chargenet $c$  with its $K$th conducting edge and $I$th and  $J$th non-conducting edges as labelled. In  Fig \ref{cgrextknib}  the vertex  structure of Fig \ref{cgrextknia} 
  is deformed along an extension of $I$th edge past $v$ and the displaced
vertex $v_I$ and the $C^0$ kinks  ${\tilde v}_J$, ${\tilde v}_K$ on the $J$th, $K$th edges are as labelled. With charge colorings similar to those described in Fig \ref{cgrkni}, 
Fig \ref{cgrextknic} shows the result of a  Hamiltonian type deformation $(i, K,\beta, \delta)$ and Fig \ref{cgrextknib} shows the result of an electric diffeomorphism deformation.
The parental vertex $v$ is doubly CGR in Fig \ref{cgrextknic} and is 4 valent and planar in  Fig \ref{cgrextknib}
}%
\label{cgrextkni}%
\end{figure}

\begin{figure}
\begin{subfigure}[h]{0.3\textwidth}
 \includegraphics[width=\textwidth]{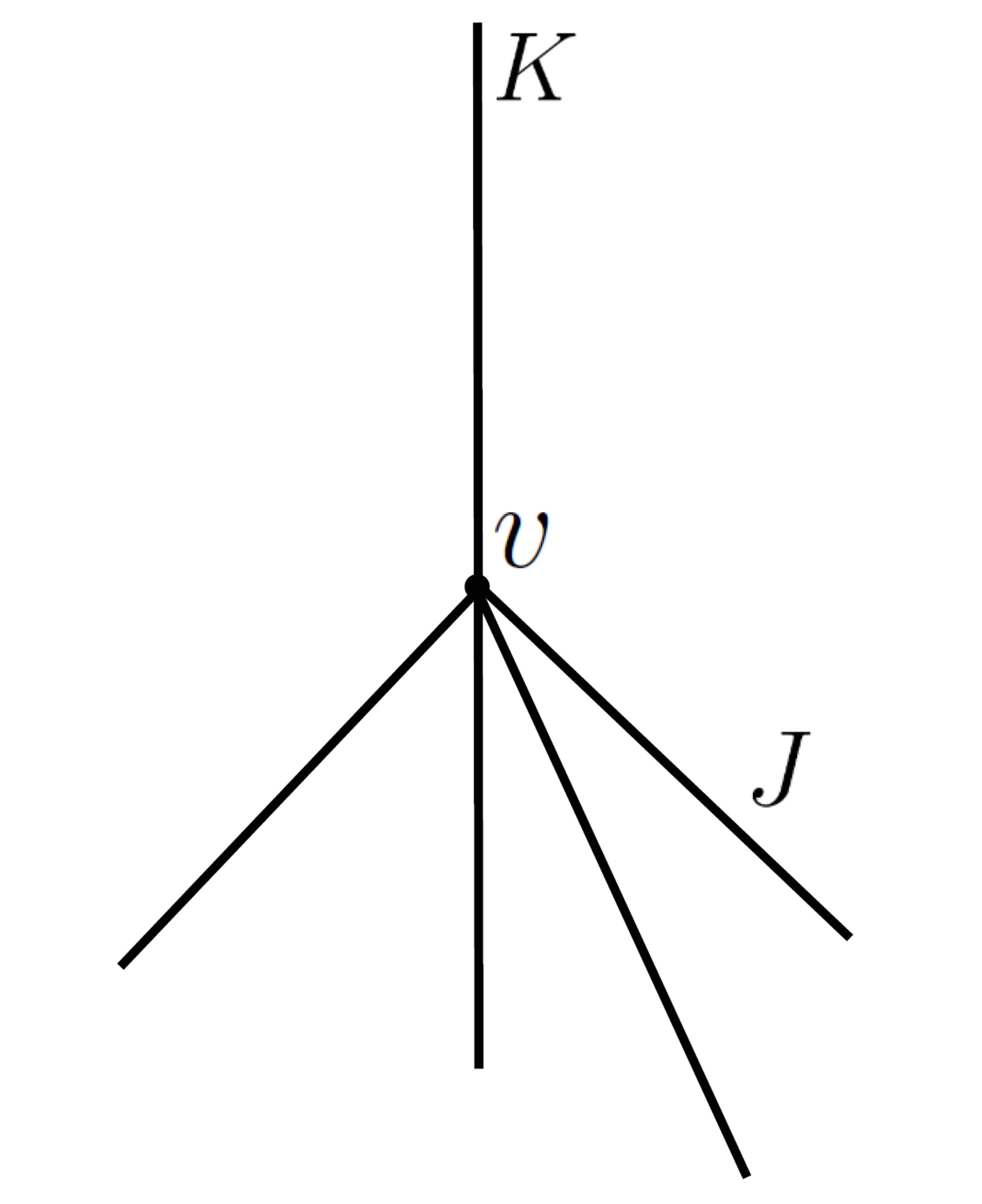}
  \caption{}
   \label{cgrextk=ia}
 \end{subfigure}
  \begin{subfigure}[h]{0.3\textwidth}
    \includegraphics[width=\textwidth]{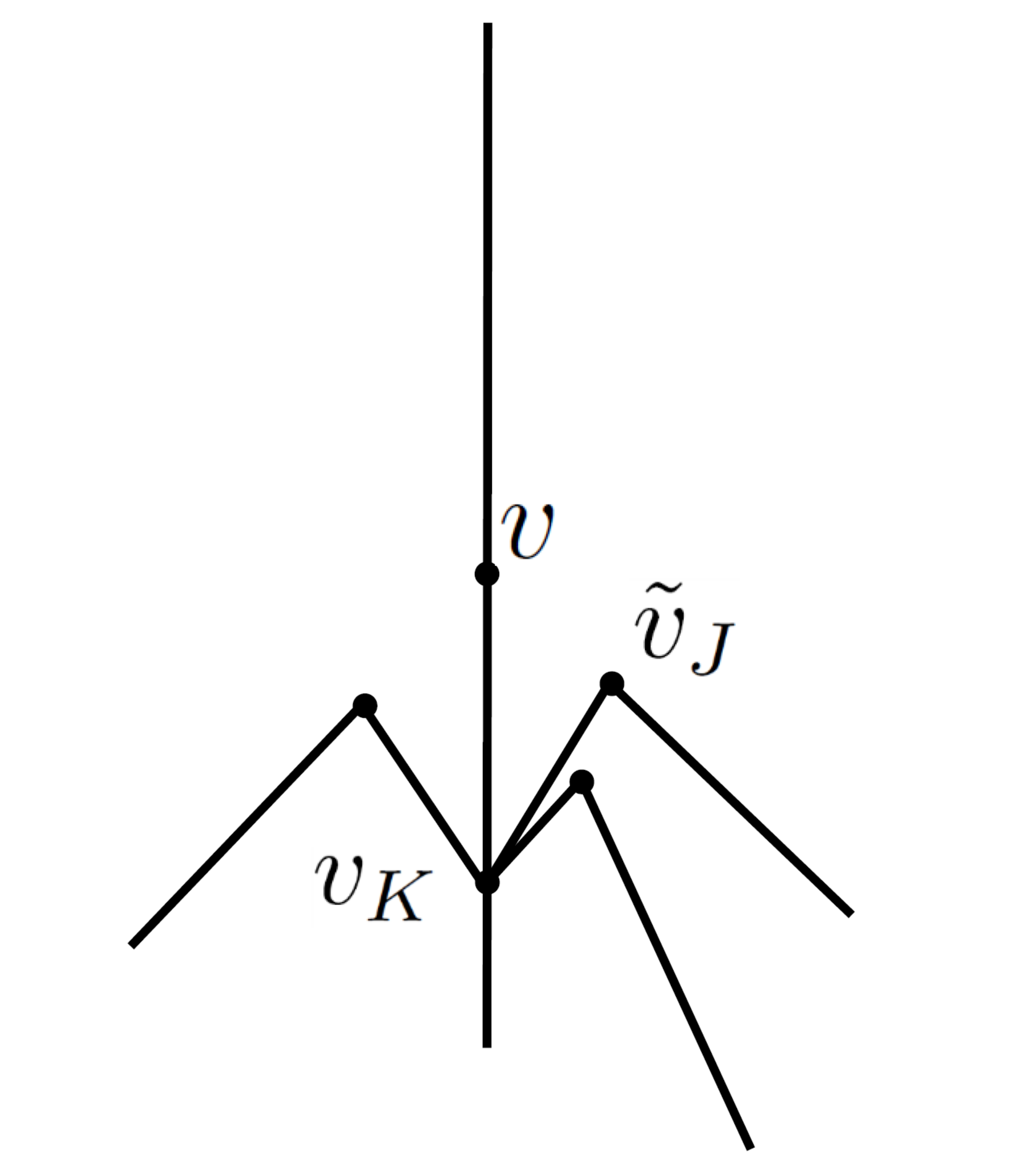}
    \caption{}
    \label{cgrextk=ib}
  \end{subfigure}
\begin{subfigure}[h]{0.3\textwidth}
    \includegraphics[width=\textwidth]{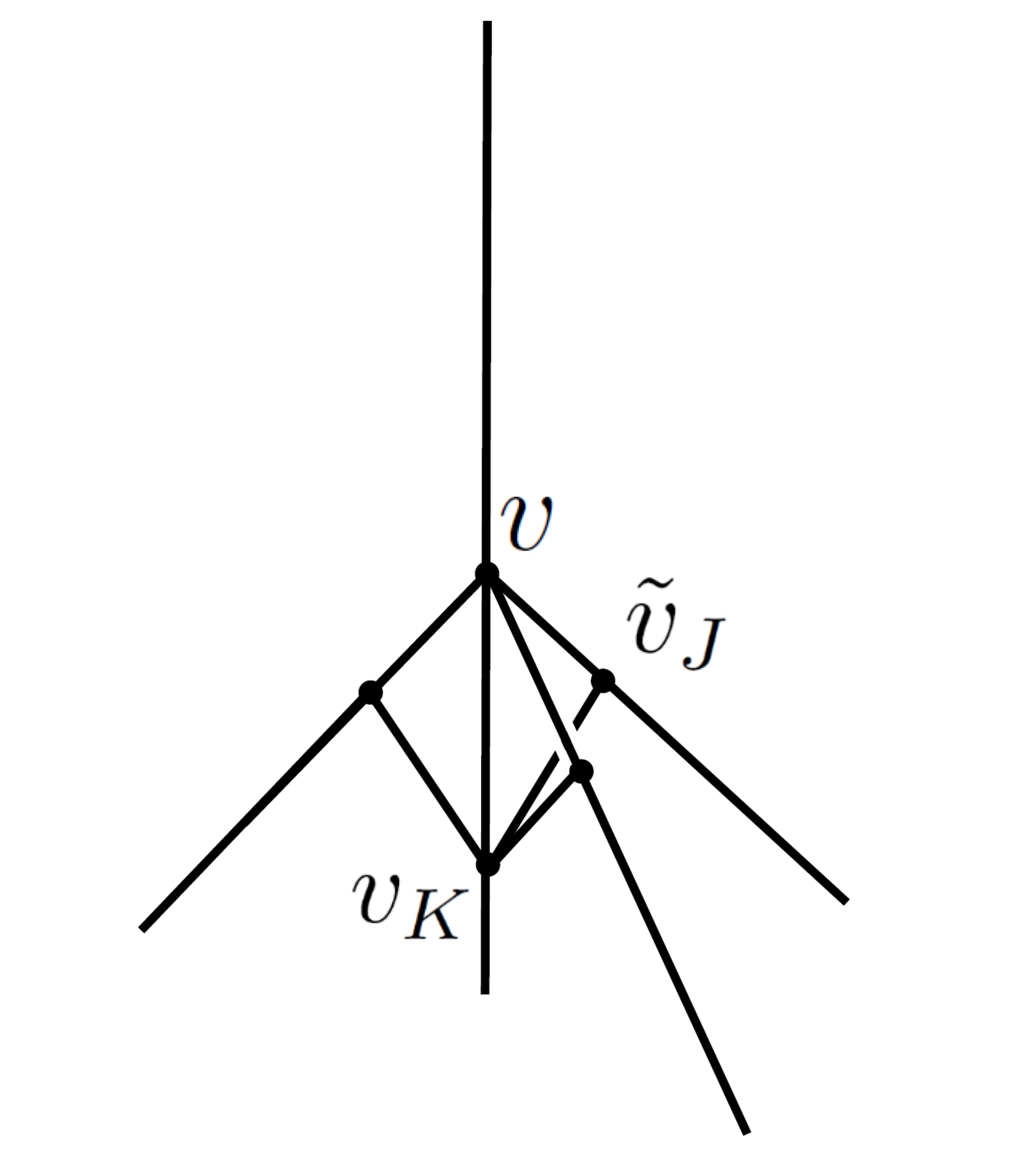}
    \caption{}
    \label{cgrextk=ic}
  \end{subfigure}
  \caption{Fig \ref{cgrextk=ia} shows an undeformed CGR vertex $v$ of a chargenet $c$  with its $K$th conducting edge and $J$th non-conducting edge as labelled. In  Fig \ref{cgrextk=ib}  the vertex  structure of Fig \ref{cgrextk=ia} 
  is deformed along an extension of $K$th edge past $v$ and the displaced
vertex $v_K$ and the $C^0$ kink  ${\tilde v}_J$  on the $J$th edge are as labelled. With charge colorings similar to those described in Fig \ref{cgrk=i}, 
Fig \ref{cgrextk=ic} shows the result of a  Hamiltonian type deformation $(i, K,\beta, \delta)$ and Fig \ref{cgrextk=ib} shows the result of an electric diffeomorphism deformation.
These deformations are isomorphic to those in Fig \ref{cgrk=i} as can be ascertained by viewing them `upside down'.
}%
\label{cgrextk=i}%
\end{figure}

\subsection{\label{secneg4} Summary and Discussion}

From our discussion in section \ref{secneg4a} and Conclusions 1 and 2, Appendix \ref{acolor}, it follows that the only possibly non-degenerate vertices which are generated
by the action of the constraints  on a nondegnerate linear GR or CGR vertex are also GR or CGR. Sections \ref{secneg1}- \ref{secneg3} 
specify the deformation of chargenets with such vertices provided the vertex structures are characterised by the kink structures discussed in sections \ref{secneg2.1} and \ref{secneg3.1}. As we
shall see in section \ref{sec4}, the chargenets of interest will have a single non-degenerate linear GR or CGR vertex with a kink structure of the type discussed.
Denoting such a chargenet of interest  with such a vertex  $v$ by $c$ and its deformed child by the deformation $(i,I, \beta, \delta)$ by $c_{(i,I,\beta,\delta)}$ where the deformed chargenets 
$c_{(i,I,\beta,\delta)}$ for all choices of $I,i,\beta$ and sufficiently small $\delta$ have been constructed in sections \ref{secneg1} -\ref{secneg3}, the action of 
discrete approximants to the Hamiltonian and electric diffeomorphism constraints is expressed in equations (\ref{ham}) and (\ref{dn}).
We shall continue to refer to these two equations with the understanding that they implement the detailed choices discussed in sections \ref{secneg1} -\ref{secneg3}.

The reason we use criteria (1)- (4)  rather than simply choose ${\vec V}_I$ to be in the direction of the outgoing tangent vector ${\vec {\hat e}}_I$ 
is that the former choice yields
anomaly free continuuum limit commutators whereas the latter does not. To see this requires a detailed study of double deformations of a charge net by 2 constraint actions which will be done
in sections \ref{sec4} - \ref{sec8}. Nevertheless we attempt to provide a brief explanation here for the choice of ${\vec V}_I$ as opposed to ${\vec {\hat e}}_I$.
The reader is urged to peruse this explanation once again after reading the entire paper as it may, at this stage, be quite opaque
.
Each double deformation generated by 2 discrete constraint actions on a chargenet is composed of a pair of single conical deformations 
Each such single deformation is along some edge of a parent state and yields deformed offspring which are conically deformed along a cone whose axis is determined by the 
direction of the parental edge.  The continuum limit involves shrinking 2 of these single deformations away from `grandchild' to `immediate parent' to `grand parent'.
It turns out that 
for certain delicate recombinations of terms to occur as a result of this process so as to generate an anomaly free result, the edge directions of the parent and the grandparent must be  correlated (and, as will be seen, in a precise
sense, identical).
To ensure that this happens we must ensure a consistent choice of edge tangent directions
in the child-parent-grandparent genealogy. This choice, it turns out, is exactly that of ${\vec V}_I$ (which clearly depends on the `genetic trace' provided by the $C^1,C^2$ kink placement), as opposed
to the choice of outgoing edge tangent $\vec{{\hat{e}}}_{I}$  (which would be a purely `local' choice independent of lineage).  
Finally, note that the use of (1)- (4) is tantamount to the  replacement of  $q^i_I\vec{{\hat{e}}}_{I}$ by $q^i_{I, net}\vec{{\hat{V}}}_{I}$ (with $q^i_{I,net}$ being the net outgoing charge along the edge $I$, see equation (\ref{defqnet})
above) in the heuristically motivated equation (\ref{defd1cdefq}). Thus there is a tension; we require the choice of ${\vec V}_I$ with the net outgoing charge 
for anomaly free commutators but the argumentation of sections \ref{sec2} and \ref{sec3} imply that we must use the 
outgoing tangent vectors with the outgoing charges. In order to remove this tension it is necessary to use  the intervention $h_{\bar l}$ of section \ref{secneg2.2} so as to ensure that criteria (1)- (4) are 
implemented through the use of valid approximants to the constraints.

\section{\label{sec4}Discrete Action of Constraint Operator Products}
In the last two sections we did not specify the choice of coordinates with respect to which the deformations generated by the discrete action of the constraints were defined. In this section we specify these
coordinates as well as the action of constraint operator products of interest along the lines sketched in section \ref{sec1}. It turns out that in view of the `single vertex' anomaly free states studied in this paper, the detailed specification
of this action only needs to be made for a certain set of kets, which we shall refer to as the Ket Set. This set  corresponds to all the kets which are obtained by multiple actions of the type
(\ref{ham}), (\ref{dn}) on certain `primordial' kets which themselves are not generated by any such action on any other state. In section \ref{sec4.1} we generalise the notation of section \ref{secgr.3} to describe  the `multiply deformed' kets which 
are generated by such multiple actions.
In section \ref{sec4.2}  we define 
the Ket Set. 
In section \ref{sec4.3} we choose a reference ket in each diffeomorphism class of kets in the Ket Set and  a set of reference diffeomorphisms such that each distinct ket in the diffeomorphism class of a reference ket
is the image of the reference ket by a unique reference diffeomorphism. 
We also define certain key structures known as Contraction Diffeomorphisms which play a crucial role in defining the continuum limit by `contracting'
the deformations away. 

In section \ref{sec4.4} we define the discrete action of  products of contraint operators on any ket in the Ket Set through 
multiple applications of equations (\ref{ham}) and (\ref{dn}). These multiple actions generate multiply deformed kets as discussed in section \ref{sec4.1}. It remains to specify the coordinates with respect to
which these deformations are defined. We do so  through slightly involved manipulations of the structures developed in sections \ref{sec4.1},  \ref{sec4.2} and \ref{sec4.3}. The end result of these manipulations
is a  specification
of the  coordinates with respect to which the deformations are defined together with  a definition of the discrete action of products of constraint operators on  any ket in the Ket Set for arbitrarily small values of the discretization parameters.
The corresponding dual action can then be defined on states in the algebraic dual space. The continuum limit of this action on anomaly free states (which reside in the algebraic dual space to the 
space of finite linear combinations of charge nets) will be evaluated in sections \ref{sec8} and \ref{sec9}.

%

\subsection{\label{sec4.1}Notation for multiply deformed states.}

Let $c$ be a state with a single non-degenerate vertex $v$, this vertex being either a linear GR or linear CGR vertex with respect to some choice of coordinates around $v$.
\footnote{In addition, as shall become clear in sections \ref{sec4.2} - \ref{sec4.4}, the kink structure of the state $c$ under consideration as well as of the states  generated from $c$ via multiple
applications of (\ref{ham}), (\ref{dn}) conforms to those
alluded to in section \ref{secneg}.}

The action of a single discrete constraint operator at discretization parameter $\delta$  on $c$ is given by (\ref{ham}), (\ref{dn}).
From Conclusion 2 of Appendix \ref{acolor}  it follows that the deformed states $c_{(i,I,\beta, \delta)}$  have at most  a single nondegenerate  GR or CGR vertex and that this corresponds to the displaced vertex in each of these states. 
We shall assume that $c$ is such that the displaced vertex in each of the deformed states $c_{(i,I,\beta, \delta)}$ is non-degenerate and that our choice of coordinates around each displaced vertex is such that the vertex is linear
with respect to this choice.
Hence these `singly' deformed states  $c_{(i,I,\beta, \delta)}$ are all single nondegenerate linear GR or CGR vertex states.  The action (\ref{ham}), (\ref{dn}) on $c$ yields these singly deformed states as well as $c$ itself.

The action (\ref{ham}), (\ref{dn}) on each of {\em these} states (namely $c_{(i,I,\beta, \delta)}$, $c$)  is then well defined because each of these states is a single linear GR or CGR state.
 Since the actions (\ref{ham}), (\ref{dn}) correspond to the discrete action of a single constraint, it follows that an action of one of (\ref{ham}) or (\ref{dn}) followed by a second
action of either (\ref{ham}) or (\ref{dn})  on $c$   corresponds to that of a
discrete approximant to  the product of two constraints and creates `doubly deformed' states,  singly deformed states and the undeformed state. 
From section \ref{secneg} it follows that  each of the doubly deformed states has a (doubly displaced)  vertex which is once again, either GR or CGR.
We shall assume that this 
vertex is non-degenerate and that the associated coordinate system is such that this vertex is linear; from Conclusion 2, the doubly deformed states are then again single, nondegenerate, linear GR or CGR vertex states.
As a result the  action (\ref{ham}), (\ref{dn}) is well defined on {\em these} states as well. In this manner any combination of 3 actions of the type (\ref{ham}) or (\ref{dn}) 
yields triply deformed states, doubly deformed, singly deformed states and the undeformed state.

Continuing on and making an appropriate non-degeneracy and linearity assumption at every stage we find that the action of a product of 
$n$ constraint operators can be approximated as $n$ applications of the type (\ref{ham}) or (\ref{dn}) and that this results in states which are $m$ - deformed, $m=0,1,..,n$ with $m=0$  corresponding to the undeformed state $c$.
The continuum limit involves contracting these deformations away; it turns out that $m$th deformation is contracted away first, then the $m-1$th one and so on all the way  to the first deformation. Hence we shall be interested
in multiple deformations such that the parameter associated with the size of each successive deformation is smaller than its predecessors.

We now develop appropriate notation  and `genealogical' language related to multiply deformed states.
We shall refer to $c$ as a parent state. As noted in section \ref{secgr.3} any  deformation of $c$ can be specified through the information $(i,I,\beta,\delta)$ where, as in that section, for the reasons explained there, we have
suppressed information about the coordinate patch used to define the deformation. In the language of section \ref{secgr.3}, this deformation yields the $(i,I,\beta,\delta)$- deformed child 
$c_{(i,I,\beta,\delta)}$. Generalising this notation, we can specify a sequence of $m$ deformations by 
\be
[ (i_{m-1}, I_{m-1},\beta_{m}, \epsilon_m), (i_{m-2}, I_{m-2}, \beta_{m-1}, \epsilon_{m-1}),.....,(i_1,I_1, \beta_2, \epsilon_2), (i,I,\beta_1,\epsilon_1 )] \;\;\;\epsilon_{i} <\epsilon_{j}< \;\;{\rm for}\;\; i>j
\label{mdef}
\ee
and denote the resulting `$m$th generation' child of the parent $c$ by 
\be
{ c}_{   [ (i_{m-1}, I_{m-1},\beta_{m}, \epsilon_m), (i_{m-2}, I_{m-2}, \beta_{m-1}, \epsilon_{m-1}),.....,(i,I, \beta_1, \epsilon_1) ]}.
\label{mchild}
\ee
This signifies that the child is obtained from the parent $c$ through  the sequence:
\begin{eqnarray}
{ c} \; \xrightarrow{i,I ,\beta_1,\epsilon_1} \;{ c}_{i,I,\beta_1,\epsilon_1} \xrightarrow{i_1, I_1, \beta_2, \epsilon_2} { c}_{ [(i_1, I_1, \beta_2, \epsilon_2),    (i,I,\beta_1,\epsilon_1)]}&\nonumber\\
..........\xrightarrow{i_{m-1}, I_{m-1}, \beta_m, \epsilon_m}{ c}_{   [ (i_{m-1}, I_{m-1},\beta_m, \epsilon_m), (i_{m-2}, I_{m-2}, \beta_{m-1}, \epsilon_{m-1}),.....,(i,I, \beta_1, \epsilon_1) ]} &
\label{mseq}
\end{eqnarray}
Above, we have assumed that the displaced vertex in each $i$th generation child with $1\leq i\leq m$ is non-degenerate and that the specification of the coordinate system around this vertex is such that 
the edges there appear as straight lines, so that the vertex is linear as well.  Further, we have chosen to enumerate the edges of each charge net in this sequence in such a way that 
the enumeration of edges in a child and in its immediate parent are related as follows. Consider the charge net ${ c}_{(i,I,\beta_1,\epsilon_1)}$ obtained by deforming its immediate parent $c$.
Each non-conducting edge emanating from  the non-degenerate vertex in ${ c}_{(i,I,\beta_1,\epsilon_1)}$ is obtained by deforming, a corresponding edge in $c$ which  emanates from the non-degenerate vertex of $c$, the two edges
meeting at a $C^0$-kink. We assign these corresponding edges the same number i.e. if  $e_{I_1}\in { c}_{(i,I,\beta_1,\epsilon_1)}$  is the deformation of the edge $e_I$ in $c$ then we have $I_1=I$.
Since there are $N-1$ such pairs of edges, one in the parent and one in the child,  the remaining edge in the parent and in the child also bear the same number. 
Clearly, we can extend this enumeration scheme so that the numbering of edges of any child and immediate parent in (\ref{mseq}) are so related. This immediately implies that given the sequence (\ref{mseq}), the enumeration scheme of 
any chargenet in the sequence is uniquely fixed by the enumeration scheme in $c$.
\footnote{In the  case of  CGR vertices   we use this numbering for non-conducting edges and the conducting {\em line}; as seen in (\ref{hamfinal}), (\ref{dnfinal}), we do not need to count the  
the upper  and lower conducting edges seperately so that this correspondence continues to hold  and the indices $I_m$ for all $m$ as well as the index $I$, all run from $1,..,N$}.

Finally, where it creates no confusion, we  will find it convenient to use the notation:\\
\be
[i,I,\beta,\epsilon]_m:= [ (i_m, I_m,\beta_m, \epsilon_m), (i_{m-1}, I_{m-1}, \beta_{m-1}, \epsilon_{m-1}),.....,(i_1,I_1, \beta_1, \epsilon_1) ]
\label{mabbrev}
\ee
so that the state in (\ref{mchild}) can be written as $c_{[i,I,\beta , \epsilon]_m}$.

\subsection{\label{sec4.2}Primordial States and The Ket Set}

We think of primordial states as being states which cannot be obtained by a Hamiltonian or electric diffeomorphism type deformation of any state.
Rather than provide a precise definition, we shall work with concrete examples and leave a more precise and complete definition of primordiality to future work.
Consider any state with a  single nondegenerate GR $N$ valent vertex. Let all other  vertices of the (coarsest)  graph underlying the state be degenerate and let no such vertex have valence $2,3,4, N,N+1$ or  $N+2$.
Let there exist some coordinate patch around the non-degenerate vertex with respect to which the edges at this vertex 
are straight lines in a small neighbourhood of the vertex i.e. let the vertex be linear with respect to some choice of coordinates. 
Such a state cannot be created by the discrete action of a constraint because, notwithstanding that we have defined this action in detail only on a restricted class of states, we visualise the action of 
the deformation maps (see section 2 of this paper as well as P1, P2) to only create  vertices of valence $2,3,4,N,N+1, N+2$ . We shall call such a state as primordial state provided it is subject to four additional
restrictions described below.

First, we  restrict attention to the case where the nondegenerate vertex is $N$ valent for some fixed {\em even} integer $N$.  This is for certain technical reasons.
Note that GR and nondgeneracy restrictions imply that $N\geq 4$.
We shall return to this point in our final section.
In order to articulate our second (mild) restriction, consider the $U(1)^3$ charge obtained by the action of a $(\beta= \beta_1, i=i_1)$ flip of the $U(1)^3$ edge charge label ${\vec q}_I:= (q^1_I, q^2_I,q^3_I)$ in $c$. We may subject the flipped
charge set to yet another flip $(\beta_2, i_2)$. Let us denote the charge obtained by $m$ such flips $[(\beta_m, i_m), (\beta_{m-1}, i_{m-1}), ..., (\beta_1, i_1)]$  as ${\vec q}_{[\beta, i]_m, I}$
with $\beta_{j} \in \{-1,1\}$. Using this notation we require that 
the charges on each edge  of $c$  satisfy
\be 
\sum_{k=1}^3 q^k_I \neq 0, \;\; \sum_{k=1}^3 q_{[\beta,i]_m,I}^k \neq 0 \; \forall [\beta, i]_m, \forall I=1,..,N
\label{sumq}
\ee
so that the sum of the 3 $U(1)$ charges on each edge as well as the sum of any `multiply flipped' image of these charges is non-vanishing.  
We also require that 
\be
q^k_I \neq 0,    \;\;\; q_{[\beta,i]_m,I}^k \neq 0 \forall [\beta, i]_m, \;\forall I=1,..,N, \forall k=1,2, 3
\label{qneq0}
\ee
It is convenient to extend the notation for $(\beta, i)$- flips to the case that $\beta=0$. Consistent with the fact that no flipping is associated with an electric diffeorphism type transformation,
we define a $(\beta, i)$ flip to be the identity operation when $\beta=0$. In this case the index $i$ is redundant but we retain it for convenience in articulating the following definitions which will be 
useful for future purposes:
\be
q_{min} = \min_{I,[\beta, i]_m \forall m,I, [\beta,i]_m} | \sum_{k=1}^3 q_{[\beta,i]_m,I}^k |, 
\label{defqmin}
\ee
\be
q_{min,1} = \min_{I,[\beta, i]_m \forall m,I,k, [\beta,i]_m} | q_{[\beta,i]_m,I}^k |,
\label{defqmin,1}
\ee
and 
\be 
q^{primordial}_{max} = \max_{(i=1,2,3),(I=1,..,N)} |q^i_{I}|.
\label{defqprimmax}
\ee
Since there are only a finite number of flipped images of the charges on each edge, equations (\ref{sumq}), (\ref{qneq0}) are well defined and  imply that $q_{min}, q_{min,1}>0$.
Note also that since the charges are integers we have that $q_{min}, q_{min,1}\geq 1$. Finally, note that while (\ref{defqprimmax}) seems identical to (\ref{defqmax}), these two definitions are in general
distinct in that the charges on the right hand side of  (\ref{defqprimmax})  are the edge charges on the primordial chargenet at its nondegenerate vertex whereas those  in the right hand side of  (\ref{defqmax}) are the edge charges for the 
edges of the (not necessarily primordial) chargenet under consideration in sections \ref{secgr} and \ref{secneg}.

Third, we restrict attention to states which exhibit linearity with respect to a particular choice of coordinate patches as follows. 
Fix a point $p_0$ on the 
Cauchy slice and a chart $\{x_0\}$ in some neighbourhood 
of $p_0$.  We  require that any state  under consideration be such that it is diffeomorphic to some state which  has a non-degenerate vertex at $p_0$ and which is linear with respect to $\{x_0\}$. 
\footnote{It seems plausible to us that any state with a single nondegenerate vertex which is linear with respect to some choice of coordinate patch must be diffeomorphic to one which is linear with 
respect to any prescribed patch. If this is indeed true, this third restriction does not actually constitute a genuine restriction. We leave an investigation of this issue to future work.}
The coordinates $\{x_0\}$ will be referred to as {\em Primary Coordinates}.

%
Fourth, we restrict attention to states which satisfy the following requirement of {\em eternal nondegeneracy}: From section \ref{sec4.1}, any multiple deformation of state yields a state with a multiply displaced
vertex. We require that any primordial state be such that any multiple deformation of the primordial state yields a state whose
 multiply displaced vertex is non-degenerate.
This `eternal nondegeneracy' is a strong and non-trivial restriction. The implementation of anomaly freedom in this paper does not go through
if this condition is not satisfied. The classical analog of this condition is the requirement that the determinant of the 3 metric stay non-zero throughout its evolution. Clearly, if this condition is violated
at any instant (i.e. anywhere on a Cauchy slice), we cannot compute the classical constraint algebra. 
In the Appendix \ref{anondeg} we show that the simplest GR vertex, namely one with 4 edges in conical configuration, 
provides an example of a state which satisfies these restrictions.

Consider the entire set of states subject to the above restrictions.  We shall call these states as primordial states. Clearly, the  set  $S_{primordial}$ of these  primordial states is closed under diffeomorphisms. 
Next, within each diffeomorphism class of these primordial states, fix a `reference' primordial state $c_{P0}$  which has a non-degenerate vertex at $p_0$ and which is linear with respect to the Primary Coordinates $\{x_0\}$. 
%
%
Consider all multiple deformations of each of these reference primordial states, these multiple deformations being   a sequence of single deformations of 
the type discussed in section \ref{sec4.1}. More in detail, consider first some primordial reference state $c_{P0}$, a neighbourhood of its vertex at $p_0$ being covered by $\{x_0\}$.
Any single deformation of $c_{P0}$ for sufficiently small deformation parameter is chosen to  be upward or downward conical according to the criteria of section \ref{secneg2} (in this case we use (4) of section \ref{secneg2.1} together with 
the sign of the edge charge labels as discussed in that section to deform upward or downward). Using the detailed constructions of Appendix \ref{acone}, and of sections \ref{sec3.1}, \ref{secgr.1} and \ref{secneg1}, these deformations
are defined for all sufficiently small values of deformation parameter 
such that the deformation is confined to the interior of a  coordinate sphere $B_{\Delta_0} (p_0)$ of some size $\Delta_0$ with  $B_{\Delta_0} (p_0)$ in the domain of $\{x_0\}$.
It follws that the resulting deformed children have displaced vertices which are  in the domain of the chart $\{x_0\}$. 
These vertices (as mentioned earlier) are GR or CGR and (by assumption) non-degenerate. They are also {\em linear} with respect to $\{x_0\}$ because of the straight line edge structure of the cones
in the vicinity of these vertices (see Appendix \ref{acone} for downward conical deformations; that a similar linearity holds for upward deformations is clear from their detailed construction in section \ref{secneg1.1}).

It is easy to check that the criteria of section \ref{secneg2} and \ref{secneg3} can be applied to these children and that {\em their} (appropriately chosen) upward or downward conical deformations can again be defined
for small enough values of deformation parameter such that the deformation is confined to the interior of $B_{\Delta_0} (p_0)$ , and that each of {\em their} children have a single nondgenerate GR or CGR vertex. The detailed construction of the 
deformation for small enough values of deformation 
parameter  implies that the primary coordinate system $\{x_0\}$ covers a small enough neighbourhood of each of these vertices in which  the edges at each such vertex are straight lines so that the vertex is linear with respect to the 
primary coordinates. Continuing in this way one can define multiply deformed states for all sufficiently small deformation parameter sets associated with the multiple deformation such that the multiple deformation lies in the interior
of $B_{\Delta_0} (p_0)$.
The set of all these  deformed children of $c_{P0}$  together  with $c_{P0}$  will be said to form a primary  family and each element  of such a family
will be called a {\em primary}.

By letting $c_{P0}$ vary over the set of all distinct reference primordials  we obtain the  set of all primaries, $S_{primary}$,  with the multiple deformation which generates any primary from a  reference primordial being confined
to the interior of $B_{\Delta_0} (p_0)$.
Finally consider the set of all diffeomorphic images of all primaries. This set is the Ket Set $S_{Ket}$.

To summarise:
We let $c$ range over all reference primordial states in  equation (\ref{mseq}).  In that equation we let $m$ range from $1...\infty$ and let the deformation sequence range over all possible choices of 
deformation specifications for all possible small enough deformation parameter sets such that  the deformations can be defined through Appendix \ref{acone} and section \ref{secneg} with respect to $\{x_0\}$
and such that the non-degenerate vertex of every deformed ket  in the sequence is covered by $\{x_0\}$. 
Our definition of primordiality ensures that the resulting set of $m$-deformed children is such that each child in this set has a    
a single nondegenerate vertex. 
The set of all these multiply deformed children together with their primordial reference ancestors comprise the set of primaries. The Ket Set $S_{Ket}$ comprises of the set of all diffeomorphic images of all primaries.

Note that since each element of the Ket Set is a diffeomorphic image of some primary, its nondegenerate vertex is linear with respect to the corresponding diffeomorphic image of $\{x_0\}$.
We note again that from the considerations of section \ref{secneg} it follows that this vertex is  a (linear) non-degenerate GR or CGR vertex and that from Conclusion 2 of Appendix \ref{acolor} this is the 
only non-degenerate vertex of that element. Finally, it is straightforward to check, using the deformations detailed in section \ref{secneg} that each element is such that the 
criteria of section \ref{secneg2.1} and \ref{secneg3.1} can be applied so that any further deformation of this element with respect to 
an appropriately specified coordinate patch is well defined. 
We develop the specification of this  coordinate patch for any given element of the Ket Set in sections \ref{sec4.2} -\ref{sec4.4}.

\subsection{\label{sec4.3}Reference states, Reference diffeormorphisms and Contraction diffeomorphisms}

Within each diffeomorphism class of elements of $S_{Ket}$ choose  one state as a {\em reference} state subject to the restriction that the state must be a primary i.e. the reference state must lie in $S_{primary}$.
A charge net label with subscript $0$ indicates a reference charge net. For the case of the diffeomorphism class of primordial states we choose the reference state to be as in section \ref{sec4.2}.
Next for each distinct element $c$ of each diffeomorphism class $[c_0]$ of a reference charge net $c_0$ choose a {\em reference} diffeomorphism  $\alpha$ such that $\alpha$ maps $c_0$ to $c$
i.e. in `ket' notation we have 
\be
{\hat U}(\alpha) |c_0\ket = |c\ket
\label{c0calpha}
\ee
where  ${\hat U}(\alpha) $ is the unitary operator representing the action of $\alpha$.

Next we define {\em contraction} diffeomorphisms. To do so, consider a ket $c$ in the Ket Set with some linear coordinate system $\{y\}$ at its nondegenerate vertex $v$. Let us deform it by the deformation $(i, I, \beta, \delta_0 )$ 
where the detailed nature of 
the deformation is  as in Appendix \ref{acone} and section \ref{secneg}. In particular, the coordinate patch used to specify the deformation (and the deformation parameter $\delta_0$) is $\{y\}$  (i.e. we set $\{x\}= \{y\}$ in Appendix \ref{acone}
and in section \ref{secneg}) 
and
the displaced vertex  and the $C^0$ kink vertices created by the deformation are each  at a coordinate  distance $|{q^i_{I,net}  }|\delta_0$ from the parent vertex
(here $q^i_{I,net}$ is the net charge as defined in (\ref{defqnet}) and Appendix \ref{acolor}).
We would like to `contract' the deformation away so that the displaced vertex 
and these kinks 
approach the parent vertex $v$ in a prescribed manner. Further, we would like the cone angle for the deformation at
the displaced vertex
to become narrower in line with our visualization of the deformation being that of a singular pulling of these edges along the $I$th edge (see section \ref{sec2}).

This `contraction' is achieved through the action of the 
 contraction diffeomorphism $\Phi^{\delta, Q, L,M,p_1,p_2,p_3 }_{c, \{y\}(i, I, \beta, \delta_0 )}$, defined for small enough $\delta_0$, for which the following properties hold:\\
\noindent (i) The contraction diffeomorphism is a semianalytic diffeomorphism connected to identity.\\
\noindent (ii) It moves the displaced vertex $v_{i,I,\delta_0}$  
along the straight line (in the coordinates $\{y\}$)  between  $v_{i,I,\delta_0}$ and $v$ to the point $v_{i,I,\delta}$  located at a  coordinate distance $|q^{i}_{I,net}| \delta <<  |q^{i}_{I,net}|\delta_0$ from the parent vertex $v$ 
\\ 
\noindent (iii) (a) The 
 $C^1, C^2$ kinks  in  $c_{(i,I,\beta,\delta_0 ) }$ 
have an area $\alpha_0^2 << \delta_0^2$ (see Appendix \ref{acone2}).
The contraction diffeomorphism 
 shrinks the area of these  kinks to $\alpha <<\delta^2$.\\
(iii)(b) It moves the   $C^0$ kink  ${\tilde v}_{L}, L\neq I$ along the edge $e_{L}$ of $c$  to a distance $\delta^{p_1}$ from the parent vertex $v$ . It moves the
$C^0$ kink  ${\tilde v}_{M}, M\neq I, M\neq  L$ along the edge $e_{M}$ of $c$  to a distance $Q\delta^{p_2}$ from $v$ (for some $Q>0$ which we specify later). It moves 
each of the remaining (N-3) kinks along its  non-conducting edge to a distance $\delta^{p_3}$ from $v$.\\
\noindent (iv) In a small vicinity of the (new position of) the displaced vertex it narrows the cone angle between the edges at that vertex by a {\em linear} deformation  generated by the diffeomorphism $G$ defined below.
\\
\noindent (v) It maps $c$ to itself and maps the straight line from $v$ to $v_{i,I,\delta_0}$  (in the coordinates $\{y\}$) to itself.
\\
\noindent (vi) It is identity outside a sphere of size $2|q^{i}_{I,net}|\delta_0$ around $v$.
\\
The construction of the  contraction diffeomorphism  is along the lines sketched in P1, P2. We proceed as follows. For convenience let us rotate the coordinate system $\{y\}$ so that $y^3$ runs (and increases) along the line from the parent vertex $v$ to 
the displaced vertex $v_{i,I,\delta_0}$.
Let the segment  of this line between $v_{i,I,\delta_0}$ and $v_{i,I,\delta}$ be $l_{\delta_0,\delta, I}$.
Let $l_{\epsilon}$ be a  straight line which contains $l_{\delta_0,\delta, I}$ and whose end points $a_{\epsilon}, b_{\epsilon}$ lie at a distance $\epsilon$ from $v_{i,I,\delta_0}, v_{i,I,\delta}$ respectively,
$\epsilon <<\delta_0, \delta$. Consider a small cylinder $C_{\epsilon, \tau}$ with axis $l_{\epsilon}$ and radius $\tau$, $\tau <<\delta, \delta_0$. Consider 2 such cylinders with parameters $\epsilon_1, \epsilon_2$ and $\tau_1, \tau_2$
with $\epsilon_1> \epsilon_2, \tau_1>\tau_2$ and with $\epsilon_1,\tau_1$ small enough  that $C_{\epsilon_1, \tau_1}$ does not intersect any edge emanating from $v$ apart from the $I$th one between $v$ and $v_{i,I,\delta_0}$. Consider the vector field $\xi^a = (\frac{\partial}{\partial y_3})^a$. Let $f$ be a function compactly supported in $C_{\epsilon_1, \tau_1}$ such that it is unity 
in $C_{\epsilon_2, \tau_2}$. Let $\phi (f\xi, t)$ be the 1 parameter set of diffeomorphisms generated by the vector field $f\xi^a$. Clearly, for an appropriate value of $t=t_0$ the diffeomorphism
$\phi (f\xi, t_0)\equiv \phi_{i,I,\delta,\delta_0}$
translates
$v_{i,I,\delta_0}$  to $v_{i,I,\delta}$ so that property (ii) is achieved.  This diffeomorphism also respects properties (v), (vi).

Next, note that within  $C_{\epsilon_2, \tau_2}$ this is a rigid translation so that the translated  edges at $v_{i,I,\delta}$ are straight lines in a small neighbourhood of $v_{i,I,\delta}$.
Hence within a small enough neighbourhood of $v_{i,I,\delta}$  we can now apply the `scrunching' diffeomorphism $G$ of  equation C.8, Appendix C4, P1. From that work we have that within a small neighbourhood of
$V_{i,I,\delta}$ of  $v_{i,I,\delta}$, $G$ acts as:
\begin{align}
(y^1(G(p))-y^1(v_{i, I, \delta }))  &  =\delta^{q-1}(y^1(p)-y^1(v_{i, I, \delta}))\nonumber\\
(y^2(G(p))-y^2(v_{i, I, \delta}))  &  =\delta^{q-1}(y^2(p)-y^2(v_{i, I,\delta }))\nonumber\\
(y^3(G(p))-y^3(v_{i, I,\delta }))  &  =(y^3(p)-y^3(v_{i, I,\delta })).
\label{scrunch}%
\end{align}
Here $q>>1$, $p$ is a point in  $V_{i,I,\delta}$, $y^i(p)$ refers to the $i$th coordinate value at $p$, $G(p)$ is the image of $p$ by $G$ and as mentioned above we have rotated our coordinates  so that $y^3$ runs along the line joining 
$v$ to $v_{i,I,\delta_0}$. Thus property (iv) is achieved. 
In addition, 
from P1, $G$ is identity outside a small neighbourhood of  $v_{i,I,\delta}$, and in particular is identity at all  the  edges of $c$ other than  the $I$th one at $v$,
maps the $I$th one at $v$ to itself (if $v$ is CGR in $c$ and if the $J\neq I$ edges are non-conducting  in $c$, then it maps the upper and lower $I$th conducting edges
to themselves) and  is identity in a 
neighbourhood of $v$. In addition, from 5., Appendix C, P1 the vector field generating $G$  (a) is supported only in an small neighborhood of  $v_{i,I,\delta}$   and (b) when  restricted to  the straight line from $v$ to  $v_{i,I,\delta_0}$ 
always points along this line wherever it is non-vanishing. Hence $G$  respects  properties (v), (vi).

Property (iii)(b) can be achieved in a similar way as (ii) by considering $\xi_J$ to be along the appropriate edge $e_J, {J\neq I}$  of $c$,  constructing suitable neighbourhoods of segments of this edge, smearing $\xi_J$ with 
suitable functions of compact support and using the finite diffeomorphisms $\phi_J$ generated by the resulting vector field to achieve the required result. Clearly these diffeomorphisms also respect properties (v), (vi). 
Property (iii)(a) can be achieved through the action of a diffeomorphism $\phi_{\alpha}$ which we shall construct at the end of this section.
The product of all the semianalytic diffeomorphisms $(\prod_{J\neq I}\phi_J) G\phi_{i,I,\delta,\delta_0}\phi_{\alpha}$ yields the required semianalytic diffeomorphism satisfying (i)-(v).
so that we have
\ba
\Phi^{\delta, Q, L,M,p_1,p_2,p_3  }_{c, \{y\}(i, I, \beta, \delta_0 )} &=& (\prod_{J\neq I}\phi_J) G\phi_{i,I,\delta,\delta_0}\phi_{\alpha},
\label{condifdef}
\\
|c_{(i,\beta,\delta)}\ket & = & {\hat U}(\Phi^{\delta, Q, L,M,p_1,p_2,p_3  }_{c, \{y\}(i, I, \beta, \delta_0 )}) |c_{(i,\beta,\delta_0)}\ket .
\label{contractket} 
\ea

Before we construct $\phi_{\alpha}$, it is useful for future purposes to derive equation (\ref{jydelta-y}) below. First note that from Appendix \ref{acone} and section \ref{secneg}, the displaced vertex is in a region covered by the coordinates $\{y\}$.
Next, consider the coordinate system $\{y^{(\delta)}\}$ obtained by the pushforward of the coordinates $\{y\}$ by the 
contraction diffeomorphism:
\be
\{y^{(\delta)}\} = (\Phi^{\delta, Q, L,M,p_1,p_2,p_3  }_{c, \{y\}(i, I, \beta, \delta_0 )})^* \{y\}.
\ee
From (\ref{contractket}) it follows that $\{y^{(\delta)}\}$ provides a coordinate patch around the displaced vertex 
of $c_{(i,I,\beta,\delta)}$. From the fact that $\phi_{\alpha}$ is the identity in a neighbourhood of the vertex $v_{i,I, \delta_0}$ (see the end of this section) together with the fact that
the displaced vertex $v_{i,I,\delta}$  and its immediate vicinity is obtained by a rigid translation 
followed by the linear transformation (\ref{scrunch}),  we have that the Jacobian
between the $\{y^{(\delta)}\}$ and $\{y\}$ coordinates at the displaced vertex is:
\ba
\frac{\partial y^{(\delta)\mu}(p)}{\partial y^\nu(p)}|_{p=v_{i,I,\delta}}  &= & \delta^{\mu}_{\nu} \delta^{-(q-1)} \;\;\; \mu =1,2\nonumber\\
                                                    &=&\delta^{\mu}_{\nu} \;\;\; \mu=3 .
\label{jydelta-y}
\ea                                                    
 Recall  that  the $\{y\}$ coordinates at $v$ are such that the $I$th edge at $v$ in $c$ runs along the 3rd coordinate direction.
To free us from this assumption let the coordinates at $v$ be $\{x\}$ with $\{x\}$ related to $\{y\}$  by a rotation $R_c$  which points $y^3$ along the $I$th edge. Then it is straightforward to see 
that the Jacobian between the coordinates $\{x^{(\delta)}\}:=   (\Phi^{\delta, Q, L,M,p_1,p_2,r  }_{c, \{y\}(i, I, \beta, \delta_0 )})^* \{x\}$ and the coordinates $\{x\}$ is:
\be
\frac{\partial x^{(\delta)\mu}(p)}{\partial x^\nu(p)}|_{p=v_{i,I,\delta}}   = (R_c G R_c^{-1})^{\mu}_{\;\;\nu}
\label{jxdelta-x}
\ee      
where 
\ba
G^{\mu}_{\;\;\nu} &= & \delta^{\mu}_{\nu} \delta^{-(q-1)} \;\;\; \mu =1,2\nonumber\\
                                                    &=&\delta^{\mu}_{\nu} \;\;\; \mu=3
\label{defG}
\ea
Note that  from property (v)  and from the fact that the $y^3$ coordinate direction coincides with the straight line joining $v$ to  $v_{i,I,\delta_0}$ (in the $\{y\}$ coordinates),
it follows that when restricted to  this straight line,   the 3rd coordinate of the coordinate system $\{y^{(\delta)}\}$ 
also points along this line (Indeed for the subset of this line lying within $V_{i,I,\delta}$, this fact can be explicitly verified from (\ref{scrunch}.).

Finally, we construct $\phi_{\alpha}$.  Let a  $C^1$ or $C^2$ kink  nearest to $v_{i,I,\delta_0}$ in $c_{(i,I,\beta, \delta_0)}$  be located at some ${\bar v}$. 
Consider 2  small spheres of radii $3\alpha_0, 2\alpha_0$ around this kink and  a semianalytic 
function which vanishes outside the larger sphere and is unity inside the smaller sphere. Smear the  dilatation vector field $\sum_{i=1}^3(y^i-  y^i({\bar v}))(\frac{\partial}{\partial y^i})^a$ with this function and
exponentiate the action of this vector field to obtain a 1 parameter family of semianalytic diffeomorphisms. Clearly for an appropriate  parameter value the size of the kink can be shrunk as required in (iii)(a)
to $\alpha$. Similarly shrink the second $C^1$ or $C^2$ kink if present. Let the diffeomorphism which shrinks these kinks be $\psi_{\alpha}$.
It is straightforward to see that the application of this diffeomorphism confines the departure from linearity,  of the edge carrying the kink,  to a sphere of radius $2\alpha$ around the kink. We shall 
choose $\alpha$ as required below.

Next, we need to ensure that the action of
 $\phi_{i,I,\delta,\delta_0}$ (which acts immediately after $\phi_{\alpha}$ in the contraction process) preserves the size of these kinks. Clearly we need only focus on any such kink if it is present
 at some ${\bar v}$ between $v_{i,I\delta_0}$ and $v$.
\footnote{Here we assume that we have chosen $\epsilon_1, \alpha$ small enough that any nearest $C^2$ or $C^1$ kink beyond $v_{i,I,\delta_0}$ does not intersect the cylinder $C_{\epsilon_1, \tau_1}$ which is used to define
 $\phi_{i,I,\delta,\delta_0}$.}
 If such a kink is present we use a construction similar to that for $\phi_{i,I,\delta,\delta_0}$ to move the kink to a distance $\bd <<\epsilon_2$ from  $v_{i,I\delta_0}$ through a rigid translation
 along the straight line joining $v$ to $v_{i,I, \delta_0}$, where $\epsilon_2$ has been defined above in the construction of $\phi_{i,I,\delta,\delta_0}$. 
 
More in detail, let ${\bar v}_1$ be on the straight line segment from $v$ to ${\bar v}$ at a distance $3\alpha$ from ${\bar v}$ with $\alpha <<\bd$.
\footnote{We choose $\bd$ to be much smaller than the distance between the kink ${\bar v}$ and $v_{i,I,\delta_0}$.}
Let $v_1$ be at a distance $\bd - 3\alpha$ from $v_{(i,I,\delta_0)}$ on the straight line segment from $v$ to  $v_{i,I,\delta_0}$.
Let the straight line  segment from ${\bar v}_1$ to $v_1$ be ${ l_1}$. Let ${l}_{1{\bar \epsilon}}$ be a straight line which contains  ${ l_1}$ and whose  end points ${\bar a}_{\bar{\epsilon}}, {\bar b}_{\bar{\epsilon}}$
lie at a distance ${\bar \epsilon}$ from ${\bar v}_{1}, v_{1}$ respectively. 
Consider a small cylinder $C_{{\bar \epsilon}, {\bar \tau}}$ with axis ${l}_{1{\bar \epsilon}}$ and radius ${\bar \tau}$. Consider 2 such cylinders with parameters ${\bar \epsilon}_1, {\bar \epsilon}_2$ and
${\bar \tau}_1, {\bar \tau}_2$ 
with ${\bar \epsilon}_1> {\bar \epsilon}_2, {\bar \tau}_1>{\bar \tau}_2$.   We shall further restrict ${\bar \epsilon}_1<<{\alpha}<<{\bar \tau}_2 <<\tau_2$.
Choose ${\bar \epsilon}_1, {\bar \tau}_1$ to be small enough that $C_{{\bar \epsilon}_1, {\bar \tau}_1}$ doesnt not intersect the graph underlying 
$c_{(i,I,\beta,\delta_0)}$ except along its edge  from $v$ to $v_{(i, I, \beta, \delta_0)}$.
Consider the vector field $\xi^a = (\frac{\partial}{\partial y_3})^a$. Let ${\bar f}$ be a function compactly supported in $C_{{\bar \epsilon}_1, {\bar \tau}_1}$ such that it is unity 
in $C_{{\bar \epsilon}_2, {\bar \tau}_2}$. Let $\phi ({\bar f}\xi, t)$ be the 1 parameter set of diffeomorphisms generated by the vector field $f\xi^a$. Clearly, for an appropriate value of $t=t_0$ the diffeomorphism
$\phi ({\bar f}\xi, t_0)\equiv {\bar \phi}$  translates the kink to its desired position. We set $\phi_{\alpha}:= {\bar \phi}\circ \psi_{\alpha}$. 


\subsection{\label{sec4.4} Discrete Action of Product of Operators }

\subsubsection{\label{sec4.4.1}Action on elements of the Ket Set}

Consider the operator product $\prod_{ {\rm i}=1 }^n {\hat O}_{ {\rm i} }(  N_{\rm i})$ where ${\hat O}^{\rm i}(N_{\rm i})$ is either a Hamiltonian  constraint operator or electric diffeomorphism constraint operator
smeared with Lagrange multiplier $N_{\rm i}$, with operators ordered such that ${\hat O}^{\rm i}(N_{\rm i})$ is to the  left of ${\hat O}^{\rm i}(N_{\rm j})$ if ${\rm i}<{\rm j}$ in the string of operators corresponding to the product.
We are interested in the action of a discrete approximant to this operator product on a state $c$ in the Ket Set.
The discrete approximant  we use is $\prod_{ {\rm i}=1 }^n {\hat O}_{ {\rm i} ,\delta_{\rm i}}(  N_{\rm i})$  where the discretization parameters $\delta_{\rm i}$ are such that  $\delta_{\rm i}< \delta_{\rm j}$ for ${\rm i}<{\rm j}$
and the action of ${\hat O}_{ {\rm i} ,\delta_{\rm i}}(  N_{\rm i})$ is given by (\ref{ham}) or (\ref{dn}) depending on whether 
${\hat O}_{ {\rm i} }(  N_{\rm i})$ is a Hamiltonian or Electric diffeomorphism constraint.
Recall that we did not adequately specify the coordinates with respect to which these individual  discrete actions were defined.
Here we shall do so indirectly through a number of steps. At the end of this multistep  procedure we shall have  a complete definition of:
\be
(\prod_{ {\rm i}=1 }^n {\hat O}_{ {\rm i} ,\delta_{\rm i}}(  N_{\rm i})) |c\ket
\label{prodact}
\ee
including a specification of the coordinates used.

\paragraph{\label{Step1}Step 1: (\ref{prodact}) as a weighted sum of Deformed States}

We define each discrete operator action in the product through (\ref{ham}) or (\ref{dn}) keeping the choice of coordinates as yet unspecified. Clearly the result is a weighted sum of deformed kets.
Our task in this step  is to find these weights. We shall use the  notation for deformed kets developed in section \ref{sec4.1}. 
Any deformed ket takes the form of an $m$th generation child of $c$, $1\leq m\leq n$. The deformation operation which produces this child from $c$ is specified by  the deformation sequence:
\ba
{   [ (i_{m-1}, I_{m-1},\beta_{\rm j_{m}}, \delta_{\rm j_{m}})
(i_{m-2}, I_{m-2}, \beta_{\rm j_{m-1}}, \delta_{ \rm j_{m-1}  }),.....,  (i_1,I_1, \beta_{\rm j_{2}}, \delta_{\rm j_{2}})
(i,I, \beta_{\rm j_{1}}, \delta_{\rm j_{1}}  ) ]}&  \nonumber\\
{\rm j_{k}} <  {\rm j_{l}} \;{\rm iff}\; k>l , \;\;\; {\rm j_{i}} \in \{1,..,n\} &
\label{m-childseq}
\ea
where if $m=1$ we only have the deformation $(i,I, \beta_{\rm j_1}, \delta_{\rm j_1})$. 
The $k$th deformation in this sequence is $(i_{k-1}, I_{k-1}, \beta_{\rm j_{k}}, \delta_{ \rm j_{k}  })$. It corresponds to the deformation generated by 
the operator ${\hat O}_{ {\rm j_{k}} ,\delta_{\rm  j_{k}}}(  N_{\rm j_{k}})$  on the $k-1$th generation child:
\be
c_{   [ (i_{k-2}, I_{k-2}, \beta_{\rm j_{k-1}}, \delta_{ \rm j_{k-1}  }),.....,(i,I, \beta_{\rm j_{1}}, \delta_{\rm j_{1}}) ]}
\label{m-child}
\ee
where if this operator is a Hamiltonian constraint we have chosen the flip $\beta_{\rm j_{k}}$ to define its action.
As is implicit in the discussion of section \ref{sec4.1}, given the deformation sequence (\ref{m-childseq}),
 the edges and internal charge indices at the nondegenerate vertex of the $k$th generation children are denoted with a subscript $k$ and the edges and internal charge indices of the parent vertex in  $c$
by $i,I$. The numbering scheme used is also that discussed in section \ref{sec4.1} so that 
the enumeration of edges of any child is related to that  for $c$. 

As emphasized before we have not yet made explicit our choices of coordinates with respect to which the deformations are defined. 
Let us see in more detail as to exactly where we need these choices to be made so as to provide a complete specification of  the deformed child (\ref{m-child}). 
Consider the deformation sequence (\ref{m-childseq}). The sequence
starts with the right most deformation $(i,I, \beta_{\rm j_{1}}, \delta_{\rm j_{1}}  )$ acting on the parent $c$. The singly deformed child it generates acts as the parent state for the next deformation
$(i_1,I_1, \beta_{\rm j_{2}}, \delta_{\rm j_{2}})$. In this way
proceeding from right to left, each successive deformation acts on the deformed state generated by the sequence to its right and produces a parent state for the deformation to its left.
Therefore, in order to specify each of these deformations we need to specify the coordinate patch for the non-degenerate vertex of the state produced by the deformation sequence to its right.
Hence  in order to specify the deformed child (\ref{m-child}) we need to specify coordinate patches for each of the deformed states in the `lineage' connecting (\ref{m-child}) to 
$c$.

The deformed states produced in (\ref{prodact}) consist of states  of the form (\ref{mchild}) for all choices of index sets $\{ {\rm i_1, i_2,.., i_m} \}$ such that the inequalities in the second line of (\ref{mchild}) hold, and for all 
${\rm m} \in \{1,..,n\}$. 
From our discussion in the previous paragraph, it follows that
for a complete specification of all the deformed states in this set, we need
a specification of a coordinate patch around each non-degenerate vertex for each $m$th generation child with $m$ ranging from $1,..,n-1$. In addition we must also, of course, specify the coordinate patch
around the nondegenerate vertex of $c$.  We shall see that in the final step of our procedure (see section \ref{Step3} below), we will have a specification of all these coordinate patches.

The notation (\ref{m-childseq}) is a cumbersome one. Hence, similar to (\ref{mabbrev}),
if there is no confusion in doing so, we will often find it convenient to abbreviate the deformation sequence in (\ref{mchild}) through:
\be
[i,I,\beta,\delta]_m\equiv {   [ (i_{m-1}, I_{m-1},\beta_{\rm j_{m}}, \delta_{\rm j_{m}}),..., (i,I, \beta_{\rm j_{1}}, \delta_{\rm j_{1}}  ) ]} .
\label{iibd}
\ee
Thus the deformed states  produced in (\ref{prodact}) consist of the states $c_{[i,I,\beta,\delta]_m}$ for all choices of $[i,I,\beta,\delta]_m$ and $m$ such that $ 1\leq m\leq n$. 
Clearly, (\ref{prodact}) can be expanded out as a sum over all these states so that we have 
\be
(\prod_{ {\rm i}=1 }^n {\hat O}_{ {\rm i} ,\delta_{\rm i}}(  N_{\rm i})) |c\ket = (\prod_{i=1}^n \delta_{\rm i})^{-1}\sum_{[i,I,\beta,\delta]_m, m=1,..,n} C_{[i,I,\beta,\delta]_m} |c_{[i,I,\beta,\delta]_m}\ket + C_0|c\ket
\label{sumCc}
\ee
Here the coefficients $C_{[i,I,\beta,\delta]_m}$ can be  computed using (\ref{ham}), (\ref{dn}) in  (\ref{prodact}). We do not need the explicit form of these coefficients here
so we refrain from displaying them. Instead we restrict ourselves to a few remarks regarding their structure. Each coefficient  is constructed out of the various factors which appear in each 
application of (\ref{ham}) or  (\ref{dn})  in (\ref{prodact}). In particular each coefficient has in it a product over all the `$n$' lapse functions in (\ref{prodact}). 
Each lapse is evaluated at a vertex of one of the  states in the lineage defined by the sequence using the coordinate patch specified at that vertex.
Note that this is the only  coordinate choice dependent feature of the coefficients. The remaining contributions come from various sign factors and overall $\hbar$ dependent numerical factors in (\ref{ham}), (\ref{dn}).
The sign factors  arise 
from the $\beta$ factors in (\ref{ham}) and from the fact that some of the actions of the constraints come from the `$-{\bf 1}$' term (see section \ref{sec3.2.1}) in  (\ref{ham}), (\ref{dn}).

To summarise: The discrete action of the operator product of interest on any ket  in the Ket Set can be written as a weighted sum over all its deformed children.  The weights (i.e. the coefficients
$C_{[i,I,\beta,\delta ]_m}, C_0$)  in this sum can be 
explicitly computed but we do not need an explicit computation in all generality for our purposes here. A complete evaluation of the coefficients and a complete specification of the deformed children requires
a choice of coordinate patch around each nondegenerate vertex of each child $c_{[i,I,\beta,\delta ]_m}, \forall [i,I, \beta, \delta]_m, m=1,..n$ as well as around the vertex of $c$.  
The coordinate choice dependence of each coefficient derives solely from its dependence on the density weighted lapse functions. 

In the next step we shall define each of the deformed children  of $c$ 
as the image of a corresponding deformation of the reference state $c_0$ by the reference diffeomorphism which maps $c_0$ to $c$.
Since we are interested in the continuum limit, it will suffice to define these deformations for small enough $\{\epsilon_{\rm i}, i=1,..,n\}$ where 
$\epsilon_{\rm i} <<\delta_{\rm i}, i=1,..n$ and  $\epsilon_{\rm i}< \epsilon_{\rm j}\: {\rm for}\;{\rm i} <{\rm j}$ from which we  define
\be
(\prod_{ {\rm i}=1 }^n {\hat O}_{ {\rm i} ,\epsilon_{\rm i}}(  N_{\rm i})) |c\ket = (\prod_{i=1}^n \epsilon_{\rm i})^{-1}\sum_{[i,I,\beta,\epsilon]_m, m=1,..,n} C_{[i,I,\beta,\epsilon]_m} |c_{[i,I,\beta,\epsilon]_m}\ket + C_0|c\ket
\label{sumCce}
\ee


\paragraph{\label{Step2}Step 2: Contraction  of deformed Reference States}

Each of the  deformed states   appearing on the right hand side of (\ref{sumCc}) is labelled by some deformation sequence $[i,I,\beta, \delta]_m$.
Replace  each such deformation sequence $[i,I,\beta, \delta]_m$,
\be
[i,I,\beta,\delta]_m\equiv {   [ (i_{m-1}, I_{m-1},\beta_{\rm j_{m}}, \delta_{\rm j_{m}}),..., (i,I, \beta_{\rm j_{1}}, \delta_{\rm j_{1}}  ) ]}
\label{cseqm} 
\ee
by the corresponding sequence:
\be
[i,I,\beta,\delta_0]_m\equiv {   [ (i_{m-1}, I_{m-1},\beta_{\rm j_{m}}, \delta_{0 \rm j_{m}}),..., (i,I, \beta_{\rm j_{1}}, \delta_{0 \rm j_{1}}  ) ]}, 
\label{c0seqm}
\ee
where we have chosen the partameters $\{\delta_{0{\rm j}}, {\rm j}=1,..,m\}$ to be suffciently small in a sense to be described below. Next, let $c_0$ be the reference state for $c$.
and consider the set $S_{\{\delta_{0\rm i}\},c_0}$  of all descendants  of $c_0$  obtained by deforming $c_0$ by all such correspondent sequences:
\be
 S_{\{\delta_{0\rm i}\},c_0}        :=  \{ c_{0[i,I,\beta,\delta_0]_m} \forall  [i,I,\beta,\delta_0]_m, m=0,1,..,n\}.
\label{setdeformd0}
\ee
Here the parameters $\{\delta_{0{\rm j}}, {\rm j}=1,..,m\}$ have been chosen sufficiently small that every element of $S_{\{\delta_{0\rm i}\},c_0}$ is a primary.   
Now, each element of the above set (apart from $c_0$) is some $m$th generation child of $c_0$. We define the coordinates with respect to which the multiple deformation sequence 
$[i,I,\beta,\delta_0]_m$ is constructed to be $\{x_0\}$. As discussed in the construction of primaries in section \ref{sec4.2}, for sufficiently small deformation parameters $\delta_{0{\rm i}}, {\rm i}= 1,..n$, these
deformations are well defined and the coordinate patch $\{x_0\}$ can be used as a linear coordinate patch for every non-degenerate vertex of every element of $S_{\{\delta_{0\rm i}\},c_0} $.

Next consider a set of deformation parameters $\{\epsilon_{\rm i}\}$ such that each $\epsilon_{\rm i} << \delta_{0{\rm i}}$  and   $\epsilon_{\rm i}< \epsilon_{\rm j}\: {\rm for}\;{\rm i} <{\rm j}$. 
Let us fix some particular deformation sequence

\be
[i,I,\beta,\delta_0]_m\equiv {   [ (i_{m-1}, I_{m-1},\beta_{\rm j_{m}}, \delta_{0\rm j_{m}}),..., (i,I, \beta_{\rm j_{1}}, \delta_{0\rm j_{1}}  ) ]}.
\label{fixseqd0}
\ee
and the corresponing sequence 
\be
[i,I,\beta,\epsilon]_m\equiv {   [ (i_{m-1}, I_{m-1},\beta_{\rm j_{m}}, \epsilon_{\rm j_{m}}),..., (i,I, \beta_{\rm j_{1}}, \epsilon_{\rm j_{1}}  ) ]}.
\label{fixseqe}
\ee
We now construct a contraction diffeomorphism which maps $c_{0[i,I,\beta,\delta_0]_m}$ to $c_{0[i,I,\beta,\epsilon]_m}$. This diffeomorphism will be constructed as a product of contraction diffeomorphisms
of the type defined in section 4.3.  We shall use the index notation as explained in Step 1 so that the subscript $k$ attached to the edge index
signifies that the edge in question is one which is obtained as 
a result of $k$ successive deformations; similary this subscript attached to the internal index of a $U(1)^3$ charge signifies that the  charge in question labels such a generation $k$ edge. 
Additionally we shall refer to the part of the deformation sequence from the 1st deformation to the $k$th one within the 
specific deformation sequence (\ref{fixseqd0}) as $[i,I,\beta, \delta_0]^k_m$ so that
\be
[i,I,\beta,\delta_0]^k_m\equiv {   [ (i_{k-1}, I_{k-1},\beta_{\rm j_{k}}, \delta_{0 \rm j_{k}}),..., (i,I, \beta_{\rm j_{1}}, \delta_{0 \rm j_{1}}  ) ]}, 
\ee
with the $k$th generation child produced in this sequence from the ancestor $c_0$  denoted as
\be
c_{0[i,I,\beta, \delta_0]^k_m }\equiv c_{0   [ (i_{k-1}, I_{k-1},\beta_{\rm j_{k}}, \delta_{0 \rm j_{k}}),..., (i,I, \beta_{\rm j_{1}}, \delta_{0 \rm j_{1}}  ) ]}.
\label{kmcodelta0}
\ee
The states $c_{0[i,I,\beta, \delta_0]^k_m }, k=1,..,m$  will be said to form the  lineage for the sequence (\ref{fixseqd0}). The states  $c_{0[i,I,\beta, \delta_0]^{k-1}_m }, c_{0[i,I,\beta, \delta_0]^k_m }$ will be called
`successive' with   $c_{0[i,I,\beta, \delta_0]^{k-1}_m }$        being the immediate parent of $c_{0[i,I,\beta, \delta_0]^{k}_m }$.
We shall use a similar notation  and language 
in relation to the deformation sequence (\ref{fixseqe}). 

Finally we introduce a hatted index notation as follows. Consider the $k$th transition \\$(i_{k-1}, I_{k-1}, \beta_{j_{\rm k}}, \delta_{0j_{\rm k}})$ which produces the child 
$c_{0[i,I,\beta, \delta_0]_m^k}$ from the parent $c_{0[i,I,\beta, \delta_0]_m^{k-1}}$. The edge indices at the non-degenerate vertex of the child are distinguished by the subscript $k$ and at that of the parent by $k-1$.
The parental edge along which the transition occurs is the $I_{k-1}$th one. Consider the edges $e_{J_k}$ in the child with $J_k\neq I_{k-1}$ (recall that the numbering of the edges of the child is correlated with that of the 
parent as described in section \ref{sec4.1}).  We shall denote such indices with a hat so that ${\hat J}_k$ signifies ${\hat J}_k\neq I_{k-1}$ in the above transition. Clearly, hatted indices index those edges which are 
non-conducting in $c_{0[i,I,\beta, \delta_0]_m^k}$  and any such edge connects the nondegenerate vertex of the child with a  $C^0$- kink.

Next, fix some $p>>1$, recall that $q>>1$ is defined in equation (\ref{scrunch}), and  proceed iteratively as follows:\\

\noindent (i)First consider the contraction diffeomorphism $\Phi^{\delta, Q, L,M,p_1,p_2,p_3  }_{c, \{y\}(i, I, \beta, \delta_0 )}$  and perform the replacements
\ba
\delta \rightarrow \epsilon_{\rm j_{1}}, & c \rightarrow c_0,  & \{y\} \rightarrow \{x_0\}  \;\;\;L, M \rightarrow {\hat J}_1, {\hat K}_1,\nonumber \\
p_1 \rightarrow \frac{2}{3}(q-1)  {\rm j_{1}}  p, & p_2 \rightarrow   \frac{2}{3}(q-1)  {\rm j_{1}} (p+1),   & p_3 \rightarrow \frac{2}{3}(q-1)  {\rm j_{1}} (p+1) + \frac{4}{3}(q-1){\rm j_{1}}
\label{1stpi}
\ea
We shall specify the factor $Q$ as we go along. $Q$ will depend on\\
(a) the operator sequence $S_{   {\rm j_1}  }   = \{ {\hat O_1} (N_1),.., {\hat O}_{\rm j_{1}}(N_{\rm j_{1}}\}$ starting from the 1st leftmost  operator  in the operator
product (\ref{prodact}) and terminating at the $j_{\rm 1}$th one,\\
(b) the charges of the child $c_{0(i,I, \beta_{\rm j_1}, \epsilon_{{\rm j_1}})}\equiv c_{0[i,I,\beta,\epsilon]_m^1}$ at  its nondegenerate vertex.\\
(c)the charges of the parent  $c_{0}$ at  its nondegenerate vertex.\\

Denoting $c_0$ by $c_{[i,I,\beta,\epsilon]_m^0}$ 
we denote this dependence through  
\be
Q\equiv Q(  c_{0[i,I,\beta,\epsilon]^{0,1}_m},                     S_{\rm j_{1}}, ).
\label{qnot}
\ee
Accordingly, we replace the label $Q$ by the label $S_{\rm j_{1}}$ and rewrite the contraction diffeomorphism as:
\be
\Phi^{\epsilon_{\rm j_{1}},  \{x_0\},    {\hat J}_1, {\hat K}_1 }_{c_{0[i,I,\beta,\epsilon]^{0}_m}, (i, I, \beta, \delta_0 ),     S_{\rm j_{1}}  }
\label{phi1}
\ee
The notation indicates that 
(a) the parent $c_0\equiv c_{   0[i,I,\beta,\epsilon]^{0}_m}$ is deformed through  $(i, I, \beta, \delta_0 )$ and the deformation parameter $\delta_0$ of the 
resulting child is contracted to the value $\epsilon_{j_1}$, both parameters being measured by the parental coordinates system $\{x_0\}$, (b) the $Q$ factor is that determined
by the charges of this child, its parent and the operator sequence  $S_{\rm j_{1}}$ (c) the kinks at ${\hat J_1}, {\hat K_1}$ are placed in accordance with the values of $p_1,p_2$ in (\ref{1stpi}).
Here to avoid notational clutter we have suppressed the labels $p_1,p_2,p_3$.

Note that the parental coordinate system $\{x_0\}$  covers a neighborhood of the displaced vertex of the uncontracted child $c_{0[i,I,\beta,\delta_0]^{1}_m}$. The diffeomorphism (\ref{phi1}) maps this displaced
vertex to its counterpart in the contracted child $c_{0[i,I,\beta,\epsilon]^{1}_m}$. Hence the 
  push forward of the parental coordinate system $\{x_0\}$  yields a coordinate system  around the displaced vertex in 
the child $c_{( i,I,\beta_{\rm j_{1}}, \epsilon_{ \rm j_{1}}) }\equiv c_{[i,I,\beta,\epsilon]^1_m}$.
Suppressing various dependences to avoid notational clutter and keeping in mind that we are discussing the contraction of the  {\em specific} deformation sequence (\ref{fixseqd0}) to that in (\ref{fixseqe}) 
we denote this coordinate system by $\{x_0^{\epsilon_{\rm j_{1}}}\}$:
\be
\{x_0^{\epsilon_{\rm j_{1}}}\}:= (\Phi^{\epsilon_{\rm j_{1}}, \{x_0\}, {\hat J}_1, {\hat K}_1}_{ c_{0 [i,I,\beta,\epsilon]^0_m},  (i, I, \beta, \delta_0 )         S_{\rm j_{1} }   })^* \{x_0\}
\label{x1}
\ee

We shall use this coordinate system associated with this first generation child to define the next transition in which this child acts as the parent for a second generation child in (ii) below.
\\

\noindent (ii)In  the contraction diffeomorphism $\Phi^{\delta, Q, L,M,p_1,p_2,p_3  }_{c, \{y\}(i, I, \beta, \delta_0 )}$  replace:
\ba
\delta \rightarrow \epsilon_{\rm j_{2}}, & c \rightarrow c_{0[i,I,\beta, \epsilon]^1_m}    ,  & \{y\} \rightarrow \{x_0^{\epsilon_{\rm j_{1}}}   \}  \;\;\;L, M \rightarrow {\hat J}_2, {\hat K}_2,\nonumber \\
p_1 \rightarrow \frac{2}{3}(q-1)  {\rm j_{2}}  p,  & p_2 \rightarrow   \frac{2}{3}(q-1)  {\rm j_{2}} (p+1),   & p_3 \rightarrow \frac{2}{3}(q-1)  {\rm j_{2}} (p+1) + \frac{4}{3}(q-1){\rm j_{2}}
\label{subii}
\ea
Similar to (i)  $Q$ depends on the operator sequence $S_{   {\rm j_2}  }   = \{ {\hat O_1} (N_1),.., {\hat O}_{\rm j_{2}}(N_{\rm j_{2}}\}$ starting from the 1st leftmost  operator  in the operator
product (\ref{prodact}) and terminating at the $j_{\rm 2}$th one, as well on the charges of the 
$[i,I, \beta, \epsilon]^2_m$  and $[i,I, \beta, \epsilon]^1_m$  children  of $c_0$ at their nondegenerate vertices so that 
 $Q\equiv Q(  c_{0[i,I,\beta,\epsilon]^{2,1}_m},S_{\rm j_{2}}) $ and we rewrite the contraction diffeomorphism as:
\be
\Phi^{ \epsilon_{\rm j_{2}}, \{x_0^{\epsilon_{\rm j_{1}}}     \}, {\hat J}_2, {\hat K}_2  }_{ c_{0[i,I,\beta,\epsilon]^{1}_m},   (i_1, I_1, \beta_{\rm j_2}, \delta_{0\rm j_2} ),        S_{\rm j_{2} }   }.
\label{phi2}
\ee
From the  substitution  $\{y\} \rightarrow \{x_0^{\epsilon_{\rm j_{1}}}   \} $ in (\ref{subii}) above, it follows that
the coordinate system with respect to which the discretization parameters $    \delta_{0 {\rm j_2} },   \epsilon_{  {\rm j_2} } $ are measured in the transition 
from the parent $c_{0[i,I,\beta,\epsilon]^1_m}$ to the child  $c_{0[i,I,\beta,\epsilon]^2_m}$ is the parental coordinate system $\{x_0^{\epsilon_{\rm j_{1}}}\}$ defined in (\ref{x1}).

More in detail, consider the uncontracted
images of this parent and child; these are the states $c_{0[i,I,\beta,\delta_0]^1_m}$, $c_{0[i,I,\beta,\delta_0]^2_m}$ of equation (\ref{kmcodelta0}).
Consider the image of {\em both} of these states by the contraction diffeomorphism  equation (\ref{phi1}). The image of the parent simply yields the parent $c_{0[i,I,\beta,\epsilon]^1_m}$  at 
parameter $\epsilon_{j_1}$ as described in (i). Clearly, by virtue of the properties of diffeomorphic images,
the image of the child $c_{0[i,I,\beta,\delta_0]^2_m}$ by this diffeomorphism defines a state which bears the same relation to its parent $c_{0[i,I,\beta,\epsilon]^1_m}$ as 
$c_{0[i,I,\beta,\delta_0]^2_m}$ bears to {\em its} parent $c_{0[i,I,\beta,\delta_0]^1_m}$. It follows that the image of $c_{0[i,I,\beta,\delta_0]^2_m}$ by (\ref{phi1}) is
a child at parameter $\delta_{0j_1}$  of $c_{0[i,I,\beta,\epsilon]^1_m}$ where the parameter $\delta_{0j_1}$  is now measured by 
{\em  the push forward coordinate system $\{x_0^{\epsilon_{\rm j_{1}}}\}$} of (\ref{x1}).  It follows that this child is obtained through the deformation $(i_1, I_1, \beta_{\rm j_2}, \delta_{0\rm j_2} )$
of its parent $c_{0[i,I,\beta,\epsilon]^1_m}$ with respect to the coordinates $\{x_0^{\epsilon_{\rm j_{1}}}\}$. The contraction diffeomorphism (\ref{phi2}) acts on this child, contracts the  parameter value
$\delta_{0\rm j_2}$ and 
produces the child $c_{0[i,I,\beta,\epsilon]^2_m}$ at parameter value $\epsilon_{j_2}$  with $\delta_{0\rm j_2},\epsilon_{j_2}$  measured by the parental coordinates $\{x_0^{\epsilon_{\rm j_{1}}}\}$
associated wuth the parent $c_{0[i,I,\beta,\epsilon]^1_m}$.

Clearly one can now define a coordinate 
around the nondegenerate vertex of $c_{0[i,I,\beta,\epsilon]^2_m}$ as the pushforward of this coordinate patch $\{x_0^{\epsilon_{\rm j_{1}}}\}$ by the contraction diffeomorphism of (\ref{phi2}) i.e. we define 
\ba
\{x_0^{\epsilon_{\rm j_{1}}   \epsilon_{\rm j_{2}}      }\} &:= &
(\Phi^{\epsilon_{\rm j_{2}}, \{x_0^{\epsilon_{\rm j_{1}}}      \}, {\hat J}_2, {\hat K}_2}_{ c_{0 [i,I,\beta,\epsilon]^{1}_m},  (i_1, I_1, \beta_{\rm j_2}, \delta_{0\rm j_2} ),            S_{\rm j_{2} }   }
)_* \{x_0^{\epsilon_{\rm j_{1}}}    \}\nonumber \\
&=&(
\Phi^{\epsilon_{\rm j_{2}}, \{x_0^{\epsilon_{\rm j_{1}}}      \}, {\hat J}_2, {\hat K}_2}_{ c_{0 [i,I,\beta,\epsilon]^{1}_m},  (i_1, I_1, \beta_{\rm j_2}, \delta_{0\rm j_2} ),            S_{\rm j_{2} }   }
\Phi^{\epsilon_{\rm j_{1}},  \{x_0\},    {\hat J}_1, {\hat K}_1 }_{c_{0[i,I,\beta,\epsilon]^{0}_m}, (i, I, \beta, \delta_0 ),     S_{\rm j_{1}}  }
)^* \{x_0\}
\label{x2}
\ea
This coordinate patch in  turn is used to define  the next transition in the sequence. We can then iterate this procedure. The structure obtained at the $k$th step is described in (iii).
\\

\noindent (iii) At the $k$th step 
the arguments of  $\Phi^{\delta, Q, L,M,p_1,p_2,p_3  }_{c, \{y\}(i, I, \beta, \delta_0 )}$   are replaced as:
\ba
\delta \rightarrow \epsilon_{\rm j_{k}}, & c \rightarrow c_{0[i,I,\beta, \epsilon]^{k-1}_m}    ,  & \{y\} \rightarrow \{x_0^{\epsilon_{\rm j_{1}}... \epsilon_{\rm j_{k-1}}}   \}  \;\;\;L, M \rightarrow {\hat J}_k, {\hat K}_k,\nonumber \\
p_1 \rightarrow \frac{2}{3}(q-1)  {\rm j_{k}}  p, & p_2 \rightarrow   \frac{2}{3}(q-1)  {\rm j_{k}} (p+1),   & p_3 \rightarrow \frac{2}{3}(q-1)  {\rm j_{k}} (p+1) + \frac{4}{3}(q-1){\rm j_{k}}
\label{replk}
\ea
with $Q$ depending  on the operator sequence $S_{   {\rm j_k}  }   = \{ {\hat O_1} (N_1),.., {\hat O}_{\rm j_{k}}(N_{\rm j_{k}})\}$ 
and on the charges of  $c_{0[i,I, \beta, \epsilon]^k_m}$,  $c_{0[i,I, \beta, \epsilon]^{k-1}_m}$  at their nondegenerate vertices so that 
 $Q\equiv Q(  c_{0[i,I,\beta,\epsilon]^{k,k-1}_m}, S_{\rm j_{k}})$ and we rewrite the contraction diffeomorphism as:
\be
\Phi^{\epsilon_{\rm j_{k}}, \{x_0^{\epsilon_{\rm j_{1}}.. \epsilon_{\rm j_{k-1}}         }     \}, {\hat J}_k, {\hat K}_k}_{ c_{0,  [i,I,\beta,\epsilon]^{k-1}_m},    (i_{k-1}, I_{k-1}, \beta_{\rm j_k}, \delta_{0{\rm j_k}}),  S_{\rm j_{k} }   }
\label{phik}
\ee
with the coordinate patch around the nondegenerate vertex of $c_{0[i,I,\beta,\epsilon]^k_m}$ defined to be
\ba
\{x_0^{\epsilon_{\rm j_{1}} ..  \epsilon_{\rm j_{k}}      }\} &:= &
(\Phi^{\epsilon_{\rm j_{k}}, \{x_0^{\epsilon_{\rm j_{1}}.. \epsilon_{\rm j_{k-1}}         }     \}, {\hat J}_k, {\hat K}_k}_{ c_{0,  [i,I,\beta,\epsilon]^{k-1}_m},    (i_{k-1}, I_{k-1}, \beta_{\rm j_k}, \delta_{0{\rm j_k}}),  S_{\rm j_{k} }   }
)^* \{x_0^{\epsilon_{\rm j_{1}} .. \epsilon_{\rm j_{k-1}}    }    \}\nonumber \\
&=&(\Phi^{\epsilon_{\rm j_{k}}, \{x_0^{\epsilon_{\rm j_{1}}.. \epsilon_{\rm j_{k-1}}         }     \}, {\hat J}_k, {\hat K}_k}_{ c_{0,  [i,I,\beta,\epsilon]^{k-1}_m},    (i_{k-1}, I_{k-1}, \beta_{\rm j_k}, \delta_{0{\rm j_k}}),  S_{\rm j_{k} }   }...
\Phi^{\epsilon_{\rm j_{1}},  \{x_0\},    {\hat J}_1, {\hat K}_1 }_{c_{0[i,I,\beta,\epsilon]^{0}_m}, (i, I, \beta, \delta_0 ),     S_{\rm j_{1}}  }
)^* \{x_0\}
\label{xk}
\ea
\\

\noindent (iv) Finally after the $m$th step we define the desired composite contraction diffeomorphism:
\be
\Phi^{ \epsilon_{\rm j_{1}} ..  \epsilon_{\rm j_{m}}, ({\hat J}_{1},{\hat K}_1),..,({\hat J}_m,{\hat K}_m)      }_{c_{0[i,I,\beta,\delta_0]_m}  , S_{\rm j_1}     }
:= 
(\prod_{k=2}^m \Phi^{\epsilon_{\rm j_{k}}, \{x_0^{\epsilon_{\rm j_{1}}   ..  \epsilon_{\rm j_{k-1}}  }      \}, {\hat J}_k, {\hat K}_k}_{ c_{0 , [i,I,\beta,\epsilon]^{k-1}_m},    (i_{k-1}, I_{k-1}, \beta_{\rm j_k}, \delta_{0{\rm j_k}})                          S_{\rm j_{k} }   })\;\;
\Phi^{\epsilon_{\rm j_{1}},  \{x_0\},    {\hat J}_1, {\hat K}_1 }_{c_{0[i,I,\beta,\epsilon]^{0}_m}, (i, I, \beta, \delta_0 ),     S_{\rm j_{1}}  }
\label{ccdiff}
\ee
where the product is ordered from right to left in increasing $k$ and where have labelled the left hand side by the sequence $S_{\rm j_1}$ because the the sequence $S_{\rm j_1}$ contains all $S_{\rm j_k}, k>1$ 
so that the label $S_{\rm j_1}$ subsumes the set of labels  $\{S_{\rm j_k}, k \geq 1\}$.
This composite contraction diffeomorphism contracts the `$\delta_0$' parameters to their corresponding `$\epsilon$' values so that we have
\be
|c_{0[i,I,\beta,\epsilon]_m }\ket := {\hat U} ( \Phi^{   \epsilon_{\rm j_{1}} ..  \epsilon_{\rm j_{m}},   ({\hat J}_{1},{\hat K}_1),..,({\hat J}_m,{\hat K}_m)                 }_{c_{0[i,I,\beta,\delta_0]_m} ,   S_{\rm j_{1}}    }) |c_{0[i,I,\beta,\delta_0]_m}\ket
\label{keteketd0}
\ee
where ${\hat U}({\Phi})$ refers to the unitary implementation of the diffeomorphism $\Phi$. The superscripts $\epsilon_{\rm j_{1}} ..  \epsilon_{\rm j_{m}}$ indicate that the deformations have been contracted down from
$\delta_{0\rm j_{1}}, .. , \delta_{0\rm j_{m}}$ in the deformation sequence (\ref{fixseqd0}) to $\epsilon_{\rm j_{1}}, .. ,  \epsilon_{\rm j_{m}}$  in the deformation sequence (\ref{fixseqe}).
The action of the deformation sequence (\ref{fixseqe}) on $c_0$ creates a series of $C^0$ kinks    in $c_{0[i,I,\beta,\epsilon]_m}$, one set of $(N-1)$ kinks for each deformation. The superscript $({\hat J}_{1},{\hat K}_1),..,({\hat J}_m,{\hat K}_m) $ 
in (\ref{keteketd0}) refers to the 2 preferred $C^0$ kinks created by each  such deformation. The preferred kinks $({\tilde v}_{{\hat J}_k}, {\tilde v}_{{\hat K}_k})$ created by the $k$th deformation 
are brought to the specific coordinate distances  specified by (iii) above (see also (iii) of section \ref{sec4.3}).
\footnote{\label{fnkink}The choice of these preferred kinks is made, at the moment, arbitrarily; in section \ref{sec6} we shall sum over these choices.}
These coordinate distances are measured by the coordinate system
$\{x_0^{ \epsilon_{\rm j_{1}}..  \epsilon_{\rm j_{k-1}}}\} $ associated with the non-degenerate vertex of the deformed state obtained by the action of the deformation $[i,I,\beta, \epsilon]_m^{k-1}$ on $c_0$, this vertex serving as the
parent vertex for the next deformation $(i_k,I_k,\beta_{\rm j_{k}}, \epsilon_{\rm j_{k}})$ in the sequence $[i,I,\beta, \epsilon]_m$. We shall see in section \ref{sec8} and \ref{sec9} that the placement of these kinks plays a
key role in obtaining an anomaly free algebra.


\paragraph{\label{Step3} Step 3:Deformed States as images of Contracted Reference States}

The reference state $c_0$ of Step 2 above is related to the state $c$ by the action of some  reference diffeomorphism $\alpha$ via the equation (\ref{c0calpha}).
We define:
\be
|c_{[i,I,\beta,\epsilon]_m }\ket := {\hat U}(\alpha)|c_{0[i,I,\beta,\epsilon]_m }\ket 
\label{alphachild}
\ee
This provides a definition of all the charge nets on the right hand side of equation (\ref{sumCce}).
Here the coordinate patch around the nondegenerate vertex of the state  $c_{[i,I,\beta,\epsilon]_m }$   
obtained through the action of  the deformation sequence $[i,I,\beta,\epsilon]_m$ on $c$ 
is defined  to be 
the image of the coordinate patch around the nondegenerate vertex of $c_{0[i,I,\beta,\epsilon]_m}$ (obtained by setting  $k=m$ in  (\ref{xk}))
by the diffeomorphism $\alpha$. 

%
Recall that  the coefficient $C_{[i,I,\beta,\epsilon]_m}$  in (\ref{sumCce}) acquires a coordinate dependence solely from the dependence of this coefficient on the lapse functions.
Each lapse function is evaluated at some nondegenerate vertex of one of the states  which define  the lineage of $c_{[i,I,\beta,\epsilon]_m}$. Since we have provided a unique choice of
coordinate patches for every such  vertex, every coefficient on the right hand side of (\ref{sumCce}) can be evaluated.

Having provided a unique and well defined evaluation of every coefficient in (\ref{sumCce}), we have a complete specification of the 
action of the operator product $\prod_{ {\rm i}=1 }^n {\hat O}_{ {\rm i} ,\epsilon_{\rm i}}(  N_{\rm i})$ for sufficiently small discretization parameters $\{\epsilon_{\rm i}, i=1,..,n\}$


%
\subsubsection{\label{sec4.4.2} Summary}

In order to define the discrete action of multiple constraint operators of a state $c$ in the Ket Set of section \ref{sec4.2}, it is necessary to define multiple deformations of $c$.
This is done in 3 stages. In the first stage, multiple deformations of the reference state $c_0$ are defined with respect to the reference coordinate coordinate system $\{x_0\}$ at
small enough parameter values as measured by $\{x_0\}$. These deformations are built out of a sequence of single deformations each constructed in detail along the lines
of Appendix \ref{acone} and section \ref{secneg1}.  
In the second stage, these parameters and the associated deformations are contracted through the action of contraction diffeomorphisms. This process involves the iterated  action of
individual contraction diffeomorphisms. The deformation which yields each  contracted child in a deformation sequence is then a deformation which is defined with respect to the coordinates
associated with the contracted parent. In the third stage, all the contracted children, now obtained at any small enough set of parameter values, are imaged by the reference diffeomorphism connecting $c_0$ to $c$
and these images define the desired multiple deformations of $c$.

In the second stage described in section \ref{sec4.4.1}, the deformation of the parental state $c_{0[i,I,\beta,\epsilon]^{k-1}_m}$ in a deformation sequence $[i,I,\beta,\epsilon]_m$ of (\ref{fixseqe}) by the deformation 
$(i_{k-1}, I_{k-1}, \beta_{\rm j_k}, \epsilon_{\rm j_k})$ yields the child $c_{0[i,I,\beta,\epsilon]^{k}_m}$. The deformation is defined with respect to the coordinate system $\{x_0^{ \epsilon_{\rm j_1},.., \epsilon_{\rm j_{k-1}}} \}$
associated with the parental state. 
In this manner the deformation which yields any child $c_{0[i,I,\beta,\epsilon]_m}$ through the specific deformation sequence ${[i,I,\beta,\epsilon]_m}$ applied to $c_0$ is 
uniquely and completely well defined in terms of the sequence of coordinate systems  $\{x_0^{\epsilon_{\rm j_1},.., \epsilon_{\rm j_{k}} }\}, k =1,2,..m-1,$ together with $\{x_0\}$.
Further,  the contraction procedure also results in the non-degenerate vertex of the child  $c_{0[i,I,\beta,\epsilon]_m}$ being equipped with the coordinates $\{x_0^{\epsilon_{\rm j_1},.., \epsilon_{\rm j_{m}} }\}$.
As is easy to check, the procedure used in the second stage to construct these coordinate patches for any deformation of $c_0$ only depends on the deformation sequence which defines the deformation.
Thus given $c_0$ and any deformation sequence, the deformed state comes equipped with a coordinate patch which is a pushforward of the reference coordinate patch $\{x_0\}$ associated with $c_0$ by
an appropriately constructed composite contraction diffeomorphism, this diffeomorphism being uniquely fixed by the specification of the deformation sequence (including the specification of the preferred set of $C^0$ kinks, 
see Footnote \ref{fnkink}).

In the third stage the images of each of these coordinate systems by the reference diffeomorphism which maps $c_0$ to $c$ yield  coordinate systems which provide  
a clear coordinate interpretation for the deformations generated by the discrete action of the 
operator product $\prod_{ {\rm i}=1 }^n {\hat O}_{ {\rm i} ,\epsilon_{\rm i}}(  N_{\rm i})$. 
In particular this procedure yields  a unique coordinate patch associated with the non-degenerate vertex of each state in the expansion (\ref{sumCce}).
It is useful to give these coordinate patches a name to distinguish them from
other coordinate patches we shall encounter.  We shall refer to the  coordinate patches associated with the non-degenerate vertex of each
of the deformed states which occur on the right hand side of (\ref{sumCce})  as {\em Contraction Coordinates} because of the role of contraction diffeomorphisms in their definition.
In section \ref{sec6} we shall encounter a different set of coordinates which we shall call {\em Reference Coordinates}.

Recall that the only coordinate dependence of the coefficients in the expansion (\ref{sumCce})  arises from the evaluation of lapse functions. The occurrence of these  lapse functions
traces back to the dependence of  the quantum shift on the lapse, this lapse being  evaluated with respect to the coordinates associated with the vertex of the state on which the quantum shift operator acts. 
Indeed, it is these coordinates in terms of which the deformations generated by the quantum shift are defined.
From this it is straightforward to see that the 
evaluation  of such a lapse function must be done in terms of the contraction coordinates associated with its argument. 

Next, let us discuss how the mapping, via contraction diffeomorphisms  in section  \ref{Step2}  and reference diffeomorphisms in section \ref{Step3}, of deformed Reference States can be interpreted 
as the discrete action of contraints. 
First consider the discrete action of  the constraint product of interest on a state $c=c_0$ which is itself a reference state
so that $\alpha ={\bf 1}$. 
Focus on some contracted child-parent pair $c_{0[i, I,\beta, \delta]^{m-1,m}_m}$ and the corresponding `primary' child-parent pair
$c_{0[i, I,\beta, \delta_0]^{m-1,m}_m}$. Recall from sections \ref{Step1} and \ref{Step2} that 
$c_{0[i, I,\beta, \delta]^{m-1}_m}, c_{0[i, I,\beta, \delta]_m}$ are the images of $c_{0[i, I,\beta, \delta_0]^{m-1}_m}, c_{0[i, I,\beta, \delta_0]_m}$
by appropriate composite contraction diffeomorphisms of the form (\ref{ccdiff}). We refer to these diffeomorphisms here, respectively,  as $\phi_{m-1}, \phi_{m}$. Also recall that  the actions of $\phi_m, \phi_{m-1}$ 
are related by that of a single contraction diffeomorphism constructed in section \ref{sec4.3}, which we denote by $\phi_1$ so that $\phi_m= \phi_1\phi_{m-1}$. We now show that the child 
$c_{0[i, I,\beta, \delta]^{m-1}_m}$ can be viewed as being generated by the discrete action of a constraint on the parent $c_{0[i, I,\beta, \delta]_m}$.  In our arguments below we shall initially refrain from creating and placing any 
$C^1,C^2$ kinks around the vertex of the child. We shall also set $\phi_{\alpha}= {\bf 1}$ in the defintion of the contraction diffeomorphism (see (\ref{condifdef})) which we have denoted here by $\phi_1$. We shall return to a discussion 
of the  placement of these kinks and justify this initial `switching off' of $\phi_{\alpha}$ after we complete our argumentation below.

First, let the parental (nondegenerate) vertex be GR without any need
for an intervention. That  the contracted child is created by the discrete action of a constraint on its parent in this case, follows immediately  from sections \ref{Step1} and \ref{Step2} and our discussion of contraction coordinates above. 
To see this we note the following using obvious notation:\\
(i) $c_{0[i, I,\beta, \delta]^{m-1}_m}  =\phi_{m-1}( c_{0[i, I,\beta, \delta_0]^{m-1}_m})$ is the parent of interest.\\
(ii) $\phi_{m-1}( c_{0[i, I,\beta, \delta_0]_m})$ is the deformed child generated by the discrete action of the appropriate constraint (Hamiltonian, if $\beta_{\rm j_m}\neq 0$ and electric diffeomorphism if $\beta_{\rm j_m}=0$) at parameter
$\delta_{0{\rm j_m}}$ with this parameter measured by the contraction coordinates the parent in (i).\\
(iii) $c_{0[i, I,\beta, \delta]_m}= \phi_m ( c_{0[i, I,\beta, \delta_0]_m}) = \phi_1(\phi_{m-1}( c_{0[i, I,\beta, \delta_0]_m}))$ is the contracted child obtained by contracting the child in (ii) from parameter value 
$\delta_{0{\rm j_m}}$ down to $\delta_{{\rm j_m}}$ where these parameters are measured by the contraction coordinates of the parent in (i).\\
It is important to note, from the construction of the contraction diffeomorphism in section \ref{sec4.3} that $\phi_1$ preserves the parent in (i) so that the process in (iii) can be viewed as a contraction of 
the child while preserving the identity of the parent.

Next consider the case where an intervention is required so that the parental vertex is either  CGR or GR of the type conforming to Case 2 section \ref{secneg2.1}.
As seen  in section \ref{secneg}, the transition from primary parent to primary child now requires an intervention. Let the intervention holonomy $h_{l_0}$ be based on the loop $l_0$. Then this transition 
unfolds through the following steps:\\
a) holonomy intervention by $h_{l_0}$  on the primary parent $c^{}_{ 0[i, I,\beta, \delta_0]^{m-1}_m }$ yielding
the parental state $c^{(l_0)}_{ 0[i, I,\beta, \delta_0]^{m-1}_m }$  with a  GR vertex, \\
(b) generation of the child $c^{(l_0)}_{0[i, I,\beta, \delta_0]_m}$, \\ 
(c) multiplication  by $h_{l_0}^{-1}$. \\
Recall  that we want to show that the  parent  $c_{0[i, I,\beta, \delta]^{m-1}_m}$ and child  $c_{0[i, I,\beta, \delta]_m}$ are related through the discrete action of a constraint. Such an action requires a holonomy intervention.  Since the parent
is the image of the primary parent by $\phi_{m-1}$, it follows that the loop $l$ labelling such an intervention must be the image of $l_0$ by the same diffeomorphism so that $l:= \phi_{m-1}(l_0)$. We may then view the child
$c_{0[i, I,\beta, \delta]_m}$ as being generated from its parents through the following steps, analogous to (a)- (c) above:\\
(a') A holonomy intervention by $h_{l}$ on the contracted parent  $c_{0[i, I,\beta, \delta]^{m-1}_m}= \phi_{m-1} (c_{0[i, I,\beta, \delta_0]^{m-1}_m})$   with $l=\phi_{m-1} (l_0)$. This intervention yields the 
state $\phi_{m-1}( c^{(l_0)}_{ 0[i, I,\beta, \delta_0]^{m-1}_m })$ with a GR vertex.\\
(b'1) Generation of the `$\delta_0$'  child $\phi_{m-1} ( c^{(l_0)}_{0[i, I,\beta, \delta_0]_m})$ from its parent $\phi_{m-1}( c^{(l_0)}_{ 0[i, I,\beta, \delta_0]^{m-1}_m })$, where 
$\delta_0$ is measured by the contraction coordinates $\phi_{m-1}^*\{x_0\}$ associated with the contracted parent $c_{0[i, I,\beta, \delta]^{m-1}_m}$. \\
(b'2)Contraction of this `$\delta_0$' child by the action of $\phi_1$ resulting in 
the  `$\delta$'  child $\phi_{m} ( c^{(l_0)}_{0[i, I,\beta, \delta_0]_m})$.\\
(c') multiplication by the inverse holonomy $h^{-1}_{l}$.\\
Clearly,  the  steps  (a')-(c') above can be viewed as corresponding to the discrete action of a constraint provided  the contraction diffeomorphism $\phi_1$ preserves the parental state
 $\phi_{m-1}( c^{(l_0)}_{ 0[i, I,\beta, \delta_0]^{m-1}_m })$ while contracting its child. If this is so  and if $\phi_1$ preserves $l$, it is easy to check that the steps (a')-(c') yield  
the child $c_{0[i, I,\beta, \delta]_m}$. 
\footnote{To see this note that $c_{0[i, I,\beta, \delta]_m}= \phi_m(     c_{0[i, I,\beta, \delta_0]_m} )=
\phi_m (h_{l_0}^{-1} c^{(l_0)}_{0[i, I,\beta, \delta_0]_m})
= h_{\phi_1\phi_{m-1} (l_0)}^{-1}( \phi_1 \phi_{m-1}( c^{(l_0)}_{0[i, I,\beta, \delta_0]_m}))= h^{-1}_{\phi_1(l)}( \phi_1 \phi_{m-1}( c^{(l_0)}_{0[i, I,\beta, \delta_0]_m}))$.}
Both of  these are ensured if  we choose the `cylinder' supports of the various diffeomorphisms  
constructed in section \ref{sec4.3}, whose product (\ref{condifdef}) yields the single contraction diffeomorphism denoted here by $\phi_1$, 
to be small enough that $\phi_1$  preserves the graph underlying $c_{0[i, I,\beta, \delta]^{m-1}_m}$  as well as the intervention loop $l$.
It is straightforward to check that these supports can be so chosen and we so choose them.

It only remains to discuss the placement of $C^1, C^2$ kinks. Any such kink, if present in the child, is either on an edge between the parental and displaced vertex or `beyond' the displaced vertex.
If there is a segment  beyond the displaced vertex this segment  must belong to the parental graph, and if an intervention is required, also belong to the straight line part of the intervention loop.
The contraction diffeomorphism (specifically the diffeomorphism $\psi_{\alpha}$ of section \ref{sec4.3})  preserves this part of the parental edge and the intervention loop.
The straight line joining the parental vertex to the displaced vertex must either be present or absent in its entirety in each of the following elements: the parental graph, the straight line
part of the intervening loop, the deformed graph prior to the kink placement. In each case the contraction diffeomorphism (specifically $\phi_{\alpha}$ of section \ref{sec4.3}) preserves this straight line.
Hence we may, as above, first  consider the deformations without kink placements (in which case $\phi_{\alpha}$ behaves as if it were the identity)   and then at the end place these kinks so as to mimic the 
result of imaging the primary child by $\phi_m$.
Since  $\phi_1$  also contracts the kink sizes, these kinks can be thought of as being placed by an appropriate holonomy which is an adequate approximant to identity to leading order in the {\em contracted} parameter value
as measured by the parental contraction coordinates.
This is why we had to demand and implement property (iii)(a) of the contraction diffeomorphism in section \ref{sec4.3}.
In this manner,  the procedure of constructing a  contracted child from its parent   can be thought of as being implemented by a discrete constraint action. 
Finally, in the case that $\alpha \neq {\bf 1}$ it is easy to see, by taking  the $\alpha$ image of the contracted child-parent pair, that the child can be thought of 
being generated by the  action of a discrete constraint action on the parent where the parental  coordinates are the $\alpha$ image of the contracted parent as in section \ref{Step3}.

{\bf Note}: We have slightly abused our notation for multiple deformations. The notation was set up so that each individual deformation was defined as in Appendix \ref{acone} and section \ref{secneg1}.
These individual deformations do place the displaced vertex at the correct location. However the $C^0$ kinks are positioned differently (they lie at distances of order of the deformation paramter $\delta$ rather than at 
the positions detailed in (iii), section \ref{sec4.3}). In this section 
(i.e. section \ref{sec4})  the contraction diffeomorphisms have been used not only to contract the displaced vertex to its desired poistion  but to also place the kinks at their desired positions (see (iii) of section \ref{sec4.3}) as well
as to `stiffen' the cone angle (see (\ref{defG}). Indeed such positioning and stiffening is more in line with the picture developed in P1,P2  of the deformation map $\varphi(q^i_I\vec{{\hat{e}}}_{I},\delta)$ (see (\ref{defd1cdefq}))
as a `singular' diffeomorphism which pulls the edges along the $I$th one.  
In section \ref{sec6} we shall augment the notation used in this section so as to include the specification of the preferred kink locations; the stiffening will be implicitly assumed  without recourse to explicit notation.

\subsection{\label{sec4.5}Action of Constraint Operators on state not in the Ket Set}

Since the Ket Set is closed under diffeomorphisms, any ket not in this set must have some diffeomorphism invariant characteristic which distinguishes it from elements of the Ket Set
We would like to define the action of constraint operators on such a ket so that this diffeomorphism invariant characteristic is preserved. However, since we do not explicitly know
what this characteristic is given such a ket, we use a blunt and inelegant definition of the deformations generated by the constraints on such a ket.
This definition deforms kets in such a way that the deformed offspring kets are in the complement of the Ket Set if the parent kets being deformed are also in the complement.
This can be done , for example, by defining the deformation map (see last line of section \ref{sec4.3} for a definition of the deformation map ) to  nontrivially knot one (or more) of the deformed edges 
at the offspring vertex. Another possibility would be to define the deformations to be `off edge' as in P1, P2.

In the remainder of the paper we assume that {\em some} such definition has been employed so as to ensure that the discrete action of constraint operators preserves the complement of the Ket Set.

\section{\label{sec5} The Anomaly Free Domain of States}
 
A state in the anomaly free domain resides in the algebraic dual space to the space of finite linear combinations of charge nets.
Such a state can be represented as a kinematically non-normalizable  sum over charge net bras. The anomaly free domain will be constructed as the linear span of basis states.
To each basis state we associate a set of bras such that the basis state is a sum over bras in this associated `Bra Set'. 
The set of kets corresponding to this Bra Set is a subset of the Ket Set we constructed in section \ref{sec4.2}. 
We discuss our choice of Bra Set in section \ref{sec5.1} and we
construct basis states in section \ref{sec5.2}. In what follows we often  denote the bra $\bra c|$ by $c$ to avoid notational clutter.

\subsection{\label{sec5.1}Bra Set}

Let $c_{P0}$ be the bra correspondent of  some primordial reference ket in the Ket Set of section \ref{sec4.2}. Consider the set of $N$ edges at the nondegenerate vertex $p_0$ of  $c_{P0}$ and the (unordered) set of 
$N$ $U(1)^3$ charge labels, one for each of these edges. 
Next consider the set  $S_{primordial,P0}$ of all primordial reference states each of whose elements  satisfy either of the restrictions below on their edge charge sets at $p_0$: \\
(i) the unordered set of edge charge labels at the vertex  $p_0$ of each such state is identical  to the corresponding set for $c_{P0}$. \\
(ii) there exists some flip $[i,\beta]_m$ such that the set of (unordered) edge charge labels at the vertex  $p_0$ of each such state is the flipped  image of  the corresponding set for $c_{P0}$ 
by this flip (here we have used the notation of (\ref{sumq}) for charge  flips).
\\

Recall that any primordial is subject to the restrictions detailed in section \ref{sec4.2}. Hence only those  states   
which have the prescribed
unordered edge charge sets of type (i) or (ii)  {\em and} satisfy these restrictions are elements of  $S_{primordial,P0}$. 
Next, fix an element $c_{{\bar P}0}$ of $S_{primordial,P0}$ and consider the set $S_{prim,{\bar P}0}$ of all its primaries (i.e. all  its children and itself) together with all  their  diffeomorphic images.
Consider the set  $B_{P0}$   of all elements of $S_{prim,{\bar P}0}$ as we vary $c_{{\bar P}0}$ over  $S_{primordial,P0}$. This set constitutes our Bra Set.

The set has the following property. 
Let $c\in B_{P0}$ and let $c_0$  be its reference state (we use the bra correspondents of the reference kets of section \ref{sec4.2} to define reference bras).  
Let $c_{P^{\prime}0}$ be a primordial reference state such that $c_0$ is a multiple deformation of $c_{P^{\prime}0}$.
Then the property of $B_{P0}$ alluded to is that $c_{P{\prime}0} \in S_{primordial,P0}$.

To see this, note the following. Since $c_0$ is diffeomorphically related to $c$, we have that $c_0$ is also in $B_{P0}$. Recall from section \ref{sec4.2} that  $c_0$ must be a primary because it is a reference state.
Hence it must be obtained as some multiple deformation of some reference primordial in the Ket Set.
From Appendix \ref{acolor}, the unordered {\em net} edge charge set at the non-degenerate vertex of $c_0$  is the same as some multiply flipped image of the unordered edge charge set of any reference primordial ancestor
whose multiple deformations give rise to $c_0$. By construction (see (i) and (ii) above), any such ancestor is in $B_{P0}$.


To appreciate the kind of situations covered by the proof  let us suppose that we  drop (i) and (ii) and choose the Bra Set to be composed of all diffeomorphic images of the primary family (including $c_{P0}$) emanating from $c_{P0}$.
As before the reference state $c_0$ for any element $c$ of this Bra Set must be a primary and hence obtained by some multiple deformation of some reference primordial in the Ket Set. Let this primordial be $c_{P^{\prime}0}$.
Consider the case where 
$c_0$ is obtained as a single electric diffeomorphism type deformation of $c_{P^{\prime}0}$. 
Next note that by construction it must be the case that  $c_0$ is diffeomorphic to a primary $c_{prim}$ emanating from $c_{P0}$.
Note also that from the kink structure of  $c_{prim}$, it must be the case that $c_{prim}$ is also a single  electric diffeomorphism deformation of $c_{P0}$.
\footnote{$c_{prim}$ must have $N-1$ $C^0$ kinks; any state with $m(N-1)$ such kinks is an $m$ deformation of a primordial.  Since $c_{prim}$ has only a single vertex of valence $N$ and none of valence $N+1$, this deformation
is of electric diffeomorphism type.}
If we could use this fact that $c_0$ is diffeomorphic to $c_{prim}$ to conclude that $c_{P^{\prime}0}$ is diffeomorphic to $c_{{ P}0}$  then, from the definition of (primordial) reference states, 
we could conclude that $c_{P^{\prime}0}$ and $c_{{ P}0}$ are identical.
Note however  that because the deformation is of an electric diffeomorphism type, the non-degenerate  vertex of $c_{P0}$ as well as the vertex structure in a small vicinity
of this vertex is absent from  the graph underlying $c_{prim}$ and, similarly, 
the non-degenerate  vertex of $c_{P^{\prime}0}$ as well as the vertex structure in a small vicinity
of this vertex is absent from  the graph underlying $c_0$. 
Hence we cannot directly conclude that the diffeomorphism which maps $c_0$ to $c_{prim}$ necessarily maps
$c_{P^{\prime}0}$  to $c_{{ P}0}$.
\footnote{We do not rule out that it may be possible to do so using a more involved argument; 
since we have not constructed any such argument, we prefer to cover the adverse consequences, sketched below,  of the possible absence of such an argument through our construction of $B_{P0}$
in the first paragraph.}

In the context above, 
the property $c_{P^{\prime}0} \in S_{primordial,P0}$ is crucial for the well defined-ness of the dual action of an electric diffeomorphism operator on anomaly free states associated with $B_{P0}$. As shall be apparent in sections \ref{sec8} and \ref{sec9},
for this  action to be  well defined, it must be the case that the discrete action of this operator on any charge net $c$  is such that either the (bra correspondents of) $c$  and all its single electric diffeomorphism deformations are absent in $B_{P0}$ or 
$c$ and all its deformations are {\em all} present in $B_{P0}$.  
If in the above example involving an electric diffeomorphism, we had that $c_{P^{\prime}0}$ above was not in $B_{P0}$, the fact that its first electric diffeomorphism deformation was in $B_{P0}$
would then lead to an ill-defined-ness  of the dual  action of an electric diffeomorphism operator on a typical anomaly free state associated with $B_{P0}$.

More in general the  restrictions (i), (ii) of the edge charge set of the primordials in $B_{P0}$  can be used to conclude  that all possible ancestors  of  any $c \in B_{P0}$ (by a possible ancestor we mean state
on which  multiple constraint actions lead to the creation of $c$) and all possible children of $c$ (by which we mean all  multiple deformations of $c$ generated by multiple constraint actions) are in $B_{P0}$ (here we freely switch between  ket and bra correspondents of the state $c$). 
To see this, note that by construction (see section \ref{sec4}) all possible ancestors and offspring of $c$ are in the Ket Set.
Recall again that all reference states must be primaries and 
consider for $c\in B_{P0}$ any ancestor $c_a$  of $c$ and its reference state $c_{a0}$. By definition of ancestry it must be true that 
deformations of this ancestor reference state  (with respect to $\{x_0\}$) yield a state diffeomorphic to the reference state, $c_0$, for $c$. 
It follows from Appendix \ref{acolor} that 
any  reference primordial for the reference state $c_{a0}$ of the ancestor must 
have an edge charge set related to that  of any  reference primordial ancestor of  $c_0$ by (i) or (ii). Since any  reference primordial for $c_0$ is in $B_{P0}$  so must any
reference primordial state for the ancestor reference state $c_{a0}$. It follows from the construction of $B_{P0}$ that the ancestor reference state and, hence the ancestor, must also be in $B_{P0}$.
Finally, note that by construction, if $c$ is in $B_{P0}$ then all its offspring are also in $B_{P0}$. This follows directly from the fact that by definition  any such offspring is diffeomorphic to 
a primary which emanates from the same primordial reference state as  one which yields the reference state $c_0$ for $c$.
The fact that all possible ancestors and offspring of any element of $c$ are necessarily in $B_{P0}$  ensures the well defined-ness of the dual actions, on anomaly free states associated with $B_{P0}$,
of those constraints which are necessary for a demonstration of anomaly free commutators.
\footnote{Of course we could have chosen the (bra correspondent) of entire Ket Set as our Bra Set as it would obviously satisfy the required property. However this would unnecessarily cut down on the size of the space of
anomaly free states.}

\subsection{\label{sec5.2} Basis States}

Let $f$ be a density weight $-\frac{1}{3}$ semianalytic function on the Cauchy Slice $\Sigma$ and let $h_{ab}$ be a semianalytic Riemannian metric such that $h_{ab}$ has no conformal symmetries.
Let $g$ be a function on $\Sigma^{m(N-1)}$ of the type specified in Appendix \ref{ag}. As detailed in Appendix \ref{adefg}, this function is determined by the network of geodesic distances, as defined by $h_{ab}$, between 
every pair of its arguments. Thus $g$ is determined once $h_{ab}$ is specified.

%
%
%
%

A basis state $\Psi_{f,h_{ab}, P_0}$ associated with the Bra Set $B_{P0}$ is constructed as a sum over all the elements of $B_{P0}$ where 
the coefficient of each such element  ${\bar c}$ in this sum is determined by $f,h_{ab}$ as follows.
Let the reference state for ${\bar c}$ be ${\bar c}_0$ and let the reference diffeomorphism which maps ${\bar c}_0$ to ${\bar c}$ be ${\bar \alpha}$ so that 
\be
|{\bar c}\ket = {\hat U} ({\bar \alpha} ) |{\bar c}_0\ket
\label{barcbarc0}
\ee
Since  the coordinate patch $\{x_0\}$ is associated with the non-degenerate vertex of ${\bar c}_0$, we define the coordinate patch associated with the non-degenerate vertex of ${\bar c}$  
to be:
\be
\{x_{{\bar \alpha}}\} :={\bar \alpha}^*\{x_0\} .
\label{defxalpha}
\ee
We shall refer to this coordinate  patch as a {\em Reference Coordinate Patch} to distinguish it from the {\em Contraction Coordinate Patches} defined at the end of  section \ref{sec4.4.2}.

Next, note that ${\bar c}_0$ is a primary and hence is either identical to , or obtained by, some multiple deformation of some reference primordial in $B_{P0}$.
While  it is possible, in principle, that this reference primordial is not unique
\footnote{While it may indeed be unique, we have not investigated the matter and hence must allow for this possibility.}, the number `$m$' of deformations of any primordial ancestor which 
yields ${\bar c}_0$ {\em is} unique. To see this, note that 
from the nature of the deformations detailed in sections \ref{sec3}, \ref{secgr} and \ref{secneg}, each single deformation generates a set of $N-1$ $C^0$ kinks. Hence the number of $C^0$ kinks in  ${\bar c}_0$,
and hence ${\bar c}$, must be $m(N-1)$ for some whole number $m$ which corresponds to the number of deformations of an appropriate reference primordial which yields ${\bar c}_0$ (If $m=0$, ${\bar c}$ is primordial).

Next, with respect to  the reference coordinates (\ref{defxalpha})  let us denote the (outward) unit edge tangents at the nondegenerate vertex $v_m$ of ${\bar c}$ by $\{{\hat e}^{a}_{I_m}, I_m=1,..,N\}$ where `unit' is with respect to  the (reference) 
coordinate norm. 
As discussed earlier if $v_m$ is CGR we shall count the upper and lower conducting edges as a single edge, where the notion of upper and lower is fixed from the kink structure in the vicinity of $v_m$ as outlined in section \ref{secneg}. 
For the conducting edge we may choose the (outward pointing) upper conducting  edge tangent.
\footnote{Since (\ref{defHi}) depends on the edge tangent norm, the choice of these orientations dont matter; we provide the above choices for concreteness.}
Define 
\be
H_{I_m} :=|| \vec{{\hat e}}_{I_m}|| := \sqrt{h_{ab}(v_m){\hat e}^{a}_{I_m}{\hat e}^{b}_{I_m}}, 
\label{defHi}
\ee
\be
h_{I_m}= \sum_{J_{m}, K_m\neq I_m} \frac{|| \vec{{\hat e}}_{J_m}||}{ || \vec{{\hat e}}_{K_m}|| }, 
\label{defhi}
\ee
and let $f(v_m)$ be the evaluation of the density weighted function $f$ at the vertex $v_m$ in the reference coordinates (i.e. in the coordinate system $\{x_{{\bar \alpha}}\}$).
Next,  consider the $m(N-1)$ $C^0$ kinks on ${\bar c}$. We evaluate $g_{\bar c}$ on the arguments corresponding to these kinks where we have defined $g_{\bar c}$ in Appendix \ref{ag}.  
Then the coefficient multiplying ${\bar c}$ in the sum over state representation of $\Psi_{f,h_{ab}, P_0}$ is:
\be
(\Psi_{f,h_{ab}, P_0}| {\bar c}> =   g_{{\bar c}}\;( \sum_{I_m} h_{I_m}H_{I_m}) \;f(v_m)
\label{psifghc}
\ee
where for any element of the algebraic dual $\Psi$, we write its action on a chargenet state $|b\ket$ as $(\Psi|b\ket$
\footnote{Recall an element of the algebraic dual is a linear map from the finite span of charge net states to the complex numbers.
We shall use the notation $\Psi$ or $(\Psi|$ for such elements depending on our convenience.}.
The formal sum over states representation of the  state $(\Psi_{f,h_{ab}, P_0}|$ is:
\be
(\Psi_{f,h_{ab}, P_0}| = \sum_{\bra {\bar c}| \in B_{P0} }  \big(g_{{\bar c}}\;( \sum_{I_m} h_{I_m}H_{I_m})\; f(v_m)\big)\bra {\bar c}|, 
\label{psifghsum}
\ee
where we have implicitly used equations (\ref{barcbarc0}), (\ref{defxalpha})  to evaluate  the quantities $g_{\bar c}, h_{I_m}, H_{I_m}, f(v_m) $ on the right hand side.

Finally we note  the following  {\em key property} of the right hand side of (\ref{psifghc}):\\

\noindent{\bf Invariance Property}:  Let the coordinates appropriate to the evaluation of the right hand side of (\ref{psifghc}) be defined through (\ref{defxalpha}) i.e. let the right hand side of (\ref{psifghc})  be evaluated with respect to
the Reference Coordinates for ${\bar c}$. Consider a second coordinate system $\{y\}$ around the 
non-degenerate vertex $v_m$ of ${\bar c}$ such that the Jacobian matrix
$J(\{x_{  {\bar \alpha}}  \}, 
\{y\})^{\mu}_{\nu} := \frac{\partial x_{\bar \alpha}^{\mu}}{\partial y^{\nu}}$ be such that its evaluation at $v_m$ is a constant times a rotation i.e.
\be
J(\{x_{\bar\alpha}\}, \{y\})^{\mu}_{\nu})|_{v_m} = C R^{\mu}_{\;\nu}
\ee
for some $C>0$ and some $SO(3)$ matrix $R$. Then the evaluation of the right hand side is the same whether the coordinates used are $\{x_{\bar \alpha}\}$ or $\{y\}$.\\

This is easily verified by inspection. It is straightforward to check that , in obvious notation, 
\be 
f(v_m)|_{\{y\}} = C^{-1} f(v_m)|_{\{x_{\bar \alpha}\}}, \;\; \;\; H_{I_m}|_{\{y\}}     = CH_{I_m}|_{\{x_{\bar\alpha}\}}.
\ee
The first equality follows from the density $-1/3$ nature of $f$ and the second from that fact the coordinate vector lengths are invariant under rotations of the coordinates and scale inversely  with scaling of the coordinates.
Further since $h_{I_m}$ involves ratios of norms of coordinate tangents it is invariant under such a transformation. Finally, from its definition in Appendix \ref{ag},  $g_{\bar c}$ is coordinate independent.

\section{\label{sec6}Continuum Limit: Final form and Contraction  behavior on Anomaly Free Domain}
In section \ref{sec4} we defined the contraction of deformations of states from larger discretization parameter to smaller ones using contraction diffeomorphisms.
The contraction moves the non-degenerate vertex from a larger coordinate distance  from its immediate parent vertex (in appropriate coordinates as explained in section \ref{sec4.4.2})  to a smaller one.
However the contraction also has a `fine structure' involving the positioning of the $C^0$ kinks generated by the transformation which produces  the state in question from its parent. 
Each choice of this fine structure yields an acceptable discrete approximant for the constraint action.  
In section \ref{sec6.1} we democratically sum over these fine structures and display our final choice of discrete approximant for  the action of a single constraint
in equations (\ref{hamsum}), (\ref{dnsum}) which replace equations (\ref{ham}), (\ref{dn}). Constraint product actions can then easily be defined
based on the machinery developed in section \ref{sec4}. In section \ref{sec6.2} we display the dual action of the constraint product on basis states in the anomaly free domain
and define its continuum limit. The contraction of deformations of kets is then transferred to that of the bras in the Bra Set of section \ref{sec5.1} and thence to the coefficients
which multiply these bras (see section \ref{sec5.2}). The evaluation of the continuum limit then depends on the contraction behavior of these coefficients. We detail this
behaviour  in section \ref{sec6.3}. A complete specification of the contraction behaviour requires a specification of the `$Q$' factors in the definition of the contraction 
diffeomorphism of section \ref{sec4.3}. We discuss this in section \ref{sec6.4}.
With this, we are ready to compute the continuum limit action of constraint products in sections \ref{sec7} and \ref{sec8}.

\subsection{\label{sec6.1}Final form of discrete constraint action}

The discrete action of the constraint product (\ref{sumCce}) is based on the single constraint actions (\ref{ham}),(\ref{dn}) at sufficiently small parameter value $\epsilon$ so that the single constraint actions are:
\be
\hat{C}[N]_{\epsilon}c  = \beta\frac{\hbar}{2\mathrm{i}}\frac{3}{4\pi}N(x(v))\nu_{v}^{-2/3}\sum_{I}\sum_{i}  
\frac{c_{(i,I,\beta, \epsilon)}- c}{\epsilon} ,
\label{hame}
\ee
\be
\hat{D}_{\epsilon}[\vec{N}_{i}]c   =\frac{\hbar}{\mathrm{i}}\frac{3}{4\pi}%
N(x(v))\nu_{v}^{-2/3}\sum_{I}   \frac{1}{\epsilon
}(   c_{   (  i, I, \beta=0, \epsilon   )  }  -   c   ).
\label{dne}
\ee
The deformed kets 
in (\ref{sumCce}) arise as a result of repeated  applications of (\ref{hame}), (\ref{dne}). These kets are defined through the contraction of
their images at larger discretization parameter values as explained in section \ref{sec4}. The contraction procedure involves a contraction of kink vertices to precisely defined locations.
These locations are specified by a choice of 2  edges in the child, $c_{   (  i, I, \beta, \epsilon   )  }$, each of which is distinct from the edge along which the child vertex is displaced
(this is reflected in the dependence of the contraction diffeomorphism on  the `hatted' indices   in, for example, equation (\ref{keteketd0}).
As a result, a deformed ket $c_{(i,I,\beta, \epsilon)}$  is characterised not only by the labels $(i,I,\beta, \epsilon)$ which describe the main features of the deformation
such as the location of the displaced vertex 
but also the labels ${\hat J_1}, {\hat K_1}$ which describe the fine structure of the location of the 
kinks.
\footnote{The subscript $1$ refers to the fact that the hatted indices number non-conducting edges of the child  $c_{   (  i, I, \beta=0, \epsilon   )  }$ which, here, is obtained by a {\em single} deformation its parent $c$;
see the discussion after equation (\ref{fixseqd0}) for the definition of  hatted indices.}
Hence a more complete notation  replaces the label set $(i,I,\beta, \epsilon)$ by  $(i,I,  {\hat J}_1, {\hat K}_1,   \beta, \epsilon )$. Of course a complete set of labels pertinent to multiply deformed kets is, for example,
that in  equation (\ref{keteketd0}). However, to display the single constraint actions in a more complete way
than in  (\ref{hame}), (\ref{dne}) 
it suffices to use
the abbreviated set of symbols $(i,I,{\hat J}_1, {\hat K}_1, \beta, \epsilon  )$ so that equations (\ref{hame}), (\ref{dne}) take the form:
\be
\hat{C}[N]_{\epsilon}c  = \beta\frac{\hbar}{2\mathrm{i}}\frac{3}{4\pi}N(x(v))\nu_{v}^{-2/3}\sum_{I}\sum_{i}  
\frac{c_{(i,I,{\hat J}_1, {\hat K}_1,\beta, \epsilon )}- c}{\epsilon} ,
\label{hamejk}
\ee
with $\beta=+1$ or $\beta =-1$, and
\be
\hat{D}_{\epsilon}[\vec{N}_{i}]c   =\frac{\hbar}{\mathrm{i}}\frac{3}{4\pi}%
N(x(v))\nu_{v}^{-2/3}\sum_{I}   \frac{1}{\epsilon
}(   c_{   (  i, I, {\hat J}_1, {\hat K}_1,\beta=0, \epsilon )  )  }  -   c   ).
\label{dnsejk}
\ee
Since each choice of hatted indices provides an acceptable discrete action which is derivable from the heuristics of section \ref{sec2}, summing over these choices also yields an acceptable 
discrete action. Accordingly we may repeat the considerations of section \ref{sec4} based  on the following  form of  single constraint actions:
\be
\hat{C}[N]_{\epsilon}c  = \beta\frac{\hbar}{2\mathrm{i}}\frac{3}{4\pi}N(x(v))\nu_{v}^{-2/3}\sum_{I}\sum_{{\hat J}_1, {\hat K}_1}\frac{1}{(N-1)(N-2)}   \sum_{i}  
\frac{c_{(i,I, {\hat J}_1, {\hat K}_1,\beta, \epsilon   )}- c}{\epsilon} ,
\label{hamsum}
\ee
with $\beta=+1$ or $\beta =-1$, and
\be
\hat{D}_{\epsilon}[\vec{N}_{i}]c   =\frac{\hbar}{\mathrm{i}}\frac{3}{4\pi}%
N(x(v))\nu_{v}^{-2/3}\sum_{I} \sum_{{\hat J}_1, {\hat K}_1}\frac{1}{(N-1)(N-2)}   \frac{1}{\epsilon
}(   c_{   (  i, I, {\hat J}_1, {\hat K}_1,\beta=0, \epsilon   )  }  -   c   ).
\label{dnsum}
\ee
The  $N(N-1)$ factors stem from the choice of 2 of the $N-1$ edges which bear the $C^0$ kinks created in the deformation of the parent. Equations (\ref{hamsum}), (\ref{dnsum}) are the final form of the single constraint actions which we shall use.
Once again, similar to  section \ref{sec4.4} we can expand the action of the constraint operator product 
$(\prod_{ {\rm i}=1 }^n {\hat O}_{ {\rm i} ,\epsilon_{\rm i}}(  N_{\rm i}))$ on the state $c$ through repeated applications of (\ref{hamsum}), (\ref{dnsum}) to obtain an expression of the form  (\ref{sumCce}) 
except that the label set must now, in obvious notation, be embellished
by the specification of the hatted indices so that we have 
\be
(\prod_{ {\rm i}=1 }^n {\hat O}_{ {\rm i} ,\epsilon_{\rm i}}(  N_{\rm i})) |c\ket = (\prod_{i=1}^n \epsilon_{\rm i})^{-1}\sum_{[i,I, {\hat J}, {\hat K}, \beta ,\epsilon ]_m, m=1,..,n} 
C_{[i,I,\beta,{\hat J} , {\hat K},\epsilon ]_m} |c_{[i,I,{\hat J}, {\hat K},\beta,\epsilon ]_m}\ket + C_0|c\ket
\label{sumCcehat}
\ee 
where we have abbreviated:
\be
[i,I,{\hat J}, {\hat K},\beta,\epsilon]_m\equiv {   [ (i_{m-1}, I_{m-1},\beta_{\rm j_{m}},  {\hat J}_{m}, {\hat K}_{m}, \epsilon_{\rm j_{m}}  ),..., 
(i,I, {\hat J}_1, {\hat K}_1 , \beta_{\rm j_{1}}, \epsilon_{\rm j_{1}} ) ]}.
\label{fixseqehat}
\ee
Each $c_{[i,I,{\hat J}, {\hat K},\beta,\epsilon ]_m}$ is defined exactly as in section \ref{sec4.3}.
Thus, each $c_{[i,I,{\hat J}, {\hat K},\beta,\epsilon ]_m}$ is 
the $\alpha$ image (where as before  $\alpha$ maps the reference ket $c_0$ to $c$) of the state 
$c_{0[i,I,{\hat J}, {\hat K},\beta,\epsilon ]_m}$  and each $c_{0[i,I, {\hat J}, {\hat K}, \beta,\epsilon]_m}$ is obtained as the image of the state 
$c_{0[i,I,\beta,\delta_0]_m}$ 
through  equation (\ref{keteketd0}), the state $c_{0[i,I,\beta,\delta_0]_m}$ being defined by repeated conical deformations with respect to the coordinate system $\{x_0\}$ , each of the type described in Appendix \ref{acone} and section \ref{secneg}. 
Note that the deformations  of Appendix \ref{acone} and section \ref{secneg} do not have a further fine structure labelled by `hatted indices' so that $c_{0[i,I,\beta,\delta_0]_m}$ is defined as the result of the deformation 
$[i,I,\beta,\delta_0]_m$ of equation (\ref{c0seqm}) applied to the reference state $c_0$.

The discussion of the coordinates underlying the deformed states $c_{[i,I,{\hat J}, {\hat K},\beta,\epsilon]_m}$ is exactly that of  sections \ref{Step3} and  \ref{sec4.4.2}. 
The coefficient $C_{ [i,I,{\hat J}, {\hat K},\beta,\epsilon]_m}$  in (\ref{sumCcehat}) acquires a coordinate dependence solely from the dependence of this coefficient on the lapse functions.
Each lapse function is evaluated at some nondegenerate vertex of one of the states  which define  the lineage of $c_{[i,I, {\hat J}, {\hat K}, \beta ,\epsilon ]_m}$. 
Since the considerations of section \ref{sec4.4}  (see especially sections \ref{Step3} and \ref{sec4.4.2}) have provided a unique choice of
coordinate patches for every such  vertex, every coefficient on the right hand side of (\ref{sumCcehat}) can be evaluated.

\subsection{\label{sec6.2}Dual Action on Anomaly free domain}
The dual action of $(\prod_{ {\rm i}=1 }^n {\hat O}_{ {\rm i} ,\epsilon_{\rm i}}(  N_{\rm i}))$ on a basis state $\Psi_{f,h_{ab}, P_0}$
is defined as 
\be
(\Psi_{f,h_{ab}, P_0}|
(\prod_{ {\rm i}=1 }^n {\hat O}_{ {\rm i} ,\epsilon_{\rm i}}(  N_{\rm i})) |c\ket
\label{dual}
\ee
The action of the operator product $(\prod_{ {\rm i}=1 }^n {\hat O}_{ {\rm i} }(  N_{\rm i}))$ 
is then defined by the continuum limit:
\be
(\lim_{\epsilon_n\rightarrow 0}(\lim_{\epsilon_{n-1}\rightarrow 0}...(\lim_{\epsilon_1\rightarrow0}
(\Psi_{f,h_{ab}, P_0}|
(\prod_{ {\rm i}=1 }^n {\hat O}_{ {\rm i} ,\epsilon_{\rm i}}(  N_{\rm i})) |c\ket)..))
\label{contlim}
\ee
The continuum limit action exists if equation (\ref{contlim}) holds for all charge net states $|c\ket$.
From section \ref{sec4.5}, this limit vanishes for all $c$ which are not in the Ket Set. Indeed, it follows from the discussion in section \ref{sec5.1}  that this limit also vanishes if (the bra correspondent of)  $c$ is not in the 
Bra Set $B_{P_0}$ associated with the anomaly free state $\Psi_{f,h_{ab}, P_0}$. Hence we need only analyse the continuum limit for states  $c$ (whose bra correspondents) are in the Bra Set
$B_{P_0}$.  For such states we expand out the  discrete operator product action as in (\ref{sumCcehat}), so that we have 
\ba
&(\Psi_{f,h_{ab}, P_0}|(\prod_{ {\rm i}=1 }^n {\hat O}_{ {\rm i} }(  N_{\rm i}))c\ket  \lim_{\epsilon_n\rightarrow 0}...(\lim_{\epsilon_1\rightarrow0}
(\Psi_{f,h_{ab}, P_0}|
(\prod_{ {\rm i}=1 }^n {\hat O}_{ {\rm i} ,\epsilon_{\rm i}}(  N_{\rm i})) |c\ket).. ) \nonumber \\
&= \lim_{\epsilon_n\rightarrow 0}(\lim_{\epsilon_{n-1}\rightarrow 0}...(\lim_{\epsilon_1\rightarrow0}
(\prod_{i=1}^n \epsilon_{\rm i})^{-1} \nonumber\\
&\big(\sum_{[i,I,{\hat J}, {\hat K},\beta,\epsilon]_m, m=1,..,n} 
C_{[i,I,{\hat J}, {\hat K},\beta,\epsilon ]_m}
(\Psi_{f,h_{ab}, P_0}
            |c_{[i,I,{\hat J}, {\hat K}\beta,\epsilon]_m}\ket + C_0   (\Psi_{f,h_{ab}, P_0}      |c\ket \big) \;)..) .
\label{contsum}
\ea
Clearly, in order to compute this limit we need to know the limiting behaviour of the coefficients  $C_{[i,I,{\hat J}, {\hat K}\beta,\epsilon ]_m}$  and of the `amplitudes' 
$(\Psi_{f,h_{ab}, P_0}
            |c_{[i,I,{\hat J}, {\hat K}, \beta,\epsilon ]_m}\ket$.  The  limiting behaviour of the coefficients  stems from the dependence of the coefficients on the coordinate dependent lapse function evaluations, these coordinates being
            dependent on the $\epsilon$- parameters. 
The limiting behaviour of the amplitudes can be computed from that of the expression (\ref{psifghc}) and the limiting behaviour of the functions $f,g$   and   the (reference) coordinate dependent unit edge tangents.
In the next section we compute this  limiting  `contraction' behavior of the amplitudes. 

\subsection{\label{sec6.3}Contraction behaviour of Amplitudes}

In this section we are interested in the behavior of
\be
(\Psi_{f,h_{ab}, P_0}|c_{[i,I, {\hat J}, {\hat K}, \beta ,\epsilon ]_m}\ket  
\label{amp}
\ee
for small $\epsilon_{\rm j_{m}}$.  We shall restrict attention to the particular deformation sequence $[i,I, {\hat J}, {\hat K}, \beta ,\epsilon ]_m$  in this section. As in sections \ref{sec4}, \ref{sec6.1}  the deformed state
$c_{[i,I, {\hat J}, {\hat K}, \beta ,\epsilon ]_m}$ will be assumed to have been generated by the discrete  action of the operator product $(\prod_{ {\rm i}=1 }^n {\hat O}_{ {\rm i} ,\epsilon_{\rm i}}(  N_{\rm i})) $
on the state $c$ (see (\ref{sumCcehat})).
Hence the sizes of the contraction parameters are defined with respect to the {\em Contraction Coordinates} of section \ref{sec4} (see the end of section \ref{sec4.4.2}).
More in detail, the contraction coordinates which specify the magnitude of $\epsilon_{\rm j_m}$  are those associated with the immediate parent  $c_{[i,I,\beta,\epsilon, {\hat J}, {\hat K}]^{m-1}_m}$ of the state $c_{[i,I, {\hat J}, {\hat K}, \beta ,\epsilon ]_m}$.
On the other hand, the amplitude (\ref{amp}) is evaluated  using the {\em Reference Coordinates} associated with ${\bar c}\equiv c_{[i,I, {\hat J}, {\hat K}, \beta ,\epsilon ]_m}$  in (\ref{psifghc}).

Therefore we proceed as follows. First  we transit  from the reference coordinates associated with $c_{[i,I, {\hat J}, {\hat K}, \beta ,\epsilon ]_m}$ to the contraction coordinates 
associated with this state. It turns out that the evaluation (\ref{psifghc}) is the same irrespective of which one of these coordinate systems we use. This is a {\em key} result and can be 
traced back to the definition of deformations developed in section \ref{sec4}. Next, using the fact (\ref{xk}) that contraction coordinates for a deformed state and its immediate parent
are related by a contraction diffeomorphism, we are able to compute the amplitude (\ref{amp}) in terms of the contraction coordinates of the immediate parent. Since the 
the size of the parameter $\epsilon_{\rm j_{m}}$ is measured by these coordinates, we are able to evaluate the small  $\epsilon_{\rm j_{m}}$ behaviour of this amplitude.

As noted in section \ref{sec6.2},  if $c \notin B_{P_0}$ then  all amplitudes on the right hand side of (\ref{contsum}) vanish. 
Hence hereon we  focus on the nontrivial case  $c \in B_{P_0}$ so that $c_{[i,I, {\hat J}, {\hat K}, \beta ,\epsilon ]_m} \in B_{P_0}$ .
\\

\noindent{\bf Step1: Transition from Reference to Contraction coordinates of $c_{[i,I, {\hat J}, {\hat K}, \beta ,\epsilon ]_m}$}:\\ 
Let the reference ket for the state $c_{[i,I, {\hat J}, {\hat K}, \beta ,\epsilon ]_m}$ be $(c_{[i,I, {\hat J}, {\hat K}, \beta ,\epsilon ]_m})_0$
Let the reference diffeomorphism be $\alpha_{[i,I, {\hat J}, {\hat K}, \beta ,\epsilon ]_m}$ so that similar to (\ref{c0calpha}) we have 
\be
 |c_{[i,I, {\hat J}, {\hat K}, \beta ,\epsilon ]_m}\ket := {\hat U}(\alpha_{[i,I, {\hat J}, {\hat K}, \beta ,\epsilon ]_m}) |   (c_{[i,I, {\hat J}, {\hat K}, \beta ,\epsilon ]_m})_0                \ket , 
\ee 
so that the associated reference coordinate system  around the non-degenerate vertex $v_m$ of $c_{[i,I, {\hat J}, {\hat K}, \beta ,\epsilon ]_m}$ is :
\be
(\alpha_{[i,I, {\hat J}, {\hat K}, \beta ,\epsilon ]_m})^*\{x_0\} .
\label{refxm}
\ee

We now turn to the contraction coordinates for $c_{[i,I, {\hat J}, {\hat K}, \beta ,\epsilon ]_m}$
Let the reference ket for $c$ be $c_0$ and let the reference diffeomorphism which maps $c_0$ to $c$ be $\alpha$
so that (\ref{c0calpha}) holds. Note that we have not restricted  $c_0$ to be a primordial.
The state  $c_{[i,I, {\hat J}, {\hat K}, \beta ,\epsilon ]_m}$ is obtained as the image of the state 
$c_{0[i,I, {\hat J}, {\hat K}, \beta ,\epsilon ]_m}$ by $\alpha$.   Recall that $c_{0[i,I, {\hat J}, {\hat K}, \beta ,\epsilon ]_m}$  is obtained through the action of 
a composite contraction diffeomorphism on the state 
$c_{0[i,I,\beta,\delta_0]_m}$  as in equation  (\ref{keteketd0}).
\footnote{In the more complete notation introduced in  section \ref{sec6.1} the left hand side states  in these equations  would also have a hatted indice specification.}
The state $c_{0[i,I,\beta,\delta_0]_m}$ is a primary which is obtained by deforming the reference state $c_0$  $m$ times, each these deformations being defined with respect to the reference coordinates $\{x_0\}$ 
and each of these deformations being of the type detailed in 
Appendix \ref{acone} and section \ref{secneg}.
\footnote{These deformations do not have the additional specification of hatted indices because the placement of the associated $C^0$ kinks
in Appendix \ref{acone} and section \ref{secneg} does not require this additional specification.} 
It follows  from (iv), section \ref{Step2} that the contraction coordinates around the nondegenerate
vertex of $c_{0[i,I, {\hat J}, {\hat K}, \beta ,\epsilon ]_m}$ are obtained as the image of the primary coordinates $\{x_0\}$ around the nondegenerate vertex of $c_{0[i,I,\beta,\delta_0]_m}$ by the
composite contraction diffeomorphism of  (\ref{keteketd0}) defined by (\ref{ccdiff}). We denote the contraction coordinates for $c_{0[i,I, {\hat J}, {\hat K}, \beta ,\epsilon ]_m}$ by
$\{x_0^{\epsilon_{\rm j_1}..     \epsilon_{\rm j_{m}}}        \}$.
\footnote{This is similar to the notation used in (\ref{xk}). Recall that (\ref{xk}) was defined for $k<m$. Extending the notation in (\ref{xk}) for $k=m$, it can be easily checked that 
(\ref{ccdiff}) together with (\ref{xk}) for $k=m-1$ imply equation (\ref{xk}) for $k=m$.}
We then have that:
\be
\{x_0^{\epsilon_{\rm j_1}..     \epsilon_{\rm j_{m}}}        \} := (\Phi^{ \epsilon_{\rm j_{1}} ..  \epsilon_{\rm j_{m}}, ({\hat J}_{1},{\hat K}_1),..,({\hat J}_m,{\hat K}_m)      }_{c_{[i,I,\beta,\delta_0]_m}  ,S_{\rm j_1}    })^*\{x_0\}
\label{xm}
\ee
%
and that the contraction coordinates for $c_{[i,I, {\hat J}, {\hat K}, \beta ,\epsilon ]_m}$ are:
\ba
\{x_{\alpha}^{\epsilon_{\rm j_1}..     \epsilon_{\rm j_{m}}}        \} & :=& \alpha^*\{x_0^{\epsilon_{\rm j_1}..     \epsilon_{\rm j_{m}}}        \} \nonumber \\
&=& \alpha^*(\Phi^{ \epsilon_{\rm j_{1}} ..  \epsilon_{\rm j_{m}}, ({\hat J}_{1},{\hat K}_1),..,({\hat J}_m,{\hat K}_m)      }_{c_{[i,I,\beta,\delta_0]_m}  , S_{\rm j_1}    })^*\{x_0\}
\label{conxm}
\ea
%

Our task is to compute the Jacobian between the reference coordinates (\ref{refxm}) and the contraction coordinates (\ref{conxm}). This is computed in the Appendix \ref{ajacob}
wherein it is shown that  the Jacobian between the 2 coordinate systems takes the form of a constant times a rotation matrix.
From  the Invariance Property of section \ref{sec5.2} it then  follows that we can as well evaluate (\ref{amp}) using the  contraction  coordinates (\ref{conxm}).\\

\noindent{\bf Step 2:Transition to contraction coordinates of immediate parent}:\\

The immediate parent of $c_{[i,I, {\hat J}, {\hat K}, \beta ,\epsilon ]_m}$ is $c_{[i,I,\beta,\epsilon, {\hat J}, {\hat K}]^{m-1}_m}$. The contraction coordinates for this immediate parent are
the $\alpha$ image of those for the state $c_{0[i,I,\beta,\epsilon, {\hat J}, {\hat K}]^{m-1}_m}$. Accordingly,  taking the $\alpha$ image of  (\ref{xk} ) with $k=m-1$,  we have that
these contraction coordinates are:
\be
\{x_{\alpha}^{\epsilon_{\rm j_1}..     \epsilon_{\rm j_{m-1}}}        \} :=\alpha^*\{x_0^{\epsilon_{\rm j_1}..     \epsilon_{\rm j_{m-1}}}        \} \nonumber \\
\label{conxm-1}
\ee
The relationship between  the contraction coordinates of the child -parent pair  $c_{0[i,I, {\hat J}, {\hat K}, \beta ,\epsilon ]_m}$, \\$c_{0[i,I,\beta,\epsilon, {\hat J}, {\hat K}]^{m-1}_m}$
can be readily inferred from
equations (\ref{xk}) (with $k=m-1$), (\ref{xm}),(\ref{ccdiff}), (\ref{conxm}) and (\ref{conxm-1}), so that we have that 
\ba
&\{x_{\alpha}^{\epsilon_{\rm j_1}..     \epsilon_{\rm j_{m}}}        \}=
\alpha^*\{x_0^{\epsilon_{\rm j_1}..     \epsilon_{\rm j_{m}}}        \} & \label{xm1}\\
&=\alpha^*
(\Phi^{\epsilon_{\rm j_{m}}, \{x_0^{\epsilon_{\rm j_{1}}.. \epsilon_{\rm j_{m-1}}         }     \}, {\hat J}_m, {\hat K}_m}_{ c_{0 [i,I,\beta,\epsilon]^{m-1}_m}, (i_{m-1}, I_{m-1}, \beta_{\rm j_m}, \delta_{0{\rm j_m}})    , S_{\rm j_{m} }   })^*
\{x_0^{\epsilon_{\rm j_1}..     \epsilon_{\rm j_{m-1}}}        \}.&
\label{xm2}
\ea
The above equation simply expresses the ($\alpha$ image of the) fact that  the contraction coordinates of any deformed state and its immediate parent are related by the action of  a contraction diffeomorphism defined in section \ref{sec4.3}.
Indeed the iterative procedure used in section \ref{Step2} implements this very fact. Next,
from the fact that for any diffeomorphism $\gamma$ and any coordinate systems $\{x\}, \{y\}$ we have that 
\be
\frac{\partial (\gamma^*x)^{\mu}}{\partial (\gamma^*y)^{\nu}}|_{\gamma (p)}=: \frac{\partial x^{\mu}}{\partial y^{\nu}}|_{(p)}, 
\ee
it follows that the Jacobian between the contraction coordinates of offspring and immediate parent is given exactly by that of equation (\ref{jxdelta-x}) with 
the identifications 
\be
x^{\delta}\equiv x_{\alpha}^{\epsilon_{\rm j_1}..     \epsilon_{\rm j_{m}}}        \;\;\;\; x\equiv x_{\alpha}^{\epsilon_{\rm j_1}..     \epsilon_{\rm j_{m-1}}} \;\;\;
\delta \equiv  \epsilon_{\rm j_{m}}  
\label{appnote1}
\ee
Recall from the discussion at the beginning of this subsection as well as from section \ref{sec4.4.2} that the contraction  parameter $\epsilon_{\rm j_{m}}$ is measured with respect to the parental contraction coordinates 
$x_{\alpha}^{\epsilon_{\rm j_1}..     \epsilon_{\rm j_{m-1}}}$. Hence the contraction behaviour of the amplitude can be inferred from the behaviour of 
the quantities  $h_{ I_{\rm j_m}},H_{ I_{\rm j_m} },f,g_{c_{[i,I, {\hat J}, {\hat K}, \beta ,\epsilon ]_m}}$ (see section \ref{sec4}) in these parental contraction coordinates.
This a straightforward though lengthy exercise and we relegate it to the Appendices. Specifically, we compute the contraction behaviour of the first three quantities   in Appendix \ref{aconb} 
using the Jacobian in equation (\ref{jxdelta-x})  and that of the last quantity in Appendix \ref{acong}.

\subsection{\label{sec6.4}Specification of $Q$ factors} 

Recall that $Q$ is one of the parameters specifying a contraction diffeomorphism (see (iii), section \ref{sec4.3}). We had briefly discussed its specification in Step 2 of section \ref{Step2}.
Here we summarise its dependences (see (\ref{qnot})) in a bit more detail. Our explicit choices for $Q$ will be displayed in section \ref{sec7} and \ref{sec8} wherein the rationale for these choices
will be self evident.

Recall that we are interested in a state which is produced by the action of some specific  product of Hamiltonian and electric diffeomorphism constraint operators (\ref{prodact}) on a state $c$.
This state is produced through some deformation sequence applied to its ancestor $c$. We are interested in the contraction of this state to its immediate parent in this deformation sequence.
Using the notation of section \ref{sec4.3}, let the state be an $m$th generation offspring  $c_{[i,I,{\hat J}, {\hat K}, \beta, \delta]_m}$ where we have used the augmented notation with the `hatted' indices as 
explained in (\ref{fixseqehat}), and let its parent be  $c_{[i,I,{\hat J}, {\hat K}, \beta, \delta]^{m-1}_m}$ and let the parameter being contracted away be $\delta_{\rm j_m}$.
Then $Q$ depends on the {\em net} edge charges of the child and the parent at their non-degenerate vertices
as well as the  sequence of operators starting from the first operator to the  ${\rm j_m}$th one i.e. on the sequence:
\be
S_{\rm j_m}= \prod_{i=1}^{\delta_{\rm j_m}} {\hat O}_{{\rm i},\delta_{\rm i}}(N_i) .
\label{defsq1}
\ee
Since the charges at the vertices of $c_{[i,I,{\hat J}, {\hat K}, \beta, \delta]_m}$, $c_{[i,I,{\hat J}, {\hat K}, \beta, \delta]^{m-1}_m}$  are the same as the charges on their (diffeomorphically related)
`$\delta_0$' counterparts, we express the dependence $Q$ for this contraction in the following equivalent notations:
\be 
Q(  c_{[i,I,\beta,\delta ]^{m-1,m}_m},S_{\rm j_{m}} )\equiv 
Q(  c_{0[i,I,\beta,\delta_0 ]^{m-1,m}_m},S_{\rm j_{m}} )\equiv 
Q(   c_{0[i,I,\beta,\delta_0]_m} ,            c_{0[i,I,\beta,\delta_0]^{m-1}_m}, S_{\rm j_m})
\label{qnot1}
\ee

The individual constraint operators in the above sequence can be of 2 types, namely a Hamiltonian constraint or an electric diffeomorphism constraint in the $k$th $U(1)^3$ direction with $k=1,2,3$.
Let us denote these constraint types as $h$ and  $d_k, k=1,2,3$ respectively so that the constraint type of an operator ${\hat C}(N)$ is $h$ and that of ${\hat D}(N_k)$ is $d_k$.
We shall say that the constraint type $t_{\rm i}$  of the operator ${\hat O}_{{\rm i}}(N_i)$ (or of its discrete approximant ${\hat O}_{{\rm i},\delta_{\rm i}}(N_i)$ is either $h$ or $d_k$.
It then turns out that the information in $S_{\rm j_m}$ relevant for the specification of $Q$ is the sequence of {\em constraint types} $(t_1,...t_{\rm j_m})$.
We shall re-designate the symbol $S_{\rm j_m}$ to denote the ordered set of constraint types $(t_1,...t_{\rm j_m})$: 
\be
S_{\rm j_m}\equiv (t_1,...t_{\rm j_m}).
\label{defsqt}
\ee
Henceforth we shall interpret  $S_{\rm j_m}$ in   (\ref{qnot1}) through (\ref{defsqt}). 

$Q$ also depends on the cone angle $\theta$ which characterises the conical deformations of reference states, these deformations being constructed using the Primary Coordinates $\{x_0\}$.
This cone angle is fixed for all deformations of reference states in the Bra Set which labels an anomaly free state. To avoid notational clutter we will suppress explicit notational reference to the dependence of 
$Q$ on $\theta$.

\section{\label{sec7} Anomaly free product of 2 Hamiltonian constraints} 

In section \ref{sec7.1} we compute the continuum limit action of a product of two Hamiltonian constraints. In section \ref{sec7.2} we compute the commutator between two electric diffeomorphism constraints
and thereby demonstrate the anomaly free nature of the commutator between the pair of Hamiltonian constraints whose product is computed in section \ref{sec7.1}.
The computations are long but straightforward. We shall only highlight the main steps. 

\noindent{\bf NOTE}: In the remainder of the main body of the paper, unless mentioned otherwise, all the edge charges considered will be {\em net} charges where, as in Appendix \ref{acolor} we define the net charge as follows:
\\

\noindent{\em Definition: Net Edge Charge:} The {\em net charge}  $q_{net\;I}^i$ on a conducting edge $e_I$  at the nondegenerate vertex of a charge net is  the {\em sum} of 
of the {\em outgoing} upper  and lower conducting charges; 
if the edge $e_I$  is non-conducting we shall define its 
lower conducting charge to be zero so that the net charge $q^i_{net\;I}$ on such an edge is just its outgoing charge $q^i_I$.
\\

In what follows we shall drop the `$net$' subscript; all charges henceforth, unless mentioned otherwise, will be net charges and the net charge associated with an $I$th edge will be denoted simply by $q^i_I$.

\subsection{\label{sec7.1}Product of 2 Hamiltonian constraints.}
\subsubsection{\label{sec7.1.1}Notation}
We  compute the continuum limit (\ref{contlim}) when $n=2$ and ${\hat O}_{\rm i}(N_{\rm i}), {\rm i}=1,2$ are Hamiltonian constraint operators.
We restrict attention to  the case that  $c$  in (\ref{contlim})  is in the Bra Set because, as mentioned in section \ref{sec6.1}, for $c$ not in the Bra Set, the dual action vanishes.
In equation (\ref{contlim}), we set:
\ba
&N_1 \equiv M \;\; \beta_1\equiv\beta_M \;\; N_2 \equiv N\;\;\; \beta_2\equiv \beta_N &\nonumber \\
&\epsilon_1 \equiv \bd  \;\;\;  \epsilon_2 \equiv \delta &
\label{not71}
\ea
As we shall see, the discrete action of this Hamiltonian constraint operator product generates the doubly deformed states $c_{[i,I,{\hat J}, {\hat K},\beta,\delta]_2}$, the singly deformed states 
$c_{[i,I,{\hat J}, {\hat K}\beta, \delta]_1}$ and $c_{(j,J,{\hat R}_1, {\hat S}_1,\beta_M , \bd)}$,  and $c$. Here we have defined the transitions:
\ba
[i,I,{\hat J}, {\hat K} ,\beta, \delta]_2 &=& [(i_1, I_1, {\hat J}_2, {\hat K_2}, \beta_M, \bd), (i, I, {\hat J}_1, {\hat K}_1, \beta_N, \delta )] \label{c2not}
\\
\;   [ i,I,{\hat J}, {\hat K}, \beta,\delta]_1 &= & (i, I, {\hat J}_1, {\hat K}_1, \beta_N, \delta ) 
\label{c1not}
\ea
The singly deformed state $c_{(j,J,{\hat R}_1, {\hat S}_1,\beta_M , \bd)}$  is distinct from the singly deformed state (\ref{c1not}) and is obtained through the deformation 
$(j,J,{\hat R}_1, {\hat S}_1,\beta_M , \bd)$ of $c$. In particular, the parameter for this transformation is $\bd$ whereas that for (\ref{c1not}) is $\delta$.

The above transitions are exactly those described in sections \ref{sec4.4} augmented with the `hatted indices' of (\ref{fixseqehat}). To see this, use (\ref{not71}). It is then straightforward to see that
by setting , in (\ref{fixseqehat}), 
\ba 
m=2,\; j_2=1, \;  j_1=2, &  \;\;{\rm we \;\;obtain\;} &\; [i,I,{\hat J}, {\hat K}, \beta, \delta]_2 ,\label{m=2} \\ 
m=1, \; j_1= 2 ,& \;\;{\rm we \;\;obtain\;} &\; [i,I,{\hat J}, {\hat K},\beta, \delta]_1 , \label{j1=2}\\
m=1 ,\; j_1= 1, \;i=j,\;I=J,\; {\hat J_1}= {\hat R}_1, {\hat K_1} = {\hat S_1}, & \;\;{\rm we \;\;obtain\;} \; &(j,J,{\hat R}_1, {\hat S}_1,\beta_M , \bd).
\label{j1=1}
\ea
The contraction coordinates associated with the states obtained by applying the deformations (\ref{m=2}) - (\ref{j1=1}) are, denoted respectively (in abbreviated notation)  in section \ref{sec4.3} and in Step 2 of section \ref{sec6.3}
by $\{x_{\alpha}^{\epsilon_2,\epsilon_1}\}, \{x_{\alpha}^{\epsilon_1}\},\{x_{\alpha}^{\epsilon_2}\}$ and the coordinates for $c$ by $\{x_{\alpha}\}$. Here we set
\be
\{x_{\alpha}\} \equiv \{x\},  \;\;\;\;\;
\{x_{\alpha}^{\epsilon_1}\} \equiv \{x^{\bd}\}, \;\;\;\;\;
\{x_{\alpha}^{\epsilon_2}\} \equiv  \{x^{\delta}\}, \;\;\;\;\;
\{x_{\alpha}^{\epsilon_2,\epsilon_1}\}\equiv  \{x^{\delta, \bd}\} .
\label{xnot}
\ee
The notation we use  for the non degenerate vertex of:
\ba
&c \;\;{\rm is}\;\;v,  \;\;\;\;\;\; c_{(j,J,{\hat R}_1, {\hat S}_1,\beta_M , \bd)} \;\;{\rm is}\;\;v_{(j,J,\bd)}, &\nonumber\\
&c_{[i,I,{\hat J}, {\hat K}\beta, \delta]_1}\;\;{\rm is}\;\;v_{[i,I,\delta]_1}, \;\;\;\;\;\;\;c_{[i,I,{\hat J}, {\hat K},\beta,\delta]_2} \;\;{\rm is}\;\;v_{[i,I,\beta,\delta]_2}.& \label{vnot}
\ea
Wherever required explicitly, we denote the density weighted object $B$ evaluated at point $p$ in the coordinate system $\{y\}$ by $B(p, \{y\})$.


\subsubsection{\label{sec7.1.2}Calculation}

From (\ref{hamsum}) we have:
\be
\hat{C}[N]_{\delta}c  = \beta_N\frac{3\hbar}{8\pi\mathrm{i}}N(v, \{x\})\nu_{v}^{-2/3}\sum_{i,I,{\hat J_1},{\hat K_1}}  
\frac{c_{[i,I, {\hat J}, {\hat K}, \beta,\delta]_1}- c}{(N-1)(N-2)\delta} ,
\label{hh1}
\ee
\be
           \hat{C}[M]_{\bd}              \hat{C}[N]_{\delta}c  = \beta_N\frac{3\hbar}{8\pi\mathrm{i}}N(v, \{x\})\nu_{v}^{-2/3}\sum_{i,I,{\hat J_1},{\hat K_1}}  
\frac{ \hat{C}[M]_{\bd} c_{[i,I, {\hat J}, {\hat K}, \beta,\delta]_1}-  \hat{C}[M]_{\bd} c}{(N-1)(N-2)\delta} ,
\label{hh2}
\ee
Using (\ref{hamsum}) again,
\be
\hat{C}[M]_{\bd} c_{[i,I, {\hat J}, {\hat K}, \beta,\delta]_1} = 
\beta_M\frac{3\hbar}{8\pi\mathrm{i}}M(v_{[i,I,\delta]_1}, \{x^{\delta}\})\nu_{v_{   [i,I,\delta ]_1 }}^{-2/3}\sum_{i_1,I_1,{\hat J_2},{\hat K_2}}  
\frac{ c_{[i,I, {\hat J}, {\hat K}, \beta,\delta]_2}    -  c_{[i,I, {\hat J}, {\hat K}, \beta,\delta]_1}  }{(N-1)(N-2)\bd} ,
\label{hh3}
\ee
\ba
\Rightarrow
(\Psi_{f,h_{ab}, P_0}|\hat{C}[M]_{\bd} c_{[i,I, {\hat J}, {\hat K}, \beta,\delta]_1}\ket=
\beta_M\frac{3\hbar}{8\pi\mathrm{i}}M(v_{[i,I,\delta]_1}, \{x^{\delta}\})\nu_{v_{   [i,I,\delta ]_1 }}^{-2/3} &\nonumber\\
\sum_{i_1,I_1,{\hat J_2},{\hat K_2}} 
\frac{ (\Psi_{f,h_{ab}, P_0}|c_{[i,I, {\hat J}, {\hat K}, \beta,\delta]_2}\ket    -  (\Psi_{f,h_{ab}, P_0}|c_{[i,I, {\hat J}, {\hat K}, \beta,\delta]_1}\ket  }{(N-1)(N-2)\bd} &
\label{hh4}
\ea
Using (\ref{psifghc}),
\be
(\Psi_{f,h_{ab}, P_0}|c_{[i,I,{\hat J}, {\hat K},\beta, \delta]_2}\ket   = g_{ c_{[i,I,{\hat J}, {\hat K},\beta, \delta]_2}  }  f(v_{[i,I,\beta,\delta]_2}, \{x^{\delta, \bd}\} )   \;( \sum_{L_2} h_{L_2}H_{L_2})
\label{hh5}
\ee
where we have used  Step1 and 2, section \ref{sec6.3} to evaluate the amplitude  with respect to the contraction coordinates at $v_{[i,I,\beta,\delta]_2}$. Next, we evaluate its contraction behaviour.

From Appendix
\ref{acong}, and using $q>>1$, we have, as $\bd\rightarrow 0$:
\be
\sum_{{\hat J_2},{\hat K_2}}g_{   c_{[i,I,{\hat J}, {\hat K},\beta, \delta]_2}        }= g_{ c_{[i,I, {\hat J}, {\hat K}, \beta,\delta]_1}     }
(\bd)^{\frac{2}{3}(q-1)}
Q(c_{[i,I,\beta, \delta_0]^{2,1}_2}, S_{1}) h_{I_1}(1 +O(\bd^2))
\label{hhgcon}
\ee
where we have used (\ref{not71}), (\ref{m=2}) to set $j=1$ in equation (\ref{p2-13-2}).

From Appendix \ref{aconb}, a straightforward computation yields:
\be
\sum_{L_2} h_{L_2}H_{L_2} = [(N-1)(N-2)(2+\cos^2\theta + (N-3)|\cos \theta |)] \sqrt{h_{ab}(v_{[i,I,\beta,\delta]_2}) {\hat V}^{(\delta)a}_{I_1} {\hat V}^{(\delta)b}_{I_1}} + O(\bd^2)
\label{hhhHcon}
\ee
where, as in Appendix \ref{aconb}, ${\hat V}^{(\delta)a}_{I_1}$ is the constant extension in the chart $\{x^{\delta}\}$ of the unit upward direction for the  $I_1$th edge of the immediate parent $c_{[i,I, {\hat J}, {\hat K}, \beta,\delta]_1}$.
Here $c_{[i,I, {\hat J}, {\hat K}, \beta,\delta]_1}$ is the immediate parent of $c_{[i,I,{\hat J}, {\hat K},\beta, \delta]_2}$. This  immediate parent has contraction coordinates $\{x^{\delta}\}$ from (\ref{xnot})
and unit upward direction for its $I_1$th edge ${\hat V}^{(\delta)a}_{I_1}$. This vector is extended in an open neighbourhood of the parental vertex $v_{[i,I,\delta]_1}$ by defining its components in the chart $\{x^{\delta}\}$ at 
any point $p$ in this neighbourhood to be the same as its components at the parental vertex, this neighbourhood being large enough to contain the child vertex $v_{[i,I,\beta,\delta]_2}$.
Finally we have, from (\ref{afnecon})that:
\be
f(v_{[i,I,\beta,\delta]_2}, \{x^{\delta, \bd}\} ) =(\bd)^{-\frac{2}{3}(q-1)}f(v_{[i,I,\beta,\delta]_2}, \{x^{\delta}\} )
\label{f2hh}
\ee
We choose the $Q$ factor above to be:
\be
Q(c_{0[i,I,\beta, \delta_0]^{2,1}_2}, S_{1}) := \frac{N(N-1)(N-2)}{[(N-1)(N-2)(2+\cos^2\theta + (N-3)|\cos \theta |)]}.
\label{defq2hh}
\ee
Clearly, $Q>0$ as required.
From (\ref{hh5}), (\ref{hhgcon})- (\ref{defq2hh}), 
and setting 
\be
\sqrt{h_{ab}(v_{[i,I,\beta,\delta]_2}) {\hat V}^{(\delta)a}_{I_1} {\hat V}^{(\delta)b}_{I_1}} \equiv ||{\vec {\hat V}}^{(\delta)}_{I_1}||_{v_{[i,I,\beta,\delta]_2} }
\ee
we have:
\ba
&\sum_{i_1,I_1,{\hat J}_2,{\hat K}_2}(\Psi_{f,h_{ab}, P_0}|c_{[i,I,{\hat J}, {\hat K},\beta, \delta]_2}\ket =&\nonumber\\ 
&N(N-1)(N-2) g_{ c_{[i,I, {\hat J}, {\hat K}, \beta,\delta]_1}} \sum_{i_1,I_1} h_{I_1} ||{\vec {\hat V}}^{(\delta)}_{I_1}||_{v_{[i,I,\beta,\delta]_2} }f (v_{[i,I,\beta,\delta]_2}, \{x^{\delta}\} )
+ O(\bd^2 )&
\label{hh6}
\ea

From ( \ref{psifghc}), the second amplitude in (\ref{hh4}) is:
\be
(\Psi_{f,h_{ab}, P_0}|c_{[i,I, {\hat J}, {\hat K}, \beta,\delta]_1}\ket  = g_{ c_{[i,I,{\hat J}, {\hat K},\beta, \delta]_1}  }  f(v_{[i,I,\beta,\delta]_1}, \{x^{\delta}\} )   \;( \sum_{L_1} h_{L_1}H_{L_1})
\label{hh7}
\ee

From (\ref{hh6}), (\ref{hh7}) and the fact that  $v_{[i,I,\beta,\delta]_2}$ is displaced by an amount $q^{i_1}_{I_1}\bd$ in the direction ${\vec {\hat V}}^{(\delta)}_{I_1}$  from $v_{[i,I ,\delta]_1}$, we have that:
\ba
\sum_{i_1,I_1,{\hat J}_2,{\hat K}_2}(\Psi_{f,h_{ab}, P_0}|c_{[i,I,{\hat J}, {\hat K},\beta, \delta]_2}\ket =     3N(N-1)(N-2)(\Psi_{f,h_{ab}, P_0}|c_{[i,I, {\hat J}, {\hat K}, \beta,\delta]_1}\ket &\nonumber\\
+\;\bd N(N-1)(N-2) g_{ c_{[i,I, {\hat J}, {\hat K}, \beta,\delta]_1} } \sum_{i_1,I_1} h_{I_1}   q^{i_1}_{I_1}{\hat V}^{(\delta)a}_{I_1}
(\partial_a   ||{\vec {\hat V}}^{(\delta)}_{I_1}||_p f (p, \{x^{\delta}\} ))|_{p= v_{[i,I ,\delta]_1}} + O(\bd^2) &
\label{hh8}
\ea
where similar to Appendix \ref{aconb}, we have set:
\be
||{\vec {\hat V}}^{(\delta)}_{I_1}||_p= \sqrt{h_{ab}(p) {\hat V}^{(\delta)a}_{I_1}(p) {\hat V}^{(\delta)b}_{I_1}(p)}
\ee
with  ${\hat V}^{(\delta)a}_{I_1}(p)$ being the constant extension of  ${\hat V}^{(\delta)a}_{I_1}$ at $v_{[i,I ,\delta]_1}$. The partial derivative $\partial_a$ can be taken with respect to any coordinates
as its tangent space index $a$ is contracted with that of  ${\hat V}^{(\delta)a}_{I_1}$. If we take it to be the coordinate derivative with respect to $\{x^{\delta}\}$ then it passes through 
 ${\hat V}^{(\delta)a}_{I_1}(p)$ and only acts on $h_{ab}, f$.
Using (\ref{hh8}) in (\ref{hh4}) and taking the limit $\bd \rightarrow 0$, we have that:
\ba
\lim_{\bd\rightarrow 0}(\Psi_{f,h_{ab}, P_0}|\hat{C}[M]_{\bd} c_{[i,I, {\hat J}, {\hat K}, \beta,\delta]_1}\ket =
\beta_M\frac{3\hbar N}{8\pi\mathrm{i}}M(v_{[i,I,\delta]_1}, \{x^{\delta}\})\nu_{v_{   [i,I,\delta ]_1 }}^{-2/3}&
\nonumber\\
%
g_{ c_{[i,I, {\hat J}, {\hat K}, \beta,\delta]_1} } \sum_{i_1,I_1} h_{I_1}   q^{i_1}_{I_1}{\hat V}^{(\delta)a}_{I_1}
(\partial_a   ||{\vec {\hat V}}^{(\delta)}_{I_1}||_p f( p, \{x^{\delta}\} ))|_{p= v_{[i,I ,\delta]_1}}
\label{hh9}
\ea
Using the notation of Appendix \ref{aconb}, we may write this concisely as:
\be
\lim_{\bd\rightarrow 0}(\Psi_{f,h_{ab}, P_0}|\hat{C}[M]_{\bd} c_{[i,I, {\hat J}, {\hat K}, \beta,\delta]_1}\ket =
\beta_M\frac{3\hbar N}{8\pi\mathrm{i}}\nu_{v_{   [i,I,\delta ]_1 }}^{-2/3}
g_{ c_{[i,I, {\hat J}, {\hat K}, \beta,\delta]_1} } \sum_{i_1,I_1} h_{I_1}   q^{i_1}_{I_1}(H^1_{I_1} (p= v_{[i,I ,\delta]_1}))
\label{hh10}
\ee

Next consider the term   $\hat{C}[M]_{\bd} c$ in (\ref{hh2}).  We have, 
using (\ref{hamsum}),
\be
\hat{C}[M]_{\bd} c  = 
\beta_M\frac{3\hbar}{8\pi\mathrm{i}}M(v, \{x\})\nu_{v}^{-2/3}\sum_{j,J,{\hat R_1},{\hat S_1}}  
\frac{ c_{(j,J, {\hat R}, {\hat S}, \beta_M,\bd)}    -  c }{(N-1)(N-2)\bd} ,
\label{hh11}
\ee
A similar analysis yields:
\ba
\lim_{\bd\rightarrow 0}(\Psi_{f,h_{ab}, P_0}|\hat{C}[M]_{\bd} c\ket =
\beta_M\frac{3\hbar N}{8\pi\mathrm{i}}M(v, \{x\})\nu_{v}^{-2/3}&
\nonumber\\
%
g_{ c} \sum_{j,J} h_{J}   q^{j}_{J}{\hat V}^{a}_{J}
(\partial_a   ||{\vec {\hat V}}_{J}||_p f( p, \{x\} ))|_{p= v}
\label{hh12}
\ea
which can be written in  the notation of Appendix \ref{aconb} as
\be
\lim_{\bd\rightarrow 0}(\Psi_{f,h_{ab}, P_0}|\hat{C}[M]_{\bd} c\ket =
\beta_M\frac{3\hbar N}{8\pi\mathrm{i}}\nu_{v}^{-2/3}
g_{ c } \sum_{j,J} h_{J}   q^{j}_{J}(H^1_{J} (p= v))
\label{hh13}
\ee
In the above calculation 
the $Q$ factor is the same as that in (\ref{defq2hh}):
\be
Q(c_{0(j,J,\beta_M, \delta_0)},c, S_{1}) := \frac{N(N-1)(N-2)}{[(N-1)(N-2)(2+\cos^2\theta + (N-3)|\cos \theta |)]},
\label{defq11hh}
\ee
and we have  used equation (\ref{p2-13-2}) in conjunction  with equations (\ref{not71}), (\ref{j1=1}).
Note that the $Q$ factor in (\ref{defq11hh}) and the $Q$ factor in (\ref{defq2hh}) are labelled by the same sequence label $S_1 =h$ (see section \ref{sec6.4} for a discussion of this labelling). It follows  from this fact,
together with the charge independence of the $Q$ factor in (\ref{defq11hh}) and the discussion in section \ref{sec6.4} that the $Q$ factors in (\ref{defq11hh}) and (\ref{defq2hh}) {\em must necessarily} be identical.

Finally, we need to the compute the contraction limit $\delta\rightarrow 0$ of (\ref{hh10}). 
From Appendix
\ref{acong}, and using $q>>1$, we have, as $\delta\rightarrow 0$:
\be
\sum_{{\hat J_1},{\hat K_1}}g_{   c_{[i,I,{\hat J}, {\hat K},\beta, \delta]_1}        }= g_{ c    }
(\delta)^{\frac{4}{3}(q-1)}
Q(c_{0[i,I,\beta, \delta_0]^{0,1}_1}, S_{2}) h_{I}(1 +O(\delta^2))
\label{hhgcon2}
\ee
where we have used (\ref{not71}), (\ref{j1=2}) to set $j=2$ in equation (\ref{p2-13-2}).
Using equations (\ref{hicon}), (\ref{dmHncon}) from Appendix \ref{aconb}, as well as (\ref{rot=1}),  we have that as $\delta\rightarrow 0$ :
\ba
&\sum_{i_1,I_1} h_{I_1}   q^{i_1}_{I_1}(H^1_{I_1} (p= v_{[i,I ,\delta]_1}))
= &
%
\nonumber \\
&\delta^{-\frac{4}{3}(q-1)}\big\{ 
          \big( M(v_{[i,I ,\delta]_1} , \{x\})      {{\hat V}^{a}}_{I}\partial_{a} (f(v_{[i,I ,\delta]_1}       , \{x\}) 
               \sqrt{h_{ab}(v_{[i,I ,\delta]_1})  {{\hat V}^{a}}_{I}  {{\hat V}^{b}}_{I}})  &
\nonumber\\
 &\;[(N-1)(N-2) q^{i_1}_{I_1=I} +  (\sum_{I_1\neq I} q^{i_1}_{I_1})(N-2) (1+ \cos\theta (1+ \cos^2 \theta + (N-3)|\cos \theta | )] \big)&
\nonumber\\
& + O(\delta^{2})
\big\}&\nonumber\\
&=
\delta^{-\frac{4}{3}(q-1)}
\big\{ 
          \big(M(v_{[i,I ,\delta]_1} , \{x\})     q^{i_1}_{I_1=I} {{\hat V}^{a}}_{I}\partial_{a} (f(v_{[i,I ,\delta]_1}       , \{x\}) 
               \sqrt{h_{ab}(v_{[i,I ,\delta]_1})  {{\hat V}^{a}}_{I}  {{\hat V}^{b}}_{I}}) &
\nonumber\\
 &[(N-1)(N-2) - (N-2)(\cos\theta )(1+ \cos^2 \theta + (N-3)|\cos \theta| )]\big) + O(\delta^{2}) 
\big\} &
\label{hh14}
\ea
where we have used gauge invariance applied to the  {\em net} charges to go from the first equality to the second.

Next, we choose the $Q$ factor in (\ref{hhgcon2}) to be:
\be
Q(c_{0[i,I,\beta, \delta_0]^{0,1}_1}, S_{2}) = \frac{\nu_{v}^{-2/3}}{\nu_{v_{   [i,I,\delta ]_1 }}^{-2/3}}\frac{3N(N-1)(N-2)}{[(N-1)(N-2) - (N-2)(\cos\theta )(1+ \cos^2 \theta + (N-3)|\cos \theta| )]}
\label{defq1h}
\ee
It is straight forward to check that (using the facts that $N\geq 4$, $|\cos \theta|<1)$, as required, this $Q$ factor is positive. Note that for $Q$ to be well defined, we need the non-degenaracy condition 
$ \nu_{v_{   [i,I,\delta ]_1 }}\neq 0$ (see the relevant discussion in the beginning of  section \ref{sec4.2}).

Next, using (\ref{hhgcon2})- (\ref{defq1h}) in (\ref{hh10}) yields:
\ba
&\sum_{{\hat J}_1, {\hat K}_1}\lim_{\bd\rightarrow 0}(\Psi_{f,h_{ab}, P_0}|\hat{C}[M]_{\bd} c_{[i,I, {\hat J}, {\hat K}, \beta,\delta]_1}\ket =
3(N)(N-1)(N-2)\beta_M\frac{3\hbar N}{8\pi\mathrm{i}}\nu_{v}^{-2/3}&
\nonumber\\
 & g_{ c} h_I \sum_{i_1}    q^{i_1}_{I_1=I} (M(v_{[i,I ,\delta]_1} , \{x\})      {{\hat V}^{a}}_{I}\partial_{a} (f(v_{[i,I ,\delta]_1}       , \{x\}) 
               \sqrt{h_{ab}(v_{[i,I ,\delta]_1})  {{\hat V}^{a}}_{I}  {{\hat V}^{b}}_{I}}) ) +O(\delta^2 ) &
\label{hh15}
\ea

In the above equation note that $q^{i_1}_{I_1=I}$ refers to the charge on the $I_1=I$th edge of $c_{[i,I, {\hat J}, {\hat K}, \beta,\delta]_1}$. This charge is related to the charge on the $I$th edge of $c$
by an $(i,\beta_M)$ flip. Hence, depending on whether $\beta_M =\pm1$  we have from (\ref{defchrgeflip}), (\ref{defchrgeflip-}) that:
\begin{equation}
q^{i_1}_{I_1=I} = \left.  ^{(i)}\!q_I^{i_1}\right.  =\delta^{ii_1}q_I^{i_1}\mp%
{\textstyle\sum\nolimits_{k}}
\epsilon^{ii_1k}q_I^{k} \label{hhchrgeflip}%
\end{equation}
Two identities, {\em key} to the anomaly free result follow from the above equation:
\be
\sum_{i,i_1} q^{i_1}_{I_1=I}  = \sum_{i} q^{i}_I
\label{qid1}
\ee
and 
\be
\sum_{i,i_1} q^{i_1}_{I_1=I}  q^i_I  = \sum_{i} (q^{i}_I)^2
\label{qid2}
\ee

Using  (\ref{hhchrgeflip}) in (\ref{hh15}) we have:
\ba
&\sum_{{\hat J}_1, {\hat K}_1}\lim_{\bd\rightarrow 0}(\Psi_{f,h_{ab}, P_0}|\hat{C}[M]_{\bd} c_{[i,I, {\hat J}, {\hat K}, \beta,\delta]_1}\ket =
3(N)(N-1)(N-2)\beta_M\frac{3\hbar N}{8\pi\mathrm{i}}\nu_{v}^{-2/3} g_c &
\nonumber\\
 & h_I\sum_{i_1}     \left.  ^{(i)}\!q_I^{i_1}\right.        (M(v_{[i,I ,\delta]_1} , \{x\})      {{\hat V}^{a}}_{I}\partial_{a} (f(v_{[i,I ,\delta]_1}       , \{x\}) 
               \sqrt{h_{ab}(v_{[i,I ,\delta]_1})  {{\hat V}^{a}}_{I}  {{\hat V}^{b}}_{I}}) ) +O(\delta^2 ) &
\label{hh16}
\ea
Next, we expand the second line of (\ref{hh16}) in a Taylor approximation and  sum over $i, I$ to obtain
\ba
  & \sum_{I}h_I\sum_i\sum_{i_1}     \left.  ^{(i)}\!q_I^{i_1}\right.        (M(v_{[i,I ,\delta]_1} , \{x\})      {{\hat V}^{a}}_{I}\partial_{a} (f(v_{[i,I ,\delta]_1}       , \{x\}) 
               \sqrt{h_{ab}(v_{[i,I ,\delta]_1})  {{\hat V}^{a}}_{I}  {{\hat V}^{b}}_{I}}) ) &              
\nonumber\\
&  =\sum_{I}h_I\big\{
\sum_{i, i_1}     \left.  ^{(i)}\!q_I^{i_1}\right.        (M(v, \{x\})      {{\hat V}^{a}}_{I}\partial_{a} (f(v       , \{x\}) 
  \sqrt{h_{ab}(v)  {{\hat V}^{a}}_{I}  {{\hat V}^{b}}_{I}}) ) &
\nonumber \\
& + \delta \sum_{i,i_1} \left.  ^{(i)}\!q_I^{i_1}\right.    \big(q^{i}_I  {\hat V}^b_I \partial_b    (M(p, \{x\}) (     {{\hat V}^{a}}_{I}\partial_{a} (f(p       , \{x\}) 
  \sqrt{h_{ab}(p)  {{\hat V}^{a}}_{I}  {{\hat V}^{b}}_{I}}) )\big)|_{p=v}
  \big\}    +O(\delta^2)  &
\nonumber\\
& =  \sum_{I}h_I \big\{  
\sum_{i}    q^i_I       (M(v, \{x\})      {{\hat V}^{a}}_{I}\partial_{a} (f(v       , \{x\}) 
  \sqrt{h_{ab}(v)  {{\hat V}^{a}}_{I}  {{\hat V}^{b}}_{I}}) ) &
\nonumber \\
& + \delta \sum_{i}\big(  (q^{i}_I)^2  {\hat V}^b_I \partial_b    (M(p, \{x\}) (     {{\hat V}^{a}}_{I}\partial_{a} (f(p       , \{x\}) 
  \sqrt{h_{ab}(p)  {{\hat V}^{a}}_{I}  {{\hat V}^{b}}_{I}}) )\big)|_{p=v}  
  \big\}  +O(\delta^2) &
\label{hh17}
\ea
Here we have used the identities (\ref{qid1}), (\ref{qid2}) to obtain the second equality from the first.

Next, consider the dual action of (\ref{hh2}) on the anomaly free state in the limit $\bd\rightarrow 0$:
\ba
&\lim_{\bd\rightarrow 0}(\Psi_{f,h_{ab}, P_0}|    \hat{C}[M]_{\bd}              \hat{C}[N]_{\delta}c\ket  = \beta_N\frac{3\hbar}{8\pi\mathrm{i}}N(v, \{x\})\nu_{v}^{-2/3}  \frac{1}{(N-1)(N-2)\delta}&
\nonumber\\
&\big(\sum_{i,I,{\hat J_1},{\hat K_1}}  \lim_{\bd\rightarrow 0} (\Psi_{f,h_{ab}, P_0}|\hat{C}[M]_{\bd} c_{[i,I, {\hat J}, {\hat K}, \beta,\delta]_1}\ket) &
\nonumber\\
&-         \sum_{i,I,{\hat J_1},{\hat K_1}}                    (\Psi_{f,h_{ab}, P_0}|\hat{C}[M]_{\bd} c\ket \big) &
\label{hh18}
\ea
The second line of (\ref{hh18})  can be evaluated using (\ref{hh17}).  
The  zeroth order term in $\delta$ in this expansion  is 
precisely $3(N)(N-1)(N-2)$ times the right hand side of (\ref{hh12}). In the term on the 3rd line of (\ref{hh18}), the amplitude is exactly that of (\ref{hh12}) and the indices 
$i,{\hat J}_1,{\hat K}_1$ are dummy indices for this amplitude so that the amplitude is simply multiplied by a factor of $N$ (coming from the sum over $I$),  $(N-1)(N-2)$ (from the sum over the hatted indices)
and $3$ (from the sum over $i$). Hence the zeroth order term of the second line cancels the contribution from the third line and this what allows the $\delta\rightarrow 0 $ limit of the 
left hand side in the first line to exist.
\\

\noindent {\bf NOTE}:{\em This cancellation is  precisely due to the $-{\bf 1}$ structure introduced in section \ref{sec3}, this structure  being motivated by considerations of `propagation'.}
\\

Finally taking the $\delta \rightarrow 0$ limit of (\ref{hh18}), we obtain 
\ba
&(\Psi_{f,h_{ab}, P_0}|    \hat{C}[M]              \hat{C}[N]c\ket  =  3\beta_N\beta_M(\frac{3\hbar N}{8\pi\mathrm{i}})^2N(v, \{x\})\nu_{v}^{-4/3} g_c&
\nonumber\\
&\big\{\sum_{i, I}h_I\big(  (q^{i}_I)^2  {\hat V}^b_I \partial_b    (M(p, \{x\}) (     {{\hat V}^{a}}_{I}\partial_{a} (f(p       , \{x\}) 
  \sqrt{h_{ab}(p)  {{\hat V}^{a}}_{I}  {{\hat V}^{b}}_{I}}) )\big)|_{p=v}  
  \big\} &
\label{hheval}
\ea

Next, as in Appendix \ref{aconb3}, it is convenient to define the  quantity   $H^l_{L_m}(N_1,N_2,..,N_3;p)$ associated with 
equation (\ref{dual}) as follows.  Let $c$  in (\ref{dual})
be in the Bra Set (as in this section). Let the  
contraction coordinate associated with the non-degenerate vertex of its $m$th  generation descendant, $c_{[i,I,{\hat J}, {\hat K}, \beta, \epsilon]_m}$ 
be denoted by $\{z\}$ so that $\{z\}:= \{ x^{ \epsilon_{\rm j_1}..\epsilon_{\rm j_m}}\}$. Then we define:
\be 
H^{l}_{L_m}(N_1,..,N_l;p):= \prod_{i=1}^l  N_{l-i+1}( p, \{z\})  {\hat V}^{a_{l-i+1}}_{L_m}(p)\partial_{a_{l-i+1}} (f( p, \{z\})\sqrt{h_{ab}(p)      {\hat V}^{ {a } }_{L_m} (p) {   {\hat V}^{b}  }_{L_m}(p)    })
\label{defHlmp}
\ee
where the product is ordered from left to right in order of increasing $i$ and the point $p$ is in a small enough neighbourhood of the nondegenerate vertex of $c_{[i,I,{\hat J}, {\hat K}, \beta, \epsilon]_m}$
wherein the unit (with respect to $\{z\}$) upward direction ${\vec {\hat V}}_{L_m}    $ associated with the $L_m$th edge at this vertex admits the  constant extension  ${\vec  {\hat V}}_{L_m}  (p) $    as discussed in Appendix \ref{aconb3}. 
\footnote{ It is straightforward to check that if we express equation (\ref{defHlmp}) in terms of the notation and the `right to left' ordering convention for `$\prod$' used in Appendix \ref{aconb3} that (\ref{defHlmp}) 
takes the form of (\ref{defdmHp}).}

Making contact through (\ref{not71}) with (\ref{defHlmp}), equation (\ref{hheval}) can be written succintly as:
\be
(\Psi_{f,h_{ab}, P_0}|    \hat{C}[M]              \hat{C}[N]c\ket  =  3\beta_N\beta_M(\frac{3\hbar N}{8\pi\mathrm{i}})^2\nu_{v}^{-4/3}g_c \sum_i\sum_I (q^i_I)^2h_I H^2_I(M,N;p=v), 
\label{hhconcise}
\ee
from which the commutator can be written as:
\be
(\Psi_{f,h_{ab}, P_0}|  [  \hat{C}[M]  ,            \hat{C}[N]] c\ket  =  3\beta_N\beta_M(\frac{3\hbar N}{8\pi\mathrm{i}})^2\nu_{v}^{-4/3}g_c \sum_i\sum_I (q^i_I)^2h_I (H^2_I(M,N;p=v)  -  H^2_I(N,M;p=v))  .
\label{h,hconcise}
\ee


\subsection{\label{sec7.2}Electric diffeomorphism commutator}

\subsubsection{\label{sec7.2.1}Notation}
We  compute the continuum limit  of the electric diffeomorphism commutator. Accordingly we consider the action of (\ref{dual})  on a state $c$ when $n=2$ and ${\hat O}_{\rm i}(N_{\rm i}), {\rm i}=1,2$ 
are electric diffeomorphism constraint operators. We 
compute the commutator from the ensuing product of discrete approximants and then take the continuum limit. 
Similar to  section \ref{sec7.1.1}, and for the reason articulated there  we  restrict attention to  the case that  $c$  is in the Bra Set. 
For this section we use notation similar to that in section \ref{sec7.1.1}. 
However, the notation denotes transitions and their associated structures  which are appropriate to the action of the electric diffeomorphism constraints and hence are often  distinct from  those
appropriate to the Hamiltonian constraint in section \ref{sec7.1.1}.

We use the notation (\ref{not71}) in (\ref{dual}) so that once again we have:
\ba
&N_1 \equiv M \;\; \beta_1\equiv\beta_M \;\; N_2 \equiv N\;\;\; \beta_2\equiv \beta_N &\nonumber \\
&\epsilon_1 \equiv \bd \;\;\;  \epsilon_2 \equiv \delta &
\label{dnot71}
\ea
The discrete action of the electric diffeomorphism  constraint operator product generates the doubly deformed states $c_{[i,I,{\hat J}, {\hat K}, \delta]_2}$, the singly deformed states 
$c_{[i,I,{\hat J}, {\hat K}, \delta]_1}$ and $c_{(i,J,{\hat R}_1, {\hat S}_1, \bd)}$ and $c$. Here we have defined the transitions:
\ba
[i,I,{\hat J}, {\hat K},\delta]_2 &=& [(i_1, I_1, {\hat J}_2, {\hat K_2}, \beta=0 , \bd), (i, I, {\hat J}_1, {\hat K}_1, \beta=0 , \delta )] \label{d2not}
\\
\;   [ i,I,{\hat J}, {\hat K}, \delta]_1 &= & (i, I, {\hat J}_1, {\hat K}_1, \beta=0, \delta )
\label{d1not}
\\
(i,J,{\hat R}_1, {\hat S}_1, \bd) &=& (i,J,{\hat R}_1, {\hat S}_1, \beta=0, \bd)
\label{d1bdnot}
\ea

We set in (\ref{fixseqehat}):
\ba 
m=2,\; j_2=1, \;  j_1=2, &  \;\;{\rm to \;\;obtain\;} &\; [i,I, {\hat J}, {\hat K}, \delta]_2 ,\label{dm=2} \\ 
m=1, \; j_1= 2 ,& \;\;{\rm to \;\;obtain\;} &\; [i,I,  {\hat J}, {\hat K},    \delta]_1 , \label{dj1=2}\\
m=1 ,\; j_1= 1, \;I=J,\; {\hat J_1}= {\hat R}_1, {\hat K_1} = {\hat S_1}, & \;\;{\rm to \;\;obtain\;} \; & (i ,J,{\hat R}_1, {\hat S}_1,  \bd).
\label{dj1=1}
\ea
The contraction coordinates associated with the states obtained by applying the deformations (\ref{dm=2}) - (\ref{dj1=1}) are, denoted respectively (in abbreviated notation)  in section \ref{sec4.3} and in Step 2 of section \ref{sec6.3}
by $\{x_{\alpha}^{\epsilon_2,\epsilon_1},\}, \{x_{\alpha}^{\epsilon_2}\},\{x_{\alpha}^{\epsilon_1}\}$ and the coordinates for $c$ by $\{x_{\alpha}\}$. Similar to (\ref{xnot}), we set:
\be
\{x_{\alpha}\} \equiv \{x\},  \;\;\;\;\;
\{x_{\alpha}^{\epsilon_2}\} \equiv \{x^{\delta}\}, \;\;\;\;\;
\{x_{\alpha}^{\epsilon_1}\} \equiv  \{x^{\bd}\}, \;\;\;\;\;
\{x_{\alpha}^{\epsilon_2,\epsilon_1}\}\equiv  \{x^{\delta, \bd}\} .
\label{dxnot}
\ee
The notation we use  for the non degenerate vertex of:
\ba
&c \;\;{\rm is}\;\;v,  \;\;\;\;\;\; c_{(i,J,{\hat R}_1, {\hat S}_1, \bd)} \;\;{\rm is}\;\;v_{(i,J,\bd)}, &\nonumber\\
&c_{[i,I,{\hat J}, {\hat K}, \delta]_1}\;\;{\rm is}\;\;v_{[i,I,\delta]_1}, \;\;\;\;\;\;\;c_{[i,I,{\hat J}, {\hat K},\delta]_2} \;\;{\rm is}\;\;v_{[i,I,\delta]_2}.& \label{dvnot}
\ea

\subsubsection{\label{sec7.2.2}Calculation}

Applying (\ref{dnsum}) to $c$  we obtain:
\be
\hat{D}_{\delta}[\vec{N}_{i}]c   =\frac{\hbar}{\mathrm{i}}\frac{3}{4\pi}%
N(x(v))\nu_{v}^{-2/3}\sum_{I} \sum_{{\hat J}_1, {\hat K}_1}\frac{1}{(N-1)(N-2)}   \frac{1}{\delta
}(   c_{   [  i, I, {\hat J}, {\hat K}, \delta  ]_1  }  -   c   ).
\label{dd1}
\ee
so that:
\be
 \hat{D}_{\bd}[\vec{M}_{i}]     \hat{D}_{\delta}[\vec{N}_{i}]c   =\frac{\hbar}{\mathrm{i}}\frac{3}{4\pi}%
N(x(v))\nu_{v}^{-2/3}\sum_{I} \sum_{{\hat J}_1, {\hat K}_1}\frac{1}{(N-1)(N-2)}   \frac{1}{\delta
}(  \hat{D}_{\bd}[\vec{M}_{i}] c_{   [ i, I, {\hat J},  {\hat K}  \delta  ]_1}  -  \hat{D}_{\bd}[\vec{M}_{i}] c   ).
\label{dd2}
\ee
Using (\ref{dnsum}) again,
\be
\hat{D}_{\bd}[\vec{M}_{i}] c_{   [ i, I, {\hat J},  {\hat K}  \delta  ]_1} = 
\frac{3\hbar}{4\pi\mathrm{i}}M(v_{[i,I,\delta]_1}, \{x^{\delta}\})\nu_{v_{   [i,I,\delta ]_1 }}^{-2/3}\sum_{I_1,{\hat J_2},{\hat K_2}}  
\frac{ c_{[i,I, {\hat J}, {\hat K}, \delta]_2}    -  c_{[i,I, {\hat J}, {\hat K}, \delta]_1}  }{(N-1)(N-2)\bd} ,
\label{dd3}
\ee
\ba
\Rightarrow
(\Psi_{f,h_{ab}, P_0}|\hat{D}[{\vec M}_i]_{\bd} c_{[i,I, {\hat J}, {\hat K}, \delta]_1}\ket=
\frac{3\hbar}{4\pi\mathrm{i}}M(v_{[i,I,\delta]_1}, \{x^{\delta}\})\nu_{v_{   [i,I,\delta ]_1 }}^{-2/3} &\nonumber\\
\sum_{I_1,{\hat J_2},{\hat K_2}} 
\frac{ (\Psi_{f,h_{ab}, P_0}|c_{[i,I, {\hat J}, {\hat K}, \delta]_2}\ket    -  (\Psi_{f,h_{ab}, P_0}|c_{[i,I, {\hat J}, {\hat K},\delta]_1}\ket  }{(N-1)(N-2)\bd} &
\label{dd4}
\ea
Using (\ref{psifghc}),
\be
(\Psi_{f,h_{ab}, P_0}|c_{[i,I,{\hat J}, {\hat K}, \delta]_2}\ket   = g_{ c_{[i,I,{\hat J}, {\hat K}, \delta]_2}  }  f(v_{[i,I,\delta]_2}, \{x^{\delta, \bd}\} )   \;( \sum_{L_2} h_{L_2}H_{L_2})
\label{dd5}
\ee
where we have used  Step1 and 2, section \ref{sec6.3} to evaluate the amplitude  with respect to the contraction coordinates at $v_{[i,I, \delta]_2}$. Next, we evaluate its contraction behaviour.

From Appendix
\ref{acong}, and using $q>>1$, we have, as $\bd\rightarrow 0$:
\be
\sum_{{\hat J_2},{\hat K_2}}g_{   c_{[i,I,{\hat J}, {\hat K}, \delta]_2}        }= g_{ c_{[i,I, {\hat J}, {\hat K}, \delta]_1}     }
(\bd)^{\frac{2}{3}(q-1)}
Q({c_{0[i,I, \delta_0]^{2,1}_2}, S_{1}}) h_{I_1}(1 +O(\bd^2))
\label{ddgcon}
\ee
where we have used (\ref{dnot71}), (\ref{dm=2}) to set $j=1$ in equation (\ref{p2-13-2}).

From Appendix \ref{aconb}, a straightforward computation identical to that used in deriving (\ref{hhhHcon}) yields
\be
\sum_{L_2} h_{I_2}H_{L_2} = [(N-1)(N-2)(2+\cos^2\theta + (N-3)|\cos \theta |)] \sqrt{h_{ab}(v_{[i,I,\delta]_2}) {\hat V}^{(\delta)a}_{I_1} {\hat V}^{(\delta)b}_{I_1}} + O(\bd^2)
\label{ddhHcon}
\ee
where, as in Appendix \ref{aconb}, ${\hat V}^{(\delta)a}_{I_1}$ is the constant extension in the chart $\{x^{\delta}\}$ of the unit upward direction for the  $I_1$th edge of the immediate parent $c_{[i,I, {\hat J}, {\hat K},\delta]_1}$.
Similar to (\ref{f2hh}), from (\ref{afnecon}), we have that:
\be
f(v_{[i,I,\delta]_2}, \{x^{\delta, \bd}\} ) =(\bd)^{-\frac{2}{3}(q-1)}f(v_{[i,I,\delta]_2}, \{x^{\delta}\} )
\label{f2dd}
\ee
We choose the $Q$ factor for this electric diffeomorphsim type transition  to be identical that of (\ref{defq2hh}) so that:
\be
Q(c_{0[i,I, \delta_0]^{2,1}_2}, S_{1}) := \frac{N(N-1)(N-2)}{[(N-1)(N-2)(2+\cos^2\theta + (N-3)|\cos \theta |)]}.
\label{defq2dd}
\ee
where $c_{0[i,I, \delta_0]^{2,1}_2}$ denote  the `$\delta_0$' images (see (\ref{c0seqm}), (\ref{setdeformd0}) and section \ref{sec6.4})  of the `electric diffeomorphism' children
$c_{[i,I, {\hat J}, {\hat K}, \delta ]^{2,1}_2}$. Note that in principle the $Q$ factors in (\ref{defq2dd}) and (\ref{defq2hh}) could be chosen to be distinct from each other because in the former
case the sequence label $S_1$ corresponds to $d_i$ whereas in the latter case $S_1=h$.

From (\ref{dd5}), (\ref{ddgcon})- (\ref{defq2dd}), 
and setting, similar to section \ref{sec7.1.2}, 
\be
\sqrt{h_{ab}(v_{[i,I, \delta]_2}) {\hat V}^{(\delta)a}_{I_1} {\hat V}^{(\delta)b}_{I_1}} \equiv ||{\vec {\hat V}}^{(\delta)}_{I_1}||_{v_{[i,I, \delta]_2} }
\ee
we have:
\be
\sum_{I_1, {\hat J}_2, {\hat K}_2 }(\Psi_{f,h_{ab}, P_0}|c_{[i,I,{\hat J}, {\hat K},  \delta]_2}\ket = 
N(N-1)(N-2) g_{ c_{[i,I, {\hat J}, {\hat K}, \delta]_1}} \sum_{I_1} h_{I_1} ||{\vec {\hat V}}^{(\delta)}_{I_1}||_{v_{[i,I,\delta]_2} }f (v_{[i,I,\delta]_2}, \{x^{\delta}\} )
\label{dd6}
\ee

From ( \ref{psifghc}), the second amplitude in (\ref{dd4}) is:
\be
(\Psi_{f,h_{ab}, P_0}|c_{[i,I, {\hat J}, {\hat K},  \delta]_1}\ket  = g_{ c_{[i,I,{\hat J}, {\hat K},  \delta]_1}  }  f(v_{[i,I, \delta]_1}, \{x^{\delta}\} )   \;( \sum_{L_1} h_{L_1}H_{L_1})
\label{dd7}
\ee

Similar to the derivation of (\ref{hh8}), from  (\ref{dd6}), (\ref{dd7}), we  have that:
\ba
\sum_{I_1, {\hat J}_2, {\hat K}_2 }(\Psi_{f,h_{ab}, P_0}|c_{[i,I,{\hat J}, {\hat K}, \delta]_2}\ket =     N(N-1)N-2)(\Psi_{f,h_{ab}, P_0}|c_{[i,I, {\hat J}, {\hat K}, \delta]_1}\ket &\nonumber\\
+\;\bd N(N-1)(N-2) g_{ c_{[i,I, {\hat J}, {\hat K},\delta]_1} } \sum_{I_1} h_{I_1}   q^{i_1=i}_{I_1}{\hat V}^{(\delta)a}_{I_1}
(\partial_a   ||{\vec {\hat V}}^{(\delta)}_{I_1}||_p f( p, \{x^{\delta}\} ))|_{p= v_{[i,I ,\delta]_1}} + O(\bd^2) &
\label{dd8}
\ea
where 
\be
||{\vec {\hat V}}^{(\delta)}_{I_1}||_p := \sqrt{h_{ab}(p) {\hat V}^{(\delta)a}_{I_1}(p) {\hat V}^{(\delta)b}_{I_1}(p)}
\ee
with  ${\hat V}^{(\delta)a}_{I_1}(p)$ being the constant extension of  ${\hat V}^{(\delta)a}_{I_1}$ at $v_{[i,I ,\delta]_1}$. As in section \ref{sec7.1.2}, the partial derivative $\partial_a$ can be taken with respect to any coordinates
as its tangent space index $a$ is contracted with that of  ${\hat V}^{(\delta)a}_{I_1}$; if we take it to be the coordinate derivative with respect to $\{x^{\delta}\}$ then it passes through 
 ${\hat V}^{(\delta)a}_{I_1}(p)$ and only acts on $h_{ab}, f$.
Using (\ref{dd8}) in (\ref{dd4}) and taking the limit $\bd \rightarrow 0$, we have that:
\ba
\lim_{\bd\rightarrow 0}(\Psi_{f,h_{ab}, P_0}|\hat{D}[{\vec M}_i]_{\bd} c_{[i,I, {\hat J}, {\hat K},  \delta]_1}\ket =
\frac{3\hbar N}{4\pi\mathrm{i}}M(v_{[i,I,\delta]_1}, \{x^{\delta}\})\nu_{v_{   [i,I,\delta ]_1 }}^{-2/3}&
\nonumber\\
%
g_{ c_{[i,I, {\hat J}, {\hat K},\delta]_1} } \sum_{I_1} h_{I_1}   q^{i_1=i}_{I_1}{\hat V}^{(\delta)a}_{I_1}
(\partial_a   ||{\vec {\hat V}}^{(\delta)}_{I_1}||_p f( p, \{x^{\delta}\} ))|_{p= v_{[i,I ,\delta]_1}}
\label{dd9}
\ea
In the  notation of Appendix \ref{aconb}, we may write this as:
\be
\lim_{\bd\rightarrow 0}(\Psi_{f,h_{ab}, P_0}|\hat{D}[{\vec M}_i]_{\bd} c_{[i,I, {\hat J}, {\hat K},  \delta]_1}\ket =
\frac{3\hbar N}{4\pi\mathrm{i}}\nu_{v_{   [i,I,\delta ]_1 }}^{-2/3}
g_{ c_{[i,I, {\hat J}, {\hat K},  \delta]_1} } \sum_{I_1} h_{I_1}   q^{i_1=i}_{I_1}(H^1_{I_1} (p= v_{[i,I ,\delta]_1}))
\label{dd10}
\ee

Next consider the term   $\hat{D}[{\vec M}_{i}]_{\bd} c$ in (\ref{dd2}). 
From (\ref{dnsum}),
\be
\hat{D}[{\vec M}_i]_{\bd} c  = 
\frac{3\hbar}{4\pi\mathrm{i}}M(v, \{x\})\nu_{v}^{-2/3}\sum_{J,{\hat R_1},{\hat S_1}}  
\frac{ c_{(i,J, {\hat R}, {\hat S},\bd)}    -  c }{(N-1)(N-2)\bd} ,
\label{dd11}
\ee
As can be seen from (\ref{dd2}), (\ref{dd11}),   the term  $\hat{D}[{\vec M}_i]_{\bd} c$      does {\em not} contribute to the commutator because it is multiplied by a product of lapse functions evaluated at the {\em same point $v$}. 
Nevertheless it is instructive to evaluate it for reasons which will become clear towards the end of this section.
A similar analysis to that involved in obtaining (\ref{dd10}) yields:
\ba
\lim_{\bd\rightarrow 0}(\Psi_{f,h_{ab}, P_0}|\hat{D}[{\vec M}_i]_{\bd} c\ket =
\frac{3\hbar N}{4\pi\mathrm{i}}M(v, \{x\})\nu_{v}^{-2/3}&
\nonumber\\
%
g_{ c} \sum_{J} h_{J}   q^{i}_{J}{\hat V}^{a}_{J}
(\partial_a   ||{\vec {\hat V}}_{J}||_p f( p, \{x\} ))|_{p= v}
\label{dd12}
\ea
which can be written in  the notation of Appendix \ref{aconb} as
\be
\lim_{\bd\rightarrow 0}(\Psi_{f,h_{ab}, P_0}|\hat{D}[{\vec M}_i]_{\bd} c\ket =
\frac{3\hbar N}{4\pi\mathrm{i}}\nu_{v}^{-2/3}
g_{ c } \sum_{J} h_{J}   q^{i}_{J}(H^1_{J} (p= v))
\label{dd13}
\ee
In the above calculation the $Q$ factor is the same as that in (\ref{defq2dd}):
\be
Q(c_{0(i,J, \delta_0)},c_0, S_{1}) := \frac{N(N-1)(N-2)}{[(N-1)(N-2)(2+\cos^2\theta + (N-3)|\cos \theta |)]},
\label{defq11dd}
\ee
and have  used (\ref{dnot71}), (\ref{dj1=1}) to set $j=1$ in equation (\ref{p2-13-2}). Note that the sequence label $S_1$ is identical for (\ref{defq2dd}), (\ref{defq11dd}). The charge independence of the $Q$ factor (\ref{defq2dd})
together with the discussion in section \ref{sec6.4} implies that the $Q$ factor for (\ref{defq11dd}) {\em must necessarily} be the same as that for (\ref{defq2dd}).

Next we compute the contraction limit $\delta\rightarrow 0$ of (\ref{dd10}). 
From Appendix
\ref{acong}, and using $q>>1$, we have, as $\delta\rightarrow 0$:
\be
\sum_{{\hat J_1},{\hat K_1}}g_{   c_{[i,I,{\hat J}, {\hat K},\delta]_1}        }= g_{ c    }
(\delta)^{\frac{4}{3}(q-1)}
Q(c_{0[i,I,  \delta_0]^{0,1}_1}, S_{2}) h_{I}(1 +O(\delta^2))
\label{ddgcon2}
\ee
where we have used (\ref{dnot71}), (\ref{dj1=2}) to set $j=2$ in equation (\ref{p2-13-2}).
From equations (\ref{hicon}), (\ref{dmHncon}), (\ref{rot=1}),   as $\delta\rightarrow 0$ we have:
\ba
&\sum_{I_1} h_{I_1}   q^{i_1=i}_{I_1}(H^1_{I_1} (p= v_{[i,I ,\delta]_1}))
= &
%
\nonumber \\
&\delta^{-\frac{4}{3}(q-1)}\big\{ 
          \big( M(v_{[i,I ,\delta]_1} , \{x\})      {{\hat V}^{a}}_{I}\partial_{a} (f(v_{[i,I ,\delta]_1}       , \{x\}) 
               \sqrt{h_{ab}(v_{[i,I ,\delta]_1})  {{\hat V}^{a}}_{I}  {{\hat V}^{b}}_{I}})  &
\nonumber\\
 &\;[(N-1)(N-2) q^{i_1=i}_{I_1=I} +  (\sum_{I_1\neq I} q^{i_1=i}_{I_1})(N-2) (1+ \cos\theta (1+ \cos^2 \theta + (N-3)|\cos \theta | )] \big)&
\nonumber\\
& + O(\delta^{2})
\big\}&\nonumber\\
&=
\delta^{-\frac{4}{3}(q-1)}
\big\{ 
          \big(M(v_{[i,I ,\delta]_1} , \{x\})     q^{i_1=i}_{I_1=I} {{\hat V}^{a}}_{I}\partial_{a} (f(v_{[i,I ,\delta]_1}       , \{x\}) 
               \sqrt{h_{ab}(v_{[i,I ,\delta]_1})  {{\hat V}^{a}}_{I}  {{\hat V}^{b}}_{I}}) &
\nonumber\\
 &[(N-1)(N-2) - (N-2)(\cos\theta )(1+ \cos^2 \theta + (N-3)|\cos \theta| )]\big) + O(\delta^{2}) 
\big\} &
\label{dd14}
\ea
where we have used gauge invariance to go from the first equality to the second.

Next, we choose the $Q$ factor in (\ref{ddgcon2}) to be:
\be
Q(c_{[i,I,  \delta_0]^{0,1}_1}, S_{2}) = \frac{\nu_{v}^{-2/3}}{\nu_{v_{   [i,I,\delta ]_1 }}^{-2/3}}\frac{A N(N-1)(N-2)}{[(N-1)(N-2) - (N-2)(\cos\theta )(1+ \cos^2 \theta + (N-3)|\cos \theta| )]}
\label{defq1d}
\ee
where we shall specify the positive constant $A$ shortly and where, as in (\ref{defq1h}), $N>3, |\cos\theta | <1$ implies that $Q$ is positive. Note that in (\ref{defq1d}), we have $S_2= (d_i, d_i)$ whereas in 
(\ref{defq1h}), $S_2=(h,h)$ so that the $Q$ factors for these 2 equations can be (and are) chosen to be distinct from each other.

Using (\ref{ddgcon2})- (\ref{defq1d}) in (\ref{dd10}) yields:
\ba
&\sum_{{\hat J}_1, {\hat K}_1}\lim_{\bd\rightarrow 0}(\Psi_{f,h_{ab}, P_0}|\hat{D}[{\vec M}_i]_{\bd} c_{[i,I, {\hat J}, {\hat K},  \delta]_1}\ket =
A(N)(N-1)(N-2)\frac{3\hbar N}{4\pi\mathrm{i}}\nu_{v}^{-2/3}&
\nonumber\\
 & g_{ c} h_I     q^{i_1=i}_{I_1=I} (M(v_{[i,I ,\delta]_1} , \{x\})      {{\hat V}^{a}}_{I}\partial_{a} (f(v_{[i,I ,\delta]_1}       , \{x\}) 
               \sqrt{h_{ab}(v_{[i,I ,\delta]_1})  {{\hat V}^{a}}_{I}  {{\hat V}^{b}}_{I}}) )  + O(\delta^2 )&
\label{dd15}
\ea
In the above equation note that $q^{i_1}_{I_1=I}$ refers to the charge on the $(I_1=I)$th edge of $c_{[i,I, {\hat J}, {\hat K}, \delta]_1}$. Since the transition involved is of electric diffeomorphism type, there is no charge flipping  so that 
this charge is equal to the  charge on the $I$th edge of $c$ so that we have:
\be
 q^{i_1=i}_{I_1=I}= q^i_I
\label{dqid}
\ee 
Using  (\ref{dqid}) in (\ref{dd15}) we have:
\ba
&\sum_{{\hat J}_1, {\hat K}_1} \lim_{\bd\rightarrow 0}(\Psi_{f,h_{ab}, P_0}|\hat{D}[{\vec M}_i]_{\bd} c_{[i,I, {\hat J}, {\hat K},\delta]_1}\ket =
A(N)(N-1)(N-2)\frac{3\hbar N}{4\pi\mathrm{i}}\nu_{v}^{-2/3} g_c &
\nonumber\\
 & h_I q^i_I        (M(v_{[i,I ,\delta]_1} , \{x\})      {{\hat V}^{a}}_{I}\partial_{a} (f(v_{[i,I ,\delta]_1}       , \{x\}) 
               \sqrt{h_{ab}(v_{[i,I ,\delta]_1})  {{\hat V}^{a}}_{I}  {{\hat V}^{b}}_{I}}) ) +O(\delta^2 ) &
\label{dd16}
\ea
Expanding the second line of (\ref{dd16}) in a Taylor approximation and   summing over $I$, we  obtain
\ba
  & \sum_{I}h_I     q^i_I        (M(v_{[i,I ,\delta]_1} , \{x\})      {{\hat V}^{a}}_{I}\partial_{a} (f(v_{[i,I ,\delta]_1}       , \{x\}) 
               \sqrt{h_{ab}(v_{[i,I ,\delta]_1})  {{\hat V}^{a}}_{I}  {{\hat V}^{b}}_{I}}) ) &              
\nonumber\\
&  =\sum_{I}h_I\big\{
q^i_I      (M(v, \{x\})      {{\hat V}^{a}}_{I}\partial_{a} (f(v       , \{x\}) 
  \sqrt{h_{ab}(v)  {{\hat V}^{a}}_{I}  {{\hat V}^{b}}_{I}}) ) &
\nonumber \\
& + \delta  (q^{i}_I)^2  {\hat V}^b_I \partial_b    \big( M(p, \{x\}) (     {{\hat V}^{a}}_{I}\partial_{a} (f(p       , \{x\}) 
  \sqrt{h_{ab}(p)  {{\hat V}^{a}}_{I}  {{\hat V}^{b}}_{I}}) )\big)|_{p=v}
  \big\}    +O(\delta^2)  &
\label{dd17}
\ea

Next, consider the dual action of (\ref{dd2}) on the anomaly free state in the limit $\bd\rightarrow 0$:
\ba
&\lim_{\bd\rightarrow 0}(\Psi_{f,h_{ab}, P_0}|    \hat{D}[{\vec M}_i]_{\bd}              \hat{D}[{\vec N}_i]_{\delta}c\ket  = \frac{3\hbar}{4\pi\mathrm{i}}N(v, \{x\})\nu_{v}^{-2/3}  \frac{1}{(N-1)(N-2)\delta}&
\nonumber\\
&\big(\sum_{I,{\hat J_1},{\hat K_1}}  \lim_{\bd\rightarrow 0} (\Psi_{f,h_{ab}, P_0}|\hat{D}[{\vec M}_i]_{\bd} c_{[i,I, {\hat J}, {\hat K},\delta]_1}\ket) &
\nonumber\\
&-         \sum_{I,{\hat J_1},{\hat K_1}}                    (\Psi_{f,h_{ab}, P_0}|\hat{D}[{\vec M}_i]_{\bd} c\ket \big) &
\label{dd18}
\ea
The second line of (\ref{dd18})  can be evaluated using (\ref{dd17}).  
The  zeroth order term in $\delta$ in this expansion  is 
$A(N)(N-1)(N-2)$ times the right hand side of (\ref{dd12}). In the term on the 3rd line of (\ref{dd18}), the amplitude is exactly that of (\ref{dd12}) and the indices 
$I, {\hat J}_1,{\hat K}_1$ are dummy indices for this amplitude so that the amplitude is simply multiplied by a factor of $N$ (coming from the sum over $I$) and $(N-1)(N-2)$ (from the sum over the hatted indices)
Hence the zeroth order term of the second line cancels the contribution from the third line {\em only if we set} $A=1$.

On the other hand, as mentioned above the term on the 3rd line of (\ref{dd18}) does not contribute to the commutator. Hence we are not restricted to the choice $A=1$ if we are only interested in the commutator.
This commutator is:
\ba
&\lim_{\delta\rightarrow}\lim_{\bd\rightarrow 0}(\Psi_{f,h_{ab}, P_0}|   ( \hat{D}[{\vec M}_i]_{\bd}              \hat{D}[{\vec N}_i]_{\delta}  - N\leftrightarrow M )| c\ket  = 
\frac{3\hbar}{4\pi\mathrm{i}}\nu_{v}^{-2/3}   &
\nonumber\\
&\lim_{\delta\rightarrow 0}\frac{1}{(N-1)(N-2)\delta}\;\;\big\{ (\sum_{I,{\hat J_1},{\hat K_1}}  \lim_{\bd\rightarrow 0} (\Psi_{f,h_{ab}, P_0}|N(v, \{x\})\hat{D}[{\vec M}_i]_{\bd} c_{[i,I, {\hat J}, {\hat K},\delta]_1}\ket) &
\nonumber\\
&-   (\sum_{I,{\hat J_1},{\hat K_1}}  \lim_{\bd\rightarrow 0} (\Psi_{f,h_{ab}, P_0}|M(v, \{x\})\hat{D}[{\vec N}_i]_{\bd} c_{[i,I, {\hat J}, {\hat K},\delta]_1}\ket)\big\} &      
\label{dd19}
\ea

Using (\ref{dd17}) in  (\ref{dd19}),   taking the $\delta \rightarrow 0$ limit and leaving $A$ undetermined (and in particular, not necessarily equal to unity),  we obtain: 
\ba
&(\Psi_{f,h_{ab}, P_0}| \sum_{i=1}^3  [ \hat{D}[{\vec M}_i]   ,           \hat{D}[{\vec N}_i ] ]c\ket  =  A(\frac{3\hbar N}{4\pi\mathrm{i}})^2\nu_{v}^{-4/3} g_c&
\nonumber\\
&\big\{\sum_{i, I}h_I  (q^{i}_I)^2  \;    [N(v, \{x\}) {\hat V}^b_I (\partial_b    M(p, \{x\}))-  M(v, \{x\}) {\hat V}^b_I (\partial_b    N(p, \{x\}))] &
\nonumber\\
&\;[     {{\hat V}^{a}}_{I}\partial_{a} (f(p       , \{x\}) 
  \sqrt{h_{ab}(p)  {{\hat V}^{a}}_{I}  {{\hat V}^{b}}_{I}}) ]|_{p=v}  
  \big\} &
\label{d,deval}
\ea

Making contact through (\ref{dnot71}) with (\ref{defHlmp}) this can be written succintly as:
\ba
&(\Psi_{f,h_{ab}, P_0}|    \sum_{i=1}^3  [ \hat{D}[{\vec M}_i]   ,           \hat{D}[{\vec N}_i ]                 c\ket  =  A(\frac{3\hbar N}{4\pi\mathrm{i}})^2(\nu_{v}^{-4/3})g_c &
\nonumber\\
&\sum_i\sum_I h_I (q^i_I)^2 
(H^2_I(M, N; p=v) - H^2_I(N, M; p=v))&
\label{d,dconcise}
\ea
Comparing this with  (\ref{h,hconcise}), we obtain:
\be
(\Psi_{f,h_{ab}, P_0}|    \sum_{i=1}^3  [ \hat{D}[{\vec M}_i]   ,           \hat{D}[{\vec N}_i ]                 c\ket  = \frac{4A}{3\beta_M\beta_N} 
(\Psi_{f,h_{ab}, P_0}|     [ \hat{C}[{ M}]   ,           \hat{C}[{ N} ]                 c\ket
\label{ddhh1}
\ee
Comparing this with  (\ref{key}), we obtain an anomaly free commutator if:\\
\noindent(a) We choose succesive actions of the Hamiltonian constraint to have opposite flips so that \\
$\beta_M= -\beta_N$ so that $\beta_M\beta_N=-1$.
\\
\noindent (b) We choose $A=\frac{1}{4}$.

Making these choices we obtain the desired anomaly free result.
Note that because we have been obliged to choose $A\neq 1$, the continuum limit product of 2 electric diffeomorphism constraints is {\em not} defined; only their commutator is well defined.
However, the electric diffeomorphism constraint operator is not one of the constraint operators used to generate the constraint algebra; its role is restricted to the demonstration 
of an anomaly free  commutator between a pair of Hamiltonian constraints in accord with (\ref{key}). Hence the ill definedness of the product of 2 electric diffeomorphism constraint operators
is not an obstruction to our treatment of the constraint algebra.

\section{\label{sec8} Multiple products of Hamiltonian constraints}

In section \ref{sec8.1} we derive, through an inductive proof,  the expression for the action of multiple products of  Hamiltonian constraint operators on an anomaly free state. 
In section \ref{sec8.2} we show that the action derived in \ref{sec8.1} yields anomaly free single commutators (see section \ref{sec1} for our usage of the term `anomaly free single commutator'). 
Since the detailed calculations below are similar to those of section \ref{sec7}, we shall only highlight the main steps of these calculations in  our exposition.

\subsection{\label{sec8.1} Multiple products of Hamiltonian constraints: Derivation}

\subsubsection{\label{sec811}Introductory Remarks}

We note the following:\\

\noindent (1)The discrete action of a  product of  $n$ Hamiltonian constraints in (\ref{dual})  requires a choice of $\beta$ (which characterises the charge flips) for each constraint action.
In the rest of this work, we choose $\beta =\beta_{\rm i}$ for the ${\rm i}$th Hamiltonian constraint in (\ref{dual}) to be $+1$ if ${\rm i}$ is {\em odd} and $-1$ if ${\rm i}$ is {\em even}.
This is consistent with the choice made in section \ref{sec7.1}  for the case of  $n=2$.\\

\noindent (2) Note  that equation (\ref{dual}) is evaluated through the 2 steps outlined in section \ref{sec6.3} so that the coordinate patch with respect to which the amplitude of a deformed 
child generated by the operator product  is evaluated is the appropriate {\em contraction} coordinate patch. Only for the undeformed state $c$, the amplitude is evaluated with respect to
the {\em reference} coordinates associated with $c$.
\\

\noindent (3) Recall that the cone angle is acute or obtuse depending on whether the deformations are upward or downward.
Hereon we will tailor our choice of cone angle to the choice of Bra Set so that $|\cos \theta|$ is fixed and the same for all deformations of ket correspondents of members of the Bra Set
and is chosen such that: 
\be
|\cos\theta | (3N q^{primordial}_{max}) <  1
\label{thetachoice}
\ee
with $q^{primordial}_{max}$
defined as in (\ref{defqprimmax}), where the set of edge charges in that equation can be taken to be those of any primordial charge net in the Bra Set.
\footnote{The value of $q^{primordial}_{max}$ is independent of the choice of primordial charge net in the Bra Set since any such primordial has the same set of unordered edge charges (see section \ref{sec5.1}).}
Note that this  condition is equivalent to the condition:
\be
|\cos\theta | (3N q^{net}_{max}) <  1
\label{thetanetchoice}
\ee
where 
\be 
q^{net}_{max} = \max_{(i=1,2,3),(I=1,..,N)} |q^i_{I}|.
\label{defqnetmax}
\ee
where the charges $q^i_{I}$ are the {\em net} edge charges
\footnote{This notation is  consistent with the Note at the beginning of section \ref{sec7}; note that this equation is in general distinct from (\ref{defqmax}) because the charges on the right hand side of that equation
refer to the actual edge charges not the net edge charges.}
at the nondegenerate vertex of any element of the Bra Set. It is easy to check that this equivalence follows immediately from Appendix \ref{acolor} together with 
the definition of the Bra Set in section \ref{sec5.1}.

\noindent (4) Equations (\ref{defqmin}) and (\ref{defqmin,1}) are defined as conditions on primordial charges.  Appendix \ref{acolor} shows that  the net charges and primordial charges on corresponding edges are identical or flipped images of 
each other.
Hence the equations (\ref{defqmin}) and (\ref{defqmin,1}) also hold for net charges on multiply deformed children of primordial charge nets and we shall so interpret them when we refer to them hereon.

\subsubsection{\label{secsum}Summary of Choices}
It is useful to note that  from sections  \ref{sec4.2}, \ref{sec4.3},  \ref{sec4.5} and (3) of section \ref{sec811}, that the action (\ref{dual}) is fixed once the following choices have been  made:\\
\noindent (a) The Set of Primordial States $S_{primordial}$. \\
\noindent (b) A Primary Coordinate patch $\{x_0\}$ around a point $p_0$.\\
\noindent (c) A set of primordial reference states, one for each diffeomorphism class of states in $S_{primordial}$,  the non-degenerate vertex of each
such reference state being located at $p_0$ and linear with respect to $\{x_0\}$. These primordial reference states are divided into exhaustive and mutually exclusive classes
each class defining a Bra Set so that members of each class have the same set of unordered edge charges. For each class we choose a cone angle $\theta$ which satisfies (\ref{thetachoice}).
\\
\noindent (d) A reference state for each distinct diffeomorphism class of elements of $S_{primary}$, each such reference state itself being an element of $S_{primary}$.
\footnote{Recall that $S_{primary}$ is the set of primaries  generated from reference primordials through repeated conical deformations of the type constructed in 
Appendix \ref{acone} and section \ref{secneg} with respect to $\{x_0\}$ so that $S_{primary}$ is determined once (a)-(b) above are fixed.  In particular the cone angles characterizing the 
conical deformations  are fixed by (b).  Recall also 
that the Ket Set $S_{ket}$ comprises of all diffeomorphic images of elements in $S_{primary}$ and hence is also determined once (a)- (b) are fixed.}
\\
\noindent (e) A choice of reference diffeomorphism,  one for each element $c$ of the Ket Set,  which maps the reference state  $c_0$ for this element $c$,  to $c$.\\
\noindent (f) A choice of deformation such that any (single or multiple)  deformation of any element of  the complement of the Ket Set is also in this complement.   
\\

Once the choices (a)-(f) are made, the formalism is rigid in that the choice of upward/downward conical deformations is fixed as in section \ref{secneg}, the contraction 
procedure is fixed as in section \ref{sec4.3}, the discrete action of operator products is fixed as in section \ref{sec4.4} with the sign flips chosen in accord with (1) above, the anomaly free basis states are chosen as in section \ref{sec5}
and the continuum limit is defined as in equation (\ref{contlim}).\\

\subsubsection{\label{sec812}Notation}
Recall that the Cauchy manifold is a $C^{k}$- semianalytic manifold for some $k>>1$.
We  compute the continuum limit (\ref{contlim}) for arbitrary $n < k$  with ${\hat O}_{\rm i}(N_{\rm i}), {\rm i}=1,..,n$ being Hamiltonian constraint operators.
We restrict attention to  the case that  $c$  in (\ref{contlim})  is in the Bra Set because, as mentioned in section \ref{sec6.1}, for $c$ not in the Bra Set, the dual action vanishes.

We denote the non-degenerate vertex of $c$ by $v$ and its associated reference coordinate patch by  $\{x\}$.
We shall be interested in a proof by mathematical induction. In the course of that proof, it will suffice to develop notation only for singly deformed states.
The single deformations of interest will be denoted as
\be
[i,I,{\hat J}, {\hat K}, \beta, \delta ]_1= ( i, I, {\hat J}_1, {\hat K_1}, \beta, \delta )
\label{8not1}
\ee
The vertex of the singly deformed state  $c_{[i,I,{\hat J}, {\hat K}, \beta, \delta ]_1}$ is denoted by $v_{[i,I,\delta]_1}$ and its associated 
{\em contraction} coordinate patch by $\{x^{\delta}\}$. In the induction proof it will turn out that the single deformation of (\ref{8not1}) will play the role of the first of  $m+1$
deformations applied to $c$. Accordingly, in relation to the  notation  of section \ref{sec4.3},  in this section we have set:
\ba
\{x_{\alpha}\} \equiv \{x\},&  \;\;\;  j_1= m+1,&  \;\; \epsilon_{\rm j_1} \equiv \delta \nonumber\\
&\{x_{\alpha}^{\epsilon_{\rm j_1} }\} \equiv \{x^{\delta}\}& 
\label{8not2}
\ea
As usual,  wherever required explicitly, we denote the density weighted object $B$ evaluated at point $p$ in the coordinate system $\{y\}$ by $B(p, \{y\})$.
We shall also make extensive use of the notation developed in Appendix \ref{aconb}.

\subsubsection{\label{sec813}Proof by Induction}

Let  $n$ be a positive integer with $n\leq k-1$ where $k$ is the differentiability class of the semianalytic Cauchy slice.
Define $k_n$ as:
\ba
k_n &=&\frac{n-1}{2} \;\;\;{\rm if} \;\;n\;\;{\rm is \;\;odd}, \nonumber\\
k_n &=&\frac{n}{2} \;\;\;{\rm if} \;\;n\;\;{\rm is \;\;even}, 
\label{defkn}
\ea
Let $c$ be in the Bra Set and let the $i$th net charge at the $I$th edge at its nondegenerate vertex $v$ be $q^i_I$. Define:
\be
|{\vec q}_I|= \sqrt{ \sum_{i=1}^3(q^i_I)^2}
\label{normq}
\ee

\noindent{\em Claim}: The continuum limit of the dual action of a product of $n$ Hamiltonian constraints  when $n$ is even is
\be
(\Psi_{f,h_{ab}, P_0}|
(\prod_{ {\rm i}=1 }^n {\hat C}_{}(  N_{\rm i})) |c\ket = 
(-3)^{k_n} (\frac{3\hbar N}{8\pi i})^n (\nu^{-\frac{2}{3}})^ng_c 
\sum_{I} |{\vec q}_I|^n h_IH_I^n (N_1,..,N_n;v), 
\label{heven}
\ee
and the continuum limit of the dual action of a product of $n$ Hamiltonian constraints  when $n$ is odd is
\be
(\Psi_{f,h_{ab}, P_0}|
(\prod_{ {\rm i}=1 }^n {\hat C}_{}(  N_{\rm i})) |c\ket = 
(-3)^{k_n} (\frac{3\hbar N}{8\pi i})^n (\nu^{-\frac{2}{3}})^ng_c 
\sum_{I} |{\vec q}_I|^{n-1}(\sum_{i=1}^3q^i_I) h_IH_I^n (N_1,..,N_n;v), 
\label{hodd}
\ee
and where we have used equation  (\ref{defHlmp}) to define $H_I^n (N_1,..,N_n;p)$ so that:
\be
H_I^n (N_1,..,N_n;v)= 
 \big(\prod_{{\rm i}=1}^n N_{\rm n-i+1}(p, \{x\})     {\hat V}_I^{a_{\rm n-i+1}}(p)\partial_{a_{\rm n-i+1}}\big)(f(p,\{x\}) \sqrt{h_{ab}(p){\hat V}_I^a(p){\hat V}^b_I(p)})|_{p=v}
\ee
and the product is ordered from left to right in increasing $i$.
\\

\noindent {\em Proof By Induction}:\\
\noindent{\bf{Step 1}}: In section \ref{sec7.1} we have shown that (\ref{hodd}) holds for $n=1$ (see (\ref{hh13})) and that (\ref{heven}) holds for $n=2$  (see (\ref{hhconcise})).
More in detail, 
clearly, we have shown (\ref{hodd}) holds for $n=1$ and that (\ref{heven}) holds for $n=2$ {\em for any  choice of} (a)-(f), section \ref{secsum}  with $c$ being in a Bra Set resulting from these choices.
\\

\noindent{\bf Step 2} Assume that (\ref{heven}) holds for $n=m$, $m$ even, {\em for any choice of} (a)- (f), section \ref{secsum}. 
Then we show below that (\ref{hodd})  holds for $n=m+1$ for  any choice of (a) -(f), section \ref{secsum}.  We have that
\be
(\Psi_{f,h_{ab}, P_0}|
(\prod_{ {\rm i}=1 }^{m+1} {\hat C}_{ }(  N_{\rm i})) |c\ket 
= \lim_{\delta\rightarrow 0}(\Psi_{f,h_{ab}, P_0}|
(\prod_{ {\rm i}=1 }^m {\hat C}_{}(  N_{\rm i})) {\hat C}_{\delta} (N_{\rm m+1})|c\ket .
\label{he1}
\ee
Using (\ref{hamsum}) we obtain:
\ba
&(\Psi_{f,h_{ab}, P_0}|
(\prod_{ {\rm i}=1 }^m {\hat C}_{}(  N_{\rm i})) {\hat C}_{\delta} (N_{\rm m+1})|c\ket = (-1)^m
\frac{3\hbar}{8\pi \mathrm{i}}N(x(v))\nu_{v}^{-2/3}\frac{1}{(N-1)(N-2)\delta}  & \nonumber\\    
& \sum_{i,I{\hat J}_1, {\hat K}_1}   \big( (\Psi_{f,h_{ab}, P_0}| (\prod_{ {\rm i}=1 }^m {\hat C}(  N_{\rm i})) | c_{[i,I, {\hat J}, {\hat K}, \beta, \delta]_1}\ket- (\Psi_{f,h_{ab}, P_0}|(\prod_{ {\rm i}=1 }^m {\hat C}(  N_{\rm i}))|c\ket \big)&
\label{he2}
\ea
Here the $(-1)^m$ factor is unity for $m$ even in accord with (1), section \ref{sec811}. 
\footnote{\label{fnodd}The equation as it is written would also be valid for $m$ odd where from (1),  section \ref{sec811} we require an overall $-1 = (-1)^m$ factor coming from our choice 
of the $m+1$th $\beta$- flip when $m$ is odd.}
The second amplitude in the last line of (\ref{he2}) is given by (\ref{heven}) with $n=m$. The first amplitude (within the summation symbol)  looks as if  it could be evaluated through a direct application of (\ref{heven}). However from (2), 
section \ref{sec811}, the 
coordinates associated with $c_{[i,I, {\hat J}, {\hat K}, \beta, \delta]_1}$ are the contraction coordinates whereas (\ref{heven}) is applicable only if these coordinates were reference coordinates.
Recall however that we have assumed (\ref{he2}) for {\em any} choice of (a)- (f). Our strategy is then to make choices for (a)- (f) such that (\ref{heven}) is directly applicable to the first term in the context of such choices.

We proceed as follows.
Consider some fixed choice of (a)- (f) in section \ref{secsum}, for which (\ref{heven}) is used to evaluate the second term  in the last line of (\ref{he2}).
In this fixed choice, as in section \ref{sec4},  let  $c_0$ be the reference state for $c$, let the reference diffeomorphism which maps $c_0$ to $c$ be $\alpha$,
let the deformation of $c_0$ with respect to the primary coordinates $\{x_0\}$ at parameter $\delta_0$ be $c_{0(i,I,  \beta, \delta_0)}$ and  let the  contraction image of $c_{0(i,I , \beta, \delta_0)}$
by the 
appropriate contraction diffeomorphism  be $c_{ 0[i,I , {\hat J}, {\hat K}, \beta, \delta ]_1}$ so that $c_{ [i,I, {\hat J}, {\hat K}, \beta, \delta]_1}$ is the image by $\alpha$ of $c_{0[i,I, {\hat J}, {\hat K}, \beta, \delta]_1 }$.
Using (\ref{8not2}) this contraction diffeomorphism from (\ref{phi1}) is:
\be
\Phi^{\epsilon_{{\rm j_{1}}=m+1 }= \delta, \{x_0\}, {\hat J}_1, {\hat K}_1}_{ c_0, (i,I,\beta,\delta_0), S_{{\rm j_{1}}=m+1 }   } \equiv \phi
\label{he3}
\ee
where for  notational simplicity  in this Step (i.e Step 2), we have  denoted the contraction diffeomorphism on the left hand side of (\ref{he3}) by 
$\phi$. 
Now consider the choices (a')- (c') below  which are images of the fixed choice made above by the diffeomorphism $\phi$. These  `$\phi$'- choices are then as follows: 
\\
\noindent (a') The set of primordials $S_{\phi,primordial}$ is chosen to be the image of the set $S_{primordial}$ of primordials chosen in accordance with choice (a);
since $S_{primordial}$ is closed under diffeomorphisms we have that $S_{\phi,primordial}$ is equal to $S_{primordial}$. \\
\noindent (b') The Primary Coordinate patch is $\phi^*\{x_0\}$ around the  point $\phi (p_0)$.\\
\noindent (c') The  set of primordial reference states is just the set of images by $\phi$ of the fixed choice (c) of reference primordials.
The cone angles, as measured by the primary coordinates in (b'),  for deformations of primordial reference states are chosen to be identical to the choices in (c) for their diffeomorphically related 
counterparts \\

Next, note that the set of primaries, $S_{\phi, primary}$  are now generated from the reference primordials of (c') through conical deformations with respect to $\phi^*\{x_0\}$; it follows that
$S_{\phi, primary}$ consists of the images of the  elements of $S_{primary}$ by $\phi$. The Ket Set generated from $S_{\phi, primary}$ is then identical to the Ket Set $S_{Ket}$  generated from $S_{primary}$
because the Ket Set is closed under the action of diffeomorphisms.  Next, consider 
the diffeomorphism class $[c_{[i,I, {\hat J}, {\hat K}, \beta, \delta]_1}]$ of $c_{[i,I, {\hat J}, {\hat K}, \beta, \delta]_1}$. 
Clearly we have that $[c_{[i,I, {\hat J}, {\hat K}, \beta, \delta]_1}] = [c_{0(i,I,  \beta, \delta_0)}  ]$. Further, since $c_{0(i,I  \beta, \delta_0)} \in S_{primary}$, equation (\ref{he3}) and (c') above
imply that 
$c_{ 0[i,I , {\hat J}, {\hat K}, \beta, \delta ]_1}\in S_{\phi, primary}$. Hence may choose $c_{ 0[i,I , {\hat J}, {\hat K}, \beta, \delta ]_1}$
to be a reference state for $c_{[i,I, {\hat J}, {\hat K}, \beta, \delta]_1}$. Recall  that 
$c_{[i,I, {\hat J}, {\hat K}, \beta, \delta]_1}$ is the image by $\alpha$ of  $c_{ 0[i,I , {\hat J}, {\hat K}, \beta, \delta ]_1}$
Accordingly we choose (d') and (e') as follows:
\\
\noindent (d') We choose the reference state for $[c_{[i,I, {\hat J}, {\hat K}, \beta, \delta]_1}]$ to be $c_{ 0[i,I , {\hat J}, {\hat K}, \beta, \delta ]_1}$
and choose reference states for 
other diffeomorphism classes of elements of the Ket Set arbitrarily.\\
\noindent (e') We choose the reference diffeomorphism for $c_{[i,I, {\hat J}, {\hat K}, \beta, \delta]_1}$ to be $\alpha$ and choose the remaining reference diffeomorphisms
arbitrarily.
\\
Finally, since the Ket Set is unaltered we retain the choice of (f) i.e. we set (f') to be the same as the fixed choice (f) above. It is then easy to see that the Bra Set $B_{P_0}$ chosen with respect to (a)- (f)
is also a valid Bra Set with respect to (a')- (f').

Accordingly we consider the same Bra Set $B_{P0}$ as before and choose $h_{ab}, f$ also as before and obtain a state $\Psi^{\phi}_{f,h, P_0}$ based on its amplitude evaluations in the context of the choices  (a')- (f') above.
Our choices (a')- (f')  (especially (d'),(e') ) ensure that the {\em reference coordinates} for $c_{[i,I, {\hat J}, {\hat K}, \beta, \delta]_1}$   in the context of these choices is the same as the 
{\em contraction coordinates} for this state in the context of the fixed choices for (a)- (f) above. It is then straightforward to check that 
the contraction coordinates for any    deformed state state 
generated by 
the action of $(\prod_{ {\rm i}=1 }^m {\hat C}_{\epsilon_{\rm i}}      (  N_{\rm i}))$ on 
$c_{[i,I, {\hat J}, {\hat K}, \beta, \delta]_1}$ 
in the context of choices (a')- (f') coincides with the contraction coordinates for the same deformed state 
when it is generated by the action of $(\prod_{ {\rm i}=1 }^{m} {\hat C}_{\epsilon_{\rm i}}(  N_{\rm i}))  {\hat C}_{\delta}(N_{\rm m+1})$  on $c$.
It follows that the evaluation of (\ref{heven}) in accordance with section \ref{secsum} in the context of  the choices (a')- (f'), and , 
with $c$ replaced by $c_{[i,I, {\hat J}, {\hat K}, \beta, \delta]_1}$ and with $\Psi_{f,h, P_0}$  replaced by $\Psi^{\phi}_{f,h, P_0}$, coincides precisely
with the first term (within the summation symbol) in the last line of equation (\ref{he2}). 
It then follows from (\ref{heven}) that:
\ba
&(\Psi_{f,h_{ab}, P_0}|
(\prod_{ {\rm i}=1 }^m {\hat C}_{}(  N_{\rm i})) |c_{[i,I, {\hat J}, {\hat K}, \beta, \delta]_1}              \ket = 
(-3)^{k_m} (\frac{3\hbar N}{8\pi i})^m (\nu_{   v_{[i,I\delta ]_1}        }^{-\frac{2}{3}})^mg_{c_{[i,I, {\hat J}, {\hat K}, \beta, \delta]_1}} &\nonumber\\
&\sum_{L_1} |{\vec q}_{L_1}|^m h_{L_1}H_{L_1}^m (N_1,..,N_m;v_{ [i,I,\delta ]_1          }), &
\label{he4}
\ea
where, using the notation (\ref{8not2}),(\ref{defHlmp}), the coordinate dependent parts of $H_{L_1}^m(N_1,..,N_m; v_{[i,I, \delta]_1} $ are evaluated with respect to the coordinates $\{x^{\delta}\}$
which serve both as the reference coordinates for the state  $c_{[i,I, {\hat J}, {\hat K}, \beta, \delta]_1}$ in the choice scheme (a')-(f') or as the contraction coordinates 
in the fixed choice scheme of (a)- (f) above. We now revert back  to the latter interpretation of these coordinates.

Using the contraction behaviour of $h_{L_1}, H^{m}_{L_1}, g_{ c_{[i,I, {\hat J}, {\hat K}, \beta, \delta]_1}}$ derived in Appendices \ref{aconb}, \ref{acong} together with (\ref{8not2}),
we obtain,  as $\delta \rightarrow 0$ :
\ba
    &  \sum_{{\hat J}_1, {\hat K}_1}    (\Psi_{f,h_{ab}, P_0}| (\prod_{ {\rm i}=1 }^m {\hat C}(  N_{\rm i})) | c_{[i,I, {\hat J}, {\hat K}, \beta, \delta]_1}\ket 
= (-3)^{k_m} (\frac{3\hbar N}{8\pi i})^m (\nu_{   v_{[i,I\delta ]_1}        }^{-\frac{2}{3}})^mg_c  
Q (c_{0[i,I,\beta, \delta_0]^{1,0}_1}, S_{m+1}) h_I &  \nonumber\\
& \{|{\vec q}_{L_1=I}|^m(N-1)(N-2) + \cos^m(\theta)(\sum_{L_1\neq I}|{\vec q}_{L_1\neq I}|^m)(N-2)(1+\cos^2\theta +(N-3) |\cos \theta|\} &\nonumber\\
&\big(\prod_{{\rm i}=1}^m N^{a_{\rm m-i+1}}_{\rm m-i+1}(p, \{x\})     {\hat V}_I^{a_{\rm m-i+1}}(p)\partial_{a_{\rm m-i+1}}(f(p,\{x\}) \sqrt{h_{ab}(p){\hat V}_I^a(p){\hat V}^b_I(p)})|_{p=v_{[i,I,\delta]_1}}\big)  +\;O(\delta^2)\;\;\;&
\label{he5}
\ea

We set:
\ba
&Q (c_{0[i,I,\beta, \delta_0]^{1,0}_1}, S_{m+1} ) = &
\nonumber\\
&\frac{( \nu_v^{-\frac{2}{3}})^m  }{ (\nu_{   v_{[i,I\delta ]_1}        }^{-\frac{2}{3}})^m    }
\big(\frac{(N)(N-1)(N-2)|{\vec q}_I|^m}{\{|{\vec q}_{L_1=I}|^m(N-1)(N-2) + \cos^m(\theta)(\sum_{L_1\neq I}|{\vec q}_{L_1\neq I}|^m)(N-2)(1+\cos^2\theta +(N-3) |\cos \theta|\}}\big)&
\label{he6}
\ea
Here  ${\vec q}_I$ refers to the charge on the $I$th edge at $v$ in $c$ and ${\vec q}_{L_1}$ to the charge on the $L_1$th edge at $v_{[i,I,\delta ]_1}$ in $c_{[i,I, {\hat J}, {\hat K}, \beta, \delta]_1}$.  
Since $m$ is even, the $Q$ factor above is maniefstly positive as required.
With this choice of $Q$ we obtain:
\ba
&\sum_{{\hat J}_1, {\hat K}_1} (\Psi_{f,h_{ab}, P_0}|
(\prod_{ {\rm i}=1 }^m {\hat C}_{}(  N_{\rm i})) |c_{[i,I, {\hat J}, {\hat K}, \beta, \delta]_1}              \ket = 
(-3)^{k_m} (\frac{3\hbar N}{8\pi i})^m (\nu_{   v      }^{-\frac{2}{3}})^m g_{c}N(N-1)(N-2)h_I  |{\vec q}_I|^{m} &
\nonumber\\
&\prod_{{\rm i}=1}^m N^{a_{\rm m-i+1}}_{\rm m-i+1}(p, \{x\})     {\hat V}_I^{a_{\rm m-i+1}}(p)\partial_{a_{\rm m-i+1}}(f(p,\{x\}) \sqrt{h_{ab}(p){\hat V}_I^a(p){\hat V}^b_I(p)}|_{p=v_{[i,I,\delta]_1}}  + O(\delta^2) &
\label{he7}
\ea

As in our treatment of similar terms in section \ref{sec7}, we expand the right hand side of the above equation in a Taylor approximation in powers of $\delta$ .
\footnote{\label{tev} This Taylor expansion is valid provided $m+1\leq k-1$.}
It is easy to see that the $0$th order contribution 
exactly cancels the contribution from the second term in (\ref{he2}). The first order term provides the only contribution to (\ref{he2}) which survives in the $\delta\rightarrow 0$ limit. It is straightforward to check that 
taking this limit of (\ref{he2}),  we obtain:
\ba
&(\Psi_{f,h_{ab}, P_0}|
(\prod_{ {\rm i}=1 }^{m} {\hat C}_{}(  N_{\rm i})) {\hat C}_{\delta} (N_{\rm m+1})    |c\ket = &\nonumber \\
&(-3)^{k_m} (\frac{3\hbar N}{8\pi i})^{m+1} (\nu_{   v      }^{-\frac{2}{3}})^{m+1} g_{c}
\sum_{i,I} \big\{|{\vec q}_I|^{m} 
 N_{\rm m+1}(p,\{x\}) h_I q^{i}_I 
 & \nonumber\\
& {\hat V}^a_I(p)\partial_a  (\prod_{{\rm i}=1}^{m} N^{a_{\rm m-i+1}}_{\rm m-i+1}(p, \{x\})     
{\hat V}_I^{a_{\rm m-i+1}}(p)\partial_{a_{\rm m-i+1}}(f(p,\{x\}) \sqrt{h_{ab}(p){\hat V}_I^a(p){\hat V}^b_I(p)}))_{p=v}
\big\}&
\label{he8}
\ea
From (\ref{defkn}) it follows that when $m$ is even $k_m= k_{m+1}$. Using the notation (\ref{defHlmp}) together with this fact in (\ref{he8}) yields:
\ba
&(\Psi_{f,h_{ab}, P_0}|
(\prod_{ {\rm i}=1 }^{m+1} {\hat C}_{}(  N_{\rm i}))  |c\ket = &\nonumber \\
&(-3)^{k_{m+1}} (\frac{3\hbar N}{8\pi i})^{m+1} (\nu_{   v      }^{-\frac{2}{3}})^{m+1} g_{c} 
\sum_{i,I} \big\{|{\vec q}_I|^{m-1}  q^{i}_I h_I H^{m+1}_I (N_1,..,N_{m+1}; v)&
\label{he9}
\ea
which is the desired result (\ref{hodd}) with $n=m+1$.
\\
Since the fixed choice (a)- (f) underlying this derivation is arbitrary and since the assumed form for $n=m$ holds for any such choice, the  result (\ref{he9}) also holds for any choice of (a)- (f).
\\

\noindent{\bf Step 3} Assume that (\ref{hodd}) holds for $n=m$, $m$ odd, {\em for any choice of} (a)- (f). Then we show below that (\ref{heven})  holds for $n=m+1$ for  any choice of (a) -(g).
The first part of our analysis is identical to the first part of the analysis in Step 2. Note that in Step 2, equations (\ref{he1}) - (\ref{he3}) hold regardless of whether $m$ is odd or even 
(see Footnote \ref{fnodd} with regard to the validity of  (\ref{he1}) when $m$ is odd).
The second amplitude in the last line of (\ref{he2}) is now given by (\ref{hodd}) with $n=m$.  To apply (\ref{hodd}) to the first amplitude in the last line of (\ref{he2}) we repeat the analysis
subsequent to (\ref{he2}) till (but not inclusive of) (\ref{he4}). The choices (a')-(f') allow us to apply (\ref{hodd}) also to the first amplitide in the last line of (\ref{he2}). 
Accordingly, this term evaluates to:
\ba
&(\Psi_{f,h_{ab}, P_0}|
(\prod_{ {\rm i}=1 }^m {\hat C}_{}(  N_{\rm i})) |c_{[i,I, {\hat J}, {\hat K}, \beta, \delta]_1}              \ket = 
(-3)^{k_m} (\frac{3\hbar N}{8\pi i})^m (\nu_{   v_{[i,I\delta ]_1}        }^{-\frac{2}{3}})^mg_{c_{[i,I, {\hat J}, {\hat K}, \beta, \delta]_1}} &\nonumber\\
&\sum_{L_1} |{\vec q}_{L_1}|^{m-1}  (\sum_{i_1}q^{i_1}_{L_1}) h_{L_1}H_{L_1}^m (N_1,..,N_m;v_{ [i,I\delta ]_1          }), &
\label{ho4}
\ea
where, similar to (\ref{he4}), using the notation (\ref{8not2}),(\ref{defHlmp}), the coordinate dependent parts of $H_{L_1}^m(N_1,..,N_m; v_{[i,I, \delta]_1} )$ are evaluated with respect to the coordinates $\{x^{\delta}\}$.
Once again, using the contraction behaviour of $h_{L_1}, H^{m}_{L_1}, g_{ c_{[i,I, {\hat J}, {\hat K}, \beta, \delta]_1}}$ derived in Appendices \ref{aconb}, \ref{acong} together with (\ref{8not2}), 
we obtain,  as $\delta \rightarrow 0$ :
\ba
&  \sum_{{\hat J}_1, {\hat K}_1}    (\Psi_{f,h_{ab}, P_0}| (\prod_{ {\rm i}=1 }^m {\hat C}(  N_{\rm i})) | c_{[i,I, {\hat J}, {\hat K}, \beta, \delta]_1}\ket 
= (-3)^{k_m} (\frac{3\hbar N}{8\pi i})^m (\nu_{   v_{[i,I\delta ]_1}        }^{-\frac{2}{3}})^mg_c  &
\nonumber \\
&Q (c_{0[i,I,\beta, \delta_0]^{1,0}_1}, S_{m+1}) h_I 
 \{|{\vec q}_{L_1=I}|^{m-1} (\sum_{i_1}q^{i_1}_{L_1=I})     (N-1)(N-2) &
\nonumber\\ 
& + \cos^m(\theta)(\sum_{L_1\neq I}|{\vec q}_{L_1\neq I}|^{m-1} (\sum_{i_1}q^{i_1}_{L_1})  )(N-2)(1+\cos^2\theta +(N-3) |\cos \theta|\} &\nonumber\\
&\big(\prod_{{\rm i}=1}^m N^{a_{\rm m-i+1}}_{\rm m-i+1}(p, \{x\})     {\hat V}_I^{a_{\rm m-i+1}}(p)\partial_{a_{\rm m-i+1}}(f(p,\{x\}) \sqrt{h_{ab}(p){\hat V}_I^a(p){\hat V}^b_I(p)})|_{p=v_{[i,I,\delta]_1}} \big)&
\nonumber \\
&+\;\; O(\delta^2)&
\label{ho5}
\ea

We set:
\ba
&Q (c_{0[i,I,\beta, \delta_0]^{1,0}_1}, S_{m+1} ) = \frac{( \nu_v^{-\frac{2}{3}})^m  }{ (\nu_{   v_{[i,I\delta ]_1}        }^{-\frac{2}{3}})^m    } \times &
\nonumber\\
&
\frac{  3(N)(N-1)(N-2)|{\vec q}_I|^{m-1} (\sum_{i_1}q^{i_1}_{L_1=I})        }{     
[|{\vec q}_{L_1=I}|^{m-1}( \sum_{i_1}q^{i_1}_{L_1=I}) (N-1)(N-2) ]
+ [\cos^m\theta(\sum_{L_1\neq I}|{\vec q}_{L_1}|^{m-1}(\sum_{i_1}q^{i_1}_{L_1}) )
(N-2)(1+\cos^2\theta +(N-3) |\cos \theta|  ]  }   \;\;\;\;\;\;\;\;  
&
\label{ho6}
\ea
For $\theta$ constrained by equation (\ref{thetachoice}), (\ref{thetanetchoice}) it is straightforward to check that the sign of the denominator is that of its first term $|{\vec q}_{L_1=I}|^{m-1}(\sum_{i_1}q^{i_1}_{L_1=I}) (N-1)(N-2)$ 
which is then the same as the 
sign of the numerator so that $Q$ is positive. Note that $Q$ in (\ref{ho6}) is different from that in (\ref{he6}); this is not a problem because their associated constraint strings $S_{m+1}$ are different in that in one
case $m$ is even and in the other $m$ is odd. Note that because the transition $[i,I, {\hat J}, {\hat K}, \beta, \delta]_1$ is a Hamiltonian constraint generated one (with $\beta =-1$ in accord with (1), section \ref{sec811}),
we have that 
\begin{equation}
q^{i_1}_{I_1=I} = \left.  ^{(i)}\!q_I^{i_1}\right.  =\delta^{ii_1}q_I^{i_1} +%
{\textstyle\sum\nolimits_{k}}
\epsilon^{ii_1k}q_I^{k} \label{hoflip}%
\end{equation}
where the left hand side refers to the $I$th edge charge in the child $c_{[i,I, {\hat J}, {\hat K}, \beta, \delta]_1}$ and the right hand side to the $I$th edge charge in the parent $c$.
Using (\ref{ho6}) and (\ref{hoflip}) in (\ref{ho5}),  together with the fact that the norm of the charge vector is flip- independent we obtain:
\ba
&\sum_{{\hat J}_1, {\hat K}_1} (\Psi_{f,h_{ab}, P_0}|
(\prod_{ {\rm i}=1 }^m {\hat C}_{}(  N_{\rm i})) |c_{[i,I, {\hat J}, {\hat K}, \beta, \delta]_1}              \ket = 
(-3)^{k_m} (\frac{3\hbar N}{8\pi i})^m (\nu_{   v      }^{-\frac{2}{3}})^m g_{c}3N(N-1)(N-2)& \nonumber \\
& h_I  |{\vec q}_I|^{m-1}  (\sum_{i_1}    \left.  ^{(i)}\!q_I^{i_1}\right.  )     
\prod_{{\rm i}=1}^m N^{a_{\rm m-i+1}}_{\rm m-i+1}(p, \{x\})     {\hat V}_I^{a_{\rm m-i+1}}(p)\partial_{a_{\rm m-i+1}}(f(p,\{x\}) \sqrt{h_{ab}(p){\hat V}_I^a(p){\hat V}^b_I(p)})|_{p=v_{[i,I,\delta]_1}} &
\nonumber\\
&+ O(\delta^2) &
\label{ho7}
\ea
Expanding in a Taylor approximation subject to Footnote \ref{tev}, we obtain:
\ba
&\sum_I\sum_i\sum_{{\hat J}_1, {\hat K}_1} (\Psi_{f,h_{ab}, P_0}|
(\prod_{ {\rm i}=1 }^m {\hat C}_{}(  N_{\rm i})) |c_{[i,I, {\hat J}, {\hat K}, \beta, \delta]_1}              \ket =& \nonumber\\
&(-3)^{k_m} (\frac{3\hbar N}{8\pi i})^m (\nu_{   v      }^{-\frac{2}{3}})^m g_{c}3N(N-1)(N-2)  \sum_{i,I} h_I  |{\vec q}_I|^{m-1}  (\sum_{i_1}    \left.  ^{(i)}\!q_I^{i_1}\right. )      &
\nonumber\\
&\big\{\prod_{{\rm i}=1}^m N^{a_{\rm m-i+1}}_{\rm m-i+1}(p, \{x\})     {\hat V}_I^{a_{\rm m-i+1}}(p)\partial_{a_{\rm m-i+1}}(f(p,\{x\}) \sqrt{h_{ab}(p){\hat V}_I^a(p){\hat V}^b_I(p)})|_{p=v}  &
\nonumber\\
&+ \delta  q^i_IV^a_I \partial_a \big( \prod_{{\rm i}=1}^m N^{a_{\rm m-i+1}}_{\rm m-i+1}(p, \{x\})     {\hat V}_I^{a_{\rm m-i+1}}(p)\partial_{a_{\rm m-i+1}}(f(p,\{x\}) \sqrt{h_{ab}(p){\hat V}_I^a(p){\hat V}^b_I(p)})\big)|_{p=v}\big\}&
\nonumber\\
&+ O(\delta^2)  &
\label{ho7a}
\ea
Using (\ref{qid1}), it is straightforward to see that the contribution, to (\ref{he2}) (with $m$ odd in (\ref{he2})),   of the $0$th order term in $\delta$ in (\ref{ho7a}) cancels with the contribution, to (\ref{he2}) of the 
second term in the last line of (\ref{he2}). Hence only the first order term in $\delta$ in (\ref{ho7a}) contributes to (\ref{he2}). Using (\ref{qid2}) and  (\ref{ho7a}) in (\ref{he2}) yields:
\ba
&(\Psi_{f,h_{ab}, P_0}|
(\prod_{ {\rm i}=1 }^{m} {\hat C}_{}(  N_{\rm i})) {\hat C}_{\delta} (N_{\rm m+1})    |c\ket = &\nonumber \\
&(-3)^{k_m} (\frac{3\hbar N}{8\pi i})^{m+1} (\nu_{   v      }^{-\frac{2}{3}})^{m+1} (-3) g_{c}
\sum_{i,I} \big\{|{\vec q}_I|^{m-1} (q^{i}_I)^2 
 N_{\rm m+1}(p,\{x\}) h_I 
 & \nonumber\\
& \big({\hat V}^a_I(p)\partial_a  (\prod_{{\rm i}=1}^{m} N^{a_{\rm m-i+1}}_{\rm m-i+1}(p, \{x\})     
{\hat V}_I^{a_{\rm m-i+1}}(p)\partial_{a_{\rm m-i+1}}(f(p,\{x\}) \sqrt{h_{ab}(p){\hat V}_I^a(p){\hat V}^b_I(p)})
\big)_{p=v}
\big\}&
\label{ho8}
\ea
where the overall $(-3)$ factor  comes from the $3$ in the numerator of (\ref{ho6}) and the $(-1)^m=-1$ factor in (\ref{he2}). 
From (\ref{defkn}) it follows that when $m$ is odd  $k_{m+1}= k_{m}+1 $. Using this fact together with the definition of the charge vector norm (\ref{normq}) and the notation (\ref{defHlmp}), in (\ref{ho8})  yields:
\ba
&(\Psi_{f,h_{ab}, P_0}|
(\prod_{ {\rm i}=1 }^{m+1} {\hat C}_{}(  N_{\rm i}))  |c\ket = &\nonumber \\
&(-3)^{k_{m+1}} (\frac{3\hbar N}{8\pi i})^{m+1} (\nu_{   v      }^{-\frac{2}{3}})^{m+1} g_{c} 
\sum_{i,I} |{\vec q}_I|^{m+1}  h_I H^{m+1}_I (N_1,..,N_{m+1}; v)&
\label{ho9}
\ea
which is the desired result (\ref{heven}) with $n=m+1$.
\\
Once again, since the fixed choice (a)- (f) underlying this derivation is 
arbitrary and since the assumed form for $n=m$ holds for any such choice, the  result (\ref{ho9}) also holds for any choice of (a)- (f).

Steps 1, 2 and 3 above complete the proof of the Claim. The caveat in Footnote \ref{tev} restricts the validity of the proof to the case that $n\leq k-1$, consistent with the Claim.

\subsection{\label{sec8.2}  Anomaly free single  commutators}

In section \ref{sec821} we  summarise our notation. In section \ref{sec8d} we compute the action of a  multiple product of Hamiltonian constraints multiplied by a single electric diffeomorphism constraint. We use this
in section \ref{sec8dd} to compute the action of a  multiple product of Hamiltonian constraints multiplied by a single commutator between a pair of electric diffeomorphism constraints. 
We show that the
result is the same as that of the action of this product of Hamiltonian constraints multiplied by the appropriate commutator between a pair of Hamiltonian constraints.  Hence this  single commutator 
between a pair of Hamiltonian constraints is anomaly free in the sense that it can be replaced, within the particular string of operators under consideration, 
by the commutator between a pair of electric diffeomorphism constraints in line with (\ref{key}). 
In section \ref{sec8m} we use this result to show that each of the commutators in (\ref{eqn1}) is anomaly free in the sense that each of them can be replaced by a corresponding appropriate electric diffeomorphism 
commutator.


\subsubsection{\label{sec821}Notation}
We denote the non-degenerate vertex of $c$ by $v$ and its associated reference coordinate patch by  $\{x\}$.
As in section \ref{sec8.1} it will suffice to develop notation only for singly deformed states.
The single deformations of interest will be denoted as
\be
[i,I,{\hat J}, {\hat K}, \beta=0, \delta ]_1= ( i, I, {\hat J}_1, {\hat K_1}, \delta )
\label{d8not1}
\ee
The vertex of the singly deformed state  $c_{[i,I,{\hat J}, {\hat K},  \delta ]_1}$ is denoted by $v_{[i,I,\delta]_1}$ and its associated 
{\em contraction} coordinate patch by $\{x^{\delta}\}$. The  single deformation of (\ref{d8not1}) will play the role of the first of  $m+1$
deformations applied to $c$ in section \ref{sec8d} and the role of the first of $m+2$ deformations in section \ref{sec8dd}. Accordingly, in relation to the  notation  of section \ref{sec4.3},  in 
section \ref{sec8d} we set:
\ba
\{x_{\alpha}\} \equiv \{x\},&  \;\;\;  j_1= m+1,&  \;\; \epsilon_{\rm j_1} \equiv \delta \nonumber\\
&\{x_{\alpha}^{\epsilon_{\rm j_1} }\} \equiv \{x^{\delta}\}& , 
\label{d8not2}
\ea
and in section \ref{sec8dd} we set:
\ba
\{x_{\alpha}\} \equiv \{x\},&  \;\;\;  j_1= m+2,&  \;\; \epsilon_{\rm j_1} \equiv \delta \nonumber\\
&\{x_{\alpha}^{\epsilon_{\rm j_1} }\} \equiv \{x^{\delta}\}& , 
\label{dd8not2}
\ea

\subsubsection{\label{sec8d}Single Electric Diffeomorphism}

In this section we evaluate the action of $(\prod_{ {\rm i}=1 }^{m} {\hat C}_{}(  N_{\rm i})){\hat D}({\vec N}_{m+1\;i})$.  We have that:
\be
(\Psi_{f,h_{ab}, P_0}|
(\prod_{ {\rm i}=1 }^{m} {\hat C}_{ }(  N_{\rm i}))  {\hat D}({\vec N}_{m+1\;i})              |c\ket 
= \lim_{\delta\rightarrow 0}(\Psi_{f,h_{ab}, P_0}|
(\prod_{ {\rm i}=1 }^m {\hat C}_{}(  N_{\rm i})) {\hat D}_{\delta}({\vec N}_{m+1\;i})        |c\ket .
\label{de1}
\ee
Using (\ref{dnsum}) we obtain:
\ba
&(\Psi_{f,h_{ab}, P_0}|
(\prod_{ {\rm i}=1 }^m {\hat C}_{}(  N_{\rm i}))   {\hat D}_{\delta}({\vec N}_{m+1\;i})                |c\ket = 
\frac{3\hbar}{4\pi \mathrm{i}}N(x(v))\nu_{v}^{-2/3}\frac{1}{(N-1)(N-2)\delta}  & \nonumber\\    
& \sum_{I{\hat J}_1, {\hat K}_1}   \big( (\Psi_{f,h_{ab}, P_0}| (\prod_{ {\rm i}=1 }^m {\hat C}(  N_{\rm i})) | c_{[i,I, {\hat J}, {\hat K},  \delta]_1}\ket- (\Psi_{f,h_{ab}, P_0}|(\prod_{ {\rm i}=1 }^m {\hat C}(  N_{\rm i}))|c\ket \big)&
\label{de2}
\ea
We use (\ref{heven}), (\ref{hodd}) to evaluate the amplitudes in the last line. The second amplitude admits a direct application of these equations with some fixed choice for (a)-(f), section \ref{secsum}.
These equations are applied to the evaluation of the first amplitude with the choices (a')- (f') outlined in Step 2, section \ref{sec813} except that we set $\beta=0$ there.
The calculational details differ slightly for $m$ even and $m$ odd.
\\

\paragraph{\label{sec8de}Case A: $m$ even}

Using the contraction behaviour of various quantities in Appendices \ref{aconb}, \ref{acong} and  the notation (\ref{d8not2}),(\ref{defHlmp}), we obtain:
\ba
    &  \sum_{{\hat J}_1, {\hat K}_1}    (\Psi_{f,h_{ab}, P_0}| (\prod_{ {\rm i}=1 }^m {\hat C}(  N_{\rm i})) | c_{[i,I, {\hat J}, {\hat K},\delta]_1}\ket 
= (-3)^{k_m} (\frac{3\hbar N}{8\pi i})^m (\nu_{   v_{[i,I\delta ]_1}        }^{-\frac{2}{3}})^mg_c  
Q (c_{0[i,I, \delta_0]^{1,0}_1}, S_{m+1}) h_I &  \nonumber\\
& \{|{\vec q}_{L_1=I}|^m(N-1)(N-2) + \cos^m(\theta)(\sum_{L_1\neq I}|{\vec q}_{L_1\neq I}|^m)(N-2)(1+\cos^2\theta +(N-3) |\cos \theta|)\} &\nonumber\\
&\prod_{{\rm i}=1}^m N^{a_{\rm m-i+1}}_{\rm m-i+1}(p, \{x\})     {\hat V}_I^{a_{\rm m-i+1}}(p)\partial_{a_{\rm m-i+1}}(f(p,\{x\}) \sqrt{h_{ab}(p){\hat V}_I^a(p){\hat V}^b_I(p)})|_{p=v_{[i,I,\delta]_1}}  +\; O(\delta^2)\;\;\;\;&
\label{de5}
\ea
We set:
\ba
&Q (c_{0[i,I, \delta_0]^{1,0}_1}, S_{m+1} ) = &
\nonumber\\
&\frac{( \nu_v^{-\frac{2}{3}})^m  }{ (\nu_{   v_{[i,I\delta ]_1}        }^{-\frac{2}{3}})^m    }
\big(\frac{(N)(N-1)(N-2)|{\vec q}_I|^m}{\{|{\vec q}_{L_1=I}|^m(N-1)(N-2) + \cos^m(\theta)(\sum_{L_1\neq I}|{\vec q}_{L_1\neq I}|^m)(N-2)(1+\cos^2\theta +(N-3) |\cos \theta|)\}}\big)&
\label{de6}
\ea
For $m$ even, clearly $Q$ is positive. 
\footnote{Note that the $Q$ factors in (\ref{de6}), (\ref{ho6}) are identical functions of their associated child-parent charges. In priniciple they could have been chosen to differ from each other because their
sequence labels differ in that the $m+1$th operator type is $h$ for (\ref{ho6}) and $d_i$ for (\ref{de6}).}

Expanding the contribution of (\ref{de5}) to (\ref{de2}) in a Taylor approximation in powers of $\delta$ subject to Footnote \ref{tev}, the zeroth order contribution cancels the contribution from second term in the last line of (\ref{de2}).
Only the first order contribution remains in the $\delta \rightarrow 0$ limit in (\ref{de2}) and we obtain:
\ba
&(\Psi_{f,h_{ab}, P_0}|
(\prod_{ {\rm i}=1 }^{m} {\hat C}_{}(  N_{\rm i}))   {\hat D}({\vec N}_{m+1\;i})                   |c\ket =     \lim_{\delta\rightarrow 0}(\Psi_{f,h_{ab}, P_0}|
(\prod_{ {\rm i}=1 }^m {\hat C}_{}(  N_{\rm i})) {\hat D}_{\delta}({\vec N}_{m+1\;i}        |c\ket                                                       &\nonumber \\
&=(-3)^{k_{m}} 2(\frac{3\hbar N}{8\pi i})^{m+1} (\nu_{   v      }^{-\frac{2}{3}})^{m+1} g_{c} 
\sum_{I} |{\vec q}_I|^{m}  q^{i}_I h_I H^{m+1}_I (N_1,..,N_{m+1}; v)&
\label{deven}
\ea
where we have used the notation (\ref{defHlmp}) so that :
\be
H^{m+1}_I (N_1,..,N_{m+1}; v):= \prod_{i=1}^l  N_{l-i+1}( p, \{x\})  {\hat V}^{a_{l-i+1}}_{L_m}(p)\partial_{a_{l-i+1}} (f( p, \{x\})\sqrt{h_{ab}(p)      {\hat V}^{ {a } }_{L_m} (p) {   {\hat V}^{b}  }_{L_m}(p)    })|_{p=v}
\ee
The reader may skip to  section \ref{sec8dd}  wherein we continue on to the electric diffeomorphism commutator calculation for $m$ even  in section \ref{sec8dde}.

\paragraph{\bf Case B: $m$ odd}

Using  Appendices \ref{aconb}, \ref{acong} and  the notation (\ref{d8not2}),(\ref{defHlmp}), we obtain:
\ba
    &  \sum_{{\hat J}_1, {\hat K}_1}    (\Psi_{f,h_{ab}, P_0}| (\prod_{ {\rm i}=1 }^m {\hat C}(  N_{\rm i})) | c_{[i,I, {\hat J}, {\hat K},\delta]_1}\ket 
= (-3)^{k_m} (\frac{3\hbar N}{8\pi i})^m (\nu_{   v_{[i,I\delta ]_1}        }^{-\frac{2}{3}})^mg_c  & \nonumber\\
&Q (c_{0[i,I, \delta_0]^{1,0}_1}, S_{m+1}) h_I
\{|{\vec q}_{L_1=I}|^{m-1} (\sum_{i_1}q^{i_1}_{L_1=I})     (N-1)(N-2) &
\nonumber\\
&+ \cos^m(\theta)(\sum_{L_1\neq I}|{\vec q}_{L_1}|^{m-1}(\sum_{i_1}q^{i_1}_{L_1=I}))(N-2)(1+\cos^2\theta +(N-3) |\cos \theta|)\} &\nonumber\\
&\big(\prod_{{\rm i}=1}^m N^{a_{\rm m-i+1}}_{\rm m-i+1}(p, \{x\})     {\hat V}_I^{a_{\rm m-i+1}}(p)\partial_{a_{\rm m-i+1}}(f(p,\{x\}) \sqrt{h_{ab}(p){\hat V}_I^a(p){\hat V}^b_I(p)}\big)_{p=v_{[i,I,\delta]_1}} &
\nonumber\\
&+\;\; O(\delta^2)&
\label{do5}
\ea
We set:
\ba
&Q (c_{0[i,I, \delta_0]^{1,0}_1}, S_{m+1} ) = 
\frac{( \nu_v^{-\frac{2}{3}})^m  }{ (\nu_{   v_{[i,I\delta ]_1}        }^{-\frac{2}{3}})^m    } \times &
\nonumber\\
&\frac{(N)(N-1)(N-2)|{\vec q}_I|^{m-1} (\sum_{j}q^{j}_{I}))    }{|{\vec q}_{L_1=I}|^m(\sum_{i_1}q^{i_1}_{L_1=I})(N-1)(N-2) + \cos^m(\theta)(\sum_{L_1\neq I}|{\vec q}_{L_1}|^{m-1}(\sum_{i_1}q^{i_1}_{L_1}))(N-2)(1+\cos^2\theta +(N-3) |\cos \theta |)}
&
\label{do6}
\ea
Since the transition $[i,I, {\hat J}, {\hat K},\delta]_1$ is electric diffeomorphism type we have that
\be
q^i_{L_1=I} = q^i_{I}
\label{de6a}
\ee
Using this with (\ref{thetachoice}), (\ref{thetanetchoice}) implies that $Q>0$.
Expanding the contribution of (\ref{do5}) to (\ref{de2}) in a Taylor approximation in powers of $\delta$ subject to Footnote \ref{tev}, the zeroth order contribution cancels the contribution from second term in the last line of (\ref{de2}).
Only the first order contribution remains in the $\delta \rightarrow 0$ limit in (\ref{de2}) and we obtain:
\ba
&(\Psi_{f,h_{ab}, P_0}|
(\prod_{ {\rm i}=1 }^{m} {\hat C}_{}(  N_{\rm i}))   {\hat D}({\vec N}_{m+1\;i})                   |c\ket =     \lim_{\delta\rightarrow 0}(\Psi_{f,h_{ab}, P_0}|
(\prod_{ {\rm i}=1 }^m {\hat C}_{}(  N_{\rm i})) {\hat D}_{\delta}({\vec N}_{m+1\;i}        |c\ket                                                       &\nonumber \\
&=(-3)^{k_{m}} 2(\frac{3\hbar N}{8\pi i})^{m+1} (\nu_{   v      }^{-\frac{2}{3}})^{m+1} g_{c} 
\sum_{I} |{\vec q}_I|^{m-1}    (\sum_{j}q^{j}_{I}))           q^{i}_I h_I H^{m+1}_I (N_1,..,N_{m+1}; v)&
\label{dodd}
\ea
where, as in (\ref{deven}), we have used the notation (\ref{defHlmp}).

\subsubsection{\label{sec8dd}Electric diffeomorphism commutator}

In this section we evaluate the action of $(\prod_{ {\rm i}=1 }^{m} {\hat C}_{}(  N_{\rm i}))[{\hat D}({\vec N}_{m+1\;i}),  {\hat  D}({\vec N}_{m+2\;i})]       $.  We have that:
\ba
&(\Psi_{f,h_{ab}, P_0}|
(\prod_{ {\rm i}=1 }^{m} {\hat C}_{ }(  N_{\rm i}))  [{\hat D}({\vec N}_{m+1\;i})   ,   {\hat D}({\vec N}_{m+2\;i})]             |c\ket &\nonumber\\
&= \lim_{\delta\rightarrow 0}(\Psi_{f,h_{ab}, P_0}|
(\prod_{ {\rm i}=1 }^m {\hat C}_{}(  N_{\rm i}))     \big(  {\hat D}({\vec N}_{m+1\;i})   ,        {\hat D}_{\delta}({\vec N}_{m+2\;i})  -    {\hat D}({\vec N}_{m+2\;i})   ,        {\hat D}_{\delta}({\vec N}_{m+1\;i})\big)   |c\ket\; \; &
\label{dde1}
\ea
Using (\ref{dnsum}) we obtain:
\ba
&(\Psi_{f,h_{ab}, P_0}|
(\prod_{ {\rm i}=1 }^m {\hat C}_{}(  N_{\rm i}))  {\hat D}({\vec N}_{m+1\;i})                      {\hat D}_{\delta}({\vec N}_{m+2\;i}                |c\ket = 
\frac{3\hbar}{4\pi \mathrm{i}}N_{m+2}(v, \{x\})\nu_{v}^{-2/3}\frac{1}{(N-1)(N-2)\delta}  & \nonumber\\    
& \sum_{I{\hat J}_1, {\hat K}_1}   \big( (\Psi_{f,h_{ab}, P_0}| (\prod_{ {\rm i}=1 }^m {\hat C}(  N_{\rm i})){\hat D}({\vec N}_{m+1\;i})                 | c_{[i,I, {\hat J}, {\hat K},  \delta]_1}\ket
&\nonumber\\
&- 
(\Psi_{f,h_{ab}, P_0}|(\prod_{ {\rm i}=1 }^m {\hat C}(  N_{\rm i})) {\hat D}({\vec N}_{m+1\;i})   |c\ket \big)&
\label{dde2}
\ea
We may use (\ref{deven}), (\ref{dodd}) to evaluate the amplitudes in the last 2 lines. The second amplitude admits a direct application of these equations with some fixed choice for (a)-(f), section \ref{secsum}.
It is straightforward to check that the application of (\ref{deven}), (\ref{dodd}) to the evaluation of the second amplitude results in an expression with an overall factor $N_{m+2}(v, \{x\})N_{m+1}(v, \{x\})$.
It then follows from (\ref{dde1}) that this term does not contribute to the commutator and, hence, we disregard it.

Equations (\ref{deven}), (\ref{dodd})  may be applied to the evaluation of the first amplitude with the choices (a')- (f') outlined in Step 2, section \ref{sec813} except that, once again,  we set $\beta=0$ there.
The calculational details for this contribution differ slightly for $m$ even and $m$ odd.
\\

\paragraph{\label{sec8dde}Case A: $m$ even}
Using (\ref{deven}) as indicated above to evaluate the first amplitude in the last line of (\ref{dde2}), we obtain its contraction  limit using the Appendices \ref{aconb}, \ref{acong} together with (\ref{dd8not2}) to be:
\ba
    &  \sum_{{\hat J}_1, {\hat K}_1}    (\Psi_{f,h_{ab}, P_0}| (\prod_{ {\rm i}=1 }^m {\hat C}(  N_{\rm i}))  {\hat D}({\vec N}_{m+1\;i})          | c_{[i,I, {\hat J}, {\hat K},\delta]_1}\ket 
= (-3)^{k_m} 2(\frac{3\hbar N}{8\pi i})^{m+1} (\nu_{   v_{[i,I\delta ]_1}        }^{-\frac{2}{3}})^{m+1}g_c  
&\nonumber\\
&Q (c_{0[i,I, \delta_0]^{1,0}_1}, S_{m+2}) h_I 
 \{|{\vec q}_{L_1=I}|^{m}  q^i_{L-1=I}  (N-1)(N-2) +  &\nonumber\\
&\cos^{m+1}(\theta)(\sum_{L_1\neq I}|{\vec q}_{L_1}|^m  q^i_{L_1})(N-2)(1+\cos^2\theta +(N-3) |\cos \theta|)\} &\nonumber\\
&\big(\prod_{{\rm i}=1}^{m+1} N^{a_{\rm m-i+1}}_{\rm m-i+1}(p, \{x\})     {\hat V}_I^{a_{\rm m-i+1}}(p)\partial_{a_{\rm m-i+1}}(f(p,\{x\}) \sqrt{h_{ab}(p){\hat V}_I^a(p){\hat V}^b_I(p)})|_{p=v_{[i,I,\delta]_1}} \big)& \nonumber\\
&+\;\; O(\delta^2)&
\label{dde5}
\ea
Note that the  transition $[i,I, {\hat J}, {\hat K},\delta]_1$ is an electric diffeomorphism type deformation so that:
\be
q^i_{L_1=I} = q^i_{I}
\label{dde6a}
\ee
We set for some $A>0$:
\ba
&Q (c_{0[i,I, \delta_0]^{1,0}_1}, S_{m+2} ) =
\frac{( \nu_v^{-\frac{2}{3}})^{m+1}  }{ (\nu_{   v_{[i,I\delta ]_1}        }^{-\frac{2}{3}})^{m+1}    } \times  &\nonumber\\
&\frac{A(N)(N-1)(N-2)|{\vec q}_I|^m q^i_I}{\{|{\vec q}_{L_1=I}|^mq^i_{L_1=I}(N-1)(N-2) + \cos^{m+1}(\theta)(\sum_{L_1\neq I}|{\vec q}_{L_1}|^m q^i_{L_1})(N-2)(1+\cos^2\theta +(N-3) |\cos \theta|\}}&
\label{dde6}
\ea

Equation (\ref{dde6a}) together with (\ref{thetachoice}),(\ref{thetanetchoice}) once again implies that $Q>0$. Next, we expand (\ref{dde5}) in a Taylor expansion in powers of $\delta$. It is easy to check that the zeroth order term does not
contribute to the commutator (\ref{dde2}). Only the first order term contributes. Using this first order term in (\ref{dde2}) and taking the contraction limit,   we obtain
\ba
&(\Psi_{f,h_{ab}, P_0}|
(\prod_{ {\rm i}=1 }^{m} {\hat C}_{ }(  N_{\rm i}))  [{\hat D}({\vec N}_{m+1\;i})   ,   {\hat D}({\vec N}_{m+2\;i})]             |c\ket = (-3)^{k_m}A4 (\frac{3\hbar N}{8\pi i})^{m+2} (\nu_{   v      }^{-\frac{2}{3}})^{m+2}  g_{c}
\sum_{I} |{\vec q}_I|^{m} (q^{i}_I)^2 h_I & \nonumber\\
&\big\{N_{\rm m+2}(p,\{x\})
 {\hat V}^a_I(p)\partial_a  (\prod_{{\rm i}=1}^{m+1} N^{a_{\rm m-i+1}}_{\rm m-i+1}(p, \{x\})     
{\hat V}_I^{a_{\rm m-i+1}}(p)\partial_{a_{\rm m-i+1}}(f(p,\{x\}) \sqrt{h_{ab}(p){\hat V}_I^a(p){\hat V}^b_I(p)})&
\nonumber \\
&- N_{\rm m+2}(p,\{x\})\leftrightarrow N_{\rm m+1}(p,\{x\})
\big\}_{p=v}
&
\label{dde8}
\ea
Summing over $i$ in (\ref{dde8}) and using the notation (\ref{defHlmp}) and the definition of the charge norm (\ref{normq}), we obtain:
\ba
&(\Psi_{f,h_{ab}, P_0}|
(\prod_{ {\rm i}=1 }^{m} {\hat C}_{ }(  N_{\rm i}))  [{\hat D}({\vec N}_{m+1\;i})   ,   {\hat D}({\vec N}_{m+2\;i})]             |c\ket & \nonumber\\
&= (4A) (-3)^{k_m} (\frac{3\hbar N}{8\pi i})^{m+2} (\nu_{   v      }^{-\frac{2}{3}})^{m+2}  g_{c}& \nonumber \\
&\sum_{I} \big\{|{\vec q}_I|^{m+2}  h_I (H^{m+2}_I (N_1,..,N_{m+1}, N_{m+2}; v) - H^{m+1}_I (N_1,..,N_{m+2},N_{m+1}; v)&
\label{dde9}
\ea
On the other hand, replacing the commutator $\sum_{i}[{\hat D}({\vec N}_{m+1\;i})   ,   {\hat D}({\vec N}_{m+2\;i})] $ with $[{\hat C}({ N}_{m+1})   ,   {\hat C}({N}_{m+2})] $ and noting that $m+2$ is even,
we obtain, from (\ref{heven})
\ba
&(\Psi_{f,h_{ab}, P_0}|
(\prod_{ {\rm i}=1 }^m{\hat C}_{}(  N_{\rm i}))   [{\hat C}({ N}_{m+1})   ,   {\hat C}({N}_{m+2})]                                             |c\ket & \nonumber\\
&= 
(-3)^{k_{m+2}} (\frac{3\hbar N}{8\pi i})^{m+2} (\nu^{-\frac{2}{3}})^{m+2}g_c & \nonumber\\
&\sum_{I} |{\vec q}_I|^{m+2} h_I(H^{m+2}_I (N_1,..,N_{m+1}, N_{m+2}; v) - H^{m+1}_I (N_1,..,N_{m+2},N_{m+1}; v) )&
\label{hhe}
\ea
From (\ref{defkn}) we have that $k_{m+2} = k_m +1$. It is then easy to see that an anomaly free commutator results if we choose $A=1/4$ in (\ref{dde9})


\paragraph{\label{sec8ddo}Case B: $m$ odd}
Using (\ref{dodd}) as indicated above to evaluate the first amplitude in the last line of (\ref{dde2}), we obtain its contraction  limit using the Appendices \ref{aconb}, \ref{acong} and equation (\ref{dd8not2}) as:
\ba
    &  \sum_{{\hat J}_1, {\hat K}_1}    (\Psi_{f,h_{ab}, P_0}| (\prod_{ {\rm i}=1 }^m {\hat C}(  N_{\rm i}))  {\hat D}({\vec N}_{m+1\;i})          | c_{[i,I, {\hat J}, {\hat K},\delta]_1}\ket 
= (-3)^{k_m} 2(\frac{3\hbar N}{8\pi i})^{m+1} (\nu_{   v_{[i,I\delta ]_1}        }^{-\frac{2}{3}})^{m+1}g_c  
&\nonumber\\
&Q (c_{0[i,I, \delta_0]^{1,0}_1}, S_{m+2}) h_I 
 \{|{\vec q}_{L_1=I}|^{m-1}   (\sum_{j_1}q^{j_1}_{L_1=I})                       q^{i_1=i}_{L-1=I}  (N-1)(N-2) +  &\nonumber\\
&\cos^{m+1}(\theta)(\sum_{L_1\neq I}|{\vec q}_{L_1}|^{m-1}  (\sum_{i_1}q^{i_1}_{L_1})                     q^{i_1=i}_{L_1})(N-2)(1+\cos^2\theta +(N-3) |\cos \theta|)\} &\nonumber\\
&\big(\prod_{{\rm i}=1}^{m+1} N^{a_{\rm m-i+1}}_{\rm m-i+1}(p, \{x\})     {\hat V}_I^{a_{\rm m-i+1}}(p)\partial_{a_{\rm m-i+1}}(f(p,\{x\}) \sqrt{h_{ab}(p){\hat V}_I^a(p){\hat V}^b_I(p)})|_{p=v_{[i,I,\delta]_1}} \big)& \nonumber\\
&+\;\; O(\delta^2)&
\label{ddo5}
\ea
Note that the  transition $[i,I, {\hat J}, {\hat K},\delta]_1$ is an electric diffeomorphism type deformation so that:
\be
q^{i_1=i}_{L_1=I} = q^i_{I}
\label{ddo6a}
\ee
We set for some $A>0$:
\ba
& Q (c_{0[i,I, \delta_0]^{1,0}_1}, S_{m+2} ) =
\frac{( \nu_v^{-\frac{2}{3}})^{m+1}  }{ (\nu_{   v_{[i,I\delta ]_1}        }^{-\frac{2}{3}})^{m+1}    } \times  \nonumber\\
& \frac{A(N)(N-1)(N-2)|{\vec q}_I|^{m-1} (\sum_{j}q^{j}_{I}) q^i_I}{|{\vec q}_{L_1=I}|^{m-1}  (\sum_{i_1}q^{i_1}_{L_1=I})            q^i_{L_1=I}(N-1)(N-2) + \cos^{m+1}(\theta)(\sum_{L_1\neq I}
|{\vec q}_{L_1}|^{m-1}  (\sum_{i_1}q^{i_1}_{L_1}) q^i_{L_1})(N-2)(1+\cos^2\theta +(N-3) |\cos \theta|)}\;\;\;\;\;\;\;\;\; 
\label{ddo6}
\ea
Equation (\ref{ddo6a}) together with (\ref{thetachoice}),(\ref{thetanetchoice}) once again implies that $Q>0$. Expanding (\ref{dde5}) in a Taylor expansion in powers of $\delta$ it is easy to check that the zeroth order term does not
contribute to the commutator (\ref{dde2}). Only the first order term contributes. Using this first order term in (\ref{dde2}) and taking the contraction limit,   we obtain
\ba
&(\Psi_{f,h_{ab}, P_0}|
(\prod_{ {\rm i}=1 }^{m} {\hat C}_{ }(  N_{\rm i}))  [{\hat D}({\vec N}_{m+1\;i})   ,   {\hat D}({\vec N}_{m+2\;i})]             |c\ket = \nonumber\\
&(-3)^{k_m}A4 (\frac{3\hbar N}{8\pi i})^{m+2} (\nu_{   v      }^{-\frac{2}{3}})^{m+2}  g_{c}
\sum_{I} |{\vec q}_I|^{m-1}   (\sum_{j}q^{j}_{I})            (q^{i}_I)^2 h_I \nonumber\\
&\big\{N_{\rm m+2}(p,\{x\})
 {\hat V}^a_I(p)\partial_a  (\prod_{{\rm i}=1}^{m+1} N^{a_{\rm m-i+1}}_{\rm m-i+1}(p, \{x\})     
{\hat V}_I^{a_{\rm m-i+1}}(p)\partial_{a_{\rm m-i+1}}(f(p,\{x\}) \sqrt{h_{ab}(p){\hat V}_I^a(p){\hat V}^b_I(p)})\;\;
\nonumber \\
&- N_{\rm m+2}(p,\{x\})\leftrightarrow N_{\rm m+1}(p,\{x\})\;\;\;
\big\}_{p=v}
\label{ddo8}
\ea
Summing over $i$ in (\ref{ddo8}) and using the notation (\ref{defHlmp}) and the definition of the charge norm (\ref{normq}), we obtain:
\ba
&(\Psi_{f,h_{ab}, P_0}|
(\prod_{ {\rm i}=1 }^{m} {\hat C}_{ }(  N_{\rm i}))  [{\hat D}({\vec N}_{m+1\;i})   ,   {\hat D}({\vec N}_{m+2\;i})]             |c\ket & \nonumber\\
&= (4A) (-3)^{k_m} (\frac{3\hbar N}{8\pi i})^{m+2} (\nu_{   v      }^{-\frac{2}{3}})^{m+2}  g_{c}& \nonumber \\
&\sum_{I} \big\{|{\vec q}_I|^{m+1} (\sum_{j}q^{j}_{I})     h_I (H^{m+2}_I (N_1,..,N_{m+1}, N_{m+2}; v) - H^{m+1}_I (N_1,..,N_{m+2},N_{m+1}; v)&
\label{ddo9}
\ea
On the other hand, replacing the commutator $\sum_{i}[{\hat D}({\vec N}_{m+1\;i})   ,   {\hat D}({\vec N}_{m+2\;i})] $ with $[{\hat C}({ N}_{m+1})   ,   {\hat C}({N}_{m+2})] $ and noting that $m+2$ is odd,
we obtain, from (\ref{hodd})
\ba
&(\Psi_{f,h_{ab}, P_0}|
(\prod_{ {\rm i}=1 }^m{\hat C}_{}(  N_{\rm i}))   [{\hat C}({ N}_{m+1})   ,   {\hat C}({N}_{m+2})]                                             |c\ket & \nonumber\\
&= 
(-3)^{k_{m+2}} (\frac{3\hbar N}{8\pi i})^{m+2} (\nu^{-\frac{2}{3}})^{m+2}g_c & \nonumber\\
&\sum_{I} |{\vec q}_I|^{m+1}   (\sum_{j}q^{j}_{I})           h_I(H^{m+2}_I (N_1,..,N_{m+1}, N_{m+2}; v) - H^{m+1}_I (N_1,..,N_{m+2},N_{m+1}; v)&
\label{hho}
\ea
From (\ref{defkn}) we have that $k_{m+2} = k_m +1$. It is then easy to see that, once again,  an anomaly free commutator results if we choose $A=1/4$ in (\ref{ddo9})

\subsection{\label{sec8m}Multiple Single Anomaly free commutators}
 
Consider the action of the operator in equation (\ref{eqn1}) on the anomaly free state $\Psi_{f,h_{ab}, P_0}$.
 In this section we show this action is invariant under the replacement of each Hamiltonian constraint commutator in (\ref{eqn1})    by a corresponding electric diffeomorphism commutator, this replacement being 
a quantum implementation of the anomaly free condition (\ref{key}). We proceed as follows.

First we further develop the notation for operator sequence labels of $Q$ factors developed in section \ref{sec6.4} as follows.
Consider the sequence:
\be
(\underbrace{h,..,h}_{m_1}, t_1,t_1, \underbrace{h,..,h}_{m_2}, t_2,t_2,\underbrace{h,..,h}_{m_n}, t_n,t_n,\underbrace{h,..,h}_{p_n})
\label{as1}
\ee
where each $t_i$ is either $h$ or $d_k, k\in {1,2,3}$ and $m_i, p_n$ are whole numbers.
The $Q$ factors which we define below only depend on whether or not a $t_i$ is Hamiltonian or electric diffeomorphism; the particular component of the electric diffeomorphism doesnt matter.
Since the $\beta$ factors for the Hamiltonian constraint are such that $\beta^2=1$ and since $\beta=0$ for an electric diffeomorphism, we denote the essential part of the sequence above through
the symbol $\sigma_n$ as follows:
\be
\sigma_n (m_1, m_2,..,m_n; \beta^2_1,..,\beta^2_n;p_n )
\label{as2}
\ee
Thus, the specification of the arguments of $\sigma_n$ allow us to reconstruct the sequence (\ref{as1}) upto irrelevant (for the $Q$ factors of interest) ambiguities regarding the specific components of electric diffeomorphism
operators in such a sequence. Next, consider any operator product of the type 
${\hat O}_{\epsilon_1,.., \epsilon_{q_n}}(M_1,..,M_{q_n})$                         corresponding to a discrete approximant for the operator product
${\hat O}(M_1,..,M_{q_n})$ where each of the operators in the operator product are either Hamiltonian or electric diffeomporphism operators similar to the operator product in (\ref{prodact}).
Let the sequence associated with this operator product be ${\cal S}_{q_n}$ and any subsequence of of this sequence of operators from the first to the ${\rm j_k}$th, ${\rm j_k}\leq q_n$ be ${\cal S}_{\rm{j_k}}$.

Next consider a `big' operator product consisting of the  sequence of operators of type (\ref{as2}) followed by the operator product ${\hat O}_{\epsilon_1,.., \epsilon_{q_n}}(M_1,..,M_{q_n})$ where the 
latter occurs to the right of the former and so acts first on any charge net $c$. We denote the operator sequence for such a `big' product by
\be
S(\sigma (m_1, m_2,..,m_n; \beta^2_1,..,\beta^2_n;p_n ), {\cal S}_{q_n}) \equiv \sigma (m_1, m_2,..,m_n; \beta^2_1,..,\beta^2_n;p_n ), {\cal S}_{q_n}
\label{as3}
\ee
If, in this big operator product, we replace  ${\hat O}_{\epsilon_1,.., \epsilon_{q_n}}(M_1,..,M_{q_n})$ by an  operator consisting of the product of the  first ${\rm j_k}$  operators in ${\hat O}_{\epsilon_1,.., \epsilon_{q_n}}(M_1,..,M_{q_n})$, 
then  we denote the sequence corresponding to the new `big'  operator product  by:
\be
S(\sigma_n (m_1, m_2,..,m_n; \beta^2_1,..,\beta^2_n;p_n ), {\cal S}_{\rm j_k}) \equiv \sigma_n (m_1, m_2,..,m_n; \beta^2_1,..,\beta^2_n;p_n ), {\cal S}_{\rm j_k}
\label{as4}
\ee
We shall be interested in $Q$ factors for child-parent contractions $c_{ [i,I,{\hat J}, {\hat K}, \beta, \epsilon]^{k-1,k}_k}$
where the child and parent states are generated by the action of ${\hat O}_{\epsilon_1,.., \epsilon_{q_n}}(M_1,..,M_{q_n})$  on some charge net $c$ and 
when the  operator product sequence label for $Q$ is of the type (\ref{as2}) or (\ref{as3}). The $Q$ factor for such a situation in the case wherein all the operators in (\ref{as1}) are Hamiltonian 
(so that all the $\beta^2_i$ are unity in (\ref{as2})) is, using the above notation in conjunction with that of section \ref{sec6.4}, then:
\ba
&Q(  c_{0[i,I,\beta, \delta_0]^{k-1,k}_k}  , S(\sigma_n (m_1, m_2,..,m_n; 1,1,..1;p_n ), {\cal S}_{\rm j_k})) &\nonumber \\
&\equiv Q(  c_{0[i,I,\beta, \delta_0]^{k-1,k}_k } ; \sigma_n (m_1, m_2,..,m_n; 1,1,..1;p_n ), {\cal S}_{\rm j_k})&
\label{as5}
\ea
We define the $Q$ factors labelled by the sequence (\ref{as4}) for the  transition $c_{[i,I,\beta, \delta_0]^{k-1,k}_k}$ to be such that 
these $Q$ factors
for all choices of $\sigma_n$  in (\ref{as4})  are the same as that in (\ref{as5}):
\ba
&Q( c_{0[i,I,\beta, \delta_0]^{k-1,k}_k}; \sigma_n (m_1, m_2,..,m_n; \beta^2_1,..,\beta^2_n;p_n ), {\cal S}_{\rm j_k})& \nonumber \\
&:=Q(  c_{0[i,I,\beta, \delta_0]^{k-1,k}_k}  ; \sigma_n (m_1, m_2,..,m_n; 1,1,..1;p_n ), {\cal S}_{\rm j_k}) ,&\nonumber \\
&\forall {\rm j_k} \in \{1,..,q_n\},\;\;q_n>0, \;\;\;\forall \beta^2_i \in \{0,1\},\;\;\;\forall p_n, m_i, i=1,..,n &
\label{as6}
\ea
where  $p_n$ and  $m_i,i=1,..,n$ range over the set of whole numbers.

It is straightforward to see that the kinds of operator products  implicated  in a demonstration that operator strings of the type (\ref{eqn1}) have anomaly free single commutators are exactly  those for which the sequence labels 
are of the type (\ref{as2}). We now construct such a demonstration through an inductive proof on the index `$n$' which occurs in (\ref{as2}). Note that the index $n$ corresponds to the number of single commutators involved.

First consider the case $n=1$.  Let ${\hat O}_{\epsilon_1,.., \epsilon_{r_1}}(M_1,..,M_{r_1}), r_1>0$ be a product of $r_1$ Hamiltonian constraints.
Then using section \ref{sec8.2} together with equation (\ref{as6})  with $q_1:=r_1$  it follows that for any $r_1 >0$ we have that:
\ba
& (\Psi_{f,h_{ab}, P_0}|
(\prod_{ {\rm i}=1 }^m{\hat C}_{}(  N_{\rm i}))   [{\hat C}({ N}_{m+1})   ,   {\hat C}({N}_{m+2})]    {\hat O}_{\epsilon_1,.., \epsilon_{r_1}}(M_1,..,M_{r_1})                          |c\ket  &\nonumber\\
&=(-3) \sum_i (\Psi_{f,h_{ab}, P_0}| 
(\prod_{ {\rm i}=1 }^{m} {\hat C}_{ }(  N_{\rm i}))  [{\hat D}({\vec N}_{{ m+1}\;i})   ,   {\hat D}({\vec N}_{   { m+2}\;i})]   {\hat O}_{\epsilon_1,.., \epsilon_{r_1}}(M_1,..,M_{r_1} )       |c\ket & 
\label{as7}
\ea
From section \ref{sec8.2}, the above equation also holds if  
we replace ${\hat O}_{\epsilon_1,.., \epsilon_{r_1}}(M_1,..,M_{r_1})$ by the identity operator. 
Taking the continuum limit of (\ref{as7}) 
we see that the result on anomaly free single commutators holds for the case that $n=1$ with $p_1=r_1$ in (\ref{as2}). Similarly taking the continuum limit of 
of the equation obtained by replacing   ${\hat O}_{\epsilon_1,.., \epsilon_{r_1}}(M_1,..,M_{r_1})$ by the identity operator in (\ref{as7}), this result also holds for the case that $n=1$ and $p_1=0$.
Thus we have established the desired result for the case that $n=1$ which corresponds to the case of a single commutator.

Next let us assume that the anomaly free single commutator property holds for all operator strings with sequences (\ref{as1}) for some $n=s$.
More in detail consider an operator product consisting of $m_1$ Hamiltonian constraints and  a Hamiltonian constraint commutator followed by $m_2$ Hamiltonian constraints and a Hamiltonian constraint commutator, all the way
upto $m_s$ Hamiltonian constraints and  the $s$th Hamiltonian constraint  commutator, followed by a product of $p_s$ Hamiltonian constraints. Then the assumption is that in any such product any subset of these Hamiltonian commutators 
can be replaced by sums over electric diffeomorphism commutators in accordance with (\ref{key}).  From this assumption we now show that the same statement hold for $n=s+1$.

First define
\be
{\hat O}(\sigma_n (m_1, m_2,..,m_n; 1,1..,1;p_n )):= \prod_{j=1}^n
\big(
(\prod_{ {\rm i}=1 }^{m_j}  {\hat C}(  N_{\rm i})  )
[{\hat C}(A_j)   ,   {\hat C} (B_j)] \big) 
(\prod_{   {\rm k}=1   }^{ p_n }  {\hat C}(  F_{\rm k}  ))
\label{as8}
\ee
where the sequence on the left hand side has all its $\beta^2_i$ as unity. Next, define the operator
\be
{\hat O}(\sigma_n (m_1, m_2,..,m_n; \beta^2_1,..,\beta^2_n;p_n ))
\label{as9}
\ee
as follows. For each $i$ for which $\beta^2_i=0 $ in $\sigma_n (m_1, m_2,..,m_n; \beta^2_1,..,\beta^2_n;p_n )$, replace the $i$th Hamiltonian constraint
commutator in (\ref{as8}) by an  appropriate sum over electric diffeomorphism commutators consistent with (\ref{key}). Note that this `${\hat O}(\sigma_n )$' notation is consistent with (\ref{as2}) in that
each of the constraint operator products obtained by expanding out the commutators in (\ref{as9}) correspond to the (same) sequence $\sigma_n (m_1, m_2,..,m_n; \beta^2_1,..,\beta^2_n;p_n )$.
In this notation our assumption for $n=s$ may be written as 
\ba
&(\Psi_{f,h_{ab}, P_0}|{\hat O}(\sigma_s (m_1, m_2,..,m_s; 1..,1;p_s ))|c\ket & \nonumber \\
&=(\Psi_{f,h_{ab}, P_0}|{\hat O}(\sigma_s (m_1, m_2,..,m_s; \beta^2_1,..,\beta^2_s;p_s ))|c\ket \;\;\forall \{\beta^2_{i},m_i, i=1,..,s\}\;\;{\rm and}\;\;\forall p_s.&
\label{as10}
\ea
Next let ${\hat O}^{1}_{\epsilon_1, \epsilon_2}(G_1, G_2)$ be the discrete approximant to the Hamiltonian constraint commutator  $[{\hat C}(G_1), {\hat C}(G_2)]$ and 
let ${\hat O}^{2}_{\epsilon_1, \epsilon_2}(G_1, G_2)$ be the discrete approximant to the appropriate sum over electric diffeomorphism commutators through (\ref{key}).
Since the action of these discrete approximants is a finite linear combination of charge nets with $Q$ factors as defined in (\ref{as6}), it follows for $\alpha=1,2$ that:
\ba
&(\Psi_{f,h_{ab}, P_0}|{\hat O}(\sigma_s (m_1, m_2,..,m_s; 1..,1;p_s )) {\hat O}^{\alpha}_{\epsilon_1, \epsilon_2}(G_1, G_2)                     |c\ket & \nonumber \\
&=(\Psi_{f,h_{ab}, P_0}|{\hat O}(\sigma_s (m_1, m_2,..,m_s; \beta^2_1,..,\beta^2_s;p_s ))   {\hat O}^{\alpha}_{\epsilon_1, \epsilon_2}(G_1, G_2)                              |c\ket \;\;\;\forall \{\beta^2_{i}, i=1,..,s\}&
\label{as11}
\ea
It is then straightforward to see that taking the continuum limit of the left hand and right hand sides and applying the anomaly free single commutator result of section \ref{sec8.2} yields the result
\ba
&(\Psi_{f,h_{ab}, P_0}|{\hat O}(\sigma_{s+1} (m_1, m_2,..,m_{s+1}; 1..,1;p_{s+1}=0 ))|c\ket & \nonumber \\
&=(\Psi_{f,h_{ab}, P_0}|{\hat O}(\sigma_{s+1} (m_1, m_2,..,m_{s+1}; \beta^2_1,..,\beta^2_{s+1};p_{s+1}=0 ))|c\ket \;\;\; \forall \{\beta^2_{i}, i=1,..,s+1\}&
\label{as12}
\ea
where we have set $p_s:= m_{s+1}$.
Next consider the approximant ${\hat O}_{\vec{\epsilon}}:=\prod_{t=1}^{p_{s+1}} {\hat C}_{\epsilon_t}(P_t)$ 
to a product of  $p_{s+1}\neq 0$  Hamiltonian constraint operators. Again using (\ref{as6}) we may substitue $|c\ket$ in (\ref{as12}) by the finite linear combination of charge nets
${\hat O}_{\vec{\epsilon}}|c\ket$. Taking the continuum limits of the resulting equations, we obtain:
\ba
&(\Psi_{f,h_{ab}, P_0}|{\hat O}(\sigma_{s+1} (m_1, m_2,..,m_{s+1}; 1..,1;p_{s+1} ))|c\ket & \nonumber \\
&=(\Psi_{f,h_{ab}, P_0}|{\hat O}(\sigma_{s+1} (m_1, m_2,..,m_{s+1}; \beta^2_1,..,\beta^2_{s+1};p_{s+1} ))|c\ket \;\;\;\forall \{\beta^2_{i}, i=1,..,{s+1}\} ,&
\label{as13}
\ea
which is the desired result for $n=s+1$. This completes our 
inductive proof of an  anomaly free single commutator implementation of the algebra of Hamiltonian constraints.
It only remains to show that this implementation is diffeomorphism covariant. We show this in the next section.

\section{\label{sec9} Diffeomorphism Covariance }

We implement diffeomorphism covariance of the continuum limit action  of products of constraint operators  on any anomaly free basis state by tailoring the underlying discrete  action to the metric label of the basis state being acted upon.
This idea, of tailoring the action of a discrete approximant to    an operator to the state it , 
 acts  upon, is a familiar one in the case that the states lie in the kinematic Hilbert space of LQG (see for example \cite{ttrick,ttbook,mediffeo}, and also section \ref{sec2} of this paper).
Here we  apply this idea to the space of kinematically non-normalizable anomaly free states.

As a prelude to the  detailed technical description  in sections \ref{sec9.1} and \ref{sec9.2} below,    we  now describe the broad idea behind this  implementation.
Recall from section \ref{sec5.2} that an anomaly free basis state is labelled by a density $-1/3$ function 
$f$, a metric with no conformal symmetries $h_{ab}$ and a choice of Bra Set $B_{P0}$.   Given this state  $\Psi_{f,h_{ab}, P_0}$
we can construct all its amplitudes i.e all the complex numbers $(\Psi_{f,h_{ab}, P_0}|c\ket$ for any charge net $c$. 
It then turns out that  we can construct enough information 
about the metric $h_{ab}$  from these amplitudes so as to distinguish this metric from any of its diffeomorphic images. 
If we restrict the space of permissible metric labels for anomaly free basis states to be the space of all diffeomorphic images of $h_{ab}$,
this means that the metric label of any anomaly free basis state can be uniquely identified from the state itself through its amplitudes.
In this sense the state `knows' about its metric label. Hence it is meaningful to define the discrete action of constraint operators on this state
in such a way that this discrete action depends on the metric label of the state.  


It turns out that  the dual action of the unitary operator corresponding to a diffeomorphism $\phi$ maps a state with metric label $h_{ab}$ to one with metric label $h^{\phi}_{ab}$,  where $h^{\phi}_{ab}$ is the image of $h_{ab}$ by $\phi$.
The idea is then to use the metric label $h^{\phi}_{ab}$ to identify  the diffeomorphism $\phi$, since, due to the lack of (conformal) isometries,  any permissible metric label is uniquely associated with the diffeomorphism
which  maps  $h_{ab}$ to this label. Then for this metric label  $h^{\phi}_{ab}$ we choose the primary coordinates and reference diffeomorphisms to be appropriate images, by $\phi$ of the primary coordinates and reference diffeomorphisms
chosen for the state labelled by $h_{ab}$. These `image' structures are then used to regulate and define constraint operator products along the lines of section \ref{sec4} and \ref{secneg}.
It can then be shown that this diffeomorphism covariant choice of regulating structures  leads to a diffeomorphism covariant  continuum limit action of products of constraints.

In section \ref{sec9.1} we formulate and prove a precise statement which shows that anomaly free basis states have the requisite sensitivity to their metric labels. In section \ref{sec9.2} we use this sensitivity to define
a covariant choice of reference structures and express  the action of finite diffeomorphisms on anomaly free states in the context of this covariant choice.
In section \ref{sec9.3}  we  demonstrate that this covariant choice results in an implementation of diffeomorphism covariance of the continuum limit action of products of constraints.
For the remainder of this section we shall restrict attention to $-\frac{1}{3}$ density scalars $f$ which vanish at most at a finite number of points in $\Sigma$.
\footnote{\label{fnfinite0}This is for technical simplicity; it seems plausible to us that our considerations can be generalised for the case where $f$ is not restricted in this manner. We leave such a generalization for future work.}

\subsection{\label{sec9.1}Metric label sensitivity of an anomaly free state}

Let  $h_{0ab}$ be a metric which has no conformal symmetries.
Let ${\cal H}_{h_0}$ be the space of all diffeomorphic images of $h_{0ab}$ by all $C^k$ semianalytic diffeomorphisms.
Let $h_{1ab}, h_{2ab}$ be 2 distinct elements of  ${\cal H}_{h_0}$.  Note that the two metrics cannot be conformally related everywhere because
they are distinct, diffeomorphic to each other and have no conformal symmetries.  Hence there exists a point $a\in \Sigma$, and,  from the fact that the metrics are
$C^{k-1}$, a neighbourhood $U(a,\delta ) $  of $a$  for some small enough $\delta >0$  such that in a fixed coordinate patch  $\{y\}$ in this neighbourhood,  we have that:
\be
|  \frac{ h_{  1ij  } }{ h_1^{ \frac{1}{3} }} -\frac{ h_{2ij} }{h_2^{ \frac{1}{3} } }| > C  .
\label{h1-h2}
\ee
The above inequality holds for every point in   $U(a,\epsilon ) $,  for at least one fixed pair $i,j \in \{1,2,3\}$ and for some fixed positive constant $C$.
 Here our notation is such that  $s_{ij}$ denotes  the coordinate components of the metric  $s_{ab}$ in the chart $\{y\}$ and $s$ denotes its determinant in this chart.
Thus equation (\ref{h1-h2}) indicates that at least one  component, in any fixed semianalytic chart, of the (conformally invariant) metric densities  in this equation differ by some minimum non zero amount 
in a small enough neighbourhood of at least    
one point of the Cauchy slice.
\footnote{It is straightforward to see that transiting from one fixed coordinate chart to another only affects the value of the constant $C$ and the choice of $i,j$ in (\ref{h1-h2}).}

Next, consider an anomaly free state which is labelled by some element $s_{ab}$ of ${\cal H}_{h_0}$. We now show that at any point $p\in \Sigma$ this metric can be reconstructed, upto
an overall scaling at $p$, to arbitrary accuracy from the amplitudes of the anomaly free state .
Accordingly, fix  a point $p$ on the Cauchy slice $\Sigma$ and  
consider some choice set (a)- (f) of section \ref{sec8.1}. With this choice set consider a primordial state $c$ such that $c\in B_{P0}$ and such that $c$  has 
 its non-degenerate vertex at $p$ with   reference coordinate patch  $\{x\}$. The action of  a single 
 electric diffeomorphism  on $c$  generates the child,  $c_1 \equiv c_{(i,I,{\hat J}, {\hat K},   \beta=0,  \delta)}$  where
   $\delta$ is measured by $\{x\}$.
From section \ref{sec5}, 
the amplitude
$(\Psi_{f,s_{ab},P_0}|c_1\ket$ of the anomaly free state $\Psi_{f,s_{ab},P_0}$ for the state  $c_1$ is  
evaluated using $\{x_1\}$ where $\{x_1\}$ denotes the reference coordinate patch for $c_1$  around $v_1$.
In view of the fact that $f$ vanishes only at a finite number of points, it follows that we can choose $\delta$ such that $f$ is non-vanishing at $v_1$ and we so choose $\delta$.

Next, consider any diffeomorphism $\chi$ which is identity in some neighbourhood of $v_1$
and consider the state $c_{1\chi}$ which is the image of $c_1$ by $\chi$.  Let the reference coordinate patch for $c_{1\chi}$ be denoted by $\{x_{1\chi } \}$.
It is straightforward to see that  the Lemma in P2 implies that we can use 
the coordinates $\chi^*\{x_1\}$  to evaluate the amplitude $(\Psi_{f,s_{ab},P_0}|c_{1,\chi}\ket$ instead of the reference coordinates $\{x_{1\chi}\}$.
Note however that since $\chi$ is identity in a vicinity of $v_1$,  this is the same as evaluating the amplitude with respect to $\{x_1\}$. 
Since the coordinate dependent part of the amplitude is $f \sum h_I H_I$  (see section \ref{sec5.2})  and since this part only depends on the 
vertex structure of $c_{1\chi})$ at $v_1$ it follows  
that 
\ba
(\Psi_{f,s_{ab},P_0}|c_{1,\chi}\ket & = &Bg_{c_{1\chi}}   \label{sense1}\\
 (\Psi_{f,s_{ab},P_0}|c_{1}\ket   &=& Bg_{c_1}  
\label{psicdiff=c}
\ea
where $B= f\sum_{L_1}h_{L_1} H_{L_1}$ and the coordinate dependent  evaluation of the function  $f$ at $v_1$ and the coordinate dependent 
normalization of the edge tangent vectors 
 ${\vec {\hat e}}_{J_1}$ at $v_1$ are with respect to $\{x_1\}$ as argued above, both for $c_1$ and for $c_{1\chi}$.

Next, we construct diffeomorphisms which are identity in a neighbourhood of $v_1$ but which move the $C^0$ kinks of $c_1$ to certain desired positions.
Since these diffeomorphisms are of the type $\chi$ above,  the amplitudes for the diffeomorphic images of $c_1$ by these diffeomorphisms satisfy 
(\ref{psicdiff=c}) and, therefore, serve to evaluate the  function $g$  (see (\ref{defgcase1}), Appendix \ref{ag}) when its arguments have been placed at these desired positions.
By placing these arguments at positions close enough to $p$, we may use the contraction behaviour of $g$ (see Appendix \ref{acong}) to 
extract the information about the metric label $s_{ab}$ in the vicinity of the point $p$. Accordingly we proceed as follows.

First , note that in the state $c_1$ the $C^0$ kink  ${\tilde v}_{{\hat J}_1} $ lies at a distance $\delta^{p_1}$ from $p$ along the the $L$th edge of $c$ with $L={\hat J_1}$,    the $C^0$ kink 
${\tilde v}_{{\hat K}_1}$ lies at a distance  $Q\delta^{p_2}$ from $p$ along  $M$th edge of $c$ with  $M={\hat K_1}$.  The values of $p_2,p_1$ are given in the Appendix \ref{acong}.  The exact specification
and value of $Q$ is not needed here.  The remaning $C^0$ kinks lie within a distance $\delta^{p_3}$ of $p$ where $p_3>p_2 $ (see Appendix \ref{acong}), all these distances being measured by 
$\{x\}$. Next, consider any $\epsilon <<\delta$. Clearly we can apply diffeomorphisms of the type constructed in  (iii), section \ref{sec4.3} to move  kinks at coordinate distances $\delta^{p_1}, Q\delta^{p_2}, \delta^{p_3}$ to
coordinate distances $\epsilon^{p_1}, Q\epsilon^{p_2}, \epsilon^{p_3}$. Further, these diffeomorphisms can be constructed in such a way that they are identity in a neighbourhood of $v_1$. Let us apply these diffeomorphisms to 
$c_1$.

Next consider the region  $R_{\epsilon,\tau}$ bounded by 2 spherical shells of  of radius $Q\epsilon^{p_2} \pm \tau$ around the vertex  $p$ of $c$, with $\tau<<\epsilon^{p_3}$.
Let ${\vec \xi}_{\epsilon}$  be any semianalytic vector field which is tangent to the sphere of radius $Q\epsilon^{p_2}$ around $p$ and let $F_{\epsilon, \tau}$ be a semianalytic function which is 1
on this sphere and which vanishes outside $R_{\tau}$. By choosing $\xi_{\epsilon}$ appropriately we can use an appropriate finite diffeomorphism  generated by the vector field
$F_{\epsilon,\tau}{\vec \xi}_{\epsilon}$ to move the point  ${\tilde v}_{{\hat K}_1}$ to any desired location on the sphere of radius  $Q\epsilon^{p_2}$ while leaving the positions of the remaining kinks unaltered. 
More in detail by moving this kink by such a diffeomorphism $\phi_{u,\epsilon}$ to a position on
this sphere such that the straight  line from the origin of the sphere at $p$ to this position  has  unit tangent ${\vec {\hat u}}$, we obtain, from Appendix \ref{acong} that:
\be
g_{c_{1 \phi_{u,\epsilon}}}
 = \epsilon^{p_2- p_1} Q\frac{||{\vec{\hat u}}_{}||}{||{\vec{\hat e}}_{\hat J_1}||}(1 + O(\epsilon^{p_2-p_1})) g_{c_{}}.
\label{gconu}
\ee
where the metric norms are calculated at the point $p$.
Clearly $\phi_{u, \epsilon}$ is of the type $\chi$ in (\ref{psicdiff=c}).  It follows that as $\epsilon \rightarrow 0$, we have that:
\ba
&(\Psi_{f,s_{ab},P_0}|c_{1 \phi_{u, \epsilon}}\ket = B\epsilon^{p_2- p_1} Q\frac{||{\vec{\hat u}}_{}||}{||{\vec{\hat e}}_{\hat J}||}(1 + O(\epsilon^{p_2-p_1})) g_{c_{}} &\label{sense2}\\
&\Rightarrow             B_1 s_{ab}{\hat u}^a{\hat u}^b  (1 + O(\epsilon^{(p_2 -p_1)})      =       \frac{\big((\Psi_{f,s_{ab},P_0}|c_{1 \phi_{u, \epsilon}}\ket\big)^2}{\epsilon^{2(p_2-p_1)}}    
\label{hu}
\ea
where $B_1:=\frac{Q^2B^2}{s_{ab}{\hat e}^a_{\hat J}{\hat e}^b_{\hat J}}$ and $g_c=1$ because $c$ is primordial and has no $C^0$ kinks. Note that $B_1$ is independent of the position ${\vec {\hat u}}$.
By varying the position ${\vec {\hat u}}$ and by choosing $\epsilon$ as small as we wish, clearly we can reconstruct the metric at $p$  upto an overall factor to any 
desired accuracy. 
More in detail, let us suppress  the labels on the right hand side of (\ref{hu}) which do not vary with ${\vec {\hat u}}, \epsilon$ and set:
\be
\frac{\big((\Psi_{f,s_{ab},P_0}|c_{1 \phi_{u, \epsilon}}\ket\big)^2}{\epsilon^{2(p_2-p_1)}}  := F(  {\vec {\hat u}}, \epsilon )
\label{sense3a}
\ee
Clearly, from appropriate linear combinations of evaluations of $F$ for 6 appropriately chosen values  ${\vec {\hat u}_{\alpha }}, \alpha =1,..,6$ we can reconstruct the 6 coordinate components of the metric $s_{ab}$ upto an overall factor 
to any desired accuracy:
\be
B_1s_{\mu \nu}  + O(\epsilon^{(p_2 -p_1)})    =
\sum_{\alpha} \lambda^{\alpha}_{\mu\nu} F(  {\vec {\hat u}_{\alpha}}, \epsilon )
\label{sense3}
\ee
Since ${\vec {\hat u}_{\alpha }}, \lambda^{\alpha}_{\mu\nu},\alpha =1,..,6, \mu,\nu=1,2,3$ are fixed and independent of $\epsilon$, we retain only the $\epsilon$ dependence of the right hand side of the above equation and set
\be 
\sum_{\alpha} \lambda^{\alpha}_{\mu\nu} F(  {\vec {\hat u}_{\alpha}}, \epsilon ) =: s^{\epsilon}_{\mu \nu}
\label{smunu}
\ee
so that we have that:
\ba
B_1 s_{\mu \nu} +  O(\epsilon^{(p_2 -p_1)})    &=&  s^{\epsilon}_{\mu \nu}  \label{s13a}\\
\Rightarrow (B_1)^3 s + O(\epsilon^{(p_2 -p_1)})  &=& s^{\epsilon} \label{sense4}\\
\Rightarrow   (B_1)^{-1} s^{-\frac{1}{3}}   + O(\epsilon^{(p_2 -p_1)})     &=&  (s^{\epsilon})^{-\frac{1}{3}}\label{s13}
\ea
where we have used $s, s^{\epsilon} $ to denote the determinats of $s_{\mu \nu},  s^{\epsilon}_{\mu \nu}$.  Multiplying the left and right hand sides of equations  (\ref{s13}), (\ref{s13a}) we get:
\be
\frac{s_{\mu \nu}}{ s^{\frac{1}{3}}} = \frac{s^{\epsilon}_{\mu \nu}}{ (s^{\epsilon})^{\frac{1}{3}}} + O(\epsilon^{(p_2 -p_1)})
\label{sfinal}
\ee
Since $\{x\}$ is an admissible semianalytic chart on $\Sigma$, we can transit to any other fixed $\epsilon$ independent semianalytic chart. Since the Jacobian factors are independent of $\epsilon$, equation (\ref{sfinal})
holds in {\em any} such chart in obvious notation.
Next, taking the limit as $\epsilon\rightarrow 0$ of (\ref{sfinal}) and letting $p$ vary over $\Sigma$,  it follows that the conformally invariant metric density 
$\frac{s_{\mu \nu}}{ s^{\frac{1}{3}}}$ can be reconstructed  on all of $\Sigma$ from the set of amplitudes defined by any anomaly free state $\Psi_{f, s_{ab}, P_0}$ with metric label $s_{ab}$.
Note that this result is {\em independent} of the choice scheme used to define $\Psi_{f, s_{ab}, P_0}$ (recall that a choice of reference coordinates is needed to evaluate the amplitudes which
define $\Psi_{f, s_{ab}, P_0}$).

Next, we use the machinery developed above to prove the following statement:

\noindent{\em Statement}:
Consider a  choice scheme $S_1$ and anomaly free basis state  $\Psi_{f_1, h_{1ab}, P^1_0}$ defined in this choice scheme for  the scalar density, metric and Bra Set labels $f_1,h_{1ab}, B_{P_0^1}$ where
$f_1$ vanishes at most at a finite number of points, $h_{1ab}\in {\cal H}_{h_0}$ and  $B_{P_0^1}$ is an admissible Bra Set in the scheme $S_1$. 
Likewise consider a second choice scheme $S_2$ and anomaly free basis state  $\Psi_{f_2, h_{2ab}, P^2_0}$ with
$f_2$ vanishing at most at a finite number of points, $h_{2ab}\in {\cal H}_{h_0}$ and $B_{P_0^2}$ admissible in $S_2$ . Let $h_{1ab}\neq h_{2ab}$. Then  $\Psi_{f_1, h_{1ab}, P^1_0} \neq \Psi_{f_2, h_{2ab}, P^2_0}$ where the inequality 
indicates that the two states are distinct in the sense of distributions.\\

\noindent{\em Proof}: 
First suppose $B_{P^1_0} \neq B_{P^2_0}$.  Let $c$ be such that $c\in  B_{P^1_0}, c\notin B_{P^2_0}$. Let $f_1\neq 0$ at the nondegenerate vertex of $c$  (if it vanishes replace $c$ by some diffeomorphic
image of $c$ such that $f_1\neq 0$ at the nondegenerate vertex of this image and rename this state as $c$). Then $(\Psi_{f_1, h_{1ab}, P^1_0}|c\ket \neq 0$ but $(\Psi_{f_2, h_{2ab}, P^2_0} |c\ket= 0$
so the 2 states are different. 

Next consider the case  $B_{P^1_0} = B_{P^2_0}$ and denote  $B_{P^1_0} = B_{P^2_0} \equiv B_{P_0} $.
In what follows we shall freqently refer to the argumentation (\ref{sense1})- (\ref{sfinal}) in the first part of this section.
Consider a primordial state $c$ in $B_{P_0}$,       and the  electric diffeomorphism deformation of $c$ in scheme  $S_1$. Call the deformed ket $c_1$.
Proceed as in the first part of this section replacing $f, s_{ab}$ by $f_1, h_{1ab}$  so as to obtain (\ref{sfinal}) with $s_{ab}$ replaced by $h_{1ab}$.

Next,  consider the amplitude $(\Psi_{f_2, h_{2ab}, P_0}|c_1\ket$ evaluated in the scheme $S_2$ (we emphasize that  $c_1$ is still the deformed child produced in scheme $S_1$ from its parent $c$). 
Let the reference coordinates  for the evaluation be $\{x_2\}$. Now if $f_2$ vanishes at $v_1$ we have $(\Psi_{f_2, h_{2ab}, P_0}|c_1\ket =0$ whereas  $(\Psi_{f_1, h_{1ab}, P_0}|c_1\ket \neq 0$  so that the 
2 anomaly free states are again distinct. Next, let $f_2\neq 0$ at $v_1$. Consider again the action of a diffeomorphism $\chi$ which is identity in the vicinity of $v_1$ on $c_1$.
Once again the Lemma of P2 implies that we may continue to use $\{x_2\}$ for amplitude evaluations $(\Psi_{f_2, h_{2ab}, P^2_0}|c_{1\chi}\ket$.  It is easy to check that the subsequent analysis also holds 
so that we have (\ref{sense1}) - (\ref{sfinal}) with the replacements $f_2,h_{2ab}$ for $f,s_{ab}$ in those equations. Thus we have derived the equations:
\ba
\frac{h_{1\mu \nu}}{ h_1^{\frac{1}{3}}}   + O(\epsilon^{(p_2 -p_1)})   &=& \frac{h^{\epsilon}_{1\mu \nu}}{ (h_1^{\epsilon})^{\frac{1}{3}}} 
\label{h1final}\\
\frac{h_{2\mu \nu}}{ h_2^{\frac{1}{3}}} + O(\epsilon^{(p_2 -p_1)}) &=& \frac{h^{\epsilon}_{2\mu \nu}}{ (h_2^{\epsilon})^{\frac{1}{3}}} 
\label{h2final}
\ea
where $\frac{h^{\epsilon}_{1\mu \nu}}{ (h_1^{\epsilon})^{\frac{1}{3}}}$ is {\em exactly} the same function of the amplitudes 
$(\Psi_{f_1, h_{1ab},P_0}|c_{1 \phi_{u, \epsilon}}\ket$ as $\frac{h^{\epsilon}_{2\mu \nu}}{ (h_2^{\epsilon})^{\frac{1}{3}}}$ is of the 
amplitudes $(\Psi_{f_2, h_{2ab},P_0}|c_{1 \phi_{u, \epsilon}}\ket$ as can be seen from (\ref{sense3a}).

Next, recall that the left hand sides of (\ref{h1final}), (\ref{h2final}) are both evaluated at the vertex $p$ of $c$.
Choose  $c$  to be such that  its vertex $p$ lies in  $U(a, \delta)$ so that (\ref{h1-h2}) holds at $p$.
It directly follows that  by choosing $\epsilon$ small enough in 
(\ref{h1final}), (\ref{h2final}), the right hand sides of these equations differ. This implies that there must be at least one value of $\alpha \in \{1,2,..,6\}$ such that
the amplitudes $(\Psi_{f_1, h_{1ab},P_0}|c_{1 \phi_{u_{\alpha}, \epsilon}}\ket$, $(\Psi_{f_2, h_{2ab},P_0}|c_{1 \phi_{u_{\alpha}, \epsilon}}\ket$  of the two anomaly free states on the {\em same}
chargenet $c_{1 \phi_{u_{\alpha}, \epsilon}}$ 
differ.
Hence the two states are distinct and this concludes the proof.

The statement which we have proved above  implies that given 2 distinct metric labels $h_{ab}, h^{\prime}_{ab} \in  {\cal H}_{h_0}$, 
 we are free to choose 2 different choice schemes, one for the definition of anomaly free states  with metric label $h_{ab}$ and for the definition of the discrete action of constraints on these states, and a second for anomaly free states
with metric label $h^{\prime}_{ab}$ and for the definition of the discrete action of constraints on {\em these} states.
This freedom of choice leads to no inconsistency  because we are guaranteed that two states with distinct metric labels are distinct.
We shall use this freedom  in the next section.

\subsection{\label{sec9.2} Diffeomorphism covariant regulating choices}

In what follows we refer to a particular implementation of the  choice scheme (a)- (f) summarised in (2), section \ref{sec811} by the letter  $S$ or by  addending suitable 
symbols/subscripts to  $S$; for example ${S}_1$ or  $S'$ etc. In section \ref{sec921} we define a covariant  choice of such schemes by tying  each such choice scheme to the metric label of the anomaly free 
basis state under consideration. In section \ref{sec922} we derive the action of a diffeomorphism on an anomaly free basis state. 

\subsubsection{\label{sec921}Covariant Choice Schemes}
Consider, as in section \ref{sec9.1}  the space ${\cal H}_{h_0}$ of metrics diffeomorphic to $h_{0ab}$.  Let $h_{0ab}$ be associated with some choice scheme $S_0$. We shall use the notation of sections \ref{sec4.2}, \ref{sec4.3}
for the reference structures associated with this choice.  Accordingly,  the metric and the associated  primary coordinate patch, the  reference state for the diffeomorphism class of states of $c$, the reference diffeomorphism 
mapping this reference state to $c$ and the reference coordinate patch for $c$ are:
\be
h_{0ab}, \;\;\; \{x_0\},  \;\;\;   c_0,  \;\;\;\; \alpha , \;\;\; \alpha^*\{x_0\}.
\label{h0ref}
\ee
Let $h_{ab}\in {\cal H}_{h_0}$. Since $h_{0ab}$ has no (conformal) symmetries there exists a  unique diffeomorphism $\phi$ such that $h_{ab} = \phi^*h_{0ab}$.  We define the choice scheme ${S}_h$ associated with this
metric to be the images by $\phi$ of the choice scheme  $S_0$
\footnote{ The choice (f) in  section \ref{secsum} will be assumed to be the same for $S_h$ and $S_0$.}
so that the metric,   the associated  primary coordinate patch, the reference state for the diffeomorphism class of states of $c_{\phi}$, the  reference diffeomorphism 
mapping this reference state to $c_{\phi}$  and the reference coordinate patch for $c_{\phi}$ are:
\be
h_{ab} = \phi^*h_{0ab}, \;\;\; \phi^*\{x_0\},  \;\;\;   \phi\circ c_0 , \;\;\;\;   \phi \circ \alpha  \circ \phi^{-1} ,  \;\;\; \phi^*\alpha^*\{x_0\}.
\label{h0phiref}
\ee
where we have denoted the image of $c$ by the diffeomorphism $\phi$ by $c_{\phi}\equiv \phi\circ c$ so that ${\hat U}(\phi )|c\ket = |c_{\phi}\ket = |\phi\circ c\ket$.
Here, we have chosen  the 
cone angle as measured by $\{x_0\}$ for  conical deformations $c_{0[i,I,\beta,\delta_0}$ of any $c_0$  in the scheme $S_0$ to be  the same as that measured by $\phi^*\{x_0\}$ for conical deformations 
$\phi\circ c_{0[i,I,\beta,\delta_0}$
of $\phi\circ c_0$ in the scheme $S_h$.

Note that  the  choice schemes  $\{S_h, h_{ab}\in {\cal H}_{h_0}\}$  yield the same  the set of primordials and the same Ket Set.
Further  if we choose the  Bra Set $B_{P0}$ in scheme $S_0$ then this same Bra Set is admitted as a Bra Set in 
the choice scheme $S_h$ for any $h_{ab} \in {\cal H}_{h_0}$. 

Accordingly consider the anomaly free state  $\Psi_{f,h_{ab}, B_{P0}}$.
We shall adopt a {\em covariant regulator scheme} for the definition of the constraint operator products of sections \ref{sec7} and \ref{sec8} by which we mean that
the discrete action of any such operator 
on $\Psi_{f,h_{ab}, B_{P_0}}$ is defined with respect to the choice scheme $S_h$.
More in detail,  let 
\be
{\hat O}(\{N_{\rm i}, \epsilon_{\rm i}, i=1,..m\}) \equiv {\hat O}(\{N_{\rm i}, \epsilon_{\rm i}\}):=
(\prod_{ {\rm i}=1 }^m {\hat O}_{ {\rm i} ,\epsilon_{\rm i}}(  N_{\rm i})), \;\;\;\;  \epsilon_{\rm i} < \epsilon_{\rm j}\;\; {\rm iff} \;\;{\rm i}< {\rm j}, 
\label{defoprod}
\ee
where the product is ordered from left to right in increasing ${\rm i}$ and each 
${\hat O}_{ {\rm i} ,\epsilon_{\rm i}}(  N_{\rm i})$ is chosen to be the discrete approximant to a Hamiltonian or electric diffeomorphism constraint
operator,  so that the resulting operator product $ {\hat O}(\{N_{\rm i}, \epsilon_{\rm i}\})$ is of the  type encountered in sections \ref{sec7} and \ref{sec8}.
Then the action of this operator  on any state $\Psi_{f,h_{ab}, B_{P_0}}$ with $h_{ab}\in {\cal H}_{h_0}$, evaluated on any charge net $c$  yields the amplitude: 
\be
(\Psi_{f,h_{ab}, P_0}|{\hat O}(\{N_{\rm i}, \epsilon_{\rm i}\})         |c\ket
\label{defoc}
\ee
where this amplitude is evaluated as in sections \ref{sec7}, \ref{sec8} with respect to 
choice scheme  $S_h$. Denoting the continuum limit operator defined through the discrete approximant ${\hat O}(\{N_{\rm i}, \epsilon_{\rm i}\})$ by  
${\hat O}(\{N_{\rm i}\})$  we have that:
\be
(\Psi_{f,h_{ab}, P_0}|{\hat O}(\{N_{\rm i}\})         |c\ket :=
(\lim_{\epsilon_m\rightarrow 0}(\lim_{\epsilon_{n-1}\rightarrow 0}...(\lim_{\epsilon_1\rightarrow0}(\Psi_{f,h_{ab}, P_0}|{\hat O}(\{N_{\rm i}, \epsilon_{\rm i}\})         |c\ket
\label{defoccont}
\ee
where the discrete action amplitude on the right hand side is defined with respect to the scheme $S_h$.

It is straightforward to see that the following property holds  in this covariant regulator scheme. Consider the metric label $h_{ab} \in {\cal H}_{h_0}$ and its image $\phi^*h_{ab}$ by the diffeomorphism $\phi$.
Given a state $c$, let its reference coordinates, reference state and reference diffeomorphism mapping this reference state to $c$  for the choice scheme associated with $h_{ab}$ be:
\be
\{x\}_c^h , \;\;\;\; c_0^h,  \;\;\;\; \alpha_c^h .
\label{href}
\ee
Then the reference coordinates, reference state and reference diffeomorphism for the state  $c_{\phi}$  for the choice scheme associated with $\phi^*h_{ab}$ is:
\be
\{x\}_{c_{\phi}}^{\phi^*h}  = \phi^* \{x\}_c^h      ,            \;\;\;\; (c_{\phi }^{\phi^*h})_0= \phi\circ   c_0^h       ,                \;\;\;\;          \alpha_{c_{\phi}}^{\phi^*h} =        \phi\circ \alpha_c^h \circ\phi^{-1}.
\label{hphiref}
\ee


\subsubsection{\label{sec922} Action of finite diffeomorphisms on anomaly free states}

Hereon, we need to keep track of metric labels and associated reference structures. Accordingly, we use (\ref{href}) to rewrite (\ref{psifghc}) so that the evaluation of the amplitude $(\Psi_{f,h_{ab}, P_0}| { c}> $ in the 
choice scheme $S_h$ is:
\be
(\Psi_{f,h_{ab}, P_0}| { c}> =   g_{c}(h_{ab} ,\{\tilde v\})  \;( \sum_{I} h_{I}H_{I})|_{h_{ab},v,\{x\}_c^h} \;f(v, \{x\}_c^h).
\label{psihc}
\ee
Here  $I$ indexes the edges at the nondegenerate vertex $v$ of $c$.  The notation $g_{c}(h_{ab}, \{\tilde v\})$ tells us the function  $g$ of section \ref{ag} is evaluated  at the $C^0$ kinks of $c$ and the geodesic distances between
these kinks are determined by the metric $h_{ab}$. The subscript ${h_{ab},c}$ to the sum over $I$ indicates that the edge tangents which go into the definition of $H_I, h_I$ are unit with respect to the $\{x\}_c^h$ coordinates
and are evaluated at $v$ with respect to the metric $h_{ab}$.

In this notation we have, once again in the $S_h$ scheme that:
\be
(\Psi_{f,h_{ab}, P_0}|{\hat U}^{\dagger}(\phi )|  c\ket  =   g_{c_{\phi^{-1}}    }(h_{ab},  \{ \phi^{-1}(\tilde v)\}) , \;
( \sum_{I} h_{I}H_{I})|_{  h_{ab},\phi^{-1}(v),\{x\}^h_{c_{ {\phi^{-1}}}}  } \;f(\phi^{-1}(v), \{x\}_{c_{{\phi^{-1}}}}^h).
\label{910}
\ee
Next, from (\ref{href}), (\ref{hphiref}),  note that in the $C_{\phi^*h}$ scheme    
the reference coordinates for $c =\phi \circ c_{\phi^{-1}}  $  
are $\phi^*\{x\}_{c_{\phi^{-1}}}^h$. This implies that 
\be
(\Psi_{\phi^*f,\phi^*h_{ab}, P_0}|  c\ket  =   g_{c   }(\phi^*h_{ab},  \{ \tilde v\}) , \;( \sum_{I} h_{I}H_{I})|_{\phi^*h_{ab},v,\phi^*\{x\}_{c_{\phi^{-1}}}^h} \;f(v, \phi^*\{x\}_{c_{\phi^{-1}}}^h).
\label{911}
\ee
Using the properties of push forwards by diffeomorphisms and that fact that  $\phi\circ c_{\phi^{-1}} = c$,  we have that:
\ba
 g_{c   }(\phi^*h_{ab},  \{ \tilde v\})  &= & g_{c_{\phi^{-1}}    }(h_{ab},  \{ \phi^{-1}(\tilde v)\}) \label{91a}\\
 (\phi^*f)(v, \phi^*\{x\}_{ c_{    \phi^{-1}   }    }^h) &=& f(  \phi^{-1}(v), \{x\}_{  c_{   {\phi^{-1}   }  }}^h) 
 \label{91b}
 \ea
 It is also straightforward to see, from the properties of pushforwards and the definition of $h_I, H_I$, that:
\be
( \sum_{I} h_{I}H_{I})|_{\phi^*h_{ab},v,\phi^*\{x\}_{c_{\phi^{-1}}}^h} = ( \sum_{I} h_{I}H_{I})|_{  h_{ab},\phi^{-1}(v),\{x\}_{c_{ {\phi^{-1}}}^h}}
\label{92}
\ee
From (\ref{91a}), (\ref{91b}) and (\ref{92}) together with (\ref{910}), (\ref{911}), it follows that:
\be 
 (\Psi_{f,h_{ab}, P_0}|{\hat U}^{\dagger}(\phi )|  c\ket = (\Psi_{\phi^*f,\phi^*h_{ab}, P_0}|  c\ket
\label{93}
 \ee
This equality holds for every $c$ in the Bra Set  $B_{P0}$. Further both sides vanish for any $c \notin B_{P0}$. Hence we have the following equality of anomaly free basis states:
\be
{\hat U}(\phi ) \Psi_{f,h_{ab}, P_0} = \Psi_{\phi^*f,\phi^*h_{ab}, P_0}
\label{94}
\ee

\subsection{\label{sec9.3} Action of products of  constraints and finite diffeomorphisms}

As explained in the introduction  the Poisson bracket  relation (\ref{classhh}) between
a pair of Hamiltonian constraints is replaced by (\ref{key}) and this relation is implemented quantum theory in sections \ref{sec7}, \ref{sec8}. 
Here we are interested in the remaining Poisson bracket relations (\ref{classdd}), (\ref{classdh}) between the diffeomorphism constraints and between the diffeomorphism and Hamiltonian constraints. 
In LQG the primary operators related to diffeomorphisms
are the unitary operators which implement finite diffeomorphisms generated by the diffeomorphism constraints rather than the diffeomorphism constraints themselves. Hence in quantum theory we 
replace (\ref{classdd}), (\ref{classdh}) by the relations:
\ba
{\hat U}(\phi_1) {\hat U}(\phi_2 ) &= & {\hat U}(\phi_1\circ\phi_2) \label{ddhat}\\
{\hat U}^{\dagger}(\phi) {\hat C}[N] {\hat U}(\phi ) &=& {\hat C}[\phi_*N]
\label{dchat}
\ea
These relations are to be imposed on the algebra generated by arbitrary products of finite diffeomorphism unitaries and Hamiltonian constraint operators.
Hence we are interested in the imposition of these relations within operator products of the form
\be
\big(\prod_{i_1=1}^{m_1}{\hat U}(\psi_{i_1} )\big) {\hat O}^{(1)}
\big(\prod_{i_2=m_1+1}^{m_2}{\hat U}(\psi_{i_2} )\big) {\hat O}^{(2)}...
\big(\prod_{i_n=m_{n-1}+1}^{m_n}{\hat U}(\psi_{i_n} )\big) {\hat O}^{(n )}
\big(\prod_{i_1=1}^{m_1}{\hat U}(\psi_{i_{n+1}} )\big)
\label{diffhamprod1}
\ee
where the $\psi$'s are semianalytic diffeomorphisms and the ${\hat O}$'s are products of Hamiltonian constraint operators. 

Note also that we would like to show that the relation (\ref{key})  is also valid within each such  product of Hamiltonian contraint operators. More in detail, by considering appropriate
linear combinations of products of the type (\ref{diffhamprod1}), we may define a product where, now,  each ${\hat O}$ in (\ref{diffhamprod1})  contains products of single commutators between  pairs of Hamiltonian constraints i.e.
we may consider multiple products of single commutators of the type in (\ref{eqn1}) and section \ref{sec8m}. We may then obtain a  new operator product by replacing each of these commutators by 
appropriate electric diffeomorphism commutators as indicated by (\ref{key}) and we would like to show that the first operator products with Hamiltonian constraint commutators equals this new product obtained
by these replacements. Hence we are interested in computing the action of operator products of the form (\ref{diffhamprod1}) where each ${\hat O}$ can be (a) a product of Hamiltonian constraints,
(b) a product of the type (\ref{eqn1})  or (c)  the product in (b) with the replacement of Hamiltonian commutators by appropriate electric diffeomorphism ones.

Since LQG provides a representation of the relation (\ref{ddhat}) on its kinematic Hilbert space, it immediately follows that this relation is automatically
implemented on the space of distributions through dual action. Since the anomaly free states are distributions, it follows that this relation is already imposed.
Given that this relation is imposed it is easy to see that 
operator products of the form (\ref{diffhamprod1}) are equivalent to products of the form:
\be
{\hat U}(\phi_1)^{\dagger}  {\hat O}^{(1)}{\hat U}(\phi_1) {\hat U}(\phi_2)^{\dagger}  {\hat O}^{(2)}{\hat U}(\phi_2)...{\hat U}(\phi_n)^{\dagger}  {\hat O}^{(n)}{\hat U}(\phi_n){\hat U}(\phi_{n+1})
\label{diffhamprod2}
\ee
where the $\phi$'s are semianalytic diffeomorphisms.
The  imposition of (\ref{dchat}), (\ref{key})  on such operator products yields the relation:
\ba
& ({\hat U}(\phi_1)^{\dagger}  {\hat O}^{(1)}(\{N_{ {\rm i}_1 }, {\rm i_1}=1,..,m_1\})                    {\hat U}(\phi_1) )({\hat U}(\phi_2)^{\dagger} 
{\hat O}^{(2)}(\{N_{{\rm i}_2}, {\rm i_2}=m_1+1,..,m_2\}){\hat U}(\phi_2))&\nonumber \\
&...({\hat U}(\phi_n)^{\dagger}  {\hat O}^{(n)}(\{N_{{\rm i}_n}, {\rm i_n}=m_{n-1}+1,..,m_n\})       {\hat U}(\phi_n))\;{\hat U}(\phi_{n+1})&
\nonumber \\
&=  {\hat O}^{(1)}(\{\phi_{1*}N_{{\rm i}_1}, {\rm i_1}=1,..,m_1\})  
{\hat O}^{(2)}(\{\phi_{2*}N_{{\rm i}_2}, {\rm i_2}=m_1+1,..,m_2\})&
\nonumber \\
&.. {\hat O}^{(n)}(\{\phi_{n*}N_{{\rm i}_n}, {\rm i_n}=m_{n-1}+1,..,m_n\})\; {\hat U}(\phi_{n+1})&
\label{diffhamaf}
\ea

Note that each ${\hat O}^{(i)}$ operator is the continuum limit of some discrete approximant of the type (\ref{defoprod}), each such product being defined in some choice scheme $S^i$.
We show below that this choice scheme, and hence the (continuuum limit) action of the operator product (\ref{diffhamprod2}) is {\em uniquely} fixed from the following two inputs:\\
\noindent Input (A): Any such choice scheme $S^i$  must be consistent with the covariant choice scheme defined in section \ref{sec921}.
By this we mean that any amplitude $(\Psi_{{\bar f}, {\bar h}_{ab}, {\bar P}_0}| {\hat O}(\{ N_{\rm i}, \epsilon_{\rm i} \} )|c\ket$ with  ${\hat O}(\{N_{\rm i}, \epsilon_{\rm i}\})$ defined as in (\ref{defoprod})
must be evalauted in the choice scheme $S_{\bar h}$ ( see the discussion around (\ref{defoc})).
\\
\noindent Input (B): The discrete action of any such {\em discrete} approximant of the type (\ref{defoprod}) in any choice scheme  on a charge net $c$ 
yields a {\em finite} linear combination of chargenets  (see (a)- (f)  of section \ref{secsum}).
\\

Accordingly, in what follows we shall restrict attention to the covariant choice scheme defined in section \ref{sec921}.  In section \ref{sec931}  we prove a key identity. In section \ref{sec932}  we derive the action of
the operator product (\ref{diffhamprod2}) from the inputs (A) and (B) above together with the identity proved in section \ref{sec931}. The resulting action will be seen to implement the relation 
(\ref{diffhamaf})  on the domain of anomaly free states.

\subsubsection{\label{sec931} A key identity}

\noindent{\em Claim}: Let ${\hat O}(\{N_{\rm i}, \epsilon_{\rm i}\})$ be defined as in (\ref{defoprod}).
Then the following identity holds for all, $f,\{N_{\rm i}\}, c$ and all  $h_{ab} \in {\cal H}_{h_0}$:
\be
(\Psi_{f,h_{ab}, P_0}|{\hat U}^{\dagger} (\phi ) {\hat O}(\{N_{\rm i}, \epsilon_{\rm i}\})  {\hat U}(\phi )       |c\ket
= (\Psi_{f,h_{ab}, P_0}|{\hat O}(\{(\phi_*N_{\rm i}), \epsilon_{\rm i}\})       |c\ket   + O({\vec \epsilon})
\label{keyid}
\ee
where $O({\vec \epsilon})$ indicates a quantity which vanishes  in the continuum limit:
\be
\lim_{\epsilon_m\rightarrow 0}\lim_{\epsilon_{n-1}\rightarrow 0}...\lim_{\epsilon_1\rightarrow0} O({\vec \epsilon}) = 0
\label{vece}
\ee

\noindent{\em Proof}: From (\ref{93}), (\ref{94}) and (B) above, it follows that 
\be
(\Psi_{f,h_{ab}, P_0}|{\hat U}^{\dagger} (\phi ) {\hat O}(\{N_{\rm i}, \epsilon_{\rm i}\})  {\hat U}(\phi )       |c\ket
=(\Psi_{\phi^*f,\phi^*h_{ab}, P_0}| {\hat O}(\{N_{\rm i}, \epsilon_{\rm i}\})  {\hat U}(\phi )       |c\ket .
\label{95}
\ee
Introduce the following notation for the action,  on the anomaly free state $\Psi_{{\bar f},{\bar h}_{ab}, P_0}$,  of the continuum limit  operator ${\hat O}(\{N_{\rm i})$ 
obtained from its discrete approximant ${\hat O}(\{N_{\rm i}, \epsilon_{\rm i}\})$:
\be
(\Psi_{{\bar f},{\bar h}_{ab}, P_0}|{\hat O}(\{N_{\rm i}, {\rm i} =1,..,m\})         |s\ket 
:= {\cal A}_m({\bar f}, {\bar h}, \{N_{\rm i}\}, s, \{x\}^{\bar h}_s) .
\label{96}
\ee
Here the left hand side is defined as a continuum limit of its discrete approximant as in (\ref{defoccont}).  The right hand side is the appropriate explicitly calculated amplitude
\footnote{\label{fn9af}Recall that we have shown in section \ref{sec8m} that this amplitude is consistent with (\ref{key}).}
from (\ref{heven}), (\ref{hodd}) with the substitutions ${\bar f}, {\bar h} , s ,  \{x\}^{\bar h}_s$ for $f, h, c, \{x\}$ in those expressions.
The resulting  expression depends on the arguments ${\bar f}, {\bar h}, \{N_{\rm i}\}, s, \{x\}^{\bar h}_s$.
Since we work within the covariant choice scheme, the coordinates  $\{x\}^{\bar h}_s$ with respect to which the explicit expression is defined are determined by ${\bar h}_{ab}$ and $s$; nevertheless
despite this redundancy, it is useful  for pedagogical purposes to retain this argument in ${\cal A}_m$. 

From the definition of the continuum limit it follows that the corresponding discrete action can be written as  
\be
(\Psi_{{\bar f},{\bar h}_{ab}, P_0}|    {\hat O}(\{N_{\rm i}, \epsilon_{\rm i}\})      |s\ket =
= {\cal A}_m({\bar f}, {\bar h}, \{N_{\rm i}\}, s, \{x\}^{\bar h}_s)  + O({\vec \epsilon} ).
\label{97}
\ee
Setting $|c_{\phi}\ket :={\hat U}(\phi )       |c\ket$, equations (\ref{95}) and (\ref{97}) imply that:
\be
(\Psi_{f,h_{ab}, P_0}|{\hat U}^{\dagger} (\phi ) {\hat O}(\{N_{\rm i}, \epsilon_{\rm i}\})  {\hat U}(\phi )       |c\ket
= {\cal A}_m({\phi^* f}, {\phi^* h}, \{N_{\rm i}\}, c_{\phi} , \{x\}^{\phi^* h}_{c_{\phi}})  + O({\vec \epsilon} ).
\label{98}
\ee
From (\ref{hphiref}) it follows that:
\be
{\cal A}_m({\phi^* f}, {\phi^* h}, \{N_{\rm i}\}, c_{\phi} , \{x\}^{\phi^* h}_{c_{\phi}})
=
{\cal A}_m({\phi^* f}, {\phi^* h}, \{N_{\rm i}\}, c_{\phi} , \phi^*\{x\}^{h}_{c})
\label{99}
\ee
Using the properties of pull backs by diffeomorphisms together with definitions of the various quantities which figure in the explicit expressions (\ref{heven}), (\ref{hodd}),
it is straightforward to see that 
\be
{\cal A}_m({\phi^* f}, {\phi^* h}, \{N_{\rm i}\}, c_{\phi} , \phi^*\{x\}^{h}_{c})
= {\cal A}_m({ f}, {h}, \{\phi_*N_{\rm i}\}, c , \{x\}^{h}_{c})
\label{991}
\ee
Using the appropriate substitutions in  (\ref{97}) we have that 
\be
{\cal A}_m({ f}, {h}, \{\phi_*N_{\rm i}\}, c , \{x\}^{h}_{c}) = 
(\Psi_{f,h_{ab}, P_0}|{\hat O}(\{(\phi_*N_{\rm i}), \epsilon_{\rm i}\})       |c\ket  + O({\vec \epsilon}).
\label{992}
\ee
The claimed identity (\ref{keyid}) immediately follows from equations (\ref{98}), (\ref{99}), (\ref{991}) and (\ref{992}).
\\

\noindent{Key Identity}: As a corollary, we have the following key identity which we shall use repeatedly in the next section:
\be
(\Psi_{f,h_{ab}, P_0}|{\hat O}(\{N_{\rm i}, \epsilon_{\rm i}\})  {\hat U}(\phi )       |c\ket
= (\Psi_{\phi_*f,\phi_*h_{ab}, P_0}|{\hat O}(\{(\phi_*N_{\rm i}), \epsilon_{\rm i}\})       |c\ket   + O({\vec \epsilon})
\label{keyid1}
\ee

To see this, substitute $f,h$ by $\phi_*f, \phi_*h$ in (\ref{keyid}) to obtain
\be
(\Psi_{\phi_*f,\phi_*h_{ab}, P_0}|{\hat U}^{\dagger} (\phi ) {\hat O}(\{N_{\rm i}, \epsilon_{\rm i}\})  {\hat U}(\phi )       |c\ket
= (\Psi_{\phi_*f,\phi_*h_{ab}, P_0}|{\hat O}(\{(\phi_*N_{\rm i}), \epsilon_{\rm i}\})       |c\ket   + O({\vec \epsilon}).
\label{993}
\ee
From Input (B), ${\hat O}(\{N_{\rm i}, \epsilon_{\rm i}\})  {\hat U}(\phi )       |c\ket$ is a finite linear combination of charge nets so that we may apply  (\ref{93}) to the left hand side of (\ref{993}) and obtain:
\be
(\Psi_{\phi_*f,\phi_*h_{ab}, P_0}|{\hat U}^{\dagger} (\phi ) {\hat O}(\{N_{\rm i}, \epsilon_{\rm i}\})  {\hat U}(\phi )       |c\ket
= (\Psi_{f,h_{ab}, P_0}| {\hat O}(\{N_{\rm i}, \epsilon_{\rm i}\})  {\hat U}(\phi )       |c\ket
\label{994}
\ee
Equation (\ref{keyid1}) immediately follows from (\ref{994}) and (\ref{993}).

An alternative way to state the Claim is to dispense with  Inputs (A) and (B) and instead state that  if (a),(b) below hold then equations (\ref{keyid})  hold where (a), (b) are as follows:\\
(a) We define the amplitude evaluation of any anomaly free basis state labelled by any  metric ${\bar h}_{ab} \in {\cal H}_{h_0}$ on any state ${\bar c}$ to be with respect to the scheme $S_{\bar h}$. \\
(b) We choose the discrete action of the operator approximant ${\hat O}(\{N_{\rm i}, \epsilon_{\rm i}\})$ on the left hand side  (lhs)  of (\ref{keyid} to be evaluated in  the $S_{\phi^*h}$ scheme and  that of the operator approximant 
${\hat O}(\{(\phi_*N_{\rm i}), \epsilon_{\rm i}\})$ on the right hand side (rhs) in the  $S_h$ scheme.\\
It is straightforward to repeat the steps of the proof with inputs (a) and (b) and thereby prove the claim.
Similarly, the Corolloary can be restated as follows. Let (a) hold and let the discrete action of ${\hat O}(\{N_{\rm i}, \epsilon_{\rm i}\})$ on the lhs of (\ref{keyid1}) be in the $S_h$ scheme and that of 
${\hat O}(\{(\phi_*N_{\rm i}), \epsilon_{\rm i}\})$ on the rhs in the $S_{\phi^*h}$ scheme. Then equation (\ref{keyid1}) holds.  Once again the proof is basically a straightforward repitition of the proof of the 
Corollary sketched above.

\subsubsection{\label{sec932} Action of the operator product in equation (\ref{diffhamprod2})}

In this section we evaluate the action of the operator (\ref{diffhamprod2}) on the anomaly free state 
$\Psi_{f,h_{ab}, B_{P0}}$. This action is obtained from that of the left hand side of (\ref{diffhamaf}) on this state:
\ba
& (\Psi_{f,h_{ab}, B_{P0}}|({\hat U}(\phi_1)^{\dagger}  {\hat O}^{(1)}(\{N_{ {\rm i}_1 },\epsilon_{{\rm i}_1} {\rm i_1}=1,..,m_1\})                    {\hat U}(\phi_1) )&\nonumber \\
&({\hat U}(\phi_2)^{\dagger} 
({\hat O}^{(2)}(\{N_{{\rm i}_2},\epsilon_{{\rm i}_2}, {\rm i_2}=m_1+1,..,m_2\}){\hat U}(\phi_2))&\nonumber \\
&...({\hat U}(\phi_n)^{\dagger}  {\hat O}^{(n)}(\{N_{{\rm i}_n},\epsilon_{{\rm i}_n}, {\rm i_n}=m_{n-1}+1,..,m_n\})       {\hat U}(\phi_n))\;{\hat U}(\phi_{n+1}) |c\ket&
\label{100}
\ea
In order to evaluate this discrete action we use Input (A), (B) above iteratively as follows.
For any charge  net $s$ we have that:
\ba
&(\Psi_{f,h_{ab}, B_{P0}}|({\hat U}(\phi_1)^{\dagger}{\hat O}^{(1)}(\{N_{ {\rm i}_1 },\epsilon_{{\rm i}_1}, {\rm i_1}=1,..,m_1\}){\hat U}(\phi_1) )|s\ket&\nonumber \\
&= (\Psi_{f,h_{ab}, B_{P0}}|{\hat O}^{(1)}(\{\phi_{1*}N_{ {\rm i}_1 },\epsilon_{{\rm i}_1}, {\rm i_1}=1,..,m_1\}){\hat U}(\phi_2)^{\dagger}  |s_1\ket     +O_1({\vec \epsilon}^{\;(1)})& \nonumber\\
&= (\Psi_{\phi_2^*f,\phi_2^*h_{ab}, B_{P0}}|{\hat O}^{(1)}(\{  \phi_2^*\phi{_1*}N_{ {\rm i}_1 },\epsilon_{{\rm i}_1}, {\rm i_1}=1,..,m_1\})|s_1\ket  +  O_1({\vec \epsilon}^{\;(1)})     & \nonumber\\
&= (\Psi_{\phi_2^*f,\phi_2^*h_{ab}, B_{P0}}|{\hat O}^{(1)}(\{  \phi_2^*\phi_{1*}N_{ {\rm i}_1 },\epsilon_{{\rm i}_1}, {\rm i_1}=1,..,m_1\})
{\hat O}^{(2)}(\{N_{{\rm i}_2},\epsilon_{{\rm i}_2} {\rm i_2}=m_1+1,..,m_2\}){\hat U}(\phi_2)
|s_2\ket &\nonumber \\
&+ O_1({\vec \epsilon}^{\;(1)}) &     \nonumber\\
&= (\Psi_{f,h_{ab}, B_{P0}}|{\hat O}^{(1)}(\{  \phi_{1*}N_{ {\rm i}_1 },\epsilon_{{\rm i}_1}, {\rm i_1}=1,..,m_1\})
{\hat O}^{(2)}(\{  \phi_{2*}N_{{\rm i}_2},\epsilon_{{\rm i}_2}, {\rm i_2}=m_1+1,..,m_2\})
|s_2\ket &\nonumber\\
&+ O_2({\vec \epsilon}^{\;(2)})  +       O_1({\vec \epsilon}^{\;(1)})         &
\label{101}
\ea
where we have used
(\ref{keyid}) in the second line, (\ref{keyid1}) in the  third and fifth lines and where we
 have defined  $s_1,s_2,$ by:
\be 
|s\ket = {\hat U}(\phi_2)^{\dagger}  |s_1\ket, \;\;\; |s_1\ket:=  {\hat O}^{\;(2)}(\{N_{{\rm i}_2}, \epsilon{\rm i_2}, {\rm i_2}=m_1+1,..,m_2\}){\hat U}(\phi_2)
|s_2\ket .
\ee
The symbol $O_1({\vec \epsilon}^{\;(1)})$ denotes a term which vanishes in the partial continuum limit which sends the parameters $\epsilon_1, \epsilon_2,..\epsilon_{m_1}$ to zero (in that order) while
keeping $\epsilon_{\rm j}, {\rm j}>m_1$ fixed. Similarly the term $O_2({\vec \epsilon}^{\;(2)})$ vanishes in the partial continuum limit over $\{\epsilon_{\rm i}, {\rm i}=1,..,m_2\}$ while keeping
$\epsilon_{\rm j}, {\rm j}>m_2$ fixed.
Clearly this procedure may be iterated to obtain an expression for (\ref{100}). It is easy to check that the continuum limit of this expression yields the evaluation of the right hand side of equation (\ref{diffhamaf})
on the anomaly free state.

Note that the application of Inputs (A), (B) to the calculation above fixes the choice scheme for the definition of each of ${\hat O}^{(j)}(N_{ {\rm i}_{j}}, \epsilon_{{\rm i}_{j}}) $
in (\ref{100}) to be $S_{\phi^*_jh}$. We may also restate the result by dispensing with Inputs (A) and (B) and instead state that  if (a),(b) below hold, then (\ref{diffhamaf}) holds. Here (a) is identical to (a), section \ref{sec931}
and (b) is as follows:\\
(b) In equation (\ref{100}) let the choice scheme for the definition of the discrete action of  ${\hat O}^{(j)}(N_{ {\rm i}_{j}}, \epsilon_{{\rm i}_{j}}) $ be chosen to be   $S_{\phi^*_jh}$.\\
It is straightforward to see that a proof may be constructed by basically repeating the steps which lead us from (\ref{100}) to (\ref{101}), iterating, and then taking the continuum limit.
Viewed in this way, defining anomaly free states through (a) above, we have shown that there exist discrete actions of operator approximants whose continuum limit action lead to the 
relationship (\ref{diffhamaf}).

The considerations in this section  show that the operator actions of sections \ref{sec7} and \ref{sec8} are consistent with equations (\ref{ddhat}), (\ref{dchat}) so that we have a diffeomorphism covariant anomaly free single commutator implementation of
the constraint algebra. More in detail given any two operator strings of the type (\ref{diffhamprod1}) related by the substitution of commutators between Hamiltonian constraints
by the appropriate combination of electric diffeormorphism commutators, we (a) first convert the strings to the form (\ref{diffhamprod2})  through (\ref{ddhat}) (b) use (\ref{dchat}) to remove all the 
${\hat U}(\phi_i ), {\hat U}^{\dagger} (\phi_i), i=1,..,n-1$ operators from the string (c) appeal to the anomaly free single commutator results of sections \ref{sec7} and \ref{sec8}.
The steps (a)-(c) show that we have a diffeomorphism covariant anomaly free single commutator implementation of
the constraint algebra.



\section{\label{secresults} Brief Summary of Results}

Consider an operator  product of the form
\be
{\hat O}= \big(\prod_{i_1=1}^{m_1}{\hat U}(\psi_{i_1} )\big) {\hat O}^{(1)}
\big(\prod_{i_2=m_1+1}^{m_2}{\hat U}(\psi_{i_2} )\big) {\hat O}^{(2)}...
\big(\prod_{i_n=m_{n-1}+1}^{m_n}{\hat U}(\psi_{i_n} )\big) {\hat O}^{(n )}
\big(\prod_{i_1=1}^{m_1}{\hat U}(\psi_{i_{n+1}} )\big)
\label{r-1}
\ee
where (a) the $\psi$'s are semianalytic diffeomorphisms,  (b) the ${\hat O}^{(i)}$'s are products of Hamiltonian constraint operators of each of density weight  $4/3$ and each smeared by a lapse of
density weight $-1/3$, and (c) the total number of Hamiltonian constraints in the operator product (\ref{r-1}) is less than $k$, the Cauchy slice $\Sigma$  being a $C^k$ semianalytic manifold.
Such an operator product has a well defined dual action on any anomaly free basis state $\Psi_{f,h_{ab}, P_0}$   , this dual action being inferred from the amplitudes
\be 
(\Psi_{f,h_{ab}, P_0}|{\hat O} |c\ket, \;\;\; \forall \;\;c
\label{secr1}
\ee
where $c$ ranges over thet set of all chargenet states.
These amplitudes are  such that the relations
\ba
{\hat U}(\phi_1) {\hat U}(\phi_2 ) &= & {\hat U}(\phi_1\circ\phi_2) \label{r-2}\\
{\hat U}^{\dagger}(\phi) {\hat C}[N] {\hat U}(\phi )& = & {\hat C}[\phi_*N]
\label{r-3}
\ea
hold so that any individual operator string,   {\em within} the big operator product (\ref{r-1}),  which is of the form of the left hand side of (\ref{r-2}) or (\ref{r-3}) can be replaced, in the amplitude evaluation 
(\ref{secr1}),  by the 
corresponding right hand side and {\em vice versa}.

An anomaly free basis state $\Psi_{f,h_{ab},P_0}$ is a linear combination of charge net bras, these bras comprising the Bra Set $B_{P_0}$. The coefficients of these bras in this 
linear combination are determined by a  scalar density $f$ of weight $-1/3$ which vanishes at most at a finite number of points, and a metric $h_{ab}$ with no conformal symmetries.
The explicit action of a product of $n\leq k-1$  Hamiltonian constraints on the anomaly free basis state $\Psi_{f,h_{ab},P_0}$ is given, for $n$ even and $c\in B_{P0}$ by:
\be
(\Psi_{f,h_{ab}, P_0}|
(\prod_{ {\rm i}=1 }^n {\hat C}_{}(  N_{\rm i})) |c\ket = 
(-3)^{\frac{n}{2}} (\frac{3\hbar N}{8\pi i})^n (\nu^{-\frac{2}{3}})^ng_c 
\sum_{I} |{\vec q}_I|^n h_IH_I^n (N_1,..,N_n;v), 
\label{r-4}
\ee
and for $n$ odd and $c\in B_{P0}$ by:
\be
(\Psi_{f,h_{ab}, P_0}|
(\prod_{ {\rm i}=1 }^n {\hat C}_{}(  N_{\rm i})) |c\ket = 
(-3)^{\frac{n-1}{2}} (\frac{3\hbar N}{8\pi i})^n (\nu^{-\frac{2}{3}})^ng_c 
\sum_{I} |{\vec q}_I|^{n-1}(\sum_{i=1}^3q^i_I) h_IH_I^n (N_1,..,N_n;v), 
\label{r-5}
\ee
%
%
where:
\be
H_I^n (N_1,..,N_n;v):= 
\big( \prod_{{\rm i}=1}^n N^{a_{\rm n-i+1}}_{\rm n-i+1}(p, \{x\})     {\hat V}_I^{a_{\rm n-i+1}}(p)\partial_{a_{\rm n-i+1}}\big) \big(f(p,\{x\}) \sqrt{h_{ab}(p){\hat V}_I^a(p){\hat V}^b_I(p)}\big)|_{p=v}
\label{r-6}
 \ee
and where the products above are ordered from left to right in increasing $i$. 
Here the reference coordinate patch (around the nondegenerate vertex $v$ of $c$)  associated with the metric $h_{ab}$ is $\{x\}$. The  vertex structure is such that the edges of the charge net  $c$ in a small vicinity of $v$ are straight lines
in the $\{x\}$ coordinates. The $I$th such edge has  
unit coordinate  edge tangents ${\hat V}^a_I$ with  ${\hat V}^a_I$ pointing outward or inward from $v$ depending on the kink structure of $c$  in the vicinity of $v$. These edge tangents are extended to constant (with respect to $\{x\}$) 
vector fields at any point $p$ in the vicinity of $v$  in equation (\ref{r-6}). The $i$th edge charge on the edge $I$ is denoted by $q^i_I$ and $\nu$ is the `volume' eigen value at $v$ in $c$.
The function $g_c$ depends on the network of geodesic distances, as measured by $h_{ab}$, between all pairs of $C^0$ kinks in $c$; a $C^0$ kink is a point at the intersection of 2 edges such that the edge tangents of the two edges
at this point are not proportional to each other. For $c\notin B_{P0}$ the right hand sides of (\ref{r-4}), (\ref{r-5}) vanish.

Equations (\ref{r-5}), (\ref{r-6}) are consistent with anomaly free single commutators. By this we mean that (a) these equations 
can be used to compute the action, on an anomaly free basis state,  of any operator string of the form (\ref{eqn1}), and,  (b) each of the commutators 
in the resulting expression  can be replaced by the appropriate electric diffeomorphism commutators in accordance with  (the quantum correspondent of) equation (\ref{key}).

The action of a diffeomorphism $\phi$ on the anomaly free basis state $\Psi_{f,h_{ab},P_0}$ yields the state  $\Psi_{\phi^*f,\phi^*h_{ab},P_0}$:
\be
(\Psi_{f,h_{ab}, P_0}|{\hat U}^{\dagger}(\phi )|c\ket = (\Psi_{\phi^*f,\phi^*h_{ab}, P_0}|c\ket
\label{r-7}
\ee
where $\phi^*$ is the push forward action of $\phi$.

The explicit action of any operator product of the form (\ref{r-1}) can be obtained from (\ref{r-4}), (\ref{r-5}), (\ref{r-7})  through a judicious use of the identities (\ref{r-2}), (\ref{r-3})
and the fact that the reference coordinate patch   $\{y\}$ associated with the metric $\phi^*h_{ab}$  for the state ${\hat U}(\phi )|c\ket$ is the pushforward of the reference coordinate patch $\{x\}$ 
associated with the metric $h_{ab}$ for the state $c$ so that $\{y\}= \phi^*\{x\}$.

\section{\label{sec10} Discussion}

\noindent{\bf (1)} Characterization of  Anomaly Free Domain:\\
In the previous section we showed that the finite span of anomaly free basis states constitute an arena wherein the constraint algebra admits a diffeomorphism covariant and anomaly free 
implementation. We refer to this finite span as the anomaly free domain ${\cal D}_{AF}$. We know very little about this domain. For example given  all the amplitudes  $(\Psi|c\ket$ of a state  $\Psi \in {\cal D}_{AF}$, we do not know
of any operational way of using these amplitudes to  reconstruct the expansion of $\Psi$ in terms of anomaly free basis states. We do not even know if this expansion is unique.
On the other hand, the action of the constraint operators depends on the basis expansion by virtue of the covariant choice scheme wherein the regulation (and hence continuum limit action) of 
constraint operators depends on the metric label of the basis state being acted upon. Hence if the expansion in basis states is not unique neither is the definition of the action of the constraints.
Nevertheless {\em given any such expansion}, the Hamiltonian constraint commutators can be replaced by appropriate electric diffeomorphism constraint commutators and the action of the constraint
operator products in (\ref{diffhamprod2}) is diffeomorphism covariant {\em within the context of this particular basis state expansion}. 
If there are several such expansions then defining all operators of interest with respect to any one fixed expansion
ensures that the {\em relations}  between these operators are consistent with anomaly free commutators and diffeomorphism covariance. It is in this sense that (\ref{ddhat}), (\ref{dchat}), 
(\ref{key})  hold.
\\

\noindent{\bf (2)} Physical States and their off shell deformations:\\
The anomaly free states introduced in section \ref{sec5}  and used in sections \ref{sec6} -\ref{sec9} do not satisfy the Hamiltonian constraint as can be seen from equation (\ref{hh13}). 
They also do not satisfy the diffeomorphism constraint, as can be seen from equation (\ref{94}). Hence they are {\em off shell} states. We would like to see them as off shell deformations of on-shell 
states. The simplest way to do this is to define the distribution $\Psi_{sol,P_0}$ :
\be 
(\Psi_{sol, P_0}| = \sum_{\bra {\bar c}| \in B_{P0} } \bra {\bar c}|.
\label{psisolsum}
\ee
It is then easy to check that the action of the distribution $\Psi_{sol, P_0}$  on (\ref{hh11}) vanishes {\em independent} of which $h_{ab}\in {\cal H}_{h_0}$ 
is used to regulate the  constraint in that equation. More in detail, if we fix any $h_{ab}\in {\cal H}_{h_0}$  and use the choice scheme $S_h$, we have that the continuum limit of the resulting Hamiltonian 
constraint on $\Psi_{sol, P_0}$ vanishes. Further, by inspection,  $\Psi_{sol, P_0}$ is invariant under the (dual) action of operators which implement  finite diffeomorphisms. 
Hence $\Psi_{sol, P_0}$  is a solution to all the constraints and constitutes a physical state.

Next, consider the following one parameter family of states based on the Bra Set $B_{P0}$:
\be
\Psi_{f,h_{ab}, P_0, \tau} = \Psi_{sol, P_0} + \tau \Psi_{f,h_{ab}, B_{P0}}, \;\;\; \tau>0
\label{psioff}
\ee
Clearly, $\Psi_{f,h_{ab}, P_0, \tau}$ is an off shell state such that its action on operator products of the type (\ref{diffhamprod2})
is diffeomorphism covariant and implements anomaly free single commuators. Further, $\Psi_{f,h_{ab}, P_0, \tau}$
 can be deformed into the physical state $\Psi_{sol, P_0}$ by allowing $\tau$ to vanish.
Thus the one parameter set of states $\{ \Psi_{f,h_{ab}, P_0,  \tau} , \tau>0\} $ constitute an off shell deformation of the physical state $\Psi_{sol, P_0}$ such that 
on these states the implementation of the constraint algebra is diffeomorphism covariant and displays anomaly free single commutators.
More generally, we may consider any state $\Psi$ in ${\cal D}_{AF}$  and  construct $\Psi_{\tau} =  \Psi_{sol, P_0} + \tau \Psi$ as off shell deformations of $\Psi_{sol, P_0}$.
The comments of {\bf (1)} above then apply to the manner in which the implementation of the constraint algebra  on such states is consistent with (\ref{ddhat}), (\ref{dchat}), 
(\ref{key}).
\\

\noindent{\bf (3)} Contrast with the conventional notion of anomaly free constraint algebras:\\
As mentioned in the Introduction, the conventional notion of anomaly free constraint algebras also includes multiple (as opposed to single) anomaly free commutators.
In the absence of structure functions, this conventional notion is powerful and appropriate as it (a)typically incorporates a representation of some underlying Lie group of gauge 
tranformations (b) ensures that there is a sufficiently large space of physical states.

In contrast, in the case of gravity, as is well known, the 4d diffeomorphism group (and it Lie algbera of vector fields) is not represented through the constraint algebra
because a spatial slice with respect to one spacetime metric is generically not spatial with respect to  the image of this metric by a diffeomorphism.  Further, due to the 
presence of structure functions, the multiple Poisson brackets between constraints, while weakly vanishing, yield constraints with more and more complicated 
phase space dependendent lapses and shifts rather than simple Lie algebra like structures.   Thus property (a) of the Lie group case seems absent so that the motivation
for anomaly free multiple commutators stems in this context mainly from (b).  However, if we drop the requirement of anomaly free multiple commutators, we may nevertheless
directly check (b) i.e. the conventional notion with regard to (b) may be viewed only as a sufficient rather than necessary condition for a nontrivial physical state space.
Another reason to question the need for anomaly free multiple commutators is that they represent properties which are higher that leading order in $\hbar$ and hence
their implementation seems to be unnecessary from  a naive view of obtaining the correct classical limit.

Our view point is then as follows. While the constraints do not offer a representation of 4d diffeomorphisms, there exists a subset of constraints whose algebra is that of 
3d diffeomorphisms. Accordingly we seek quantum representation of the group of 3d diffeomorphisms and LQG provides this. Next,  even though the 4d diffeomorphism Lie algebra
is unavailable, 
one {\em can} nevertheless
interpret the {\em single} Poisson bracket (\ref{classhh}) as the representation of  4d deformations in {\em spacetime} of the 3d Cauchy Slice \cite{hkt}.
More in detail, in Reference \cite{hkt} it is shown that commutator of a pair of such infinitesmal geometric deformations  normal to the Cauchy slice to {\em leading order} exactly mirrors
the single Poisson bracket (\ref{classhh}).
\footnote{It would be of interest to see if this correspondence also holds between multiple Poisson brackets and  higher order contributions to the commutator between infinitemal geometric deformatioms;
if the correspondence breaks down due to `embedding dependence' \cite{hkt}, this would provide added justification for dropping the requirement of anomaly free multiple commutators.}
Hence we seek to represent these single Poisson brackets in an anomaly free manner through (\ref{key}).
After doing this we may then check (b) i.e we may check if we have a large enough solution space. While the work in this paper suggests that the constraint action is compatible 
with a large enough solution space, a confirmation of this suggestion rests on {\bf (7)} below.  
\\

\noindent{\bf (4)} Dependence of solution space on regulating choices:\\
The key regulating choice is that of the Primary Coordinates when the metric label is $h_0$, other choices being fixed through our covariant regulator choice requirement.
This `preferred' choice seems to lead   to the existence of `preferred' structural properties of Physical States in that such states are combinations of charge net bras which 
are multiple deformations of primordial bras, these multiple deformations being defined with respect to this choice of Primary Coordinates. Hence it would be adviseable to see if we could 
build on the work here so that our considerations yield physical states which are combinations of multiply deformed charge net bras these multiple deformations arising from  {\em all} possible choices of 
linear coordinates for primordial states. 
\\

\noindent{\bf (5)} Structural inputs in our demonstration of anomaly free commutators:\\

\noindent{\bf (a)} Interventions: The interventions by judiciously chosen holonomies in order to define deformations in sections \ref{secgr} and \ref{secneg} play an {\em essential} role in our demonstration
of the existence of anomaly free commutators. These interventions are far from obvious and could not have been arrived at without  guidance from the requirement of anomaly freedom. Thus the requirement of
anomaly free commutators plays a key role in homing in on the (hopefully!) correct choice of (discrete approximants to) the Hamiltonian constraint.
\\
\noindent{\bf (b)} Gauge invariance: $U(1)^3$ gauge invariance plays a key role in our considerations; without it, the results  of Appendix \ref{acolor} which related net charges to primordial ones would not hold. 
As a result of the interventions {\bf (a)}, it is the  properties of the {\em net} charges (as opposed to the charges themselves) which become important (see the Note and related discussion at
the beginning of section \ref{sec7}). $U(1)^3$ gauge invariance then plays a key role in our considerations; without it, the results  of Appendix \ref{acolor} which relate net charges to primordial ones would not hold
and there would no longer be a correlation between properties of primordial charges and those of net charges. This would negatively impact many important structures/concepts  such as the definition of non-degeneracy
of CGR vertices, the properties of the Bra Set discussed section \ref{sec4.2}, 
the invariance of the  inequalities (\ref{defqmin}), (\ref{defqmin,1}) under the replacement of primordial charges by net charges,  and the  equivalence of  (\ref{thetachoice}) with (\ref{thetanetchoice}).

\noindent{\bf (c)} Linearity: Linearity of chargenet vertices plays a key role in our constructions. It  allows for  unambiguous extensions of graphs, such extensions being required for the construction 
of certain conical deformations (see section \ref{secneg}. 
It also  allows for a natural  definition of `along edge' vertex displacements (see section \ref{sec3.2.2}). This definition together with the 
linearity of the scrunching diffeomorphism $G$ (\ref{defG}) leads to constant Jacobian factors which can be pulled out when analysing the contraction behavior of 
the function $H^l_m$ (\ref{dmHcon}), (\ref{dmHncon}). This contraction behaviour neatly dovetails with that of the function $g_c$. All this would be impacted if we did not have linearity.
Our proof of the  validity of the replacement of reference coordinates by contraction coordinates for amplitude evaluations relies on the invariance of the regular conicality of deformations under
rigid translations; this, too, relies crucially on linearity.

It is not clear to us if our constructions can be generalised if we drop the requirement of linear vertices; however, any such putative construction
would be incredibly baroque. The linearity property implies that higher order moduli vanish \cite{grotrovelli}; since no physically interesting  operators in LQG to date involve higher order moduli,
linearity does not seem to signify a strong physical restriction. Linearity also plays a role in intepretations of kinematic states \cite{polyhedra} and in the application of the Minkowski theorem to
our considerations in P1 \cite{jurekmink}.

\noindent{\bf (d)} Restrictions on charge labels: As mentioned earlier the `eternal nondegeneracy' restriction on all members of a primary family is a key property without which it would be difficult to 
proceed. However, it may be worthwhile to think about how this restriction may be weakened or removed. The restriction (\ref{qneq0}) seems to be an overkill and likely can be removed without damaging our final results; this
should be confirmed. The restrictions (\ref{qneq0}), (\ref{sumq}) seem to play a role only for the products involving   more than 2  constraints (see section \ref{sec8}). In this regard, see {\bf (6)} below.

\noindent{\bf (e)} Restriction on valence: We have restricted attention to the case that the valence $N$ of any primordial at its nondegenerate vertex  is even. 
The reason is that any regular conical deformation of a GR vertex with $N$ odd results in a vertex which is not GR. This follows from the fact that the projections perpendicular to the 
cone axis of the edges pointing along the cone for a regular conical configuration seperate into pairs which point opposite to each other. This in turn is due to the fact that $\pi$ is an integer multiple 
of the azimuthal angle  $\phi= \frac{2\pi}{N-1}$ between successive projections  i.e. $\pi= \frac{N-1}{2} \phi$. 

The GR property is used crucially in the proof of the Lemma in section 3 of P2, this Lemma being used
in the arguments of section \ref{sec6.3} involving the replacement of reference coordinates by contraction coordinates.  If $N$ is odd, let us assume that there do exist charge nets which satisfy 
`eternal non-dgeneracy' under repeated conical deformations. It should still be possible to replace reference by contraction coordinates by restricting attention to charge nets in the Ket set which have
no `vertex symmetries'. By this we mean that 
 the only diffeomorphism which maps any such chargenet to itself is necessarily identity in the vicinity of the nondegenerate vertex of the charge net. This can be achieved, for example, by arranging for the 
 primordial charges on distinct edges to be unequal to each other, provided the various charge restrictions in {\bf (d)} can be show to hold. A careful check must also be made that all other considerations in this
 work go through for the $N$ odd case. While these issues need careful investigation, we believe that with suitable genericity assumptions which lead to the absence of vertex symmetries, it should be possible to
 generalise our work to the case of $N$ odd.

\noindent{\bf (f)} Role of the $C^1,C^2$ kinks: The reader is urged to peruse the last paragraph of section \ref{secneg4} wherein the necessity of correlation of the `upward' direction between members of a lineage is 
emphasised. The upward directions at any parent vertex are inferred from the positioning of the $C^1, C^2$ and $C^0$ kinks about the  parent vertex and the placement of  kinks  around the child vertex is correlated
with the set upward directions at the parent vertex. While the $C^0$ kinks occur naturally from our picture of the deformations generated by the constraints as the `abrupt pulling of edges along some particular edge', 
we have introduced the $C^1, C^2$ kinks purely as diffeomorphism invariant markers for the reconstruction of 
consistent upward directions. Their presence stems from our desire to  exercise adequate control on the calculations in this work. However we feel that they constitute an inessential technical overkill and that it should be 
possible to do away with them.\\

\noindent{\bf (6)} Products of more than 2 constraints:\\ It seems unlikely to us that the treatment of products of multiple constraint products in section \ref{sec8} will go through for the $SU(2)$ case of Euclidean gravity.
This is because the analogs of an $i$th charge component is the $i$th component of a left or right invariant vector field on $SU(2)$ i.e. the analogs of these charges are gauge {\em variant} operators. Hence it seems difficult to
define the $Q$ factors in equations  (\ref{ho6}), (\ref{do6}). On the other hand all the ingredients in our treatment are fixed already by the requirements of an anomaly free commutator
for the case of 2 constraints (see section \ref{sec7}).  Given these ingredients, the `$-{\bf 1}$' structure of the constraints ensures that any solution to the constraints is of the type discussed in {\bf (2)} above.
Hence even if we manage to generalise only the considerations of section \ref{sec7} (and section \ref{sec10}) to the $SU(2)$ case, it would constitute significant progress.
\\

\noindent{\bf (7)} Multivertex States: 
The extension of our results to the multivertex case is a key open problem. It is only in the context of such an extension that we can analyse propagation in the sense of Smolin \cite{leeprop,proppft}.
In \cite{u13prop} we make reasonable assumptions on the solution space emerging from such a putative extension and analyse the issue of propagation. 
\\


\noindent{\bf (8)} Semianalyticity Assumption: We have assumed that semianalytic vector fields generate semianalytic diffeomorphisms and used this assumption in many of our constructions. An important open technical problem is
to construct a proof of the validity of this assumption. \\

\noindent{\bf (9)} Speculations on role of the metric label: \\
Anomaly free basis states have a metric label which plays a key role in our implementation of diffeomorphism covariance. 
We have restricted metric labels to have no conformal symmetries; is it possible to allow for metric labels with  (asymptotic) symmetries such as (asympotitically) flat metrics?
Can these metric labels have any other fundamental role to play (for example in coupling to matter or in considerations of Lorentz invariance or semiclassicality)?

\section*{Acknowledgements}
I thank Fernando Barbero for his comments on a draft version of this work and for his kind help with the figures.

\section*{Appendices}

\appendix

\section{\label{adefkink}Definition of $C^0,C^1,C^2$ kinks}

\noindent $C^0$ kink: Let 2 semianalytic $C^k$ edges $e$ and $f$ intersect at a point $p$. Let the edge tangent at $p$, in some parameterization $t$ of $e$,   be ${\dot e}^a$. Let the edge tangent at $p$ in some
parameterization $s$ of $f$ be ${\dot f}^a$. Then $p$ is called a $C^0$  kink if ${\dot e}^a,{\dot f}^a $ are linearly independent. Clearly this property is invariant under diffeomorphisms.
\\

Next, consider $e,f$ as above. Let the intersection point  $p$ be  the end point of $e$ and the beginning point of $f$. Consider a semianalytic coordinate patch in an open neighbourhood
of $p$. Dots will refer to derivatives of coordinate components of points of $e,f$ with respect to their respective parameters $t,s$ at the point $p$. \\

\noindent  $p$ is a $C^1$ kink iff:\\
(a1) There exists $\lambda_1 > 0$ such that ${\dot f}^a = \lambda_1 {\dot e}^a$\\
(b1) There exists no $\lambda_2$ such that ${\ddot f}^a - (\lambda_1)^2 {\ddot e}^a = \lambda_2 {\dot e}^a$.\\

\noindent $p$ is a $C^2$ kink iff:\\
(a2) There exists $\lambda_1>0$ such that ${\dot f}^a = \lambda_1 {\dot e}^a$\\
(b2) There exists  $\lambda_2$ such that ${\ddot f}^a - (\lambda_1)^2 {\ddot e}^a = \lambda_2 {\dot e}^a$.\\
(c2) There exists no $\lambda_3$ such that ${\dddot f}^a- (\lambda_1)^3 {\dddot e}^a - 3\lambda_1 \lambda_2 {\ddot e}^a = \lambda_3 {\dot e}^a$.\\

Here the conditions (a1), (a2) ensure that there exists a reparameterization of $e$ such that first order parameter derivatives  of $e,f$ coincide at $p$.
The condition (b1) implies that no reparameterization  of $e$,  for which the first order parameter derivatives  of $e,f$ coincide at $p$, is such that the second order parameter derivatives coincide.
The conditions (b2), (c2) imply that such a reparameterization exists  but no such reparameterization can also make the third order parameter derivatives of $e,f$
coincide at $p$. It is straightforwad to verify that these conditions are invariant under change of semianalytic coordinate patch around $p$ (assuming,  that the differentiability degree $k$ of the semianalytic 
manifold is greater than 3) as well as under change of parameterizations of $e,f$. A straightforward consequence is that the defining properties of  $C^1, C^2$ kinks are  diffeomorphism invariant.

\section{\label{acone}Regular Downward Conical  Deformations of linear GR vertex structure}
The deformation is constructed in 2 steps. The first, described  in section \ref{acone1}, endows the deformation with a regular cone structure with the non-conducting edges in the vicinity
of the displaced vertex lying along a `downward regular cone' with axis along the conducting edge. Here by regular we mean that if we take the outward pointing upper conducting edge as the `$z$- axis' then the non-conducting 
edges are at equispaced azimuthal angles around this axis along the cone. In the second step described in section \ref{acone2} we introduce a `$C^m$-kink' $m\in \{1,2\}$  on the upper conducting edge.
The same techniques used below can be adapted to (a) construct regular downward conical deformations of linear CGR vertex structures as discussed in  section \ref{secgr}),  (b) use (a) as in section \ref{secneg1.1} to construct
regular upward conical deformations of linear GR vertex structures,  and,  (c) use (b) to construct regular upward linear CGR vertex structures as in  section \ref{secneg1.2}.  
\subsection{\label{acone1}Step 1: Obtaining a regular cone about the conducting line}
Let $c$ be a chargenet with a single linear non-degenerate GR vertex $v$ with $N$ edges, $N$ being even. 
We are interested in deforming this charge net along its $I$th edge to obtain the deformed charge net ${\bar c}$. If the deformation is generated by the Hamiltonian constraint 
this deformed charge net ${\bar c}$ is obtained as the product of 3 chargenet holonomies; the first holonomy is based on the deformed graph depicted in Fig \ref{grb} and the second and third on the undeformed graph
of Fig \ref{gra} underlying $c$. If the deformation is generated by an electric diffeomorphism, the chargenet is based only on the deformed graph depicted in Fig \ref{grb}
but is colored differently from the first holonomy for the Hamiltonian constraint alluded to above.
The displaced vertex of  ${\bar c}$ is CGR if the deformation is generated by the Hamiltonian constraint and  GR if the deformation is generated by an electric diffeomorphism. 
Hence only in the former case do we have a conducting line and an upper conducting edge. Nevertheless, in this section we abuse
this terminology slightly and refer to the edge in ${\bar c}$ along which the deformation has taken place, variously, as the conducting line, conducting edge or upper conducting edge. 

In this section we construct the precise  deformation which leads to the deformed graph structure of Fig \ref{grb}.
Since we are exclusively concerned with graph structure near the deformed vertex, we shall not be interested in the colorings of $c, {\bar c}$ in this section. We shall use the language
`deformation of $c$' to mean `deformation of the graph underlying $c$ so as to yield the deformed part of the graph near the deformed vertex in the graph underlying ${\bar c}$'.

Let $\{x\}$ be the chosen (linear) coordinate patch around $v$ so that there is a small enough coordinate ball, $B_{2\delta}(v)$ of radius  $2\delta$ around $v$
whose surface intersects each edge emanating from $v$  only once and such that these edges within this ball are straight lines. 
In what follows all our considerations will be restricted to this ball and we shall freely use  coordinate structures (with respect to $\{x \}$ such as  (the restriction to this ball of) planes, lines, rigid rotations etc. In what follows
 we shall use the
notation $B_{\tau}(p)$ to denote a coordinate ball of radius $\tau$  around the point $p$. 

Let $e_I$ be the edge of $c$ at $v$  along which the deformation is to be constructed. We use hatted indices to denote the edges of $c$ at $v$ other than the $I$th so that such an edge is denoted  $e_{\hat J}, {\hat J \neq  I}$.
Let $B_{\delta}(v)$  intersect each $e_{\hat J}$ at the point ${\tilde v}_{\hat J}$ and $e_I$ at the point $v_I$. Join $v_I$ to each 
${\tilde v}_{\hat J}$  by the straight line $l_{I{\hat J}}$. 

Each  $l_{I{\hat J}}$ is in the coordinate plane $P_{I{\hat J}}$ spanned by the tangent vectors  ${\vec {\hat e}}_I (v), {\vec {\hat e}}_{\hat J}(v)$ at $v$ (these vectors are in the direction of the the  straight line edges $e_{\hat J}$ and 
$e_I$). Since $v$ is GR, these $N-1$ planes (one for each ${\hat J}$) only intersect along the straight line along $e_I$. Consider any such plane $P_{I{\hat J}}$ and the rotation vector field ${\vec \xi}_{I{\hat J}}$ about the axis passing through $v_I$ normal to this plane.
Consider $B_{\epsilon}(v_I)$, $\epsilon<\delta$ so that $B_{\epsilon}(v_I) \subset B_{2\delta} (v)$.
Let $f$ be a semianalytic function of compact support which is unity in $B_{\frac{\epsilon}{2}}(v_I)$ and vanishes outside $B_{\epsilon}(v_I)$. Let  $\phi(f\xi_{I{\hat J}}, t)$ be the one parameter family of diffeomorphisms generated by 
$f{\vec \xi}_{I{\hat J}}$. For an appropriate value of $t= t(\theta,{\hat J})$ apply this diffeomorphism {\em only} to the the line $l_{I{\hat J}}$ so as to rotate it rigidly within  $B_{\frac{\epsilon}{2}}(v_I)$ to the 
required cone angle $\theta$ while retaining its semianalytic character.  Performing this `rotation' for each line $l_{I{\hat J}}$  yields a downward cone structure in the vicinity of $v_{I}$. 
With a slight abuse of notation we shall continue to refer to  $\{l_{I{\hat J}}\}$  so deformed by the same symbol. 
The above structure  defines a deformation of $c$ wherein 
$\{{\tilde v}_{\hat J}\}$ are the $C^0$ kinks, $v_I$ is the displaced vertex and the edges $\{e_{\hat J}\}$ have been `abruptly dragged' at the $C^0$ kinks so as to form the curves $l_{{\hat J},I}$  which are straight lines in the 
vicinity of $v_I$ and point along the `downward' cone there with cone angle $\theta$. Further, due to the use of semianalytic diffeomorphisms the graph so obtained remains semianalytic and, due to the details of the procedure
no unwanted intersections have been created.

Our considerations hereon are restricted to $B_{\frac{\epsilon}{2}}(v_I)$.
Recall that the edges at $v_I$ in  $B_{\frac{\epsilon}{2}}(v_I) \subset B_{\epsilon}(v)$ are all straight lines.
Since in this section we have occassion to refer to both the undeformed edges of $c$ as well as their deformed counterparts in ${\bar c}$, we denote these deformed counterparts through `bars'. Accordingly,
using the same numbering for the deformed edges at $v_I$ in the deformed charge net as for their undeformed counterparts at $v$ in $c$, we denote the deformed counterpart of $e_J$  at $v_I$ in ${\bar c}$ by ${\bar e}_J$. Using the hatted
index notation, the edges $\{{\bar e}_{\hat J}\}$ are non-conducting at $v_I$ and the conducting edge is ${\bar e}_I$. 

Consider the projections of each of the non-conducting edges  transverse to the conducting edge at $v_I$ in ${\bar c}$. These projections take the form of radially directed rays in a 2 dimensional
disk $D_{\perp}$  emanating outward from its center. As shown in P1, the {\em angular order} of these projections around this center is a coordinate invariant property. We shall now further deform the structure around $v_I$ so that these
transverse projections are  at equal angles $\phi = \frac{2\pi}{N-1}$ with respect to each other while 
maintaining the downward cone angle $\theta$ of their unprojected non-conducting edge correspondents. 
 For this purpose it is useful to change our notation slightly and denote the non-conducting  edges by ${\bar e}_0, {\bar e}_1,..{\bar e}_{N-2}$ where we have numbered the edges in the angular order of the transverse projections of their tangents at $v_I$
with respect to ${\bar e}_I$ and we have arbitrarily picked ${\bar e}_0$  to be some particular non-conducting edge. Let the edges ${\bar e}_1,..{\bar e}_{N-2}$ be such that their transverse projections onto $D_{\perp}$  make angles $\phi_1,.., \phi_{N-2}$ with respect to ${\bar e}_0$.
Starting at ${\bar e}_0$ and moving anticlockwise around the cone axis in order of increasing $\phi$, let ${\bar e}_i$ be the first edge encountered such that:
\be
\phi \geq i\frac{2\pi}{N-1},  \;\;\;\;\; \phi_j <j\frac{2\pi}{N-1} \;\forall j<i  .
\label{acon11}
\ee
Assume $i>1$.
Let ${\vec \xi}_{\phi}$ be the rotational vector field about the axis ${\bar e}_I$.
Consider a semianalytic  function  of compact support $f_{i-1}$  such that $f_{i-1} =1$  on $B_{\frac{\epsilon}{4}}(v_I)$ and $f_i=0$ outside $B_{\frac{\epsilon}{2}}(v_I)$ with $f_i$ decreasing from $1$ to $0$ in the region between the boundaries of
these  2 spheres.
Let $\Phi ( f_{i-1} \xi_{\phi}, t) $ be the one parameter family of 
semianalytic diffeomorphisms generated by $f_{i-1}{\vec \xi}_{\phi}$. Then for an appropriate value of $t=t_{i-1}$ apply the diffeomorphism $\Phi ( f_i \xi_{\phi}, t_{i-1}) $
 {\em only} to ${\bar e}_{i-1}$ so that its transverse angle  with respect to the ${\bar e}_0$ in the vicinity of $v_I$ is increased to  $(i-1)\frac{2\pi}{N-1}$.
 \footnote{If ${\bar e}_{i-1}$ is already at its desired position set $t_{i-1}=0$.}
 It can be checked that this deformation does not create any additional 
 intersections between any edges. Next repeat this procedure for the edge ${\bar e}_{i-2}$ 
 replacing $f_{i-1}$ by $f_{i-2}$ which is unity in $B_{\frac{\epsilon}{8}}(v_I)$, vanishes outside
$B_{\frac{\epsilon}{4}}(v_I)$ and is between $1$ and $0$ in the region between the boundaries of these 2 spheres. This  brings ${\bar e}_{i-2}$ `forward' to its desired angular position. Repeat this procedure for all the $i-1$ edges between 
${\bar e}_i$ and ${\bar e}_0$. This leads to a situation in which we have straight line edges 
in $B_{\frac{\epsilon}{2^{i}}}(v_I)$ and we proceed to the next step.
\footnote{In the case that $i=1$ satisfies (\ref{acon11}), there are no `in between' edges and we can directly proceed to the next step.
Also note that if there is no edge $i$ which satisfies (\ref{acon11}), then all edges are either at their correct positions or need to be rotated anticlockwise to their positions. 
In such a case we start our procedure as above
by setting $i-1= N-2$ so as to first bring  the $N-2$th edge to its correct position, then $N-3$th edge all the way upto the 2nd edge (the second edge is $e_1$ since our numbering starts from $0$) so that all edges are now at their
correct positions.}

The next step is to check if ${\bar e}_i$ is already at its correct position. If so we skip this paragraph and move on to the  step outlined in the next paragraph.
If ${\bar e}_i$ is not at its correct position,
we apply a similar procedure to ${\bar e}_i$   with $f_i$ unity in $B_{\frac{\epsilon}{2^{i+1}}}(v_I)$ and vanishing outside $B_{\frac{\epsilon}{2^{i}}}(v_I)$ so as to
 rotate ${\bar e}_i$  {\em clockwise} about ${\bar e}_I$ in the vicinity of $v_I$  to its desired position. 
At the end of all this, we have created no additional 
intersections, and we have edges ${\bar e}_0,{\bar e}_1,.., {\bar e}_i$ at their desired positions with these edges being straight lines in $B_{\frac{\epsilon}{2^{i+1}}}(v_I)$.

Next,  if $i<N-1$, repeat the considerations above for the edges ${\bar e}_{j}, j>i$ by replacing the role of ${\bar e}_0$ in the above procedure by ${\bar e}_i$. Clearly the procedure terminates in a finite number of steps at the end of which 
the deformed graph remains semianalytic without any additional intersections between its edges, and, the vertex structure in a small neighbourhood of $v_I$ is exactly of the `regular conical' type required.

Let us now revert to our old notation which numbered the deformed and undeformed edge counterparts identically.
It is important to note that the following property holds for the graph deformation we have just defined. Consider the projections of the edge tangents for ${\hat J}\neq I$  transverse to the $I$th edge tangent at the vertex $v$
of the undeformed chargenet $c$.  Call this set of projections as $\{ {\vec {\hat e}}_{\hat J, \perp}\}$. 
The elements of this set can be ordered in order of increasing transverse angle $\phi$. Let this ordering be $( {\vec {\hat e}}_{\hat J_1, \perp}, {\vec {\hat e}}_{\hat J_2, \perp}, ..,{\vec {\hat e}}_{\hat J_{N-1}, \perp})$.
Recall again that each undeformed edge $e_{\hat J}$ has a unique deformed nonconducting counterpart ${\bar e}_{\hat J}$ 
which departs from the undeformed edge $e_{\hat J}$ at the $C^0$  kink ${\tilde v}_{\hat J}$ and terminates at $v_I$.   Call the set of projections of these non-conducting edge tangents transverse to the conducting edge tangent at the vertex $v_I$
as $\{ {\vec {\hat {\bar e}}}_{\hat J, \perp}\}$.   These also may be ordered in a similar manner. Then the property which holds for the deformation is that these two orderings are identical i.e.
the collection $( {\vec {\hat {\bar e}}}_{\hat J_1, \perp}, {\vec {\hat e}}^{}_{\hat J_2, \perp}, ..,{\vec {\hat {\bar e}}}_{\hat J_{N-1}, \perp})$ is also ordered in order of increasing $\phi$.

\subsection{\label{acone2}Step 2: Introducing a $C^2$ or $C^1$   kink on the conducting line}
The step above leaves us with a regular cone in some small  $\eta<<\delta$ sized neighbourhood
$B_{\eta}(v_I)$ of $v_I$ in the deformed charge net ${\bar c}$. Recall that ${\bar c}$  is the deformed charge net obtained in section \ref{acone1} above and, for a Hamiltonian constraint deformation is based on the graph depicted in Fig \ref{grc}
and for an electric diffeomorphism deformation on  Fig \ref{grb}.
We now show how to place a $C^2$  kink on the upper conducting edge of ${\bar c}$.  

Since the deformation of $c$ is along its $I$th edge, the upper conducting edge  in ${\bar c}$ is  a subset of $e_I$ and we confine our considerations to this subset below.
Let $\tau<<\eta$ and 
let $v_{\frac{\eta}{2},I}$, $v_{ \frac{\eta +\tau }{2},I}$  be at 
distances  $\frac{\eta}{2}$, $ \frac{\eta+\tau}{2}$ from $v_I$  on $e_I$ 
so that 
the outward pointing  upper conducting edge 
 runs from $v_I$ on to $v_{\frac{\eta}{2},I}$ and thence to    $v_{ \frac{\eta +\tau }{2},I}$.
 We seek to deform this edge
 so that $v_{\frac{\eta}{2},I}$
 becomes a $C^2$ -kink 
We require that in doing this:\\
\noindent (a)the segment of the edge  between $v_I$ and   $v_{\frac{\eta}{2},I}$   remains undeformed (and hence a straight line), 
\\
(b)the deformation be confined to within  a distance $\tau << \frac{\eta}{2}$ of  $v_{\frac{\eta}{2},I}$
in a small enough cylindrical (with axis along $e_I$) neighbourhood $U_{cyl}$ of the edge $e_I$ that no intersections of the deformed edge with any other edge of the graph underlying ${\bar c}$  ensue.\\

Clearly the desired deformation can be generated through the action of a small loop holonomy $h_{l_{\tau}}$ with charge $q^i_{l_{\tau}}$ where the loop consists of two semianalytic segments $l_{1\tau}, l_{2\tau}$ where:\\
(i) $l_{1\tau}$ runs from $v_{\frac{\eta}{2}I}$ along $e_I$ for a distance $\frac{\tau}{2}$  to the point $v_{ \frac{\eta +\tau }{2},I}$  
on $e_I$.\\
(ii) $l_{2\tau}$ runs from $v_{ \frac{\eta +\tau }{2},I}$  to $v_{\frac{\eta}{2}I}$ within a small  enough cylindrical neighbourhood $U_{cyl}$ of $e_I$ such that it doesnt intersect the deformed graph of Step 1 except at the two points
$v_{\frac{\eta}{2}I}$,  $v_{ \frac{\eta +\tau }{2},I}$  and such that it joins $v_{\frac{\eta}{2}I}$ at a $C^2$ kink but leaves $v_{\tau,I}$ as a semianalytic $C^k$ extension. \\
(iii) $q^i_{l_{\tau}}$ is chosen to be the negative of the charge which colors the (outgoing) upper conducting edge at $v_I$ in ${\bar c}$ (recall that the part of $e_I$ between $v_{\frac{\eta}{2}I}$ and   $v_{ \frac{\eta +\tau }{2},I}$ 
is a subset of this upper conducting edge).

Recall that the deformed charge net obtained at the end of  Step 1  in the case of Hamiltonian constraint type deformation can be thought of as being  generated by the product of appropriately defined holonomies  (see, for example, the line preceding equation (\ref{ciflipq}) and the discussion in section 2
after equation (\ref{p1cnf1})). This deformation is further modified through multiplication by the holonomy $h_{l_{\tau}}$.
By choosing $U_{cyl}$ and $\tau$ small enough, the area of the loop can be made as small as desired so that the corresponding classical holonomy is unity to $O(\eta^m)$ for any desired $m$. 
This implies that the above `$C^2$-kink modification'  of the discrete approximant constraint action still yields an acceptable approximant to the constraint action.
In a similar fashion multiplication by this holonomy  of the electric diffeomorphism constraint approximant which generates the deformation of Step 1 also yields an acceptable approximant which generates the $C^2$ kink modifed deformation.

To summarise:  The end result of our constructions is that in a small enough neighbourhood of the deformed vertex $v_I$ the non-conducting edges are straight lines which form a regular `downward' pointing cone around an
axis in the direction of the conducting edge which is also a straight line in this neighbourhood. The nonconducting edges emanating from $v_I$ meet their undeformed counterparts  in $C^0$ kinks whereas the  conducting edge emanating from $v_I$ 
is distinguished by its having  a $C^2$ kink. The `area' of the $C^2$ kink (i.e. the area of a holonomy which can create this kink) can be made as small as desired. In particular, given some $\alpha_0 <<\delta_0$
the departure of the edge from linearity can be confined to a small sphere of radius $2\alpha_0$ around the kink.

It is straightforward to see that similar constructions enable the placement of $C^2$ or $C^1$ kinks
at desired locations on the conducting line of the deformed charge nets encountered in the main text.


\section{\label{acolor}Coloring of Multiply Deformed States}

The concept of net charge plays a key role in this section.
Equation (\ref{defqnet})  defines the  {\em net charge}   on a conducting edge   to be   the sum 
of its {\em outgoing} upper  and lower conducting charges. Here we extend this definition to the case that 
the edge  is non-conducting; in such a case we define the 
the net charge $q^i_{net\;I}$ on such an edge to be its outgoing charge.

Next, let $c$ be a state with a single nondegenerate GR or CGR vertex $v$ which is linear with respect to the coordinate patch $\{x\}$. 
Let $c$ be  deformed as described in section \ref{secneg} by the deformation $(i, I, \beta, \delta)$. The detailed locations of kinks associated with this deformation
are not important here.
We have the following cases:\\

\noindent {\em Case 1}: The parent vertex is GR and there is no intervention.   From sections \ref{sec3}, \ref{secgr} and \ref{secneg} it follows that the displaced vertex in the child is either GR or CGR. We have two subcases:\\ 
\noindent {\em Case 1a}: Parent vertex is GR, Child vertex is displaced along the edge $e_I$ of the parent in the outgoing direction ${\vec {\hat e}}_I$: 
In this case, for any  $J\neq I$, the $J$th edge at the displaced vertex of the child is colored with  $(i, \beta)$-flipped images (i.e. unflipped charges if $\beta=0$ and $\beta$ times the $i$-flipped charges if
$\beta =\pm1$) of the charges on its undeformed parental counterpart. By gauge invariance the net charge on the  $I$th edge at the displaced vertex in the child is  the $(i, \beta)$- flipped image of the charge on the $I$th edge of $c$.
\\
\noindent {\em Case 1b}: Parent vertex is GR, Child vertex is displaced along the {\em linear extension} of the edge $e_I$ of the parent in the incoming direction $-{\vec {\hat e}}_I$ opposite to the 
outgoing direction: 
In this case there is  no conducting line in the child at the displaced vertex and, by construction, all edges  at the displaced  vertex of the child have $(i, \beta)$-flipped charges 
of their undeformed parental counterparts. 
\\

\noindent {\em Case 2}: Parent vertex is CGR:  
Let the conducting line through  the parent vertex $v$ in $c$  be the $K$th one.  
Due to the intervention by $h_l$ (see sections \ref{secgr}, \ref{secneg}), the parent vertex $v$ in $c$ becomes a GR vertex in $c_l$. The edges at $v$ in $c_l$ are equipped with the net charges of their counterparts at $v$ in $c$. 
We have 2 subcases:\\

\noindent {\em Case 2a}: The child vertex is is displaced along the edge $e_I$ of $c_l$  in the outgoing direction ${\vec {\hat e}}_I$. There are 2 further subcases:\\
\noindent {\em Case 2a.1}: $I\neq K$: The displaced vertex is not located on the conducting line in $c$. Hence the inverse intervention $h_l^{-1}$  leaves this vertex untouched. From Case 1 above, the 
displaced vertex in the child $c_{l(i,I,\beta, \delta)}$ is  either CGR or GR. Since this vertex is untouched by the inverse intervention, this vertex remains CGR or GR in $c_{(i,I,\beta, \delta)}$ .
Note also from Case 1a that the net charges at the 
displaced 
vertex in  the deformed child $c_{l(i,I,\beta, \delta)}$ are the $(i, \beta)$- flipped  images of the corresponding  net charges at $v$ in $c_l$. Since the inverse  intervention  does not touch this vertex,
the net charges at the displaced vertex in $c_{(i,I,\beta, \delta)}$ are also the $(i, \beta)$- flipped  images of the charges on $c_l$, these charges on $c_l$ being the same as the net charges 
at $v$ in $c$.
\\
\noindent {\em Case 2a.2}: $I=K$: The displaced vertex is located on  the conducting line passing through $v$ in $c$. From 
Case 1 it follows that the displaced vertex is CGR or GR in 
$c_{l(i,K,\beta, \delta)}$. Since the displacement is along the $K$th edge in $c_l$,  in the case that   the displaced vertex is CGR the upper and lower conducting edges at this vertex in $c_{l(i,K,\beta, \delta)}$
are also the $K$th ones. Since the inverse intervention can only affect the vertex structure at the displaced vertex in  $c_{l(i,K,\beta, \delta)}$ along this $K$th conducting line, it follows that 
the displaced vertex in $c_{(i,K,\beta, \delta)}$ is also either GR or CGR. Moreover the inverse intervention cannot change the net charges at the displaced vertex so that the net charges at the 
displaced  vertex in $c_{(i,K,\beta, \delta)}$ are the same as those at this vertex in $c_{l(i,K,\beta, \delta)}$. The latter, from Case 1a, are the $(i,\beta)$ flipped images of their counterparts at  $v$ in $c_l$, these charges in $c_l$
being the same as the net charges at $v$ in $c$.\\

\noindent {\em Case 2b}: The child vertex is displaced along the {\em linear extension} of the edge $e_I$ of the $c_l$ in the incoming direction $-{\vec {\hat e}}_I$ opposite to the 
outgoing direction. There are 2 subcases:\\
\noindent {\em Case 2b.1}: $I\neq K$: The displaced vertex is not located on the conducting line in $c$ so that the inverse intervention $h_l^{-1}$  leaves this vertex untouched. 
From Case 1b above it follows that  (i) the  displaced vertex in $c_{l(i,I,\beta, \delta)}$, and hence in $c_{(i,I,\beta, \delta)}$, is  GR (ii) the net charges at the displaced vertex 
in $c_{l(i,I,\beta, \delta)}$, and hence in  $c_{(i,I,\beta, \delta)}$,
 are the $(i, \beta)$- flipped  images of the charges on $c_l$. The charges at $v$ in $c_l$ are  the same as the net charges at $c$ because the inverse intervention cannot change net charges.
\\
\noindent {\em Case 2b.2}: $I=K$: The displaced vertex is located on  the conducting line in $c$. From Case 1b, it follows that the displaced vertex is GR in 
$c_{l(i,K,\beta, \delta)}$. The inverse intervention can only affect the vertex structure  at the displaced vertex in  $c_{l(i,K,\beta, \delta)}$    along the   $K$th edge of $c_{l(i,K,\beta, \delta)}$
at this vertex. It follows that 
the displaced vertex in $c_{(i,K,\beta, \delta)}$ can only be  GR or CGR. Moreover the inverse intervention cannot change the net charges at the displaced vertex so that the net charges at the 
displaced  vertex in $c_{(i,K,\beta, \delta)}$ are the same as those in $c_{l(i,K,\beta, \delta)}$. The latter, from Case 1b, are the $(i,\beta)$ flipped images of their counterparts at  $v$ in $c_l$ and the
charges at $v$ in $c_l$ are the same as the net charges at $v$ in $c$.
\\

\noindent{\em Case 3}: Parent vertex is GR but an intervention is required: This case is that of (2), section \ref{secneg2.1}. It is easy to check that this is identical to the case of a CGR vertex with
vanishing upper conducting charge and no new structures beyond those already encountered ensue. Since our arguments for the CGR case did not depend on the specific values of the edge charges,
we still have that {\bf Conclusion 1} below holds.


Thus we have \\
\noindent {\bf Conclusion 1}:
The displaced vertex in the child is either GR or CGR.
The net charges on the edges  of the child at its displaced vertex  are the $(i,\beta)$- flipped images of the net charges on their counterparts in the parent.
\\

Also note that:\\
\noindent (1a) In Case 1a above the undeformed parts of the edges $e_{J\neq I}$ emanating from $v$ in $c_{(i,I,\beta , \delta )}$  have vanishing $i$th charge when $\beta \neq 0$. By gauge invariance, the $I$th edge at $v$
in $c_{(i,I,\beta , \delta )}$  also has vanishing $i$th charge so that $v$ is degenerate in $c_{(i,I,\beta , \delta )}$. Note that $v$ remains GR in $c_{(i,I,\beta , \delta )}$.
If $\beta = 0$ (i.e. for electric diffeomorphism type deformations), $v$ is absent in 
$c_{(i,I,\beta , \delta )}$.\\

\noindent (1b)  In Case 1b above, similar to (1a)  the undeformed parts of the edges $e_{J\neq I}$ emanating from $v$ in $c_{(i,I,\beta , \delta )}$  have vanishing $i$th charge when $\beta \neq 0$. By gauge invariance, 
net charge along 
the $I$th edge  at $v$
in $c_{(i,I,\beta , \delta )}$   has vanishing $i$th component so that $v$ is degenerate in $c_{(i,I,\beta , \delta )}$. Note, however, that because of the necessity of a graph extension, the vertex $v$ in  
$c_{(i,I,\beta , \delta )}$ can be either CGR or GR.
If $\beta = 0$, then because of the graph extension $v$ is present (and bivalent) in 
$c_{(i,I,\beta , \delta )}$.\\

\noindent (2) The deformed chargenets $c_{l(i,I,\beta, \delta)}$ are obtained by actions of the type in Cases 1a, 1b  at the GR vertex $v$ of  $c_l$. Accordingly we have that:
\\
\noindent (2a) The transition from $c_l$ to  $c_{l(i,I,\beta, \delta)}$ is  of type Case 1a:  For $\beta \neq 0$, (1a) implies that the edges at the  vertex $v$
in $c_{l(i,I,\beta, \delta)}$ each have net charge with vanishing $i$th component. The action of the inverse intervention does not change these net charges so that the 
the net charges at $v$ in  $c_{(i,I,\beta, \delta)}$ also have vanishing $i$th component. If $v$ in $c_{(i,I,\beta, \delta)}$ remains CGR, it is then easy to see that 
independent of which edge at $v$ in $c_{(i,I,\beta, \delta)}$  we assign as upper,  due to the fact that the corresponding intervention which removes the lower conducting edge at $v$ 
does not change the net charge, we have that $v$ is unambiguosly degenerate in $c_{(i,I,\beta, \delta)}$ (see Definition 3, section \ref{secneg4}).  

If $\beta=0$ then from (1a) $v$ is absent in 
$c_{l(i,I,\beta, \delta)}$  so that  it is bivalent in $c_{(i,I,\beta, \delta}$.\\

\noindent (2b) The transition from $c_l$ to  $c_{l(i,I,\beta, \delta)}$ is of type Case 1b. Here it is important to delineate the 2 subcases, (2b.1) with $I\neq K$ and (2b.2) with $I= K$:\\

\noindent (2b.1)  From (1b), if $\beta \neq 0$, the edges at the  vertex $v$ in $c_{l(i,I,\beta, \delta)}$ each have net charge with vanishing $i$th component. 
Note however that from (1b) the vertex $v$ in $c_{l(i,I,\beta, \delta)}$ can be GR or CGR. If it is CGR, the conducting line at $v$ in $c_{l(i,I,\beta, \delta)}$ is along the $I$th edge of $c$ and its extension.
Since the inverse intervention only affects the vertex structure at $v$ along the $K$th edge in $c_{l(i,I,\beta, \delta)}$, this $I$th conducting line is also present at $v$ in 
$c_{(i,I,\beta, \delta)}$. In addition the inverse intervention restores the lower part of the $K$th conducting edge so that there are now {\em two} conducting lines through $v$ in 
$c_{(i,I,\beta, \delta)}$. Note however that the inverse intervention cannot change net charges so that the net charges on these lines still have vanishing $i$th component.
Definition 5, section \ref{secneg} then implies that this `doubly CGR' vertex is degenerate.

If $\beta=0$, then from (1b) $v$ is bivalent in $c_{l(i,I,\beta, \delta)}$  and the inverse intervention renders this vertex 4 valent but with only 2 linearly independent edge tangents. Hence the 
vertex is planar (and hence neither GR nor CGR)  and hence, degnerate, in $c_{(i,I,\beta, \delta)}$.\\

\noindent (2b.2) From (1b),  if $\beta \neq 0$, the edges at the  vertex $v$ in $c_{l(i,K,\beta, \delta)}$ each have net charge with vanishing $i$th component. 
If the vertex $v$ is CGR in $c_{l(i,K,\beta, \delta)}$ as a result of the graph extension, then the conducting line through $v$  is along the   $K$th  edge in $c_l$ and its extension.
Since the inverse intervention is also along the $K$th edge at $v$ in $c$, the vertex $v$ in $c_{(i,K,\beta, \delta)}$ is either GR or CGR but not doubly CGR. Since the inverse intervention
cannot change net charges, the net charges at $v$ in $c_{(i,K,\beta, \delta)}$ have vanishing $i$th component so that $v$ is unambiguously degenerate in $c_{(i,K,\beta, \delta)}$.
If $\beta =0$, only the $K$th line passes through $v$ in $c_{l(i,K,\beta, \delta)}$ so that $v$ is bivalent in $c_{l(i,K,\beta, \delta)}$.  The inverse intervention near $v$ is also along this line
and the vertex $v$ remains bivalent (and hence degenerate) in $c_{(i,K,\beta, \delta)}$.

\noindent{(3)}: As mentioned in Case 3 above, this is identical to the case of a CGR vertex with
vanishing upper conducting charge and no new structures beyond those already encountered ensue. Since our arguments for the CGR case did not depend on the specific values of the edge charges,
we still have that {\bf Conclusion 2} below holds.

It is straightforward to check that in all cases, leaving aside the vertex $v$ and its displaced image in the child, the only other vertices created by the deformation are of valence at most 3 and hence
degenerate.
Hence the only possibly non-degenerate vertices of $c_{(i,I,\beta , \delta )}$ are $v$ (which we have shown is degenerate)  and its displaced image. 

Thus we have \\
\noindent {\bf Conclusion 2}: The vertex $v$ (if present) in $c_{(i,I,\beta , \delta )}$ is degenerate.
\\

It then follows, if (as is assumed in the main text) the displaced vertex is nondegenerate, then the deformed child of a parent with a single linear, nondegenerate GR or CGR vertex also has 
a single GR or CGR vertex with net charges which are $(i, \beta)$ flipped images of their parental correspondents. Applying this to any $c$ in the Ket Set, it folows, by virtue of the fact that any such Ket 
arises as a multiple deformation of  some primordial ket, that (a) any $c \in S_{Ket}$ has a single nondegenerate vertex and (b)  the net charges at the nondegenerate vertex of $c \in S_{Ket}$ are identical to, or the flipped images of,  the 
charges on such a primordial ancestor.

\section{\label{anondeg}Examples of Primordial States}

Consider a 4 valent gauge invariant linear vertex which is linear with respect to some chart $\{x,y,z\}$. Let its outward pointing  edges be in either upward or downward conical conformation with respect to $\{x,y,z\}$ (by which we mean that one
edge points along the cone axis and the remaining three are arranged in a regular upward or downward configuration about this axis). We assume without loss of generality that the 4th edge $e_4$ points along the $z$ (or -$z$)
axis and that the remaining 3 edges $e_1,e_2,e_3$ point along a cone with angle $0<\theta<\pi $ about the $-z$ axis.  Let the projections of the outgoing tangents to $e_1,e_2,e_3$ be ${\vec e}_{1\perp},{\vec e}_{2\perp},{\vec e}_{3\perp}$. Let
the edges be placed such that these projections are ordered anticlockwise   about the $z$ axis as $\{{\vec e}_{1\perp},{\vec e}_{2\perp},{\vec e}_{3\perp}\}$ and let the angle between successive projections be $\frac{2\pi}{3}$ so that the
configuration is regular conical.

Let the  triplet of $U(1)$ charges on the $I$th edge be ${\vec q}_I= (q^1_I, q^2_I, q^3_I)$.  Choose ${\vec q}_I, I=1,2,3$ to be linearly independent vectors and 
set ${\vec q}_4 = -(\sum_{I=1}^3{\vec q}_I)$ so that the vertex is gauge invariant.
It is straightforward to verify that if the cone is in downward conformation (so that $e_4$ points along the $+z$ axis), the `volume' eigenvalue $\nu$ (see  (\ref{evq})) is 
\be 
\nu = 4|q| , \;\;\; q:= \frac{1}{48}\epsilon^{ijk}q^i_1q^j_2q^k_3
\label{nud}
\ee
and if the cone is in upward conformation (so that $e_4$ points along the $-z$ axis) then the volume eigenvalue  is
\be 
\nu = 2|q| 
\label{nuu}
\ee

Let us denote a primordial chargenet with a single 4 valent vertex of the type which results in (\ref{nud}) by $c_D$ and that with a single 4 valent vertex of the type which results in 
(\ref{nuu}) by $c_U$.  We note the following:\\

(1) For both these choices of primordials, the linear dependence of 3 of the 4 charge triplets together with  gauge invariance implies that these 4 charge triplets define a GR set of charge vectors i.e. any 3 of these vectors are linearly
independent.  Since we have only used linear independence of 3 charge vectors, gauge invariance and the conical conformation of the edge tangents, it follows that we could have chosen {\em any} of the 4 edges to be
along the cone axis and still obtained non-degeneracy. 
\\
(2) Let us subject the charges ${\vec q}_I, I=1,2,3,4$ to a $(\beta, i)$ flip. It is easy to see that  flipped charges also form a gauge invariant set and that the volume eigen value is invariant under
the replacement of the charges ${\vec q}_I, I=1,2,3,4$ by their flipped images.
\\
(3) From Appendix \ref{acolor} it follows that  the  net charges which color the edges at  the vertex of any  charge net obtained through multiple deformations of a primordial charge net are just multiply 
flipped images of the charges on the primordial.
\\

From (1)- (3) above in conjunction with Conclusion 2 of Appendix \ref{anondeg}, the  deformations constructed  in section \ref{secneg1}- \ref{secneg3} and Definition 3 of the non-degeneracy of CGR vertices in section \ref{secneg4}, 
it follows that any multiple deformation of either $c_D$ or $c_U$ results in a deformed charge net which has a single non-degenerate GR or CGR vertex.

Finally, it is straightforward to see that we can easily arrange for conditions (\ref{sumq}), (\ref{qneq0}) to be satisfied, for example by setting the charges on $c_D,C_U$ to be $q^i_I= M\delta^{i_I} +1   , I=1,2,3$, $M>>1$.

%


\section{\label{ajacob}Jacobian between  Reference  and Contraction Coordinates}

To avoid notational clutter we adopt the following change in notation in this section relative to Step 1, section \ref{sec6.3}. 
We set:\\
$c_{[i,I, {\hat J}, {\hat K}, \beta ,\epsilon ]_m}= s$ , $(c_{[i,I, {\hat J}, {\hat K}, \beta ,\epsilon ]_m})_0 = \sref, \;\;\;$ $\alpha_{[i,I, {\hat J}, {\hat K}, \beta ,\epsilon ]_m}= \phir,\;\;\;$
$c_{0[i,I,\beta,\delta_0]_m}= \scon$, \\
and denote the diffeomorphism which maps $\scon$ to $s$ by $\phic$ (here the action of $\phic$ is obtained by the action of an appropriate (composite) contraction diffeomorphism
followed by the diffeomorphism $\alpha$ of Step 1, section \ref{sec6.3}).
The reference coordinates 
for $s$ are $\phir^*\{x_0\}$ where $\phir^*$ is the {\em pushforward} action of $\phir$ and the contraction coordinates for $s$ are $\phic^*\{x_0\}$.
We are interested in evaluating the Jacobian between these two coordinate systems at the nondegenerate vertex $v$ of $s$. 

First note the following. The states $\scon$ and $\sref$ are diffeomorphic to $s$ and hence to each other. Hence there exists a diffeomorphism $\phi$ which maps $\sref$ to $\scon$. This diffeomorphism
must map the non-degenerate vertex $v^{ref}_0$ of $\sref$ to the nondegenerate vertex $v^{con}_0$ of $\scon$ and also map the set of nearest kinks   at these vertices to each other. In particular the nearest $C^1$ kink
if present in $\sref$ must be also be present in $\scon$ and be mapped to it and similarly for a nearest $C^2$ kink if present. Since at least one of these kinks must be present and since the upward direction inferred  from
the location of either or both of these kinks, if present, is uniquely defined (see section \ref{secneg2}), the upward direction  ${\vec V}$ at $v^{ref}$ is mapped to that at $v^{con}$. Since both $\sref$  and $\scon$
are primaries, the vertex structure in a small vicinity of their vertices must be either upward or downward conical. Note however that if the structure is upward conical  in $\sref$ at  $v^{ref}$ 
then it must be upward conical in $\scon$ at $v^{con}$, and similarly for downward conical structures. This follows from the fact that no diffeomorphism connected to identity can map an upward conical structure to a downward one.

To prove this, proceed as follows. Consider an upward or downward cone with respect to ${\vec V}$:\\
\noindent (i) From P2, it follows that the anticlockwise ordering of the projections, transverse to ${\vec V}$ in the coordinates $\{x_0\}$,  of  the {\em outward pointing} edge tangents which do not point along ${\vec V}$ is invariant under
orientation preserving changes of coordinates.
A quick way to see this is as follows.  Clearly, the  2 dimensional vector space of these  transverse projections is isomorphic to the vector space of equivalence classes of vectors where 2 vectors are defined to be equivalent if they 
differ by a vector proportional to ${\vec V}$. Let us denote the transverse projection of an edge tangent ${\vec v}$ (or equivalently its equivalence class as defined in the previous sentence) by ${\vec v}_{\perp}$
Since in the regular conical conformation with respect to $\{x_0\}$, the angle between 2 successive projections in this anticlockwise ordering is less that $\pi$, the condition that 2 
projections ${\vec v}_{1\perp }, {\vec v}_{2\perp }$, with ${\vec v}_{2\perp }$ occurring immediately after ${\vec v}_{1\perp }$ in this ordering  are successive is equivalent to the conditions that \\
(a) no other edge tangent projection can be written as a linear combination of ${\vec v}_{1\perp}, {\vec v_{2\perp}}$ with {\em positive} coefficients i.e. there exist no $\alpha ,\beta >0$ for which there exists an edge tangent  ${\vec v}_3$ such that 
$\alpha {\vec v}_{1\perp } +\beta {\vec v}_{2\perp }= {\vec v}_{3\perp }$. 
\\
(b) the  vectors ${\vec v}_1, {\vec v}_2, {\vec V}$ form a right handed triple i.e. with respect to the alternating Levi- Civita tensor $\eta_{abc}$ we have that 
$\eta_{abc}v_1^a v_2^b V^c>0$ (this condition encodes the fact that ${\vec v}_{2\perp }$ is encountered {\em after} ${\vec v}_{1\perp }$ in the anticlockwise ordering; we have implicitly assumed that $\{x_0\}$ is right handed).
\\
The conditions (a) and (b) are invariant under positive rescalings of ${\vec v}_1,{\vec v}_2, {\vec V}$. Clearly, any coordinate transformation can only rescale these vectors with positive rescalings since they refer
to coordinate invariant directions at the tangent space of the vertex in question. This proves that the ordering of these projections is defined in a coordinate invariant way.
\\
\noindent (ii) The $N-1$ edges not pointing along ${\vec V}$ are arranged in a regular cone of angle $\theta$  with respect to ${\vec V}$ when viewed in the  $\{x_0\}$ coordinates. Let ${\hat {\vec V}}$
be the unit vector along ${\vec V}$ (unit with respect to the $\{x_0\}$ coordinate norm and 
consider 3 successive unit (with respect to $\{x_0\}$) outward pointing edge tangents ${\hat {\vec v}}_i, i=1,2,3$ arranged in anticlockwise order as discussed in (a) so that the angle between 2 successive pairs in these coordinates is 
$\phi= \frac{2\pi}{N-1}$.
It is straightforward to show that 
\be
{\hat {\vec v}_1} + {\hat {\vec v}_3} - 4 \cos \theta \sin^2 \frac{\phi}{2}{\hat {\vec V}} = 2 \cos\phi {\hat {\vec v}_2}
\label{v1-3}
\ee
This implies the  relation 
\be
a {{\vec v}_1} + b { {\vec v}_3} - c \cos\theta  {\vec V} = d \cos \phi { {\vec v}_2}, \;\;\; {\rm for \;some} \; a,b,c,d >0
\ee
This implies that if $N=4$, for some $\alpha,\beta, \gamma >0$,  we have that:
\ba
-(\alpha  {{\vec v}_1} + \beta  { {\vec v}_3}  - \gamma  {\vec V}) =  { {\vec v}_2}, &\;\;\; &{\rm for \; a \; upward \; cone}
\label{v1-3upn3}\\
-(\alpha  {{\vec v}_1} + \beta  { {\vec v}_3}  + \gamma  {\vec V}) =  { {\vec v}_2}, &\;\;\; &{\rm for \; an \; downward \; cone} ;
\label{v1-3dn3}
\ea
and that if $N>4$, for some $\alpha, \beta, \gamma >0$, we have that:
\ba
\alpha  {{\vec v}_1} + \beta  { {\vec v}_3}  - \gamma  {\vec V} =  { {\vec v}_2}, &\;\;\; &{\rm for \; an \; upward \; cone} ;
\label{v1-3upn}\\
\alpha  {{\vec v}_1} + \beta  { {\vec v}_3}  + \gamma  {\vec V} =  { {\vec v}_2}, &\;\;\; &{\rm for \; a \; downward \; cone} 
\label{v1-3dn}
\ea
The above equations retain their form (as well as the  positivity properties of $\alpha,\beta,\gamma$) irrespective of the choice of coordinates because a change of coordinates only provides a positive rescaling to 
the vectors in these equations. 
\\

\noindent (iii) Clearly, any diffeomorphism $\phi$ connected to identity which maps $\sref$ to $\scon$ is such that:\\
(a) It maps outward pointing edge tangents at $v_{ref}$ in $\sref$ to outward pointing edge tangents at $v_{con}$ in $\scon$.\\
(b)It maps, as noted in the 2nd paragraph of this section, the upward direction at $v_{ref}$ to that at $v_{con}$.\\
(c) It retains the anticlockwise ordering of the $(N-1)$ edge tangents (which are not along the cone axis) around the upward direction; this immediately follows from the fact that $\phi$ 
(which is orientation preserving by virtue of its being connected to identity) preserves
conditions (i)(a),(b).\\
(d) From (a)-(c), it follows that $\phi$  preserves conditions (\ref{v1-3upn3})- (\ref{v1-3dn}) so that if any one of these conditions hold at $v_{ref}$ in $\sref$, the same condition holds at $v_{con}$ in $\scon$.\\

The statement (d) implies that an upward conical deformation cannot be mapped into a downward conical deformation (and vice versa) by $\phi$.
Next, recall that the multiple deformation which generates any primary from any  reference primordial is confined to a coordinate ball $B_{\Delta_0}(p_0)$ with respect to the Primary Coordinates $\{x_0\}$. 
We show the following Lemma.
\\

\noindent {\bf Lemma L1}: Let the vertex structures in a small vicinity of the nondegenerate vertices of $\sref,\scon$ be downward conical. Denote the coordinate ball of size $\eta$  (in $\{x_0\}$ coordinates) around a point $p$ by $B_{\eta}(p)$. Then
there exist small enough open balls  $B_{\tau}(v^{ref}_0), B_{\tau}(v^{con}_0)$,  
together with a rigid rotation $R$ 
and a rigid translation $T$  (with $R, T$ defined with respect to the  $\{x_0\}$ chart) 
such that  $B_{\tau}(v^{con}_0)= RT B_{\tau}(v^{ref}_0)$ and such that  $RT (\sref|_{B_{\tau}(v^{ref}_0)}) = \scon|_{B_{\tau}(v^{con}_0)}$ where  $c|_{U}$ refers to the restriction of the colored graph defining the charge net $c$ to the set $U$.
\\
\noindent{\em Proof}: 
Clearly, there exists small enough $\tau$  such that 
$B_{\tau}(v^{ref}_0), B_{\tau}(v^{con}_0)$ intersect $\sref, \scon$ only in their conical vertex structures. 
Thus $\sref|_{B_{\tau}(v^{ref}_0)}, \scon|_{  B_{\tau}(v^{con}_0)   }$ both comprise of regular conical structures with respect to $\{x_0\}$ i.e.
each of these restrictions comprise of a cone vertex with downward pointing non-conducting edges around the upward pointing cone axis.
Further both cones have the same angle $\theta$.  
Since $\sref, \scon$ are diffeomorphic (to $s$ and hence) to each other, there exists a diffeomorphism $\phi$ which maps the preferred upward pointing  axes to each other
and the set of downward pointing conical edges to each other. Such a diffeomorphism induces  a map between the sets of downward pointing unit edge tangents so that 
\be
\{\phi^* ({\vec{\hat e}}_{\hat J}),{\hat J}= 1,..,N-1 \} = \{ \beta_{\hat I} {\vec{\hat e}}_{\hat I}, \;\;\; \beta_{\hat I} >0, {\hat I}= 1,..,N-1\} 
\ee
where the edge tangents ${\vec{\hat e}}_{\hat J}, {\vec{\hat e}}_{\hat I}$ on the left and right hand sides
are unit with respect to the coordinates $\{x_0\}$ at $v^{ref}_0, v^{con}_0$ respectively. From (iii)(c) above it follows  that the sets of projections of these edge tangents
transverse to the cone axis have the same ordering.
Thus we may use $\phi$ to identify each  downward pointing  edge of the cone in $\sref$ with a downward pointing edge in $\scon$.

Let $T$ be the rigid translation which maps $v_0^{ref}$ to $v^{con}_0$. Clearly $T (B_{\tau}( v^{ref}_0))  = B_{\tau}(v^{con}_0)$. Next, rotate  $(T\sref)|_{   B_{\tau}(v^{con}_0)  }$ by $R_1$ so that its distinguished
`up' direction ${\vec V}$ aligns  with that of $\scon|_{U^{con}}$. Next, rotate around this preferred direction by $R_2$ so that  one of the downward pointing edges of $(R_1T\sref)|_{   B_{\tau}(v^{con}_0 )  }$  aligns with its counterpart
in $\scon|_{U^{con}}$ as identified by  $\phi$.  Since the transverse ordering is preserved, this automatically aligns all the remaing downward pointing edges of $(R_1T\sref)|_{  B_{\tau}(v^{con}_0     }$ with 
their counterparts in $\scon|_{U^{con}}$. Since $R_1,R_2$ preserve (the coordinate ball) $B_{\tau}(v^{con}_0)$
we set $R= R_2R_1$ to obtain the desired transformation $RT$.
\\

Next note that that there exists a small enough neighborhood $V$ of the nondegenerate vertex $v$ of $s$ such that $V$ is covered by the reference as well as the contraction coordinates
so that $(\phir)^{-1}V, (\phic)^{-1}V$ are in the domain of the coordinate patch  $\{x_0\}$, and such that  $s|_V$ only contains the nondegenerate vertex $v$ and segments of the edges emanating therefrom.
Also note that there exists small enough $\tau$
such that the open balls of the Lemma above are such that $   B_{\tau}(v^{con}_0) \subset   (\phic)^{-1} V$ , $   B_{\tau}(v^{ref}_0) \subset   (\phir)^{-1} V$. 
It follows from the Lemma above that that $\phic RT B_{\tau}(v^{ref}_0) \subset V$ and that this set is an open neighbourhood of $v$. Hence $(\phir)^{-1} \phic RT B_{\tau}(v^{ref}_0)$ is covered by $\{x_0\}$
and is an open neighbourhood of $v^{ref}_0$. Setting $(\phir)^{-1} \phic RT \equiv \psi$, and $\psi B_{\tau}(v^{ref}_0) = B^{\psi}_{\tau}(v^{ref}_0)$, we have that both 
$B_{\tau}(v^{ref}_0)$ and $B^{\psi}_{\tau}(v^{ref}_0)$ are open neighbourhoods of $v^{ref}_0$ and we can compute the Jacobian 
\be
J^{\mu}_{\;\nu}= \frac{\partial(\psi^*x_0)^{\mu}}{\partial x_0^{\nu}}|_{v^{ref}_0}
\label{cr1}
\ee
From the proof of the Lemma in P2 and the discussion above it follows that (a) $\psi$ maps the  set of edge tangents at $\sref$  at its vertex $v^{ref}_0$  to  itself modulo overall scaling (b) the upward  direction is mapped
to itself and the anticlockwise arrangement of the transverse projections of the downward pointing edge directions are mapped to themselves.
From the fact that the reference deformations of  Appendix \ref{acone} are regular conical, the results of (a) and (b) can be implemented through the action of linear transformation on the tangent space at $v^{ref}_0$ which takes the form of a 
constant times an $SO(3)$ matrix in the $\{x_0\}$ coordinates. Then it follows from the last part of the proof of the Lemma in P2 that the action of $\psi$ on the tangent space as expressed through the Jacobian
in the above equation must be that of constant times as rotation i.e. we have that
\be
J^{\mu}_{\;\nu} = C{\bar R}^{\mu}_{\;\nu}
\label{cr2}
\ee
for some $C>0$ and some $SO(3)$ matrix ${\bar R}$.

Next note that with $p\in \phic RT B_{\tau}(v^{ref}_0)$  and ${\bar p}:=   (RT)^{-1}(\phic)^{-1} p \in  B_{\tau}(v^{ref}_0)$ we have that 
\be
\frac{       \partial(\phir^*x_0)^{\mu}(p)      }{  \partial ((\phic)^* RT^* x_0)^{\nu}(p)   } =
\frac{\partial x_0^{\mu}((\phir)^{-1} p)}{\partial  x_0^{\nu}(  (RT)^{-1}(\phic)^{-1} p)} 
= \frac{\partial(\psi^*x_0)^{\mu}({\bar p})}{\partial x_0^{\nu}({\bar p})}.
\ee

Setting $p= v$ we have from  (\ref{cr1}), (\ref{cr2})  that 
\be
\frac{  \partial ((\phir)^*x_0)^{\mu}   (p)   }{    \partial ((\phic)^* (RT)^* x_0)^{\nu}(p)}|_{p=v} = C{\bar R}^{\mu}_{\;\;\nu}
\label{cr3}
\ee

Next note that with $p \in \phic RT B_{\tau}(v^{ref}_0)$, 
\be
\frac{  \partial     ((\phir)^*x_0)^{\mu}(p)     }{    \partial ((\phic)^*x_0)^{\alpha}(p)} =
\frac{\partial((\phir)^*x_0)^{\mu}(p)  }{  \partial ((\phic)^* (RT)^* x_0)^{\alpha}(p)   }
\;\;
\frac{    \partial((\phic)^*(RT)^*x_0)^{\alpha} (p)  }{   \partial ((\phic)^*x_0)^{\nu}(p)  },
\label{cr4}
\ee
so that it remains to evaluate the second Jacobian in this equation. Setting ${\bar p}= (\phic)^{-1} p \in B_{\tau}(v^{con}_0)$ we have that:
\be 
\frac{  \partial( (\phic)^*(RT)^*x_0)^{\mu}(p)}{  \partial ((\phic)^*x_0)^{\nu}(p)  } 
= \frac{  \partial x_0^{\mu}((RT)^{-1}{\bar p})}{\partial x_0^{\nu}({\bar p})}.
\ee
Since $R,T$ are rigid rotations and translations in $\{x_0\}$, we have that 
 $x_0^{\mu} (     (RT)^{-1} {\bar p} ) =  x_0^{\mu}( T^{-1} R^{-1}{\bar p})=  (R^{-1})^{\mu}_{\;\;\nu}   x_0^{\nu} ({\bar p})  - t^{\mu}$ where $t^{\mu}$ is a constant vector corresponding to the translation $T$
from which it follows that the Jacobian in the above equation is  the $(R^{-1})^{\mu}_{\;\;\nu}$. Together with (\ref{cr3}), (\ref{cr4}),  this implies that the Jacobian between the reference and the 
contraction coordinates is a constant times a rotation.

It is straightforward to check that an appropriate version of Lemma 1 and the following argumentation leads to the same conclusion for the upward conical case.

\section{\label{aconb}Contraction behaviour of various quantities of interest}
\subsection{\label{aconb1}Notation} 

To avoid notational clutter we adopt the following change in notation in this section relative to Step 2, section \ref{sec6.3}.
We set, similar to  equation (\ref{appnote1}):
\ba
x_{\alpha}^{\epsilon_{\rm j_1}..     \epsilon_{\rm j_{m}}}  \equiv   x^{(\delta)}   &  x_{\alpha}^{\epsilon_{\rm j_1}..     \epsilon_{\rm j_{m-1}}} \equiv  x   & \epsilon_{\rm j_m} \equiv \delta \nonumber \\
c_{[i,I, {\hat J}, {\hat K}, \beta ,\epsilon ]_m}  \equiv s_{\delta} & c_{[i,I,\beta,\epsilon, {\hat J}, {\hat K}]^{m-1}_m} \equiv s &  \beta_m\equiv\beta
\label{e1}
\ea
We are interested in the transition from the immediate parent $s\equiv c_{[i,I,\beta,\epsilon, {\hat J}, {\hat K}]^{m-1}_m}$ to its child $s_{\delta}\equiv c_{[i,I, {\hat J}, {\hat K}, \beta ,\epsilon ]_m}$. 
In this transition the parent $s$ is deformed conically along (or opposite to) its $I_{m-1}$th edge at its non-degenerate vertex $v$ so as to displace this vertex to the point $v^{\delta}$ which is the nondegenerate vertex
of the child $s_{\delta}$. In our notation (\ref{e1}), this transition is $(i_{m-1},I_{m-1}, \beta_m, \epsilon_{\rm j_m} )\equiv (i_{m-1}, I_{m-1}, \beta, \delta) $.
We  denote the upward direction at $v$ in $s$  by ${\vec V}_{I_{m-1}}$
so that ${\vec V}_{I_{m-1}}$ is parallel or antiparallel to the edge tangent ${\vec {\hat e}}_{I_{m-1}}$ at $v$ (see section \ref{secneg2}).

Finally, all the edge charges referred to below will be {\em net} charges (see the first paragraph of Appendix \ref{acolor} for the definition of net charge).

\subsection{\label{aconb2}Contraction coordinates and Choices of Upward Direction}

From (\ref{e1}),  the child $s_{\delta}$ of its immediate parent $s$ has contraction coordinates $\{x^{(\delta)}\}$. Step 1 of Section \ref{sec6.3} allows the evaluation of the coefficients which multiply the bra correspondent
of $s_{\delta}$ in these contraction coordinates. To extract their contraction behavior we need to transit to the parental coordinates $\{x\}$ in terms of which (see section \ref{sec4.4.2}) $\delta$ is measured.
Following equations (\ref{jydelta-y}), (\ref{jxdelta-x}), it is useful to rotate the $\{x\}, \{x^{(\delta) }\}$ coordinates by $R_s^{-1}$ so as to obtain $\{y\}, \{y^{(\delta)}\}$ coordinates with $y^3$ pointing along the 
the straight line passing through $v$ and $v^{\delta}$. From the last remark of section \ref{sec4.3} it follows that $y^{(\delta)3}$ also points along this direction.
Note that this direction is parallel (antiparallel)
to ${\vec V}_{I_{m-1}}$ if the deformation is downward (upward). Note also that ${\vec V}_{I_{m-1}}$ is defined, strictly speaking, only at $v$ in $s$. However due to the fact that $v$ in $s$ is linear with respect to $\{x\}$ (and, hence, $\{y\}$) 
we can naturally define 
${\vec V}_{I_{m-1}}$ at every point in the domain of these coordinates.  It is in this sense that we refer to the direction defined by ${\vec V}_{I_{m-1}}$  in this section.
Note that the direction so defined is consistent with $C^2,C^1$ kink placements generated by the transformation $(i_{m-1}, I_{m-1}, \beta, \delta)$.  

More in detail, any edge that emanates from $v_{\delta}$ in $c_{\delta}$ which bears such a ($C^1$ or $C^2$) kink is, in the vicinity of $v^{\delta}$, a straight line pointing along the $I_{m-1}$th edge emanating from $v$ in $c$
(or along the  linear extension  (with respect to 
$\{y\}$) of this edge. Hence its outward pointing edge tangent is along or opposite to ${\vec V}_{I_{m-1}}$. The placement of its nearest kink (see sections \ref{secneg2} and \ref{secneg3}) then defines the upward direction
${\vec V^{(\delta)}}_{I_m}$ for such an edge to be equal to ${\vec V}_{I_{m-1}}$.

\subsection{\label{aconb3} Contraction behaviour of $H_{L_m}, h_{L_m}, f$}
Note that the structure in the immediate vicinity of the nondegenerate vertex of $s_{\delta}$ is regular conical in terms of $\{x^{(\delta)}\}$ (or $\{y^{(\delta)}\}$) because this structure and this coordinate system are images {\em by the same
diffeomorphism} (see section \ref{sec4}, especially section \ref{sec4.4.2})  of 
reference deformations and reference coordinates in which the reference deformations are regular conical.  Recall, from section \ref{aconb2} above, that  the  upward direction at $v_{\delta}$  in $c_{\delta}$ is ${\vec V}_{I_{m-1}}$.
Let the cone angle  as measured by the  $\{y^{(\delta)}\}$ coordinates,    with respect to this  upward direction 
be $\theta$ so that $\theta <\frac{\pi}{2}$ defines an upward conical deformation and $\theta >\frac{\pi}{2}$ defines a downward conical deformation.
Consider any $J_m$th edge of $s_{\delta}$  with $J_m\neq I_{m-1}$ in the immediate vicinity of $v_{\delta}$. 
\footnote{Recall that we use  the edge enumeration convention described in section \ref{sec4.1}}
Such an edge points along the cone. Let its azimuthal angle in the $\{y^{(\delta)}\}$ coordinates be 
$\phi_{J_m}$. Thus we have that the unit (with respect to the $\{y^{(\delta)}\}$ coordinates) outward pointing edge tangent ${\vec {\hat e}}^{(\delta)}_{ J_m}$ along this edge has coordinates (in the $\{y^{(\delta)}\}$ chart): 
\footnote{Since $\{y^{(\delta)}\}$ and $\{x^{(\delta)}\}$ are related by the rotation $R_s$, normalization in both these systems is identical.}
\be
{\hat e}_{J_m}^{(\delta)\bar \mu}= (\sin \theta \cos \phi, \sin \theta \sin \phi, \pm\cos\theta ).
\ee
where the $+$ sign in front of $\cos \theta$ corresponds to downward deformations (for which $y^{(\delta)3}, y^3$ run upward) and the $-$ sign to upward deformations (for which $y^{(\delta)3}, y^3$ run downward).
Using equation (\ref{jydelta-y}), the components of ${\vec {\hat e}}^{(\delta)}_{ J_m}$     in the $\{y\}$ coordinates are
\be
{\hat e}_{J_m}^{(\delta) \mu}= (\delta^{q-1}\sin \theta \cos \phi, \delta^{q-1}\sin \theta \sin \phi, \pm \cos\theta ).
\ee
Since the $y^3$ direction is along ${\vec V}_{I_{m-1}}$ for downward deformations and opposite to ${\vec V}_{I_{m-1}}$  for upward deformations,  the above equation takes the form: 
\be
 {\hat e}_{J_m}^{(\delta) a}     =  \cos \theta {\hat V}^a_{I_{m-1}} + \delta^{q-1}w_{J_{m}}^a 
\label{adowncon}
\ee
where $w^a_{J_{m}}$ is a $\delta$- independent vector in the  $\{y\}$ (and hence in the $\{x\}$) coordinates and ${\hat V}^a_{I_{m-1}}$ is  the normalised (in the $\{x\}$ or, equivalently, $\{y\}$ coordinates) vector 
parallel to ${ V}^a_{I_{m-1}}$.  Note that these edges are such that the nearest kink is a $C^0$ kink  so that the upward direction   ${V}_{J_m}^{(\delta) a}$ is along the outward pointing  $J_m$th edge tangent 
(see section \ref{secneg}). Denoting its normalised (with respect to the $\{y^{\delta}\}$ coordinates associated with $v_{\delta}$) by  ${\hat V}_{J_m}^{(\delta) a}$, we may write (\ref{adowncon}) as:
\be
 {\hat V}_{J_m}^{(\delta) a}    =   {\hat e}_{J_m}^{(\delta) a} =       \cos \theta {\hat V}^a_{I_{m-1}} + \delta^{q-1}w_{J_{m}}^a 
\label{adownconv}
\ee
Next consider the upper conducting edge $e_{I_m=I_{m-1},u}$ (if present)  at $v^{\delta}$ in  $s_{\delta}$. 
A similar analysis shows that the unit (with respect to $\{x^{\delta}\}$)  edge tangent along this edge is   
also unit with respect to the $\{x\}$ coordinates so that we have:
\be
 {\hat e}_{I_m= I_{m-1}, u}^{(\delta) a}= {\hat V}^a_{I_{m-1}}
\label{acon+}
\ee
Similarly the lower conducting edge (if present
\footnote{Recall  (see for example Footnote  \ref {fncgrtogr}) that while the upper or lower conducting edge may be absent in $s_{\delta}$ because the upper or lower conducting charge happens to vanish, from Appendix \ref{acolor} it must 
be the case that at least one of these edges is present in the child due to the relation of the net conducting charge with the primordial charge.
}) at $v_{\delta}$ in $s_{\delta}$  has unit (with respect to $\{x^{\delta}\}$) tangent:
\be
 {\hat e}_{I_m= I_{m-1}, d}^{(\delta) a}= -{\hat V}^a_{I_{m-1}}
\label{acon-}
\ee
Note that by definition the outward upper conducting edge tangent is along the upward direction ${V}_{J_m}^{(\delta) a}$ at $v_{\delta}$, and the lower one opposite to it so that we may write (\ref{acon+}), (\ref{acon-}) as
the single equation:\footnote{Note that (\ref{aconv}) is consistent with last part of the discussion in section \ref{aconb1}.}
\be
 {\hat e}_{I_m= I_{m-1}, u}^{(\delta) a}= {\hat V}^a_{I_{m-1}}
=-{\hat e}_{I_m= I_{m-1}, d}^{(\delta) a}
={\hat V}_{I_m= I_{m-1}}^{(\delta) a}
\label{aconv}
\ee


It is then straightforward to obtain the following estimates for  the metric norms of the unit (with respect to $\{x^{(\delta)}\}$)  edge tangent vectors at $v^{\delta}$ in $s_{\delta}$:
\ba
||   {\hat e}_{J_m\neq I_{m-1}}^{(\delta) a}        ||_{v^{\delta}} &= & |\cos \theta|  \sqrt{{h_{ab}(v^{\delta}) {\hat V}^a_{I_{m-1}}  {\hat V}^b_{I_{m-1}}     }}
(1    + C_{1,J_m}(v^{\delta}) \delta^{q-1}  +     C_{2,J_m}(v^{\delta}) (\delta^{q-1})^2)^{\frac{1}{2}} \label{adnorm}\\      
 C_{1,J_m}(v^{\delta})&:=&  2 \frac{h_{ab}(v^{\delta})w_{J_{m}}^a {\hat V}^b_{I_{m-1}}}{\cos \theta{{h_{ab}(v^{\delta}) {\hat V}^a_{I_{m-1}}  {\hat V}^b_{I_{m-1}}     }}} \label{defc1}
\\ 
C_{2,J_m}(v^{\delta})&:=&  \frac{h_{ab}(v^{\delta})w_{J_{m}}^a w_{J_{m}}^b     }{\cos^2 \theta{{h_{ab}(v^{\delta}) {\hat V}^a_{I_{m-1}}  {\hat V}^b_{I_{m-1}}     }}} \label{defc2} 
\ea
\be
|| {\hat e}_{I_m= I_{m-1}, u}^{(\delta) a}   ||_{v^{\delta}} 
= || {\hat e}_{I_m= I_{m-1}, d}^{(\delta) a}   ||_{v^{\delta}} =
\sqrt{h_{ab}(v^{\delta}) {\hat V}^a_{I_{m-1}}  {\hat V}^b_{I_{m-1}}   }           
\label{acnorm}
\ee


From (\ref{acon+})- (\ref{acnorm}), we obtain the following behaviour for 
the quantities in (\ref{defHi}), (\ref{defhi}):
\ba
H_{I_m=I_{m-1}} &=& \sqrt{h_{ab}(v_{\delta})   {\hat V}^a_{I_{m-1}}  {\hat V}^b_{I_{m-1}}      },\nonumber \\
H_{J_m\neq I_{m-1}} &=& |\cos \theta | H_{I_m=I_{m-1}} + O(\delta^{q-1}), 
\label{Hicon}
\ea
\ba
h_{ I_m=I_{m-1}} & = &(N-1)(N-2) + O(\delta^{q-1}) \nonumber \\
h_{J_m\neq  I_{m-1}} &=& \frac{N-2}{|\cos\theta|}(1 +\cos^2\theta + (N-3)|\cos\theta| ) + O(\delta^{q-1} ).
\label{hicon}
\ea

Next, consider any density weight $-\frac{1}{3}$ scalar density $S$ evaluated at $v_{\delta}$. From (\ref{jxdelta-x}) it follows that:
\be
S( v_{\delta}, \{x^{\delta}\}) = \delta^{-\frac{2}{3}(q-1)} 
S( v_{\delta}, \{x\})
\label{afnecon}
\ee
where we have used the notation $S(p,\{z\})$ to signify the evaluation of $S$ at the point $p$ in the coordinate system $\{z\}$.
Setting $S=:f$, equation (\ref{afnecon}) yields the contraction behaviour of $f$.
Note also that if $\{y\}$ is related to $\{x\}$ by a rotation (say, as in equations (\ref{jydelta-y}) ,  (\ref{jxdelta-x}))
we have by virtue of the fact that the determinant of a rotation matrix is unity, that 
\be
f(p, \{x\}) = f(p, \{y\})
\label{rot=1}
\ee

Next consider the quantity $H^{l}_{L_m}$ defined by 
\be
H^{l}_{L_m} := \prod_{i=1}^lN_{i}( v_{\delta}, \{y^{(\delta)}\})    {\hat V}^{(\delta)a_i}_{L_m}\partial_{a_i}  (f( v_{\delta}, \{y^{(\delta)}\})  ||{\vec {\hat e}^{(\delta)}}_{L_m}||)
\label{defdmH}
\ee
Here $N_i,f$ are density weight $-1/3$ scalars.  The unit (with respect to the $\{y^{(\delta)}\}$ coordinates) upward direction the $L_m$th edge (or line) at $v^{\delta}$
is denoted by ${\hat V}^{(\delta)a_i}_{L_m}$, where the upward direction ${ V}^{(\delta)a_i}_{L_m}$ is chosen in accord with the criteria of section \ref{secneg}.
The product in (\ref{defdmH})  is ordered from left to right in decreasing value  of $i$  so that the factor $(N_1{\hat V}^{(\delta)a_1}_{L_m}\partial_{a_1})$ is rightmost.
The  vector ${\vec {\hat V}^{(\delta)}}_{L_m}$ and its norm with respect to the metric $h_{ab}$ are   defined only
at the vertex $v^{\delta}$ of $c_{\delta}$. In order to render (\ref{defdmH}) well defined, we extend the domain of definition of  ${\vec {\hat V}^{(\delta)}}_{L_m}$ from $v_{\delta}$ to a small
neighbourhood $U(v^{\delta})$
thereof so that the vector field on this extended domain is constant in the $\{y^{(\delta)}\}$ chart. This neighbourhood is small enough that equations (\ref{scrunch}) hold so that the 
vector field is also constant in the $\{y\}$ coordinates. Thus, for any point $p$ in this neighbourhood, we define this `constant extension'  ${\hat V}^{(\delta)a_i}_{L_m}$ from which we define: 
\be 
H^{l}_{L_m}(N_1,N_2,..N_l; p):= \prod_{i=1}^l  N_{i}( p, \{y^{(\delta)}\})  {\hat V}^{(\delta)a_i}_{L_m}(p)\partial_{a_i} (f( p, \{y^{(\delta)}\})\sqrt{h_{ab}(p)  {{\hat V}^{(\delta)a}}_{L_m} (p) {{\hat V}^{(\delta)b}}_{L_m}(p)}).
\label{defdmHp}
\ee
Then we render (\ref{defdmH}) well defined by setting 
\be
H^{l}_{L_m}:= H^{l}_{L_m}(N_1,N_2,..N_l; p=v^{\delta})
\label{dmhp=vd}
\ee
We are interested in the contraction  behaviour of $H^{l}_{L_m}$ as defined above.
Note that since the transformation between $\{y\}$ and $\{y^{(\delta)}\}$ is linear in the domain of interest, constant vector fields  in one system are also constant in the other.
It then immediately follows that with such extensions of vectors  $  {\hat e}^{(\delta)a_i}_{L_m} {\hat V}^{(\delta)a_i}_{L_m}, {\vec {\hat V}}_{I_{m-1}},  {\vec w}_{J_{m}}  $ in  equations (\ref{adowncon})- (\ref{aconv}), 
these equations continue to hold
in $U(v^{\delta})$.  It then follows that replacing these vectors by their constant extensions in  (\ref{adnorm}), (\ref{acnorm}) and replacing $h_{ab}(v^{\delta})$ by $h_{ab}(p)$ in these equations,
we obtain equations which hold in $U(v^{\delta})$.
These equations can then be used to write (\ref{defdmHp}) in terms of quantities natural to the $\{y\}$ coordinates. Evaluating this form of the equations at $v^{\delta}$ then allows us to
derive the contraction behaviour of $H^{l}_{L_m}$ as defined by (\ref{dmhp=vd}). Accordingly (\ref{adnorm}), (\ref{acnorm}) are extended to $U(v^{\delta})$ as:
\ba
||   {\hat e}_{J_m\neq I_{m-1}}^{(\delta) a}        ||_{p} =  |\cos \theta|  \sqrt{{h_{ab}(p) {\hat V}^a_{I_{m-1}}(p)  {\hat V}^b_{I_{m-1}}(p)    }}
(1    + C_{1,J_m}(p) \delta^{q-1}  +     C_{2,J_m}(p) (\delta^{q-1})^2)^{\frac{1}{2}}  & \label{adnormp}\\      
 C_{1,J_m}(p):=  2 \frac{h_{ab}(p)w_{J_{m}}^a (p) {\hat V}^b_{I_{m-1}}(p)}{\cos \theta{{h_{ab}(p) {\hat V}^a_{I_{m-1}} (p) {\hat V}^b_{I_{m-1}}(p)     }}} \;\;\;\;\;\;\;&\label{defc1p}
\\ 
C_{2,J_m}(p):=  \frac{h_{ab}(p)w_{J_{m}}^a(p) w_{J_{m}}^b(p)     }{\cos^2 \theta{{h_{ab}(p) {\hat V}^a_{I_{m-1}} (p) {\hat V}^b_{I_{m-1}}(p)     }}}\;\;\;\;\;\;& \label{defc2p} 
\ea
\be
|| {\hat e}_{I_m= I_{m-1}, u}^{(\delta) a}   ||_{p} = || {\hat e}_{I_m= I_{m-1}, d}^{(\delta) a}   ||_{p} =
\sqrt{h_{ab}(p) {\hat V}^a_{I_{m-1}} (p) {\hat V}^b_{I_{m-1}}(p)   }           
\label{acnormp}
\ee
Using (\ref{adnormp})- (\ref{acnormp}), (\ref{afnecon}), (\ref{adownconv}), (\ref{aconv}) in equation (\ref{defdmHp}), and noting that the only objects in these equations  with a non-trivial $p$ dependence are $h_{ab},N_i, f$ the other quantities being constant in $\{y\}$ coordinates, 
it is straightforward to obtain:
\ba
H^{l}_{L_m =I_{m-1}}  (N_1,N_2,..N_l; p)              &=&  
\delta^{-(l+1)\frac{2}{3}(q-1)}   \nonumber\\
&\big\{& \prod_{i=1}^l  N_{i}( p, \{y\})    {{\hat V}^{a_i}}_{I_{m-1}}(p)\partial_{a_i}) (f( p, \{y\})\sqrt{h_{ab}(p)  {{\hat V}^{a}}_{I_{m-1}} (p) {{\hat V}^{b}}_{I_{m-1}}} (p))\big\}\label{dmHconp}
\\
H^{l}_{ L_m \neq I_{m-1}} (N_1,N_2,..N_l; p)        &=&
\delta^{-(l+1)\frac{2}{3}(q-1)}  \nonumber \\
 &\big\{&   |\cos \theta|(\cos^l\theta) \prod_{i=1}^l  N_{i}( p, \{y\}) {{\hat V}^{a_i}}_{I_{m-1}}(p)\partial_{a_i}( f(p, \{y\})\sqrt{ h_{ab}(p)  {{\hat V}^{a}}_{I_{m-1}} (p) {{\hat V}^{b}}_{I_{m-1}} (p)})
\nonumber\\ 
 &+& O(\delta^{q-1}) \big\}
\label{dmHnconp}
\ea
As emphasised above,  the derivatives in (\ref{dmHconp}) and (\ref{dmHnconp})   are along constant coordinate directions in the $\{y\}$ coordinates. The only objects with  non-trivial $p$ dependence are $h_{ab}(p), N_i, f$ and so the above expressions only 
involve coordinate derivatives
of components of the metric and of the evaluations of $N_i, f$ in the $\{y\}$ coordinates. Setting $p= v^{\delta}$ after evaluating these derivatives, we write  the contraction behaviour of 
$H^{l}_{L_m}$ in a notation similar to that 
used in  (\ref{defdmH}) as:
\ba
 & H^{l}_{L_m=I_{m-1}}:= H^{l}_{L_m}(N_1,N_2,..N_l; p=v^{\delta})  \;\;\;\;\;\;\;\;\;\;\;\;\;\;\;\;\;\;     \;\;\;\;\;\;\;\;\;\;\;\;\;\;\;\;\;\;   \;\;\;\;\;\;\;\;\;\;\;\;\;\;\;\;\;\;   &
\nonumber \\
& =\delta^{-(l+1)\frac{2}{3}(q-1)}    \big\{ \prod_{i=1}^l  N_{i}( v_{\delta}, \{y\})      {{\hat V}^{a_i}}_{I_{m-1}}\partial_{a_i} (f( v_{\delta}, \{y\}) \sqrt{h_{ab}(v^{\delta})  {{\hat V}^{a}}_{I_{m-1}}  {{\hat V}^{b}}_{I_{m-1}}}) \big\}
 \;\;\;\;\;\;\;\;\;\;\;\;\;\;\;\;\;\;      &\label{dmHcon}\\
&  H^{l}_{L_m\neq I_{m-1}}:= H^{l}_{L_m}(N_1,N_2,..N_l; p=v^{\delta})   \;\;\;\;\;\;\;\;\;\;\;\;\;\;\;\;\;\;           \;\;\;\;\;\;\;\;\;\;\;\;\;\;\;\;\;\;   \;\;\;\;\;\;\;\;\;\;\;\;\;\;\;\;\;\;   &
\nonumber\\
& =\delta^{-(l+1)\frac{2}{3}(q-1)}\big\{|\cos \theta|(\cos^l\theta) \prod_{i=1}^lN_{i}( v_{\delta}, \{y\}) 
{{\hat V}^{a_i}}_{I_{m-1}}\partial_{a_i}( f( v_{\delta}, \{y\})\sqrt{h_{ab}(v^{\delta})  {{\hat V}^{a}}_{I_{m-1}}  {{\hat V}^{b}}_{I_{m-1}}} )&
\nonumber\\
&+O(\delta^{q-1})\big\}&
\label{dmHncon}
\ea


\section{\label{ag}The function $g_c$}
 
\subsection{\label{adefg}Specification of the function $g_c$.}

Recall that $g: \Sigma^{m(N-1)} \rightarrow {\bf R}$ with no smoothness restrictions and that we are interested in the specification of $g$ only when none of its arguments are coincident.

First, define the function $d(a_1,a_2)$ between any two distinct points $a_1,a_2 \in \Sigma$ as follows:\\
If  there exists a unique geodesic with length $l, l<1$ which joins $a_1$ to $a_2$ then we define  $d=l$ else we set $d=1$.
We shall refer to $d$ as a `distance' function.

Let $U_m$ be the set of  $m(N-1)$ (noncoincident) points in $\Sigma$ which serve as arguments of $g$.  
Consider the case in which the elements of $S$  can be uniquely segregated into $m-k+2$ sets of points $S_{i}, i=k-1,k,...,m-1,m$  with each $S_i, i\geq k$ containing $(N-1)$ points as follows.
Let $S_m$ be such that the distance between any 2 elements of $S_m$ is less than the distance between any element of $S_m$ and any element of $U_m$  not in  $S_m$ , as well as between any 2 elements of $U_m$ not in $S_m$.
This means that the $N-1\choose 2$ distances between points in $S_m$ are the shortest distances among the $m(N-1)\choose 2$ distances between points in $U_m$.

To define $S_{m-1}$ we remove the points belonging to $S_{m}$ from $U$. Call the resulting set of \\$(m-1)(N-1)$ points as $U_{m-1}$.  
Let $S_{m-1}$ be such that the distance between any 2 elements of $S_{m-1}$ is less than the distance between any element of $S_{m-1}$ and any element of $U_{m-1}$  not in  $S_{m-1}$ , as well as between any 2 elements of $U_{m-1}$ 
not in $S_{m-1}$. This means that the $N-1\choose 2$ distances between points in $S_{m-1}$ are the shortest distances among the $(m-1)(N-1)\choose 2$ distances between points in $U_{m-1}$.
We assume that the structure of points in $U_m$ is such that this procedure can be iterated so as to define
define $S_{m-2}, S_{m-3}...,S_{k}$ and that the procedure cannot be iterated beyond this so that the remaining $(k-1)(N-1)$ points are contained in $S_{k-1}$, where if $k=1$,  $S_0$ is the empty set.

Next, in each set $S_i, i>k-1$,  consider the $N-1\choose 2$ distances between pairs of points. Order these distances in decreasing value and denote this
ordered set by  $(d^{(i)}_{1}, d^{(i)}_2,..., d^{(i)}_{   N-1\choose 2     })$, where  $d^{(i)}_r \leq d^{(i)}_s$ iff $r>s$.  Then we define 
\be
g:= \prod_{i=k}^m \frac{d^{(i)}_{N-1}}{d^{(i)}_1}
\label{defgcase1}
\ee

If  $m>1$ and  $U_m$ is such that there exists no $k\geq 1$ for which the above segregation exists, we set $g=1$.
If $m=1$ then we define $k=1$ so that all the points are in the set $S_1$ and interpret (\ref{defgcase1}) as
\be
g:= \frac{d^{(1)}_{N-1}}{d^{(1)}_1}
\label{defgcase2}
\ee

\subsection{\label{acong}Contraction behavior of $g_c$} 
 
Consider the set of $C^0$ kinks of the child  $c_{[i,I, {\hat J}, {\hat K}, \beta ,\epsilon ]_m}$
which contract to the parent vertices $v$ of the parent $c_{[i,I,\beta,\epsilon, {\hat J}, {\hat K}]^{m-1}_m}$.  For the purposes of this section, we denote the contraction parameter $\epsilon_{j_m}$ by $\epsilon$,
the child $c_{[i,I, {\hat J}, {\hat K}, \beta ,\epsilon ]_m}$ by $c$ and the parent $c_{[i,I,\beta,\epsilon, {\hat J}, {\hat K}]^{m-1}_m}$ by $c_{par}$ and the {\em contraction} coordinates associated with $v_{par}$ in $c_{par}$ by $\{x\}$.

Then for small enough $\epsilon$ the function $g_c$ as defined in Appendix \ref{adefg} above seperates as:
$g_c = g_1 g_{c_{par}}$ where $g_1$ is a function only of the $N-1$ $C^0$ kinks which contract to the parent vertex with, from (\ref{defgcase1}), 
\be
g_1 = \frac{d_{N-1}}{d_1}
\label{defg1}
\ee
where the  $N-1\choose 2$ distances between pairs of these $C^0$ kinks are ordered in  decreasing value and denoted by 
$(d_{1}, d_2,..., d_{N-1\choose 2})$, where  $d_r \leq d_s$ iff $r>s$.

For small enough $\epsilon$ these distances are the geodesic distances between pairs of $C^0$- kinks.
Using the fact that geodesic normal coordinates are (at least) $C^{3}$ functions (recall that $\Sigma$ is a semianalytic manifold of differentiability class much larger than unity) of 
coordinate charts on $\Sigma$, it is straightforward to show that the geodesic distance $d$ between points seperated by a coordinate distance $\delta$ is estimated as:
\be
d(a_1,a_2) = \delta ||{\vec{\hat e}}_{a_1,a_2}||+ O(\delta^2)
\label{gdcd}
\ee
where ${\vec{\hat e}}_{a_1,a_2}$ is the unit (with respect to the coordinate norm) coordinate vector along the coordinate straight line connecting $a_1$ to $a_2$ and 
$||{\vec{\hat e}}_{a_1,a_2}||$ is the metric norm (with respect to $h_{bc}$ at either of the points $a_1$ or $a_2$):
\be
||{\vec{\hat e}}_{a_1,a_2}|| = \sqrt{     h_{bc}(a){{\hat e}^b}_{a_1,a_2}{{\hat e}^c}_{a_1,a_2}}, \;\;a = a_1 \;\; {\rm or} \;\; a=a_2
\ee
where the choice of $a=a_1$ or $a_2$ only affects the expression (\ref{gdcd}) at $O(\delta^2)$.
We may now use (\ref{gdcd}) to estimate the required  geodesic distances between the contracting $C^0$ kinks.

We shall use the notation in (iii), (iv) section \ref{sec4.3}. Note that for the contraction of $C^0$ kinks we have $p_1<p_2< p_3$ (see (i)-(iii), Step 2, section \ref{Step2}).
The $C^0$ kinks are situated such that\\
(a) one of them lies along the ${\hat J}$th  edge at the nondegenerate vertex $v_{par}$ in $c_{par}$ at a coordinate  distance $\epsilon^{p_1}$ from $v_{par}$,\\
(b) a second lies along the ${\hat K}$th  edge at the nondegenerate vertex $v_{par}$ in  $c_{par}$ at a coordinate  distance $Q\epsilon^{p_2}$ from $v_{par}$,\\
(c) the remaining $N-3$ kinks vertices lie at coordinate distances of size $\epsilon^{p_3}$ from $v_{par}$.\\

Clearly the largest distances among the pairs of these kinks will be those between the kink in (a) and the others. There are $N-2$ such distances. Clearly the $N-1$th distance  in the 
prescribed decreasing order will be one of the distances  between the kink in (b) and those in (c). These distances can be readily estimated using (\ref{gdcd}) and elementary plane geometry.
We obtain:
\ba
d_{N-1} &= & Q\epsilon^{p_2}||\;   {\vec{\hat e}}_{\hat K}\;  ||(1 + O(\epsilon^{p_2-p_1}))\nonumber \\
d_{1} &= &  \epsilon^{p_1}||\;   {\vec{\hat e}}_{\hat J} \;  ||(1 + O(\epsilon^{p_2-p_1}))
\label{dn-3dn-1}
\ea
where $||\;   {\vec{\hat e_{\hat L}}}\;   ||$ denotes the metric norm of the unit coordinate vector ${\vec{\hat e}}_{\hat L}$ at the point $v_{par}$ in the coordinate system $\{x\}$ associated with $v_{par}$ in $c_{par}$,
\be
||\;   {\vec{\hat e_{\hat L}}}\;   ||= \sqrt{h_{ab}(v_{par}) { {\hat e_{\hat L}}^a } { {\hat e_{\hat L}}^b }}, 
\ee
and where we have used the following inequalities which follow from (i)-(iii), Step 2, section \ref{Step2}  (see (\ref{p1p2p3}), (\ref{p2-13-2}) below):
\be
p_3>p_2>p_1, \;\;\;\; p_3-p_2 >p_2-p_1,  \;\;\;\;p_1 >p_2-p_1. 
\label{ineq}
\ee

From (\ref{defg1}) and  (\ref{dn-3dn-1}),   
we have that:
\be
g_1 = \epsilon^{p_2- p_1} Q\frac{||{\vec{\hat e}}_{\hat K}||}{||{\vec{\hat e}}_{\hat J}||}(1  + O(\epsilon^{p_2-p_1}))
\ee

Note also that during the contraction of the $N-1$ kinks created in the transition from $c_{par}$ to $c$, the position of any pre-existing kinks in $c_{par}$ 
are left unchanged by virtue of (vi), Step 2, section \ref{Step2}. 
Hence the contraction behvior of $g_c$ is
\be
g_c = \epsilon^{p_2- p_1} Q\frac{||{\vec{\hat e}}_{\hat K}||}{||{\vec{\hat e}}_{\hat J}||}(1 + O(\epsilon^{p_2-p_1})) g_{c_{par}}
\label{gcon}
\ee

Note that from  (i)-(iii), Step 2, section \ref{Step2} we have, for some $j\geq 1, p,q>>1$ that:
\be
p_1=jp\frac{2}{3}(q-1), \;\; \;\;\;\;p_2= j(p+1) \frac{2}{3}(q-1), \;\;\;\;\;\; p_3 = j(p+1) \frac{2}{3}(q-1) + j\frac{4}{3}(q-1) 
\label{p1p2p3}
\ee
so that 
\be
p_2-p_1 = j\frac{2}{3}(q-1),  \;\;\;\;p_3-p_2= j\frac{4}{3}(q-1) 
\label{p2-13-2}
\ee


\end{document}